\documentclass[prd,a4paper,nofootinbib,reprint,]{revtex4}
\pdfoutput=1 

\usepackage{blkarray}
\usepackage{graphicx}
\usepackage{amsfonts}
\usepackage{contractions}
\usepackage{soul}
\usepackage{cancel}
\usepackage{amssymb}
\usepackage{amsmath}
\usepackage{pifont}
\usepackage{slashed} 
\usepackage{enumerate}
\usepackage{color}
\usepackage{comment}
\usepackage{bbold}
\usepackage{tikz}
\usepackage{amsmath}
\usepackage{tikz}
\usepackage{empheq}
\usepackage{adjustbox}
\usepackage{multirow}
\usepackage{hyperref}
\usepackage[utf8]{inputenc}
\usepackage{adjustbox}
\usepackage{erewhon}
\usepackage{tcolorbox}
\usepackage{hyperref}
\usepackage{bm}

\definecolor{oucrimsonred}{rgb}{0.6, 0.0, 0.0}
\definecolor{persianblue}{rgb}{0.11, 0.22, 0.73}
\definecolor{forestgreen}{rgb}{0.13,0.35,0.13}
\definecolor{lightgray}{rgb}{0.83, 0.83, 0.83}
 \hypersetup{colorlinks, citecolor=oucrimsonred, linkcolor=persianblue, urlcolor=oucrimsonred}
\def\hhref#1{\href{http://arxiv.org/abs/#1}{#1}} 
\definecolor{cornellred}{rgb}{0.7, 0.11, 0.11}
\definecolor{navyblue}{rgb}{0.0, 0.0, 0.5}
\definecolor{amethyst}{rgb}{0.6, 0.4, 0.8}
\definecolor{yellow}{rgb}{1.0, 1.0, 0.0}
\definecolor{firebrick}{rgb}{0.7, 0.13, 0.13}
\definecolor{tangerineyellow}{rgb}{1.0, 0.8, 0.0}
\definecolor{deepfuchsia}{rgb}{0.76, 0.33, 0.76}
\definecolor{amber}{rgb}{1.0, 0.75, 0.0}
\definecolor{VioletRed4}{rgb}{0.55, 0.13, .32}
\definecolor{indiagreen}{rgb}{0.07, 0.53, 0.03}
\newcommand{\be}{\begin{equation}}
\newcommand{\ee}{\end{equation}}
\newcommand{\bea}{\begin{eqnarray}}
\newcommand{\eea}{\end{eqnarray}}
\newcommand{\nn}{\nonumber}

\definecolor{oucrimsonred}{rgb}{0.6, 0.0, 0.0}
\newcommand{\fd}[2]{\parbox{#1}{\includegraphics[width=#1]{figs/#2}}}

\definecolor{mtcolor}{rgb}{0.13, 0.55, 0.13}

\definecolor{violachiaro}{rgb}{1,0.6,1}
 
\definecolor{gbcolor}{rgb}{.43,.22,.12}
 
\definecolor{gbcolor2}{rgb}{.9,.2,.6}
\definecolor{gbcolor3}{rgb}{.3,.2,.6}

\definecolor{verdechiaro}{rgb}{0.6,1,0.6}
\definecolor{giallochiaro}{rgb}{1,1,0.6}
\definecolor{bluscuro}{rgb}{0.15, 0.2, 0.9}
\definecolor{verdes}{rgb}{0.1, 0.5, 0.1}%
\definecolor{tangerineyellow}{rgb}{1.0, 0.8, 0.0}

\begin{document}

\title[]{Solving peak theory in the presence of local non-gaussianities}

\date{\today}
\author{Flavio Riccardi$^{a,b}$}
\email{flavio.riccardi@sissa.it}
\author{Marco Taoso$^{c}$}
\email{marco.taoso@to.infn.it}
\author{Alfredo Urbano$^{d,e}$}
\email{alfredo.urbano@uniroma1.it}
\affiliation{$^a$SISSA, via Bonomea 265, I-34132 Trieste, Italy}
\affiliation{$^b$I.N.F.N. sezione di Trieste, SISSA, via Bonomea 265, I-34132 Trieste, Italy}
\affiliation{$^c$I.N.F.N. sezione di Torino, via P. Giuria 1, I-10125 Torino, Italy}
\affiliation{$^d$Dipartimento di Fisica, ``Sapienza'' Universit\`a di Roma, Piazzale Aldo Moro 5, 00185, Roma, Italy}
\affiliation{$^e$I.F.P.U., Institute  for  Fundamental Physics  of  the  Universe, via  Beirut  2, I-34014 Trieste, Italy.}

\begin{abstract}
\noindent  
We compute the probability density distribution of maxima for a scalar random field in the presence of local non-gaussianities. 
The physics outcome of this analysis
is the following.
If we focus on
maxima whose curvature is larger than a certain threshold for gravitational collapse, our calculations illustrate
how the 
fraction of the Universe's mass in the form of primordial black holes (PBHs) changes 
in the presence of local non-gaussianities.
We find that previous literature on the subject overestimates, by many orders of magnitude, the impact of 
local non-gaussianities on the PBH abundance. 
We explain the origin of this discrepancy, and conclude that, in realistic single-field inflationary models with ultra slow-roll, 
one can obtain the same abundance found with the gaussian approximation simply changing the peak amplitude of the curvature power spectrum
by no more than a factor of two.
We comment about the relevance of non-gaussianities for second-order gravitational waves.
\end{abstract}
\maketitle
 
\section{Introduction}\label{sec:Intro}

The possibility that the totality of dark matter in the Universe consists of primordial black holes (PBHs) still holds the stage 
even though
almost half-a-century has passed after the pioneering proposal of Hawking and Carr\,\cite{Hawking:1971ei,Carr:1974nx}.
This is especially true in the mass range $10^{18} \lesssim M_{\rm PBH}\,[{\rm g}] \lesssim 10^{21}$ in which
black holes are neither too light (otherwise they would have evaporated in the past through Hawking radiation\,\cite{Carr:2009jm}) or too heavy (otherwise they would distort space-time in a way that contradicts present bounds from lensing experiments\,\cite{Niikura:2017zjd,Katz:2018zrn}).

PBHs could have formed in the very early Universe during the radiation dominated era.\footnote{It is also possible to have PBH formation during matter domination\,\cite{Harada:2016mhb}.} 
The key ingredient that triggers the formation of a PBH is the presence of an over-fluctuation in the density of the Universe 
which, if large enough, gravitationally collapses dragging down any matter within its horizon, that is the parcel of space around any point reachable at the speed of light.

The theory of inflation provides an elegant mechanism that explains the origin of density perturbations in the Universe. 
In the inflationary picture, space-time fluctuates quantum mechanically around a background that is expanding exponentially fast. 
After the end of inflation, these curvature fluctuations are transferred to the radiation field, creating slightly overdense and under-dense regions. It is, therefore, fascinating to ask whether the formation of PBHs fits in the inflationary picture of structure formation.\footnote{This is not the only option. The formation of PBHs may have been independent of inflationary physics; PBHs may have been originated from topological defects 
 formed during symmetry breaking phase transition, for instance 
from the collapse of string loops\,\cite{Vilenkin:1981iu,Hawking:1990tx,Fort:1993zb}.} To answer this question, two (related) aspects need to be addressed.

\begin{itemize}

\item [{\it i)}] The inflaton dynamics should give rise to a peak in the power spectrum of curvature perturbations.\footnote{
An exception, where no amplification of the power spectrum is needed, are models where PBHs 
form from the collapse of domain walls created during inflation\,\cite{Deng:2016vzb}.
} 
This translates into a large variance for density perturbations that, in turn, 
enhances the chance to create overdense regions above the threshold for gravitational collapse. 

\item [{\it ii)}] The abundance of such collapsing regions should be large enough to explain the totality of dark matter. 

\end{itemize}

In this paper we focus on simple single-field inflationary models. It is known that in order to fulfil point {\it i)} 
slow-roll conditions must be violated. The simplest option is to introduce, few $e$-folds before the end of inflation, 
an approximate stationary inflection point in the inflaton potential (see refs.\,\cite{Starobinsky:1992ts,Ivanov:1994pa,Saito:2008em} for the earliest proposal in this direction). 
When the inflaton, during its classical dynamics, crosses this region (during the so-called ``ultra slow-roll phase''), curvature perturbations, due to the presence of negative friction, get exponentially enhanced\,\cite{Leach:2000yw,Leach:2001zf,Tsamis:2003px,Kinney:2005vj,Kinney:1997ne}.\footnote{Alternatively, a parametric  amplification  of  curvature  perturbations  could be caused  by resonance  with  oscillations  in  the  sound  speed  of  their propagation\,\cite{Cai:2018tuh}. 
Another possibility is that, after  the  inflationary  phase, the  inflaton  begins  to  oscillate near the minimum of the potential 
 and fragments into oscillons which, in turn, lead to copious production of PBHs\,\cite{Cotner:2018vug}.} 
Point {\it ii)} is more subtle. 
Due of their intrinsic quantum-mechanical origin,  the way in which 
quantum fluctuations lead to a classical pattern of perturbations can be described only in a probabilistic sense. 
Consequently, the computation of the abundance of collapsing regions requires informations about the statistical distribution of density perturbations. Most of the time, for simplicity, the gaussian approximation is assumed. 
However, the very same fact that slow-roll conditions are violated as a consequence of point {\it i)} suggests that 
non-gaussianities may play a relevant role. 
Refs.\,\cite{Atal:2018neu,Taoso:2021uvl} indeed find that during an ultra slow-roll phase sizable non-gaussianities of local type are generated. 
In the rest of this paper we will dub these non-gaussianities ``primordial'' to distinguish them from non-gaussianities that arise from the non-linear relation between curvature and density perturbations.

What is the impact of primordial non-gaussianities on the gaussian approximation when computing the 
PBH abundance?
Ref.\,\cite{Franciolini:2018vbk} addressed this question in the context of threshold statistics. 
The main result of ref.\,\cite{Franciolini:2018vbk} is that the abundance of PBHs is exponentially sensitive  
to primordial non-gaussianities~\footnote{More precisely this means that in the context of threshold statistics the PBH abundance is given by eq.\,(\ref{eq:MainRiotto}), where $\mathcal{C}_n$ are the $n^{\rm th}$ normalized 
cumulants of the non-gaussian distribution.}. 
Based on this result, ref.\,\cite{Atal:2018neu} claims that, in the context of single-field inflationary models 
which feature an approximate stationary inflection point, the gaussian approximation is hardly trustable when computing the PBH abundance.

The goal of this work is to address the same question using a different computational strategy inspired by peak theory\,\cite{Bardeen:1985tr}. 
More precisely, we associate regions where the overdensity field takes values above the threshold for gravitational collapse with spiky local maxima of the curvature perturbation field, and compute the
number density of the latter using peak theory that we extend to include local non-gaussianities. 

Our main conclusion is that the impact of local non-gaussianities on the PBH abundance is far less important compared to what previously thought. 
We confirm that in models for PBH production (at least the class of models that we are going to consider), local non-gaussianities are sizeable enough to invalid the use of the the gaussian approximation to estimate their abundance.
However we find that their impact is modest when translated in terms of the amplitude of curvature power spectrum, namely it is enough to change it by a factor $\simeq2$ or smaller to obtain the same PBH abundance predicted by the gaussian calculation.
This shift can be obtained by a small change of the parameters of the inflationary model.

As a phenomenological application, we consider the impact of local non-gaussianities on the computation 
of the amplitude of the induced second-order gravitational-wave signal. 
Our main conclusion is that a careful treatment of non-gaussianities is needed in order to provide a reliable 
comparison with the expected experimental sensitivities of future gravitational-wave interferometers.  
En route, we discuss the difference between threshold statistics and peak theory, 
and we explain under which conditions (and why) peak theory gives a PBH abundance which is larger than the one computed by means of threshold statistics.

The structure of this paper is as follows.

\begin{itemize}

\item [$\ast$]
In section\,\ref{sec:Mot} we introduce the problem and present our solution strategy. 
This section is paired with  appendix\,\ref{app:InflationPrimer} where we explain in more detail 
the cosmological interpretation of all quantities involved.

\item [$\ast$]
In section\,\ref{sec:Res} we discuss our main results and we explain the discrepancy with the previous literature. 
This section is paired with appendix\,\ref{app:GaussianPeakTheory}-\ref{app:NonLin} where we collect all relevant technical details.

\item [$\ast$] We conclude in section\,\ref{sec:Con}.

\end{itemize}

\section{Problem setup and solution strategy}\label{sec:Mot}

Consider in position space
\begin{align}\label{eq:NonGauRandomField}
h(\vec{x}) = \mathcal{R}(\vec{x}) + \alpha\mathcal{R}(\vec{x})^2\,,
\end{align}
where $\alpha$ is a constant, $\mathcal{R}(\vec{x})$ is a gaussian scalar random field while $h(\vec{x})$ is non-gaussian because of the presence of the non-linear term on the right-hand side.
 In this case, non-gaussianities are called 
of local type because for a given $\vec{x}$ at which we evaluate $h$ the amount of non-gaussianity is localized at the same position. 
 We briefly discuss in appendix\,\ref{app:InflationPrimer} the physical interpretation of eq.\,(\ref{eq:NonGauRandomField}) and the limitations of this parametrization of non-gaussianities.

The first observation is that 
\begin{align}
\partial_i h(\vec{x}) = [\partial_i \mathcal{R}(\vec{x})][1+2\alpha \mathcal{R}(\vec{x})]\,,
\end{align}
meaning that stationary points of $\mathcal{R}$ are also stationary points of $h$ ($\partial_i \mathcal{R} = 0$ implies $\partial_i h = 0$). 

What is crucial, however, is that the nature (saddle points, maxima or minima) of these ``shared'' stationary points depends on the sign of the factor $1+2\alpha \mathcal{R}$.
Consider the matrix of second derivatives evaluated at a stationary point $\vec{x}_{\rm st}$ (we will further indicate 
 a minimum with $\vec{x}_{\rm m}$ and a maximum with $\vec{x}_{\rm M}$). One finds the Hessian matrix
$h_{ij}(\vec{x}_{\rm st}) =
 [\mathcal{R}_{ij}(\vec{x}_{\rm st})](1+2\alpha \mathcal{R}_{\rm st})$.\footnote{We use the short-hand notation 
 $f_{\rm k} \equiv f(\vec{x}_{\rm k})$,
 $f_{i}(\vec{x}_{\rm k}) \equiv \partial_{i}f(\vec{x})$ evaluated at $\vec{x}_{\rm k}$ and 
$f_{ij}(\vec{x}_{\rm k}) \equiv \partial_{ij}f(\vec{x})$ evaluated at $\vec{x}_{\rm k}$ for some generic function $f$. 
The flat spatial Laplacian is $\triangle f(\vec{x}) \equiv \sum_{i}f_{ii}(\vec{x})$.}
 
To fix ideas, consider the simple case of two spatial dimensions $\vec{x} = \{x,y\}$ and the case in which the stationary point $\vec{x}_{\rm m}$ is a minimum of $\mathcal{R}$. 

Minima of $\mathcal{R}$ are identified by 
two conditions. The first one, $\mathcal{R}_{xx}(\vec{x}_{\rm m})\mathcal{R}_{yy}(\vec{x}_{\rm m}) - \mathcal{R}_{xy}(\vec{x}_{\rm m})^2 > 0$ separates extrema from saddle points. The second one, 
$\mathcal{R}_{xx}(\vec{x}_{\rm m})>0$ and $\mathcal{R}_{yy}(\vec{x}_{\rm m}) >0$, separates minima from maxima. 
Since we have $h_{xx}(\vec{x}_{\rm m})h_{yy}(\vec{x}_{\rm m}) - h_{xy}(\vec{x}_{\rm m})^2 = (1+2\alpha \mathcal{R}_{\rm m})^2[\mathcal{R}_{xx}(\vec{x}_{\rm m})\mathcal{R}_{yy}(\vec{x}_{\rm m}) - \mathcal{R}_{xy}(\vec{x}_{\rm m})^2]$ 
it is obvious that the condition $\mathcal{R}_{xx}(\vec{x}_{\rm m})\mathcal{R}_{yy}(\vec{x}_{\rm m}) - \mathcal{R}_{xy}(\vec{x}_{\rm m})^2 > 0$ is also satisfied by 
$h$.

On the contrary, since $h_{xx}(\vec{x}_{\rm m}) = (1+2\alpha \mathcal{R}_{\rm m})\mathcal{R}_{xx}(\vec{x}_{\rm m})$ and $h_{yy}(\vec{x}_{\rm m}) = (1+2\alpha \mathcal{R}_{\rm m})\mathcal{R}_{yy}(\vec{x}_{\rm m})$, 
it is possible that a minimum of $\mathcal{R}$ becomes a maximum of $h$ if $1+2\alpha \mathcal{R}_{\rm m} < 0$. 
Viceversa, a maximum of $\mathcal{R}$ can become a minimum of $h$.

The argument trivially generalizes to the more realistic case of three spatial dimensions.

In peak theory, one computes the number density of maxima\,\cite{Bardeen:1985tr}.  
 We are interested in the number density of maxima of the non-gaussian variable $h$. 
 As argued before, identifying this quantity with the number density of  
 maxima of $\mathcal{R}$ (based on the observation that $h$ and $\mathcal{R}$ have the same stationary points) is not correct.  
 Let us give a quantitative argument to support this claim. 
 From the previous discussion, 
 it is clear that counting  the maxima of $\mathcal{R}$ might be not enough. On the contrary, a simple modification could be 
 the following. One should
\begin{itemize}
\item [{\it i)}] Count the maxima of $\mathcal{R}$;
\item [{\it ii)}] {\it Add} the minima of $\mathcal{R}$ that, depending on the value of $(1+2\alpha \mathcal{R}_{\rm m})$, become maxima of $h$;
\item [{\it iii)}] {\it Subtract} the maxima of $\mathcal{R}$ that, depending on the value of $(1+2\alpha \mathcal{R}_{\rm M})$, become minima of $h$.
\end{itemize}

\begin{figure}[!htb!]
\begin{center}
\includegraphics[width=.45\textwidth]{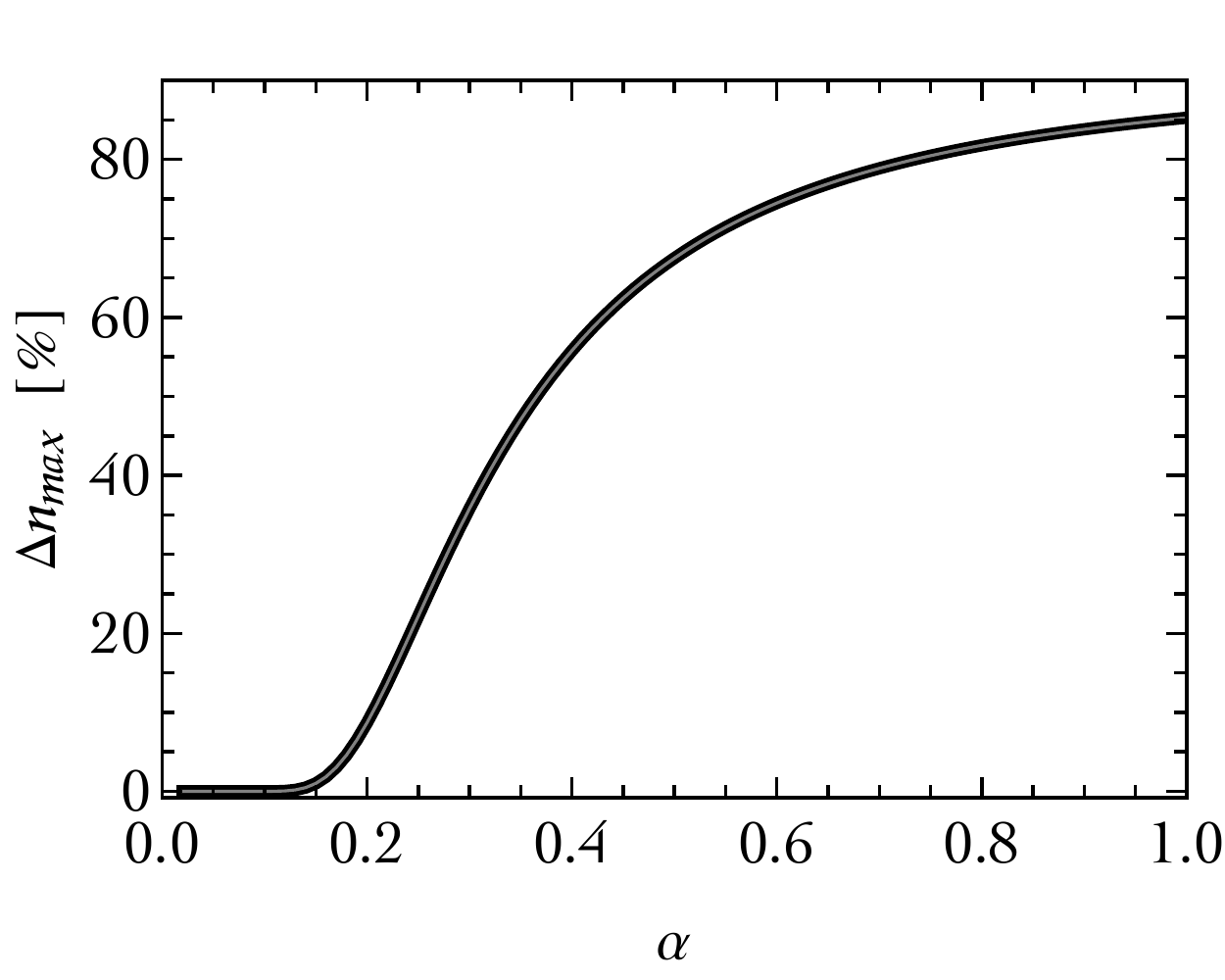}
\caption{\em \label{fig:DeltaNMax} 
Percentage increase in the number density of maxima of $\mathcal{R}$ when we {\it ii)}
add the minima of $\mathcal{R}$ that become maxima of $h$ and {\it iii)} subtract the maxima of $\mathcal{R}$ that become minima of $h$ (see eq.\,(\ref{eq:DeltaNmax})).
To make this (illustrative) plot we set $\sigma_0 = 1$ and $\gamma=3/4$ (see appendix\,\ref{app:GaussianPeakTheory} for definitions).
 }
\end{center}
\end{figure}

Ref.\,\cite{Yoo:2019pma} assumes {\it i)}.
However, the two operations {\it ii)} and {\it iii)} do not balance between each others, and a sizable correction to  {\it i)} will be introduced 
if $\alpha$ is large enough. 
In fig.\,\ref{fig:DeltaNMax} we show how the number density of maxima of the gaussian variable 
$\mathcal{R}$ changes (in percentage) as a function of $\alpha$ 
when {\it ii)} and {\it iii)} are implemented. Schematically, we compute
\begin{align}\label{eq:DeltaNmax}
\Delta n_{\rm max} = \frac{
({\rm \#\,minima\,of\,}\mathcal{R}\to{\rm maxima\,of\,}h) -
({\rm \#\,maxima\,of\,}\mathcal{R}\to{\rm minima\,of\,}h)
}{{\rm \#\,maxima\,of\,}\mathcal{R}}\,.
\end{align}
We obtain fig.\,\ref{fig:DeltaNMax} using gaussian peak theory (implementing 
the results of ref.\,\cite{Bardeen:1985tr}, see appendix\,\ref{app:GaussianPeakTheory}). 
If we take $\alpha \ll 1$, {\it ii)} and {\it iii)} do not alter the estimate of {\it i)}. 
However, for sizable $\alpha \gtrsim 0.2$ the change in the number density of maxima of $\mathcal{R}$ becomes  evident.

In situations of cosmological interest, the issue is further complicated by the fact that we are not really interested in {\it all} maxima of $h$ but only in those which are ``spiky enough.'' 
The reason is that the quantity which is relevant is the density contrast $\delta(\vec{x},t)$ (also dubbed overdensity field in the following) whose relation 
with $h(\vec{x})$ (assuming radiation dominated epoch) reads\,\cite{Harada:2015yda}
\begin{align}\label{eq:NonLinearDelta}
\delta(\vec{x},t) =  
-\frac{4}{9}\left(
\frac{1}{aH}
\right)^2 
e^{-2 h(\vec{x})}
\bigg[
\triangle h(\vec{x}) + \frac{1}{2} h_i(\vec{x})
 h_i(\vec{x})
\bigg]\,,
\end{align}
where the time dependence comes from the scale factor $a=a(t)$ and the Hubble rate $H=H(t)$ while $h$ does not depend on time because 
eq.\,(\ref{eq:NonLinearDelta}) assumes perturbations to be on super-horizon scales. Eq.\,(\ref{eq:NonLinearDelta}) can be thought as a Poisson equation in which $h$ plays the role of gravitational potential while the density contrast 
can be written more precisely as $\delta(\vec{x},t) \equiv \delta\rho(\vec{x},t)/\rho_b(t)$ where 
$\rho_b(t)$ is the average background radiation energy density and $\delta\rho(\vec{x},t) = \rho(\vec{x},t)- \rho_b(t)$ its perturbation. 
The physics-case that is relevant for the present study is the one in which the density contrast has a peak localized in some region of space that is high enough to trigger the gravitational collapse into a black hole. If the number of these peaks above threshold is large enough, these black holes can be part of dark matter. 
In the range $10^{18} \lesssim M_{\rm PBH}\,[{\rm g}] \lesssim 10^{21}$, a population of
PBHs may account for the totality of dark matter
observed in the Universe today.

Consider a peak of the overdensity field, located at some spatial point $\vec{y}_{\rm pk}$.
\begin{align}\label{eq:Curvature2Density}
\delta(\vec{y}_{\rm pk},t) =  
-\frac{4}{9}\left(
\frac{1}{aH}
\right)^2 
e^{-2 h(\vec{y}_{\rm pk})}
\bigg[
\triangle h(\vec{y}_{\rm pk}) + \frac{1}{2} h_i(\vec{y}_{\rm pk})
 h_i(\vec{y}_{\rm pk})
\bigg] \simeq 
-\frac{4}{9}\left(
\frac{1}{aH}
\right)^2 
\triangle h(\vec{y}_{\rm pk})\,,
\end{align}
where in the second step we linearized in $h$. 
We follow here the approach of refs.\,\cite{Germani:2018jgr,Musco:2018rwt} in which the linear approximation was adopted.
 Since the peak amplitude of the overdensity must be larger than some critical value $\delta_c$, we deduce the condition 
 \begin{align}\label{eq:SpikyEnough}
- \triangle h(\vec{y}_{\rm pk}) \gtrsim \frac{9}{4}(aH)^2
\delta_c\,,
 \end{align}
 on the curvature of $h$ at the peak of $\delta$. 
 
If we assume that local maxima of $h$ coincide with peaks of $\delta$ (that is $\vec{y}_{\rm pk} \simeq \vec{x}_{\rm M}$), then the condition in eq.\,(\ref{eq:SpikyEnough}) 
tells that only maxima of $h$ which are ``spiky enough'' contribute to the formation of black holes. 
Of course, the assumption that local maxima of $h$ coincide with peaks of $\delta$ requires some care.
Ref.\,\cite{DeLuca:2019qsy} argues, both analytically and numerically, that this assumption is well justified. 
However, ref.\,\cite{DeLuca:2019qsy} only considers the case in which $h$ is gaussian, that is, $h = \mathcal{R}$ with $\alpha = 0$ in our case (but they include the presence of the non-linearities in eq.\,(\ref{eq:Curvature2Density})). Since stationary points of $\mathcal{R}$ are also stationary points of $h$, we tend to believe that 
the same conclusion holds true in the case with $\alpha \neq 0$ but of course this is an important 
point that has to be checked explicitly.

All in all, the strategy we shall follow in the course of this work is the following. 

First, we will compute the number density of maxima of the non-gaussian random field 
$h$ that are ``spiky enough'' according to 
the condition in eq.\,(\ref{eq:SpikyEnough}). This requires a generalization of the work in ref.\,\cite{Bardeen:1985tr} such as to implement local non-gaussianities.
Second, we will check that these maxima are also peaks of the overdensity field.
If this last point will turn out to be true, our computation of the number density of ``spiky enough'' maxima of $h$
will provide the abundance of
peaks of the overdensity field that are large enough to form black holes.

\section{Results and discussion}\label{sec:Res}

We present in this section the main results of our analysis. 
In section\,\ref{eq:Abubnda} we discuss how primordial non-gaussianities of local type alter the abundance of PBHs.  
In section\,\ref{sec:Comp} we compare with the existing literature.
 In section\,\ref{sec:nonlin} we (partially) include the effect of non-linearities in the relation between curvature and 
 density perturbations. 

All technical details are collected in appendix\,\ref{app:InflationPrimer} (where we discuss the origin of eq.\,(\ref{eq:NonGauRandomField}) from a cosmological viewpoint),
appendix\,\ref{app:GaussianPeakTheory} (where we discuss the gaussian limit), appendixes\,\ref{app:NonGaussianPeakTheoryExact} and \ref{app:NonGaussianPeakTheory} (where we discuss how to construct the non-gaussian part and the approximations that are involved),
appendix\,\ref{app:NonPer} (where we give formulas for computing cumulants of generic order),
 appendix\,\ref{app:Threshold} (where we discuss how to 
compute the threshold value for collapse into black holes) and appendix\,\ref{app:NonLin} (where we  discuss non-linearities).

\subsection{The abundance of PBHs in the presence of primordial non-gaussianities of local type}\label{eq:Abubnda}

The quantity of central interest is the fraction of the Universe's mass in the form of PBHs at the time of their formation. 
As customary in the literature, we indicate this quantity with $\beta$. 
The present-day fractional abundance of dark matter in the form of PBHs is given by  (for a review, see ref.\,\cite{Sasaki:2018dmp})
\begin{align}\label{eq:PresentDayAbundance}
\frac{\Omega_{\rm PBH}}{\Omega_{\rm DM}} =  
O(1) \times \left(\frac{\beta}{10^{-16}}\right)
\left[\frac{g_{*}(t_{f})}{106.75}\right]^{-1/4}\left(\frac{M_{\rm PBH}}{10^{18}\,{\rm g}}\right)^{-1/2}\,,
\end{align}
where $g_{*}(t_{f})$ is the number of relativistic degrees of freedom at the time of black hole formation (that we normalize and set to its standard model value).
Eq.\,(\ref{eq:PresentDayAbundance}) is defined modulo an overall $O(1)$ factor whose precise value depends on the detail of the gravitational collapse that leads to 
black hole formation. In this paper we consider $M_{\rm PBH} \simeq 10^{18}$ g; consequently, as an order-of-magnitude estimate, $\beta \gtrsim 10^{-16}$ 
is excluded since it would imply overclosure of the present-day Universe, $\Omega_{\rm PBH} > \Omega_{\rm DM}$.

We find the following formula
\begin{tcolorbox}[colframe=amethyst!50,arc=6pt,colback=amethyst!5,width=1.031\textwidth,
title=fraction of the Universe's mass in PBH in the presence of primordial local non-gaussianities,coltitle=red!25!black]
\vspace{-.4cm}
\begin{align}
\beta \simeq
\frac{1}{4\sqrt{2\pi(1-\gamma^2)}} \, \left[ \int_{-\frac{1}{2\alpha\sigma_0}}^{\infty} d\bar{\nu}\, \int_{x_{\delta}(\bar{\nu})}^{\infty} dx\, e^{-\bar{\nu}^2/2}\, f(x)\, e^{-\frac{(x-x_*)^2}{2(1-\gamma^2)}}
+\int_{-\infty}^{-\frac{1}{2\alpha\sigma_0}}d\bar{\nu}\, \int_{-\infty}^{x_{\delta}(\bar{\nu})} dx\, e^{-\bar{\nu}^2/2}\, f(x)\, e^{-\frac{(x-x_*)^2}{2(1-\gamma^2)}}  \right]
\label{eq:MasterFormula}
\end{align}
\end{tcolorbox}
that we derive in detail in appendix\,\ref{app:GaussianPeakTheory} (as far as the gaussian limit is concerned), appendix\,\ref{app:NonGaussianPeakTheoryExact} (where we discuss how to construct the non-gaussian part and the approximations that are involved) and appendix\,\ref{app:Threshold} (where we discuss how to 
compute the threshold value for collapse into black holes).
In short:
\begin{itemize}

\item [$\ast$] The parameter $\alpha$, already defined in eq.\,(\ref{eq:NonGauRandomField}), indicates the presence of local non-gaussianities (of quadratic type). 
The limit $\alpha\to 0$ reproduces the gaussian result. A more physical interpretation of this parameter is given in appendix\,\ref{app:InflationPrimer}. 
In concrete models of inflation which generate a sizable abundance of dark matter in the form of PBHs 
(see, for instance, ref.\,\cite{Ballesteros:2020qam}), we expect $\alpha \simeq [0.24\div 0.61]$\,\cite{Atal:2018neu,Taoso:2021uvl}.

We derive our result based on peak theory. 
More precisely, we associate regions where the overdensity field takes large values with spiky local maxima of the comoving curvature perturbation, and compute the
number density of the latter using peak theory that we extend to include local non-gaussianities. 
Within  this approach, eq.\,(\ref{eq:MasterFormula}) represents an original result.

\item [$\ast$] The spectral moments $\sigma_j^2$ are defined by (see eq.\,(\ref{eq:sigmaj2}) and discussion in appendix\,\ref{app:GaussianPeakTheory})
\begin{equation}\label{eq:HHHH}
\sigma_j^2 \equiv \int \frac{dk}{k}\mathcal{P}_{\mathcal{R}}(k)\,k^{2j}\,,
\end{equation}
where $\mathcal{P}_{\mathcal{R}}(k)$ is the dimensionless power spectrum of the gaussian random field $\mathcal{R}$.
The a-dimensional parameter $\gamma$ is defined as $\gamma=\sigma_1^2/\sigma_2\sigma_0$  and takes values $0<\gamma<1$.
In this paper we analyze two possible cases. 
In order to elucidate some intermediate results of our computational strategy, we use in appendix\,\ref{app:GaussianPeakTheory}
 a simple toy-model for the power spectrum given by the log-normal function (see eq.\,(\ref{eq:ToyPS}) and related discussion)
\begin{align}\label{eq:ToyPS2}
\mathcal{P}_{\mathcal{R}}(k) = \frac{A_g}{\sqrt{2\pi}v}\exp\left[
-\frac{\log^2(k/k_{\star})}{2v^2}
\right]\,,
\end{align}
since in this case the spectral moments can be computed analytically and they are given by $\sigma_j^2 = A_g k_{\star}^{2j}e^{2j^2 v^2}$.
 The three parameters $\{k_{\star},A_g,v\}$ in eq.\,(\ref{eq:ToyPS2}) control, respectively, the position of the peak of the power spectrum, the peak amplitude of the power spectrum and its width.
However, we remark that in single-field inflationary models the value of $\alpha$ that defines the amount of local non-gaussianities and the shape of the power spectrum are intimately related, and in general one can not take $\alpha$ as a free parameter and fix the power spectrum 
to a specific functional form like the one introduced in eq.\,(\ref{eq:ToyPS2}).
A more realistic example is the following. Consider the power spectrum defined by the piecewise function
\begin{align}\label{eq:RealisticPS}
{\rm realistic\,power\,spectrum:}~~~~~~\mathcal{P}_{\mathcal{R}}(k) = \mathcal{P}_{\mathcal{R}}(k_{\star})\times 
\left\{
\begin{array}{ccc}
\left(\frac{k}{k_1}\right)^{n_1}
\exp\left[
-\frac{\log^2(k_1/k_{\star})}{2v^2}
\right]
  & {\rm for}  & k < k_1   \\
\exp\left[
-\frac{\log^2(k/k_{\star})}{2v^2}
\right]  & {\rm for}   & k_1 \leqslant k \leqslant k_2  \\
\left(\frac{k}{k_2}\right)^{n_2}
\exp\left[
-\frac{\log^2(k_2/k_{\star})}{2v^2}
\right]  & {\rm for}   &  k > k_2 
\end{array}
\right.
\end{align}
 with two power-law behaviors for $k \ll k_{\star}$ and 
$k \gg k_{\star}$ that, for $k\approx k_{\star}$, are connected by the log-normal function in eq.\,(\ref{eq:ToyPS2}). 
In this case, it is possible to show that the spectral index of the fall-off of the power spectrum after the peak at $k=k_{\star}$ 
is related to $\alpha$ by the relation $n_2\approx -4\alpha$ that is twice the value of the Hubble parameter $\eta$ 
after the end of the ultra slow-roll phase.\footnote{This can be understood as follows. The modes that constitute 
the fall-off of the power spectrum after the peak are those for which the horizon-crossing condition $k=aH$ happens after the end of the 
ultra slow-roll phase\,\cite{Ballesteros:2020qam}; during this part of the dynamics the power spectrum can be approximated by means of the 
conventional slow-roll relation $\mathcal{P}_{\mathcal{R}}(k) = H^2/8\pi^2\epsilon$ where the Hubble parameter $\epsilon$ evolves in time according to 
$\epsilon(N) \propto e^{-2\eta_0 N}$ where $\eta_0$ (which is a negative number, $\eta_0 < 0$, see appendix\,\ref{app:InflationPrimer}) is the value of the Hubble parameter $\eta$ after the end of the ultra slow-roll phase. We neglect the contribution coming from the time-evolution of $H$, which is sub-leading.
This means that we have $\mathcal{P}_{\mathcal{R}}(k) \propto e^{2\eta_0 N}$. From $k=aH$ we have $dk/k = dN$, and we can convert the $e$-fold time-dependence into 
a $k$-dependence, $\mathcal{P}_{\mathcal{R}}(k) \propto k^{2\eta_0} = k^{-4\alpha}$ where we used that $\alpha = -\eta_0/2$ (see appendix\,\ref{app:InflationPrimer}).
} 
In our numerical analysis, therefore, we use the realistic power spectrum in eq.\,(\ref{eq:RealisticPS}) with the 
condition $n_2 = -4\alpha$ for fixed $\alpha$. This provides the above-mentioned relation between the amount of local non-gaussianities and the shape of the power spectrum. We remark that this point is often overlooked in the literature. 
\begin{figure}[!h!]
\begin{center}
$$\includegraphics[width=.45\textwidth]{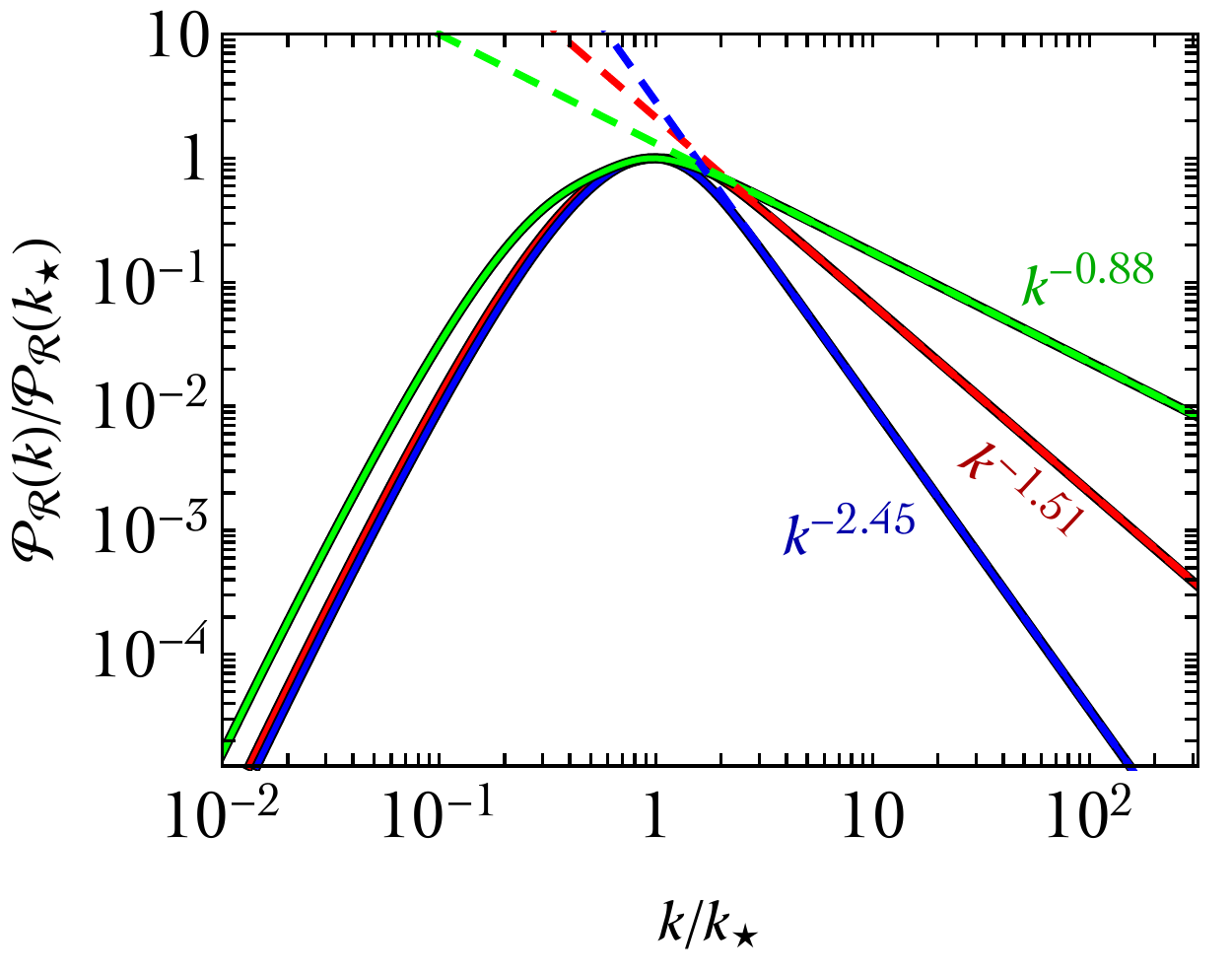}
\qquad\includegraphics[width=.45\textwidth]{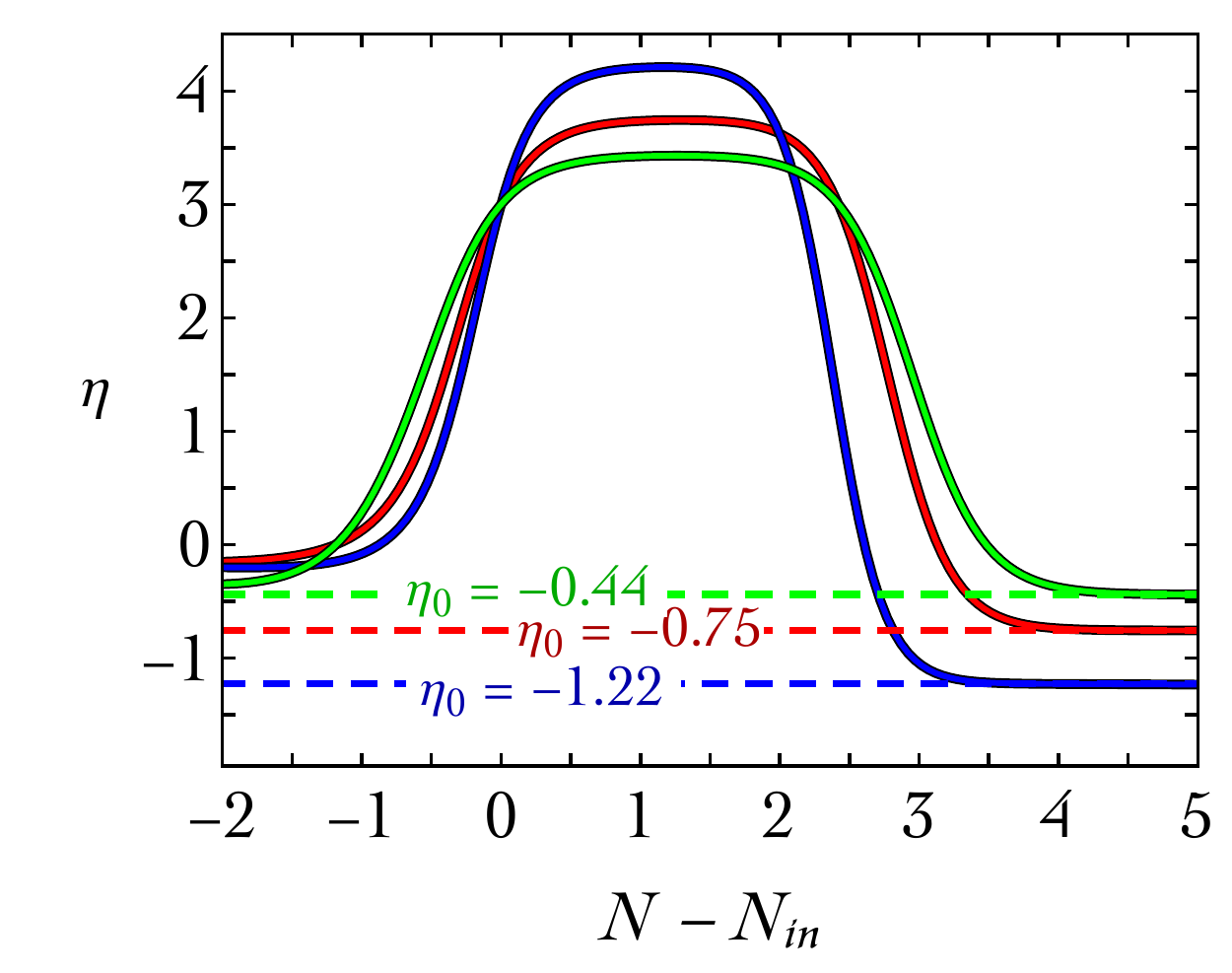}$$
\caption{\em \label{fig:Numericstuff} 
Power spectra for the model in ref.\,\cite{Ballesteros:2020qam} (red, with label $k^{-1.51}$), 
ref.\,\cite{Ballesteros:2017fsr} (blue, with label $k^{-2.45}$) 
and ref.\,\cite{Dalianis:2018frf} (green, with label $k^{-0.88}$)
computed numerically by solving the Mukhanov-Sasaki equation (left panel). 
The slope of the power-law falloff after the peak is $k^{2\eta_0}$ where $\eta_0$ is the value of the Hubble parameter 
$\eta$ (whose evolution, as function of the $e$-fold time $N$ with $N_{\rm in}$ the beginning of the ultra slow-roll phase, is shown in the right panel) after the end of the ultra 
slow-roll phase; $\eta_0$, in turn, is related to the parameter $\alpha$ that controls the size of local non-gaussianities via $\alpha = -\eta_0/2$ (see appendix\,\ref{app:InflationPrimer}).
 } 
\end{center}
\end{figure}
As a numerical check, we show in fig.\,\ref{fig:Numericstuff} (left panel) the power spectra of comoving curvature perturbations 
computed numerically for three inflationary models in which a ultra slow-roll phase takes place. 
All power spectra are well described by the analytical ansatz in eq.\,(\ref{eq:RealisticPS}). 
The slope of the power spectra after the peak is related to the value of the Hubble parameter $\eta$ (shown in the right panel 
of the same figure)
after the end of the ultra slow-roll phase which, in turn, controls the size of local non-gaussianities (see caption and appendix\,\ref{app:InflationPrimer} for more details).

As far as the values of the other parameters in eq.\,(\ref{eq:RealisticPS}) are concerned, we use $k_{\star} = 1.5\times 10^{14}$ Mpc$^{-1}$, $n_1 = 3.4$, 
$k_1 = k_{\star}/5$, $k_2 = 3k_{\star}/2$ and $v= 0.7$; the values of $k_{1,2}$, $v$ and $n_1$ are motivated by a fit of eq.\,(\ref{eq:RealisticPS}) 
done with respect to the numerical power spectrum obtained in the context of the
explicit models studied in ref.\,\cite{Ballesteros:2020qam}. In particular, notice that the spectral index $n_1$ describes the 
growth of the power spectrum that leads to the formation of the peak; its value is related to the value of the Hubble parameter $\eta$ during the ultra slow-roll phase and the duration of the latter. Semi-analytical arguments (see ref.\,\cite{Byrnes:2018txb}) suggest that $n_1 < 4$.\footnote{However, see ref.\,\cite{Ozsoy:2019lyy} for a special case (derived in the context of non-attractor inflation) in which the growth of the power spectrum can be steeper, although for a limited range of $k$.
A steeper growth can also be attained in multi-field inflationary scenarios\,\cite{Fumagalli:2020adf}.
}  
The value $k_{\star} = 1.5\times 10^{14}$ Mpc$^{-1}$ is chosen because it implies $M_{\rm PBH}\simeq 10^{18}$ 
(for which $\Omega_{\rm PBH}\simeq \Omega_{\rm DM}$ is possible, see ref.\,\cite{Taoso:2021uvl}).
We consider the peak amplitude of the power spectrum $\mathcal{P}_{\mathcal{R}}(k_{\star})$ as a free parameter.
Notice that the power spectra that we consider lead to a narrow PBHs mass function, see appendix\,\ref{app:Threshold} for details. In the following, to analyze the impact of non-gaussianities, we will focus on the PBH mass around the peak of the distribution.

\item [$\ast$] The function $f(x)$ is given in eq.\,(\ref{eq:npkxyfx}) and the quantity $x_*$ in eq.\,(\ref{eq:AQFx}).
The function $x_{\delta}(\bar{\nu})$ is defined by the relation (see eq.\,(\ref{eq:xdelta}))
\begin{align}\label{eq:ExplThreshx}
(1+2\alpha\sigma_0\bar{\nu})\, x_{\delta}(\bar{\nu}) =\frac{9 (a_m H_m)^2}{4\sigma_2}\delta_c\,,
\end{align}
where $\delta_c = O(1)$ is a threshold value above which a peak of the overdensity field collapses to form a black hole. 
The left hand side of eq.\,(\ref{eq:ExplThreshx}) corresponds to the critical curvature of $h$ in eq.\,(\ref{eq:SpikyEnough}), 
see section\,\ref{app:NonGaussianPeakTheoryExact}.
Formally, eq.\,(\ref{eq:ExplThreshx}) depends on time via the comoving Hubble radius $1/aH$. 
We evaluate eq.\,(\ref{eq:ExplThreshx}) at the time $t_m$ 
 when curvature perturbations re-enter the horizon and become causally connected (see 
 refs.\,\cite{Germani:2018jgr,Musco:2018rwt}  and appendix\,\ref{app:Threshold}) \footnote{Notice however that eq.\,(\ref{eq:NonLinearDelta}) is valid on super-horizon scales. This means that at horizon-crossing additional non linear effects are present. Recently, these corrections have been considered in ref.\,\cite{Musco:2020jjb}.}. 
 In eq.\,(\ref{eq:ExplThreshx}) we use the short-hand notation $a_m H_m \equiv a(t_m)H(t_m)$.

\item  [$\ast$] 
An important comment concerns the so-called smoothing procedure. Consider the case in which one takes a very narrow power spectrum, like the toy-model
introduced in eq.\,(\ref{eq:ToyPS2}) with a small value of $v$, say $v=0.1$. 
In this case it is not strictly necessary to introduce a smoothing procedure because the power spectrum is characterized by a well-defined 
scale in momentum space, $k=k_{\star}$. 
The realistic case introduced in eq.\,(\ref{eq:RealisticPS}), on the contrary, requires more care.
Although the power spectrum peaks at $k=k_{\star}$, the peak is broadened by the relatively large value of $v$, and it also possesses a pronounced 
power-law tail at large $k\gg k_{\star}$. 
In this situation we can not blindly apply eq.\,(\ref{eq:MasterFormula}) to compute the PBH abundance because the 
spectral moment $\sigma_2^2$ is formally ultraviolet-divergent unless the power spectrum decays fast enough, which is however not the case 
in the realistic model.\footnote{Of course, any power spectrum generated by the inflationary dynamics has an intrinsic cut-off 
set by the smallest scale (largest $k$) that exits the Hubble horizon before inflation ends. 
More precisely, therefore, with the words ``ultraviolet-divergent integral'' we mean that $\sigma_2^2$ is dominated by small scales.}
The solution to this issue (discussed in appendix\,\ref{app:Threshold}) 
is to smooth-out small scales by introducing an appropriate cut-off.  
At the operative level, we use, instead of eq.\,(\ref{eq:RealisticPS}), 
the power spectrum $\mathcal{P}_{\mathcal{R}}^{\rm cut}(k) \equiv 
\mathcal{P}_{\mathcal{R}}(k)\exp(-k^2/k_{\rm cut}^2)$ and we choose $k_{\rm cut}$ such as to minimize the threshold value
 in the right-hand side of eq.\,(\ref{eq:ExplThreshx}).
Notice that this smoothing procedure is relevant for the determination both of the threshold value and the spectral moments in eq.\,(\ref{eq:HHHH}).

Physically, the fact that the power spectrum in eq.\,(\ref{eq:RealisticPS}) does not possess a 
well-defined 
scale means that the PBHs it generates will be 
characterized by a relatively broad mass distribution (rather than sharply peaked at the value associated to $k_{\star}$). The cut-off procedure described before selects the scale (and, therefore, the value of the mass $M_{\rm PBH}$) 
at which the abundance of PBH will be the largest.

\item [$\ast$] We use the linear approximation in eq.\,(\ref{eq:NonLinearDelta}). We discuss the role of non-linearities in section\,\ref{sec:nonlin}.

\end{itemize}

We now discuss the implications of eq.\,(\ref{eq:MasterFormula}). 
\begin{figure}[!htb!]
\begin{center}
\includegraphics[width=.45\textwidth]{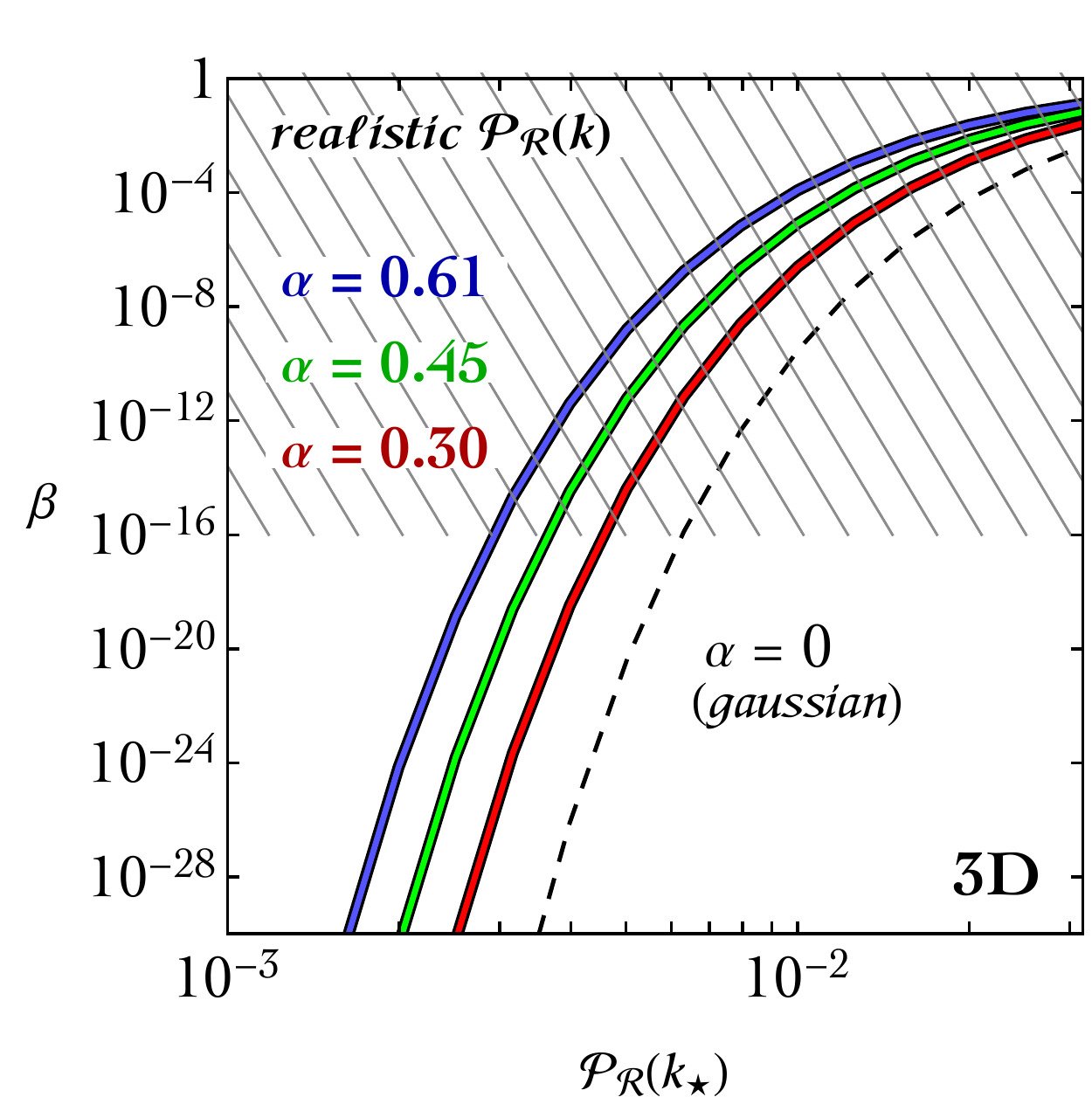}
\qquad
\includegraphics[width=.45\textwidth]{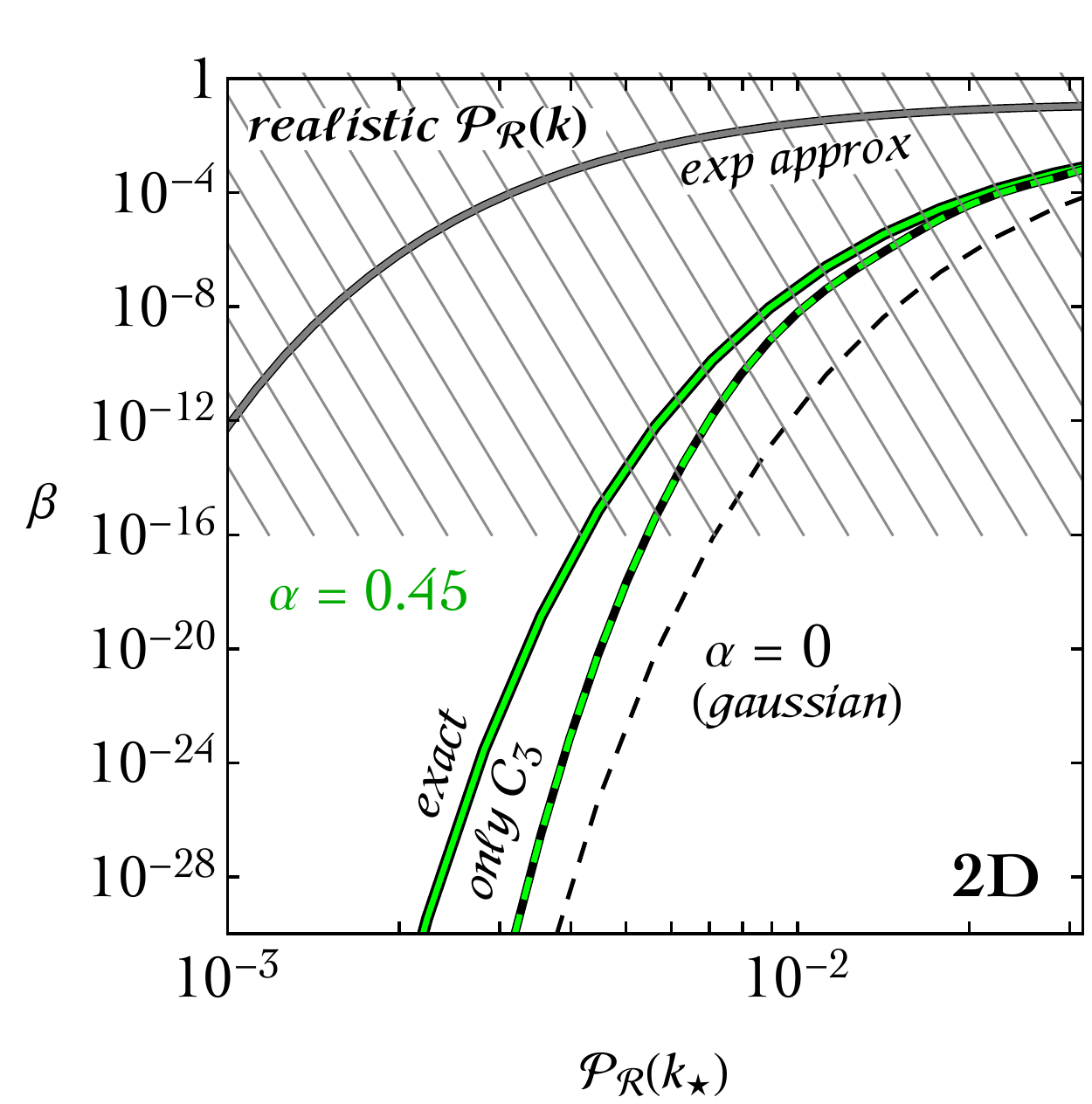}
\caption{\em \label{fig:MasterFormula} 
Left panel: Fraction of the Universe's mass in PBHs at the time of their formation computed with (solid lines with colors) and without (dashed black line) non-gaussianities as a function of the peak amplitude of the power spectrum $\mathcal{P}_{\mathcal{R}}(k_{\star})$. 
We adopt the realistic model for the power spectrum introduced in eq.\,(\ref{eq:RealisticPS}) and the abundance is computed using eq.\,(\ref{eq:MasterFormula}). 
We show the impact of non-gaussianities for different benchmark values of $\alpha$. 
 In the hatched region we have $\beta > 10^{-16}$ and the Universe is overclosed (see eq.\,(\ref{eq:PresentDayAbundance}) and related discussion).
 Right panel: we consider two spatial dimensions, and compare 
 the exact computation of $\beta$ with the approximation obtained including only the third-order cumulants. 
 We also show the value of $\beta$ obtained by means of the exponential approximation 
 in eq.\,(\ref{eq:MasterFormulaApp}); 
 the latter gives an estimate of the abundance off by many orders of magnitude compared with the actual result. 
 } 
\end{center}
\end{figure}
In the left panel of fig.\,\ref{fig:MasterFormula} we show the fraction of Universe's mass in the form of PBH 
computed according to eq.\,(\ref{eq:MasterFormula}) for increasing values of the parameter $\alpha$ starting from the gaussian case with $\alpha = 0$. 
What values of $\alpha$ are expected in concrete models?
In popular single-field models of inflation that generate a sizable abundance of dark matter in the form of PBHs, we find 
$\alpha \simeq 0.38$ (ref.\,\cite{Ballesteros:2020qam}),
$\alpha \simeq 0.37$ (ref.\,\cite{Ozsoy:2018flq}), $\alpha \simeq 0.30$ (ref.\,\cite{Cicoli:2018asa}), $\alpha \simeq 0.22$ (ref.\,\cite{Dalianis:2018frf}), $\alpha \simeq 0.61$ (ref.\,\cite{Ballesteros:2017fsr}). 
The plot shows that including local non-gaussianities of primordial origin makes the formation of PBHs easier, and the value of $\mathcal{P}_{\mathcal{R}}(k_{\star})$ required to reproduce the benchmark abundance $\beta = 10^{-16}$  turns out to be smaller than the gaussian one. 
We find that the rescaling of the peak amplitude implied by the presence of local non-gaussianities is modest, a factor a few in realistic models.
Moreover, let us mention that it can be obtained at a price of an even smaller retuning of the parameters of the inflationary models.
Similar results have been obtained in\,\cite{Taoso:2021uvl} for the calculation of the PBH abundance with threshold statistics.

\bigskip

In addition to the exact result presented in eq.\,(\ref{eq:MasterFormula}), 
we also consider a ``perturbative'' approach based on a power-series $\alpha$-expansion around the gaussian distribution. 
To this end, we work in two (instead of three) spatial dimensions. 
This simplifying assumption allows to perform most of the computations in appendix\,\ref{app:GaussianPeakTheory} and appendix\,\ref{app:NonGaussianPeakTheory} analytically, and makes possible 
 to visualize and check numerically a number of intermediate results by means of simple two-dimensional plots.

Let us first clarify the exact meaning of the word ``perturbative.''
From a statistical viewpoint, the $\alpha$-expansion corresponds to an expansion in cumulants 
of the joint non-gaussian probability distribution 
according to the schematic summarized in table\,\ref{tab:Cum} (see appendix\,\ref{app:GaussianPeakTheory}, appendix\,\ref{app:NonGaussianPeakTheory} and appendix\,\ref{app:NonPer} for details).
\begin{table}[!h!]
\begin{center}
\begin{tabular}{||c|c|c|c|c|c|c|c||}\hline\hline
\multirow{2}{*}{} & Gaussian & \multirow{2}{*}{$O(\alpha)$} &  \multirow{2}{*}{$O(\alpha^2)$} &  \multirow{2}{*}{$O(\alpha^3)$} 
&  \multirow{2}{*}{$O(\alpha^4)$} & \multirow{2}{*}{$O(\alpha^5)$} &  \multirow{2}{*}{$\dots$}   \\
& $\alpha = 0$ &  & & & & & \\ \hline\hline
\multirow{2}{*}{second-order cumulants $C_2$} & \multirow{2}{*}{{\color{indiagreen}{\ding{51}}}} & \multirow{2}{*}{{\color{cornellred}{\ding{55}}}} &  \multirow{2}{*}{{\color{indiagreen}{\ding{51}}}} & \multirow{2}{*}{{\color{cornellred}{\ding{55}}}} 
&  \multirow{2}{*}{{\color{cornellred}{\ding{55}}}} & \multirow{2}{*}{{\color{cornellred}{\ding{55}}}} &  \multirow{2}{*}{$\dots$}   \\
& & & & & & & \\ \hline 
\multirow{2}{*}{third-order cumulants $C_3$} & \multirow{2}{*}{{\color{cornellred}{\ding{55}}}} & \multirow{2}{*}{{\color{indiagreen}{\ding{51}}}} &  \multirow{2}{*}{{\color{cornellred}{\ding{55}}}} &  \multirow{2}{*}{{\color{indiagreen}{\ding{51}}}} 
&  \multirow{2}{*}{{\color{cornellred}{\ding{55}}}} & \multirow{2}{*}{{\color{cornellred}{\ding{55}}}} & \multirow{2}{*}{$\dots$}   \\
& & & & & & & \\ \hline 
\multirow{2}{*}{fourth-order cumulants $C_4$} & \multirow{2}{*}{{\color{cornellred}{\ding{55}}}} & \multirow{2}{*}{{\color{cornellred}{\ding{55}}}} &   \multirow{2}{*}{{\color{indiagreen}{\ding{51}}}} &  \multirow{2}{*}{{\color{cornellred}{\ding{55}}}} 
&  \multirow{2}{*}{{\color{indiagreen}{\ding{51}}}} & \multirow{2}{*}{{\color{cornellred}{\ding{55}}}} & \multirow{2}{*}{$\dots$}   \\
& & & & & & & \\ \hline
\multirow{2}{*}{fifth-order cumulants $C_5$} & \multirow{2}{*}{{\color{cornellred}{\ding{55}}}} & \multirow{2}{*}{{\color{cornellred}{\ding{55}}}} &   \multirow{2}{*}{{\color{cornellred}{\ding{55}}}} &  \multirow{2}{*}{{\color{indiagreen}{\ding{51}}}} 
&  \multirow{2}{*}{{\color{cornellred}{\ding{55}}}} & \multirow{2}{*}{{\color{indiagreen}{\ding{51}}}} & \multirow{2}{*}{$\dots$}   \\
& & & & & & & \\ \hline
\multirow{2}{*}{$\dots$} & \multirow{2}{*}{$\dots$} & \multirow{2}{*}{$\dots$} &  \multirow{2}{*}{$\dots$} &  \multirow{2}{*}{$\dots$} 
&  \multirow{2}{*}{$\dots$} &  \multirow{2}{*}{$\dots$} &  \multirow{2}{*}{$\dots$}   \\
& & & & & & & \\ \hline\hline       
\end{tabular}
\end{center}
\caption{\em  We organize the cumulants $C_n$ of the joint six-dimensional probability distribution $P(h,h_x,h_y,h_{xx},h_{xy},h_{yy})$ 
as a series expansion in $\alpha$. 
In the case $\alpha=0$, only second-order cumulants are non-vanishing (eqs.\,(\ref{eq:C21}-\ref{eq:C24}) with $\alpha = 0$), 
and we reconstruct the gaussian limit. 
At order $O(\alpha)$, the leading correction is given by third-order cumulants (eqs.\,(\ref{eq:Cumu1}-\ref{eq:Cumu7})).
At order $O(\alpha^2)$, we include 
corrections to the second-order cumulants (eqs.\,(\ref{eq:C21}-\ref{eq:C24})) and 
the leading pieces in the expression of fourth-order cumulants (eqs.\,(\ref{eq:C4_1}-\ref{eq:C4_14})).} 
\label{tab:Cum}
\end{table}%
In the gaussian approximation, all cumulants $C_{n\geqslant 3}$ vanish and the second-order ones correspond to the entries of the 
covariance matrix of the distribution. This result is valid in the limit $\alpha \to 0$. 
If $\alpha \neq 0$, all cumulants $C_{n\geqslant 3}$ are generated. 
However, the non-zero cumulants can be organized in terms of an $\alpha$-expansion as shown 
in table\,\ref{tab:Cum} (see caption, 
and appendix\,\ref{app:NonGaussianPeakTheory} and appendix\,\ref{app:NonPer} for details). 
Crudely speaking, we have
\begin{align}\label{eq:FundaScaling}
C_{n\geqslant 2} \sim O(\alpha^{n-2}) + O(\alpha^{n})\,.
\end{align}
At order $O(\alpha)$, only the leading part of the third-order cumulants $C_{3}$ appears.\footnote{More precisely, 
for each cumulant the expansion is controlled---considering for 
simplicity the log-normal power spectrum in eq.\,(\ref{eq:ToyPS2})---by the dimensionless parameter $\alpha A_g \ll 1$.}
We can, therefore, consider an expansion around the gaussian distribution including only the 
leading part of the
third-order cumulants $C_{3}$. 
The rationale for this approximation is twofold.
\begin{itemize}

\item [{\it i)}] If we compare the perturbative approach with the exact computation 
(downgraded in two spatial dimensions) we can estimate the validity of the approximation 
in which only the leading part of the third-order cumulant $C_{3}$ is included. 

This exercise is useful for the following reason. 
In the approach based on threshold statistics, one usually computes, using the tools of 
cosmological perturbation theory, the cumulants   
in the form of (the connected part of) correlators of the density perturbation
field. 
In the presence of ultra slow-roll, however, this computation is not 
simple (see refs.\,\cite{Atal:2018neu,Taoso:2021uvl}), and one typically includes only the leading term in the so-called bispectrum (that is the three-point correlator).
This approximation precisely corresponds to the one in which 
only the leading part of the third-order cumulant $C_{3}$ is included (that is 
the one proportional to $\alpha$ in eq.\,(\ref{eq:FundaScaling})). 
In the right panel of fig.\,\ref{fig:MasterFormula} we show the comparison between 
the exact computation of $\beta$ and the approximation in which only the 
leading part of $C_3$ is included; as mentioned before, we work in two spatial dimensions 
and we fix $\alpha = 0.45$. The comparison shows that truncating the expansion at the first order in the 
non-gaussian corrections does not fully capture the impact of non-gaussianities on $\beta$. 

Based on this result, we pose the attention on the fact that a similar conclusion is likely to be valid also 
when computing correlators in cosmological perturbation theory. 
The local non-gaussianity which is present at the level of the three-point correlator 
(computed, for instance, in refs.\,\cite{Atal:2018neu,Taoso:2021uvl} and used in 
 ref.\,\cite{Atal:2018neu} to estimate the impact of non-gaussianities on $\beta$) induces corrections also at higher orders.
In terms of (cosmological) Feynman diagrams, the situation can be sketched as follows
  \begin{align}\label{eq:CosmoFey}
  	\raisebox{-.5mm}{
	\begin{tikzpicture}
	\draw [bluscuro,thick, dashed] (-2,0)--(2,0); 
	\draw [thick] (-1.5,0)--(0,-1.5); 
	\draw [thick] (0,0)--(0,-1.5); 
	\draw [thick] (1.5,0)--(0,-1.5); 
	\draw[black,fill=tangerineyellow,thick] (0,-1.5)circle(6pt);
  \node[anchor=south] at (0,-2.5) {\scalebox{0.85}{third-order cumulant}};	   
	\end{tikzpicture}}
	\hspace{0.8cm}
	\raisebox{-.5mm}{
	\begin{tikzpicture}
	\draw [bluscuro,thick, dashed] (-2,0)--(2,0); 
	\draw [thick] (-1.75,0)--(-1,-1.5); 
	\draw [thick] (-0.25,0)--(-1,-1.5); 
	\draw [thick] (-1,-1.5)--(1,-1.5);  
	\draw [thick] (1-0.75,0)--(1,-1.5);
	\draw [thick] (1+0.75,0)--(1,-1.5);
	\draw[black,fill=tangerineyellow,thick] (1,-1.5)circle(6pt);
	\draw[black,fill=tangerineyellow,thick] (-1,-1.5)circle(6pt);
     \node[anchor=south] at (0,-2.5) {\scalebox{0.85}{fourth-order cumulant from (cubic interactions)$^2$}};
	\end{tikzpicture}}
	\hspace{0.8cm}
	\raisebox{-.5mm}{
	\begin{tikzpicture}
	\draw [bluscuro,thick, dashed] (-2,0)--(2,0); 
	\draw [thick] (0-1.5,0)--(0,-1.5);
	\draw [thick] (0-0.5,0)--(0,-1.5);
    \draw [thick] (0+1.5,0)--(0,-1.5);
	\draw [thick] (0+0.5,0)--(0,-1.5); 	 
	\draw[black,fill=violachiaro,thick] (0,-1.5)circle(6pt);
     \node[anchor=south] at (0,-2.5) {\scalebox{0.85}{pure fourth-order cumulant}};
     \node[anchor=south] at (1.5,0.1) {\scalebox{0.85}{{\color{bluscuro}{end of inflation}}}};
	\end{tikzpicture}}		
\end{align}
where the square of the third-order local interaction enters at the fourth-order (as well as at higher ones).\footnote{Of course, higher-order correlators 
generated by pure higher-order interactions---for instance, a pure quartic interaction in the violet vertex of the example above---could also be present but, as discussed in appendix\,\ref{app:InflationPrimer}, 
we do not consider them explicitly in this paper.} 
The comparison shown in the right panel of fig.\,\ref{fig:MasterFormula} 
suggests that, without including the contributions that 
third-order local interactions induce in higher-order cumulants, a precise computation of $\beta$ cannot be claimed. 

In addition to this observation there is a second, and by far more problematic, issue that we shall discuss next.

\item [{\it ii)}] 
When computing non-gaussian corrections in the form of 
a series expansion around the gaussian distribution, 
some care must be taken. Seemingly harmless approximations, indeed, may lead to erroneous conclusions. 
Consider the simplified  case discussed before in which 
only the leading part of the third-order cumulants $C_{3}$ is included. 
A wrong approximation in the evaluation of the non-gaussian probability distribution 
leads to the expression of the abundance that we report in eq.\,(\ref{eq:MasterFormulaApp}). 
In this compact expression, the non-gaussian correction exponentiates and alters the argument of the 
exponential function in the gaussian distribution. 
For illustration, we add in the right panel of fig.\,\ref{fig:MasterFormula} the value of $\beta$ that one gets 
by means of eq.\,(\ref{eq:MasterFormulaApp}). 
Clearly, eq.\,(\ref{eq:MasterFormulaApp}) overestimates the actual value of $\beta$ by many orders of magnitude. 
In appendix\,\ref{app:NonGaussianPeakTheory} we explain why this approximation is wrong and what 
is the correct procedure to follow.

The reader may wonder why we are wasting time discussing a wrong result. 
The reason is that it rings a bell. 
Previous studies on the impact of primordial non-gaussianities found, although in a 
slightly different statistical context, that non-gaussian corrections alter 
exponentially the gaussian value of $\beta$, 
analogously to what happens with eq.\,(\ref{eq:MasterFormulaApp}),
an expression that we just branded inaccurate.
A closer look at this literature, therefore, is mandatory. This will be the subject of the next section.

\end{itemize}

\subsection{Comparison with the literature}\label{sec:Comp}

The impact of primordial non-gaussianities on the computation of PBH abundance was discussed in 
ref.\,\cite{Franciolini:2018vbk} in the context of the so-called threshold statistics.
Ref.\,\cite{Franciolini:2018vbk} finds that primordial non-gaussianities may play a very relevant role because they 
 alter the argument of the exponential function that sets the value of the PBH abundance in the gaussian case.  
Ref.\,\cite{Atal:2018neu} applies the results of ref.\,\cite{Franciolini:2018vbk} to the case of
 single-field inflationary models which
feature the presence of an approximate stationary inflection point, and concludes that  
 the gaussian approximation does not give the correct estimate of the PBHs abundance precisely because 
 the non-gaussian correction exponentiates and drastically changes the gaussian result. 

However, this conclusion clashes with our result. 
In fact, as we are going to show, the impact of non-gaussianities on the PBH abundance is much smaller than what found in ref.\,\cite{Franciolini:2018vbk}.
To explain the discrepancy, let us first re-derive the main result of ref.\,\cite{Franciolini:2018vbk} in a simplified way.

In a nutshell, in the context of threshold statistics one computes, assuming a gaussian distribution, 
the probability to find regions where the overdensity field $\delta$ takes values above a given threshold. 
It is known that threshold statistics gives a smaller PBH abundance if compared with peak theory. 
Let us briefly explain the origin of this difference. This point is not crucial to understand the discrepancy between our
conclusions and  ref.\,\cite{Franciolini:2018vbk}
 but it 
will play an important role later when we will discuss the relation with gravitational waves. 
In this section we work directly with the density contrast $\delta$. Furthermore, we start considering the gaussian limit.  
The gaussian probability density distribution 
of $\delta$ with variance $\sigma_{\delta}$ and spectral moments $\bar{\sigma}_j$ (where $\bar{\sigma}_0 = \sigma_{\delta}$) is given in the two cases by the expressions
\begin{align}
P_{\rm threshold}(\delta) & = \frac{1}{\sqrt{2\pi}\sigma_{\delta}}e^{-\delta^2/2\sigma_{\delta}^2}\,,
\label{eq:ThresholdP}\\
P_{\rm peak}(\delta) & = \frac{1}{(2\pi)^2 \sigma_{\delta} R_*^3}
\underbrace{\bigg\{
\int_0^{\infty}dx \frac{f(x)}{\sqrt{2\pi(1-\gamma^2)}}\exp\left[
-\frac{(x- \delta\gamma/\sigma_{\delta})^2}{2(1-\gamma^2)}
\right]
\bigg\}}_{\equiv \, G(\gamma,\delta)}e^{-\delta^2/2\sigma_{\delta}^2} 
\equiv 
\frac{1}{(2\pi)^2 \sigma_{\delta} R_*^3}
\int_{0}^{\infty}dx P_{\rm peak}(x,\delta) \,,\label{eq:PeakdP}
\end{align}
where the function $f(x)$ is given explicitly in eq.\,(\ref{eq:npkxyfx}). 
The two quantities $0 \leqslant \gamma \equiv \bar{\sigma}_1^2/\bar{\sigma}_2\sigma_{\delta} < 1$ and
 $R_* \equiv \sqrt{3}\bar{\sigma}_1/\bar{\sigma}_2$ are factors depending on the power spectrum 
 of $\delta$ (notice in particular that $\gamma$ is the same defined below eq.\,(\ref{eq:HHHH}) but now written in terms 
 of the spectral moments of $\delta$ instead of $\mathcal{R}$). 
 The variable $x$ is defined by $x\equiv -\triangle\delta/\bar{\sigma}_2$. 
 The two expressions are similar but there are two differences. 
 First, $P_{\rm peak}(\delta)$ has dimension of inverse spatial volume (because of the factor $1/R_*^3$). 
 This is because $P_{\rm peak}(\delta)$ is defined in peak theory as a number density of maxima.  
 This implies that the true comparison is between the a-dimensional quantities $P_{\rm threshold}(\delta)$ and $R_*^3P_{\rm peak}(\delta)$. 
 Second, $P_{\rm peak}(\delta)$ contains an extra factor (the one in curly brackets) with respect to 
 $P_{\rm threshold}(\delta)$. This factor arises because in peak theory one starts from 
 the ten-dimensional gaussian joint probability density distribution $P(\delta,\delta_i,\delta_{ij})$ of $\delta$ and its first 
 ($\delta_i$) and 
 second ($\delta_{ij}$) spatial derivatives and imposes a number of conditions that 
 select among the stationary points those that are maxima\,\cite{Bardeen:1985tr}.  
 In threshold statistics, on the contrary, one simply integrates out all  spatial informations 
 by  reducing $P(\delta,\delta_i,\delta_{ij})$ to the one-dimensional gaussian distribution in eq.\,(\ref{eq:ThresholdP}). 
 The crucial aspect is that the function $G(\gamma,\delta)$ in eq.\,(\ref{eq:PeakdP}) 
 depends on $\delta$ in a way which is proportional to the parameter $\gamma$. 
 The latter controls the degree of correlation between $\delta$ and $x$. 
 When $\gamma = 0$, the two variables are completely uncorrelated, and 
 we find that $G(0,\delta) = (29-6\sqrt{6})/10\sqrt{10\pi}$. In this case, $G(0,\delta)$ does not depend on 
 $\delta$: peak theory and threshold statistics give the same qualitative answer (in the sense that the only $\delta$-dependence is encoded in the gaussian exponential which is in common between the two).  
 When $\gamma \to 1$, $\delta$ and $x$ are strongly correlated.
 \begin{figure}[!htb!]
\begin{center}
\includegraphics[width=.45\textwidth]{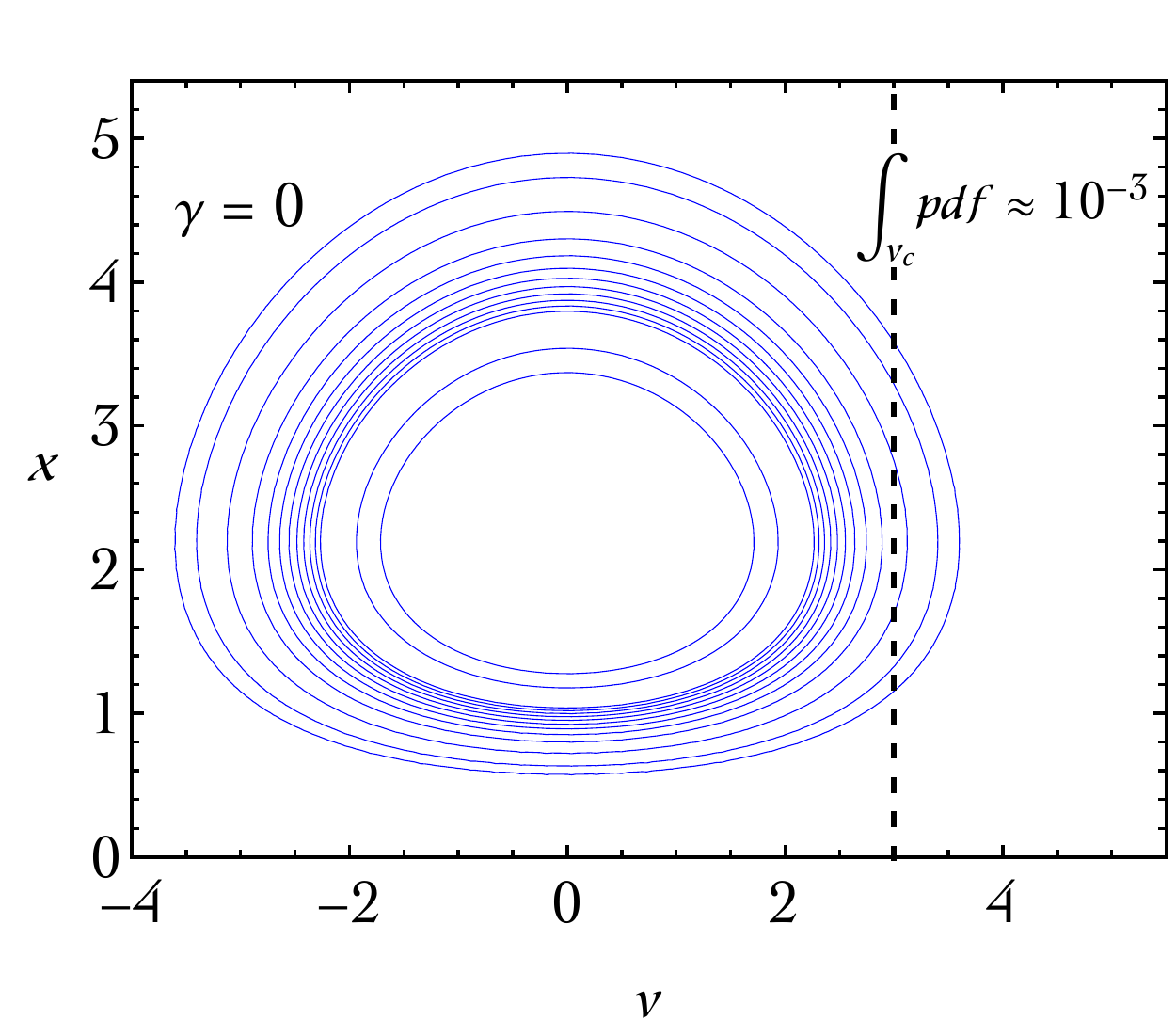}
\qquad
\includegraphics[width=.45\textwidth]{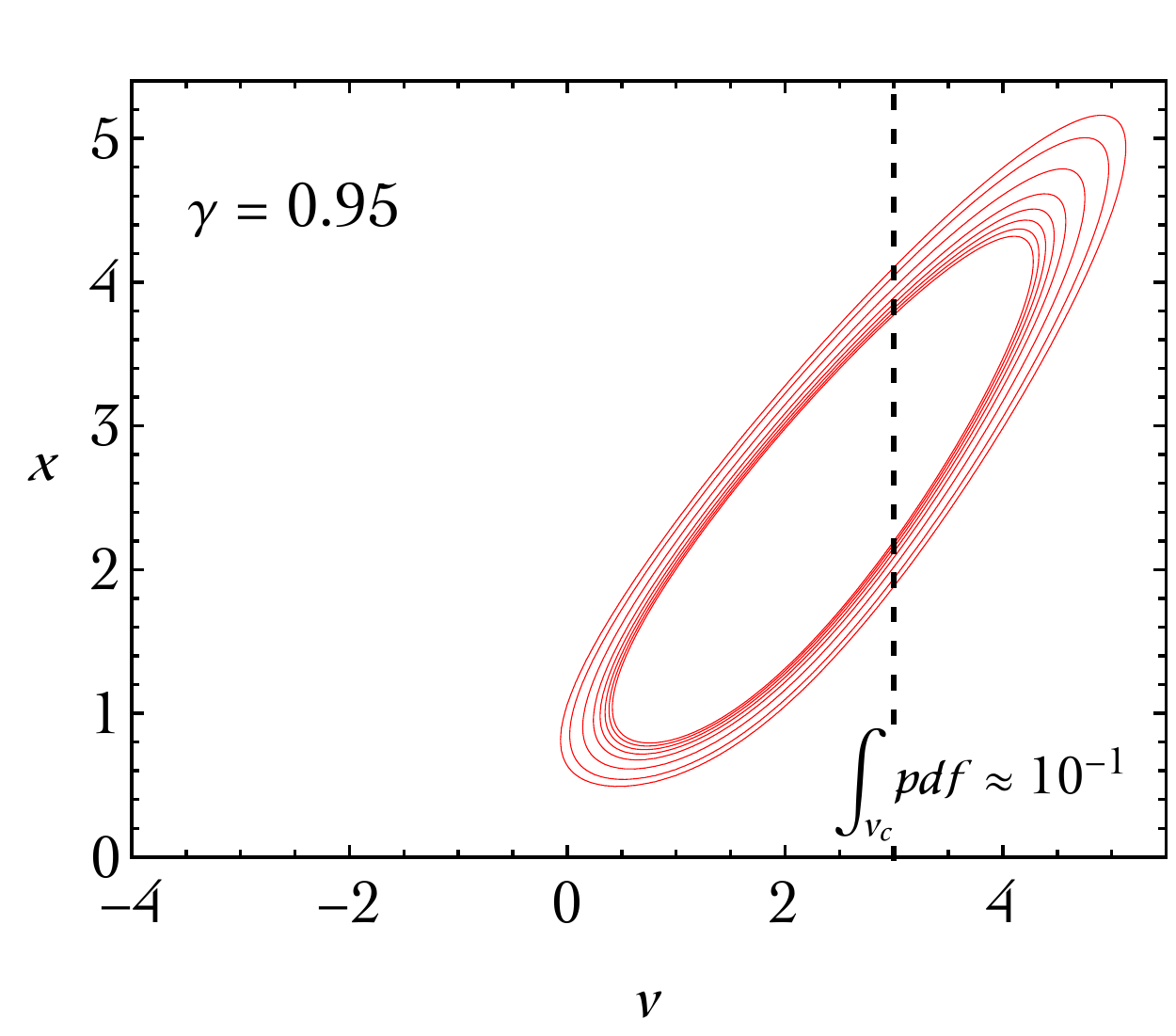}
\caption{\em \label{fig:PDFpeak} 
Isocontours of constant probability density distribution  $P_{\rm peak}(x,\delta)$ 
defined in eq.\,(\ref{eq:PeakdP}). We use $\nu \equiv \delta/\sigma_{\delta}$ and we set 
(for illustrative purposes only) the threshold at $\nu_c = 3$. 
The probability above the threshold is obtained integrating $P_{\rm peak}(x,\delta)$ for
$x\in [0,\infty)$ and $\nu\in [\nu_c,\infty)$. 
We show the case with $\gamma = 0$ (no correlation between $x$ and $\delta$, left panel) and 
$\gamma = 0.95$ (strong correlation between $x$ and $\delta$, right panel). 
 } 
\end{center}
\end{figure}
 This fact deforms the shape of the probability density distribution of $x$ and $\delta$ 
 and gives more weight to the region in which $\delta$ crosses the threshold for collapse. 
 This is illustrated schematically in fig.\,\ref{fig:PDFpeak} (using $\nu \equiv \delta/\sigma_{\delta}$). 
 In the case of strong correlation between $\delta$ and $x$, therefore, 
 it is well expected that after integrating above 
 the threshold $\nu > \nu_c$ 
 peak theory gives  a result which is larger than threshold statistics.\footnote{Notice that, for the sake of 
 simplicity, we are implicitly assuming in this example 
 the same threshold value $\nu_c$ in the case of peak theory and threshold statistics in order to highlight the 
 main difference between the two statistical approaches.} 
 In threshold statistics, this information about the correlation between $\delta$ and $x$ is completely lost since $\delta$ is treated independently from the spatial configuration. 
 We remark that the difference between threshold statistics and peak theory is 
 more and more relevant as we approach the 
 limit $\gamma \to 1$. As already pointed out, the value of $\gamma$ is dictated by the properties of the 
 power spectrum (in this case, strictly speaking, 
 the power spectrum of $\delta$ which is however related to the power spectrum of comoving curvature perturbations) 
 with a very peaked power spectrum that corresponds to the limit $\gamma\to 1$.
 
After this digression, we can back to the main point of this section. 
In the context of threshold statistics, non-gaussianities are included in the form of a  Gram–Charlier A series around the 
gaussian ansatz\,\cite{bookStat}
\begin{align}
P_{\rm NG}(\delta) & = 
P_{\rm G}(\delta)\left[1 + \sum_{n=3}^{\infty}
\frac{1}{n!2^{n/2}}B_n(0,0,\mathcal{C}_3,\dots,\mathcal{C}_n)H_n\left(
\frac{\nu}{\sqrt{2}}
\right)\right] \label{eq:GC} \\
& = P_{\rm{G}}(\delta)\bigg[
1 + \frac{\mathcal{C}_3}{3!2^{3/2}}H_3\left(\frac{\nu}{\sqrt{2}}\right)
+ 
\frac{\mathcal{C}_4}{4!2^{4/2}}H_4\left(\frac{\nu}{\sqrt{2}}\right) +
\frac{\mathcal{C}_5}{5!2^{5/2}}H_5\left(\frac{\nu}{\sqrt{2}}\right)  +
\frac{(10\mathcal{C}_3^2+\mathcal{C}_6)}{6!2^{6/2}}H_6\left(\frac{\nu}{\sqrt{2}}\right) +\dots
\bigg]\,,\nn
\end{align}
where $P_{\rm{G}}(\delta) = (1/\sqrt{2\pi}\sigma_{\delta})\exp(-\delta^2/2\sigma_{\delta}^2)$ is the gaussian probability density distribution 
of $\delta$ with variance $\sigma_{\delta}$ (that is $P_{\rm threshold}(\delta)$ introduced in eq.\,(\ref{eq:ThresholdP})) 
 and $H_n$ are the physicists' Hermite polynomials.\footnote{We remind that 
\begin{align}
\frac{d^n}{dx^n}e^{-x^2/2\sigma^2} = \frac{(-1)^n}{2^{n/2}\sigma^n}e^{-x^2/2\sigma^2}H_n\left(\frac{x}{\sqrt{2}\,\sigma}\right)\,,
~~~~~~~~~{\rm with}~~H_n(x) = (-1)^n e^{x^2}\frac{d^n}{dx^n}e^{-x^2}\,,
\end{align}
so that the reader can immediately recognize in eq.\,(\ref{eq:GC}) the structure of a derivative expansion.
}
 We define $\nu\equiv \delta/\sigma_{\delta}$ and we introduce 
the $n$-th complete exponential Bell polynomial $B_n(x_1,\dots,x_n) = \sum_{k=1}^{n}B_{n,k}(x_1,\dots,x_{n-k+1})$
with  $B_{n,k}$ the partial exponential Bell polynomials. We indicate with 
$\mathcal{C}_n$ the $n^{\rm th}$ normalized 
cumulant defined as the connected part of the $n$-point correlator (evaluated at the same point) of the overdensity field 
normalized by the $n^{\rm th}$ power of the standard deviation
\begin{align}\label{eq:NormCumu}
\mathcal{C}_n \equiv \frac{\langle\overbrace{\delta(\vec{x})\dots \delta(\vec{x})}^{n\,{\rm times}}\rangle_{\rm conn}}{\sigma_{\delta}^n}\,.
\end{align}
The explicit computation of eq.\,(\ref{eq:NormCumu}) in the presence of an ultra-slow roll phase is discussed, in the context of cosmological perturbation theory, in ref.\,\cite{Taoso:2021uvl} (see also ref.\,\cite{Atal:2018neu}). 
Notice that the quantity $\langle\delta(\vec{x})\dots \delta(\vec{x})\rangle$ is dimensionless, and so is $\mathcal{C}_n$.
We now integrate $P_{\rm NG}(\delta)$ over some threshold $\delta_c = \nu_c \sigma_{\delta}$.
We define the abundance $\beta(\nu_c) = \int_{\delta_c}^{\infty}d\delta\,P_{\rm NG}(\delta)$.
We find\footnote{
We use the property 
\begin{align}
\frac{1}{\sqrt{2\pi}\sigma}\int_{\delta_{c}}^{\infty}d\delta
e^{-\delta^2/2\sigma^2}H_n\left(
\frac{\delta}{\sqrt{2}\sigma}
\right) = \frac{1}{\sqrt{\pi}}e^{-\nu_c^2/2}H_{n-1}\left(
\frac{\nu_{c}}{\sqrt{2}}
\right)\,.
\end{align}}
\begin{align}
\beta(\nu_c) & = \frac{1}{2}{\rm Erfc\left(\frac{\nu_c}{\sqrt{2}}\right)} 
+ \frac{1}{\sqrt{2\pi}\nu_c}e^{-\nu_c^2/2}
\sum_{n=3}^{\infty}
\frac{\sqrt{2}\nu_c}{n!2^{n/2}}B_n(0,0,\mathcal{C}_3,\dots,\mathcal{C}_n)H_{n-1}\left(
\frac{\nu_c}{\sqrt{2}}
\right)\label{eq:exactbeta}\\
& \simeq \underbrace{\frac{1}{\sqrt{2\pi}\nu_c}e^{-\nu_c^2/2}}_{{\rm gaussian\,approx}\,\beta_{\rm G}(\nu_c)}
\underbrace{\left[
1 + \sum_{n=3}^{\infty}
\frac{\sqrt{2}\nu_c}{n!2^{n/2}}B_n(0,0,\mathcal{C}_3,\dots,\mathcal{C}_n)H_{n-1}\left(
\frac{\nu_c}{\sqrt{2}}
\right)
\right]}_{\rm non\,gaussian\,correction}\,,\label{eq:GCexp}
\end{align}
where in the last step we use 
$(1/2){\rm Erfc}(x/\sqrt{2}) \simeq (1/\sqrt{2\pi}x)e^{-x^2/2}$ for $x\gg 1$. 

We now adopt the approach suggested in ref.\,\cite{Franciolini:2018vbk} and approximate 
\begin{align}\label{eq:ApproxHerm}
H_n\left(
\frac{\nu_c}{\sqrt{2}}
\right) = 2^{n/2}\nu_c^n + O(\nu_c^{n-2})\,,
\end{align} 
which seems justified since $\nu_c \gg 1$. In this case eq.\,(\ref{eq:GCexp}) becomes\footnote{
The complete exponential Bell polynomial is defined by the exponential generating function\,\cite{Comtet}
\begin{align}
\exp\left(
\sum_{j=1}^{\infty}\frac{x_j t^j}{j!}
\right) = \sum_{n=0}^{\infty}B_n(x_1,\dots,x_n)\frac{t^n}{n!}\,,
\end{align}
and the first few complete Bell polynomials are $B_0 = 1$, $B_1(x_1) = x_1$, $B_2(x_1,x_2) = x_1^2 + x_2$, 
$B_3(x_1,x_2,x_3) = x_1^3 + 3x_1 x_2 + x_3$. In our case we have $x_1=x_2=0$ and the previous definition takes the form
\begin{align}
\exp\left(
\sum_{j=3}^{\infty}\frac{x_j t^j}{j!}
\right) = 1 + \sum_{n=3}^{\infty}B_n(0,0,x_3,\dots,x_n)\frac{t^n}{n!}\,,
\end{align}
which applies to our case with $t=\nu_c$ and $x_{i\geqslant 3} = \mathcal{C}_{i\geqslant 3}$ 
since the application of eq.\,(\ref{eq:ApproxHerm}) to eq.\,(\ref{eq:GCexp}) gives 
\begin{align}
\beta(\nu_c) \simeq \frac{1}{\sqrt{2\pi}\nu_c}e^{-\nu_c^2/2}\left[
1+\sum_{n=3}^{\infty}B_n(0,0,\mathcal{C}_3,\dots,\mathcal{C}_n)\frac{\nu_c^n}{n!}
\right]\,.
\end{align}
Notice that, compared to our result, ref.\,\cite{Franciolini:2018vbk} finds an additional factor 
$(-1)^n$ in eq.\,(\ref{eq:MainRiotto}) which, however, does not appear in our computation.
}
\begin{align}\label{eq:MainRiotto}
\beta(\nu_c) & \simeq \frac{1}{\sqrt{2\pi}\nu_c}
\exp\left(
-\frac{\nu_c^2}{2} + \sum_{n=3}^{\infty}\frac{\mathcal{C}_n\nu_c^n}{n!}
\right)\,.
\end{align}
We find, therefore, the main result of ref.\,\cite{Franciolini:2018vbk}: The non-gaussian correction 
exponentiates, and changes the argument of the exponential in the  gaussian distribution. 

Consider the simplified case in which only the third-order normalized cumulant (a.k.a. skewness) is non-vanishing, 
$\mathcal{C}_3 \neq 0$ and $\mathcal{C}_{n>3} = 0$. This is the approximation studied in ref.\,\cite{Atal:2018neu} in which 
the third-order cumulant is computed in the case of local 
non-gaussianities.\footnote{Notice, however, that assuming $\mathcal{C}_3 \neq 0$ and $\mathcal{C}_{n>3} = 0$ is not fully consistent 
with the local non-gaussianities computed in ref.\,\cite{Atal:2018neu}. Indeed, local non-gaussianities 
automatically generate non-vanishing cumulants of any order. This is evident in our approach, in which 
$\alpha\neq 0$ generates a whole tower of non-zero cumulants. 
From a more mathematical viewpoint, consider the following theorem.
{\it The sequence $\{\mu_n,n=0,1,2,\dots\}$ corresponds to moments of a non-negative probability density function if and only if the determinants
\begin{align}\label{eq:MomTheeo}
D_{n+1} \equiv {\rm det}
\left(
\begin{array}{ccccc}
 \mu_0 & \mu_1  & \mu_2 & \cdots  & \mu_n \\
 \mu_1 &  \mu_2 & \mu_3 & \cdots  & \mu_{n+1} \\
 \mu_2  & \mu_3  & \mu_4 & \cdots  & \mu_{n+2} \\
 \vdots  & \vdots & \vdots & \ddots &  \vdots \\
 \mu_n & \mu_{n+1} & \mu_{n+2} & \cdots & \mu_{2n}
\end{array}
\right)\,,~~~~~~{n=0,1,2,\dots}\,,
\end{align}
are {\underline{all}} non-negative\,\cite{moments}}.
Consider the illustrative case with $\mathcal{C}_{n} = 0$ for
 $n\geqslant 4$. We have $\mu_0 = 1$, $\mu_1 = 0$, $\mu_2 = \sigma_{\delta}^2$, $\mu_3 = \langle \delta^3\rangle$ and
  $\mu_n = \sigma_{\delta}^n (n-1)!!$ for $n\geqslant 4$ even ($\mu_n = 0$ for $n\geqslant 4$ odd). 
  The last condition simply means that cumulants of order higher than three vanish (and the corresponding moments purely gaussian).
The first determinant $D_3 >0$
already sets a non-trivial condition on the skewness, that is 
$-\sqrt{2} < \mathcal{C}_3 < \sqrt{2}$. 
However, there is more than this. 
It is indeed trivial to see that higher-order conditions $D_n > 0$ for $n>3$ impose increasingly strong bound on 
$\mathcal{C}_3$ so that only $\mathcal{C}_3\to 0$ is allowed if all the infinite number of constraints are implemented.
In other words, the theorem above 
implies that {\it skewness alone can not consistently parameterize a non-gaussian probability density function}.
 This result resonates with what we discussed at point {\it ii)} in section\,\ref{eq:Abubnda}. 
 Computing only the three-point correlator of the overdensity field does not fully describe non-gaussianities but 
 one should at least include the contributions that the three-point function generates at higher orders.
}
We find
\begin{align}\label{eq:WrongSumThesholdStat}
\beta(\nu_c) \simeq 
\frac{1}{\sqrt{2\pi}\nu_c}\exp\left(
-\frac{\nu_c^2}{2} + \frac{\mathcal{C}_3\nu_c^3}{6}
\right)\,,~~~~~~~{\rm with}~~~\mathcal{C}_3 \neq 0~~~{\rm and}~~~\mathcal{C}_{n>3} = 0\,.
\end{align}
Ref.\,\cite{Franciolini:2018vbk} and ref.\,\cite{Atal:2018neu} used the above equations to conclude that the gaussian estimate of the PBH abundance is hardly trustable. 
This result is based on the approximation in eq.\,(\ref{eq:ApproxHerm}). 
However, as we shall now discuss, the applicability of this approximation is not as straightforward as one may think.
Let us critically inspect the issue. 

We define, for ease of reading, the quantity $b_n(\nu_c)\equiv (\sqrt{2}\nu_c/n!2^{n/2})B_n(0,0,\mathcal{C}_3,\dots,\mathcal{C}_n)H_{n-1}(\nu_c/\sqrt{2})$ which enters in the non-gaussian correction in eq.\,(\ref{eq:GCexp}).
As a consequence of eq.\,(\ref{eq:ApproxHerm}), we have (slashed terms are neglected if we apply the approximation in eq.\,(\ref{eq:ApproxHerm}))
\begin{align}
b_3 & = \frac{\mathcal{C}_3\nu_c^3}{6}\bigg(1-\cancel{\frac{1}{\nu_c^2}}\bigg)\,,\label{eq:He1}\\
b_6 & = \frac{\mathcal{C}_3^2\nu_c^6}{72}
\bigg(
1 - \cancel{\frac{10}{\nu_c^2} + \frac{15}{\nu_c^4}}
\bigg)\,,\label{eq:He2}\\
b_9 & = \frac{\mathcal{C}_3^3\nu_c^9}{1296}
\bigg(
1 - \cancel{\frac{28}{\nu_c^2} + \frac{210}{\nu_c^4} - \frac{420}{\nu_c^6} + 
\frac{105}{\nu_c^8}}
\bigg)\label{eq:He3}\,,\\
b_{12} & = \dots
\end{align}
and so on. We see, for instance, that we neglected the term $-7\mathcal{C}_3^3\nu_c^7/324$ in eq.\,(\ref{eq:He3}) 
but we kept the term $\mathcal{C}_3\nu_c^3/6$ in eq.\,(\ref{eq:He1}). 
This is hardly justifiable given that $\nu_c \gg 1$ and we expect $\mathcal{C}_3 \sim O(1)$ 
(see ref.\,\cite{Taoso:2021uvl} for a careful computation of $\mathcal{C}_3$).  
This is enough to question the validity of the result based on  eq.\,(\ref{eq:ApproxHerm}).
Furthermore, from this simple comparison, we also see that the approximation in eq.\,(\ref{eq:ApproxHerm})---contrary to what na\"{\i}vely expected---gets worse for larger $\nu_c$.
\begin{figure}[!htb!]
\begin{center}
$$\includegraphics[width=.45\textwidth]{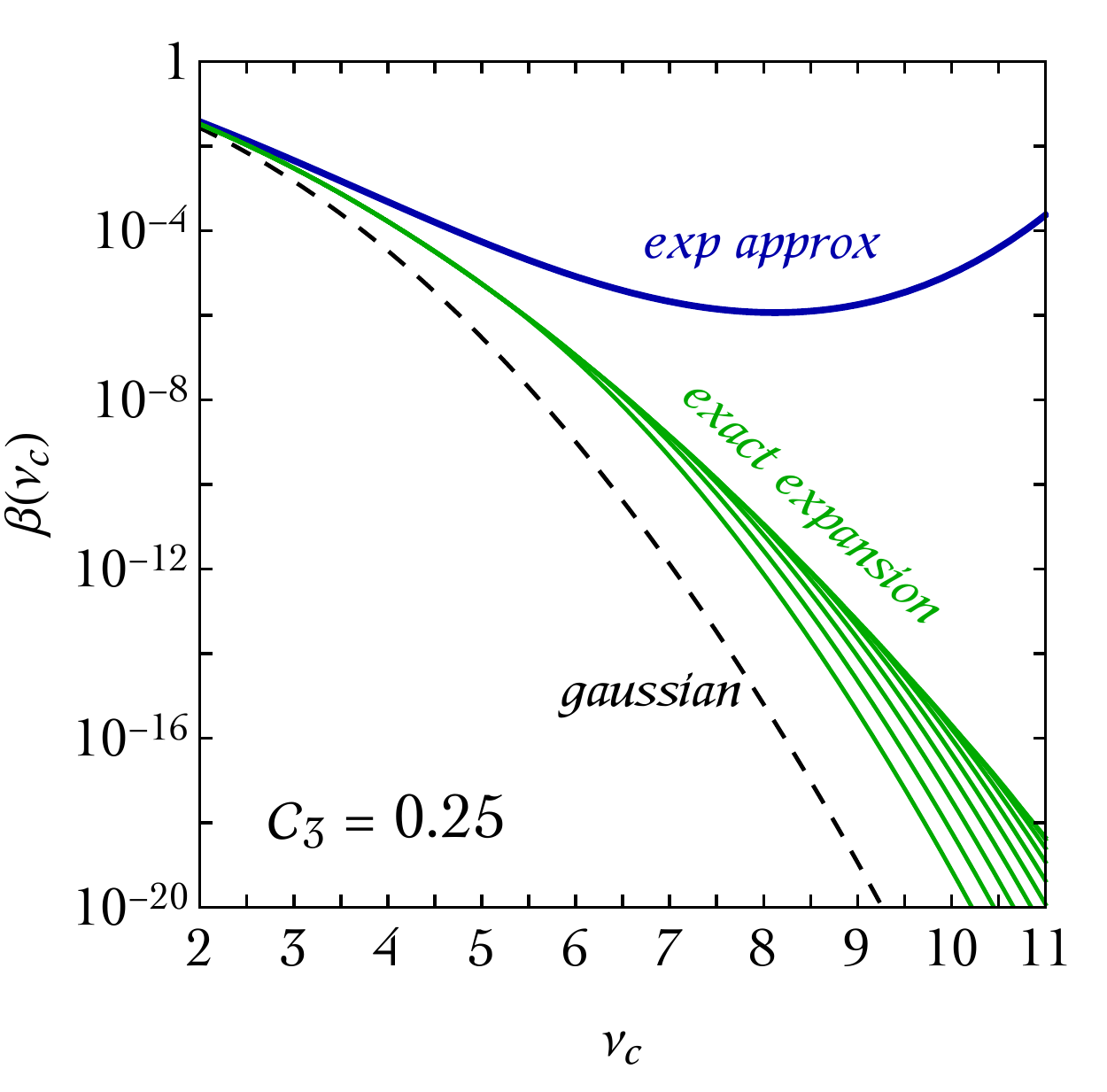}
\qquad\includegraphics[width=.45\textwidth]{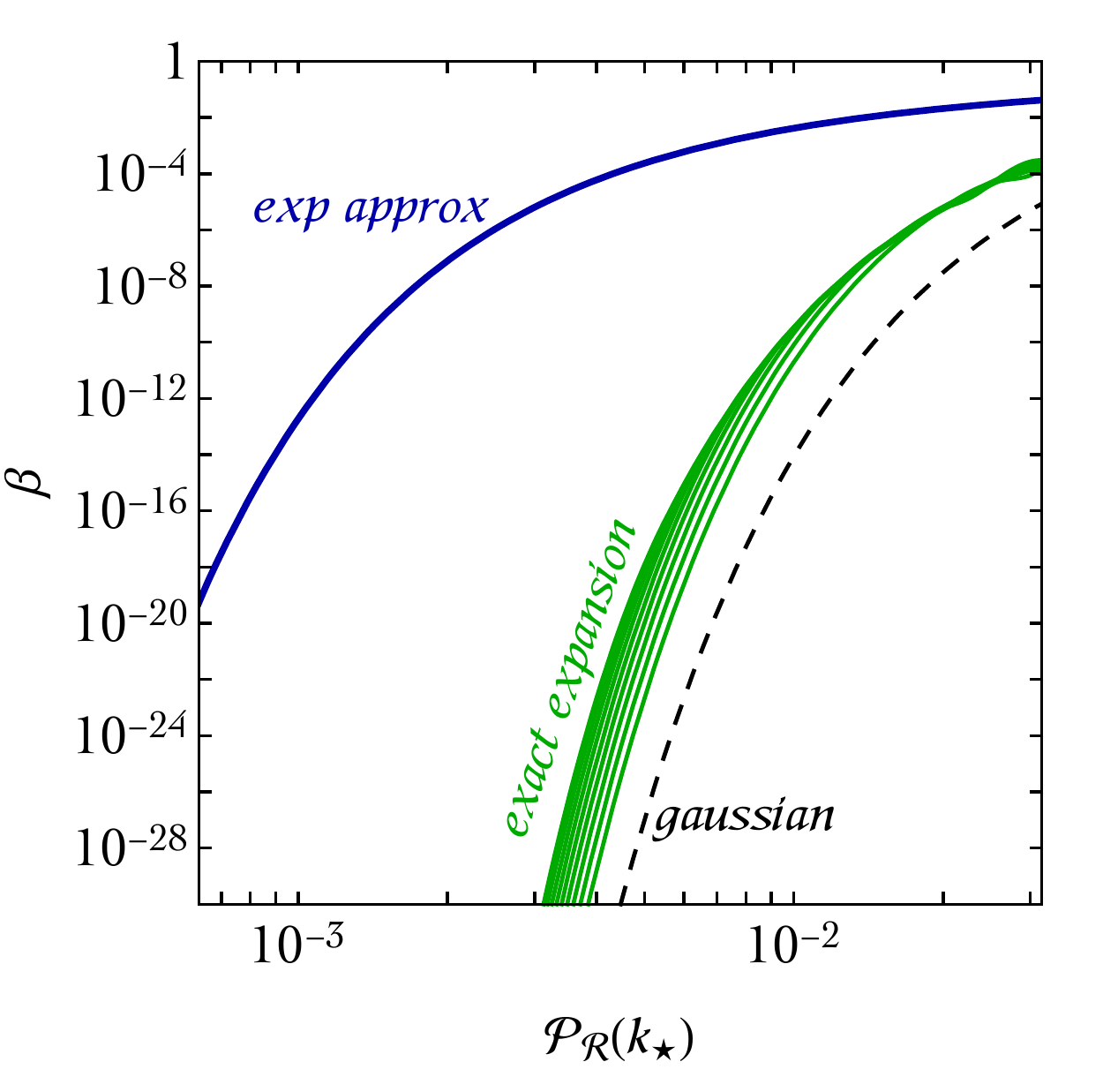}$$
\caption{\em \label{fig:WrongAppro} 
Left panel.
Numerical comparison 
between {\it i)} the gaussian result $\beta_{\rm G}(\nu_c)$ defined in eq.\,(\ref{eq:GCexp}), 
{\it ii)} the exponential approximation in 
eq.\,(\ref{eq:WrongSumThesholdStat}) and {\it iii)} the 
exact expansion of $\beta(\nu_c) = \beta_{\rm G}(\nu_c)[1+\sum_{n=3}^{n_{\rm max}}b_{n}(\nu_c)]$ 
for increasing values of $n_{\rm max}$. We take $\mathcal{C}_3 = 0.25$ and we show our results as function of $\nu_c$. 
Right panel. Same as in the left panel but using the scaling in eq.\,(\ref{eq:ScalingLeggi}) to draw 
$\beta$ as function of $\mathcal{P}_{\mathcal{R}}(k_{\star})$. We take $\delta_c = 1.2$.
 } 
\end{center}
\end{figure}

In particular, in this specific case, we can rewrite the sum (\ref{eq:exactbeta}) by using the recurrence property of the Bell polynomials:
\begin{equation}
B_{n+1}\left(x_1,\dots,x_{n+1} \right) = \sum_{k=0}^n \binom{n}{k} B_{n-k}\left(x_1,\dots,x_{n-k} \right)x_{k+1}.
\end{equation}
In our particular case $x_{k+1}=\mathcal{C}_3 \, \delta_{k+1,3}$, therefore $B_{3(k+1)}(0,0,\mathcal{C}_3,0,\dots,0)=\frac{1}{2}(3k+1)(3k+2)\mathcal{C}_3 B_{3k}(0,0,\mathcal{C}_3,0,\dots,0)$. This can be easily recasted in the form:
\begin{equation}
B_n \left(0,0,\mathcal{C}_3,0,\dots,0 \right) = \begin{cases}& 3\left(\frac{\mathcal{C}_3}{6}\right)^k \frac{(3k-1)!}{(k-1)!}, \quad \textrm{if} \quad n=3k,\\ & 0,  \quad \textrm{if} \quad n \ne 3k. \end{cases}
\end{equation}
Using this identity inside the eq. (\ref{eq:exactbeta}) we get:
\begin{equation}
\label{eq:exactseries}
\beta(\nu_c) = \frac{1}{2}\mathrm{Erfc}\left(\frac{\nu_c}{\sqrt{2}}\right)+\frac{1}{\sqrt{\pi}}e^{-\nu_c^2/2}\sum_{k=1}^\infty \frac{1}{k!}\left(\frac{\mathcal{C}_3}{12\sqrt{2}}\right)^k H_{3k-1}\left(\frac{\nu_c}{\sqrt{2}}\right).
\end{equation}

To show more explicitly the error that one makes by taking the exponential approximation in 
eq.\,(\ref{eq:WrongSumThesholdStat}) we plot in the left panel of fig.\,\ref{fig:WrongAppro} the comparison 
between {\it i)} the gaussian result $\beta_{\rm G}(\nu_c)$ defined in eq.\,(\ref{eq:GCexp}), 
{\it ii)} the exponential approximation in 
eq.\,(\ref{eq:WrongSumThesholdStat}) and {\it iii)} the 
exact expansion of $\beta(\nu_c) = \beta_{\rm G}(\nu_c)[1+\sum_{n=3}^{n_{\rm max}}b_{n}(\nu_c)]$ 
for increasing values of $n_{\rm max}$\footnote{One can argue that the power series $\sum_{k=1}^\infty \frac{z^k}{k!}H_{3k-1}(x)$ contained in eq. (\ref{eq:exactseries}) is not convergent. Nevertheless it should be regarded as an asymptotic series that gives a trustable value after an optimal truncation. Moreover it is Borel summable, and its value can be computed numerically evalutating the Borel sum $\int_0^\infty e^{-t}\left[\sum_{k=1}^\infty \frac{(tz)^k}{(k!)^2}H_{3k-1}(x)\right] dt $. This check has been done and the value obtained agrees with the one got by the optimal truncation.}. 
We take $\mathcal{C}_3 = 0.25$ and we set $\mathcal{C}_{n>3} = 0$. 
The approximation gives a wrong estimate of the actual magnitude of the non-gaussianities. 
As expected, the exponential approximation diverges from the actual result for larger values of $\nu_c$.

There is another point that is worth emphasizing. Taking $\mathcal{C}_3$ fixed and changing only $\nu_c$, as done 
in the left panel of fig.\,\ref{fig:WrongAppro} is not fully consistent since both  
$\mathcal{C}_3$ and $\nu_c$ are functions of the power spectrum $\mathcal{P}_{\mathcal{R}}$. 
More explicitly, we have the scaling $\nu_c \propto \mathcal{P}_{\mathcal{R}}(k_{\star})^{-1/2}$ and 
$\mathcal{C}_3 \propto \mathcal{P}_{\mathcal{R}}(k_{\star})^{1/2}$ (see appendix\,\ref{app:InflationPrimer}). 
This means that reducing $\mathcal{P}_{\mathcal{R}}(k_{\star})$ to decrease $\mathcal{C}_3$ does not improve on the applicability of the exponential approximation since $\nu_c$, in turn, increases. 
To better visualize this point let us consider the following benchmark scalings 
\begin{align}\label{eq:ScalingLeggi}
\sigma_{\delta} = 0.35\left[\frac{\mathcal{P}_{\mathcal{R}}(k_{\star})}{0.05}\right]^{1/2}\,,~~~~~
\mathcal{C}_3 = 0.8\left[\frac{\mathcal{P}_{\mathcal{R}}(k_{\star})}{0.05}\right]^{1/2}\,,
\end{align}
and take $\delta_c = 1.2$. 
We can now redo the comparison we did before but now as function of $\mathcal{P}_{\mathcal{R}}(k_{\star})$.
The result is shown in the right panel of fig.\,\ref{fig:WrongAppro}.
We conclude again that the exponential approximation overestimates the impact of local non-gaussianities on the PBH abundance by many orders of magnitude compared with the actual result.
Using the exponential approximation, one would wrongly conclude that, in order to fit the reference value $\beta \simeq 10^{-16}$, 
an order-of-magnitude decrease in the peak amplitude of the power spectrum is needed compared to the gaussian result.

A precise determination of the value of $\mathcal{P}_{\mathcal{R}}(k_{\star})$ that is needed to 
get $\Omega_{\rm DM}\simeq \Omega_{\rm PBH}$ has important phenomenological implications. 
Let us discuss a specific example. 
PBHs with mass $M_{\rm PBH} \simeq 10^{18}$ g have approximatively the size of the atomic nucleus 
(remember that $1\,M_{\odot} \simeq 1.48\,{\rm km}$ in the Planck unit system). 
Testing the nature of such small objects in the form of dark matter is extremely challenging. 
An interesting prospect is the following. 
The peak in the power spectrum of scalar perturbations that is responsible for the formation of PBHs also generates 
(as a second-order effect) a gravitational wave signal.  
The position of the peak amplitude of the power spectrum of curvature perturbations, the peak height in the mass distribution of PBHs and the frequency of the peak of the induced gravitational wave signal ($f_{\star}$ in the following) are related by the approximate relation
\begin{align}
\left(\frac{M_{\rm PBH}}{10^{17}\,{\rm g}}\right)^{-1/2}\simeq 
\frac{k_{\star}}{2\times 10^{14}\,{\rm Mpc}^{-1}}\simeq \frac{f_{\star}}{0.3\,{\rm Hz}}\,,
\end{align}
meaning that a peak amplitude around $k_{\star} = 10^{14}$ Mpc$^{-1}$ corresponds to a frequency range detectable by 
 future gravitational-wave interferometers like LISA, DECIGO and MAGIS-100. 
 This is, in principle, a powerful probe. 
By means of the condition $\Omega_{\rm DM}\simeq \Omega_{\rm PBH}$,  one can fix the parameters of 
 a given inflationary model that produces a sizable abundance of dark matter in the form of PBHs; in turn, this ``predicts'' the induced signal of  gravity waves since the latter is completely determined once the former step is taken. 
This is because the current energy density of gravitational waves induced by scalar perturbations is given by 
\begin{align}\label{eq:NGGW}
\Omega_{\rm GW} = \frac{c_g\Omega_r}{36}\int_{0}^{\frac{1}{\sqrt{3}}}dt\int_{\frac{1}{\sqrt{3}}}^{\infty}ds
\left[
\frac{(t^2-1/3)(s^2-1/3)}{t^2 - s^2}
\right]^{2}\left[\mathcal{I}_c(t,s)^2 + \mathcal{I}_s(t,s)^2\right]
\mathcal{P}_{\mathcal{R}}\left[\frac{k\sqrt{3}}{2}(s+t)\right]
\mathcal{P}_{\mathcal{R}}\left[\frac{k\sqrt{3}}{2}(s-t)\right]\,,
\end{align}
where $c_g\approx 0.4$, $\Omega_r$ is the current energy density of radiation 
and $\mathcal{I}_c$ and $\mathcal{I}_s$ are two functions that  can  be  computed  analytically (see, for instance, ref.\,\cite{Espinosa:2018eve} and references therein); 
$\Omega_{\rm GW}$ is completely fixed once $\mathcal{P}_{\mathcal{R}}(k)$ is given. 
In most of the analysis the power spectrum $\mathcal{P}_{\mathcal{R}}(k)$ is fixed to a specific functional form, 
and the abundance of PBHs computed assuming gaussian statistics. 
Typical functional forms for $\mathcal{P}_{\mathcal{R}}(k)$ used in the literature are the 
delta-function power spectrum (see, e.g., ref.\,\cite{Yuan:2019udt}), 
the log-normal power spectrum (see, e.g., refs.\,\cite{Kapadia:2020pir,Inomata:2018epa}) and the more realistic broken power-law in   
eq.\,(\ref{eq:RealisticPS}) (see, e.g., ref.\,\cite{Bhaumik:2020dor}); all these studies assume gaussian statistics 
for PBH formation.

The first remark that we make is that, as already anticipated, 
 taking a specific functional form for  $\mathcal{P}_{\mathcal{R}}(k)$ and 
 assuming gaussian statistics can be conceptually wrong; for instance, for single-field inflationary models of phenomenological 
 relevance, eq.\,(\ref{eq:RealisticPS}) 
 with a fixed value of $n_2 = -4\alpha$ implies the presence of local non-gaussianities, and the value of $\beta$ should be 
 computed accordingly.
The second remark is that if one takes 
the exponential approximation in eq.\,(\ref{eq:WrongSumThesholdStat}) (considering for example a situation like the one given in eq.\,(\ref{eq:ScalingLeggi})) then one would incorrectly conclude that primordial non-gaussianities imply a two orders-of-magnitude suppression of the gravitational wave signal, since the latter is proportional to $\mathcal{P}_{\mathcal{R}}(k_{\star})^2$. 
In reality (see fig.\,\ref{fig:WrongAppro}), primordial non-gaussianities reduce the value of 
$\mathcal{P}_{\mathcal{R}}(k_{\star})$ by no more than a factor of 2 compared to the gaussian result. We will come back in more detail to this point in the next section. 
Before that, let us mention some other previous analysis on the impact of primordial non-gaussianities on the PBHs phenomenology.
Ref.\,\cite{Byrnes:2012yx} approximates the calculation of the PBH abundance working with the comoving curvature perturbation instead of the density. This simplified method is discussed in appendix\,\ref{app:NonGaussianPeakTheory}, see eq.\,(\ref{eq:ExactProb}).
Ref.\,\cite{Shandera:2012ke} describes the non-gaussian probability density distribution using the Edgeworth expansion, analogous to the Gram-Charlier series in eq.\,(\ref{eq:GC}) but with a different truncation and ordering of terms.
The higher order cumulants $\mathcal{C}_{n}$ are parametrized in terms of the third order one $\mathcal{C}_{3}$ assuming scaling relations compatible with a certain ansatz for non-gaussianties (one of the two cases considered corresponds to local non-gaussianties).
Ref.\,\cite{Young:2015cyn} considers several templates for the bispectrum, which are then used to compute the non-gaussian smoothed density field.
The latter quantity is simulated numerically on a grid, and the PBHs abundance is computed counting the number of 
grid points above the threshold for collapse.
Since the formation of the PBHs is a rare event, the limitation of the computing power forces ref.\,\cite{Young:2015cyn} to focus an on large value of the PBH abundance, $\beta=10^{-4},$ order of magnitudes above the realistic ones.
As evident, the methods and the statistical contexts of these analysis are different from ours. For instance, all these works have been performed in the context of threshold statistics, instead of peak theory. 
Nevertheless, at a qualitative level, we find that their results are in agreement with ours, namely primordial non-gaussianities as those considered here increases the PBH abundance, but their effect can be compensated by a decrease of $\mathcal{P}_{\mathcal{R}}(k_{\star})$ by a modest factor.  

Finally, we shall mention that primordial non-gaussianities are constrained by the non-observations of isocurvature modes at CMB scales\,\cite{Tada:2015noa,Young:2015kda}.
These constraints rely on the influence of long wavelength fluctuations on the formation of PBHs. The role of such long modes is still under discussion in the literature, e.g.\cite{Passaglia:2018ixg,Suyama:2021adn,Matarrese:2020why}.

\subsection{Non gaussianities from the non-linear relation between density and curvature perturbations.}
\label{sec:nonlin}

In the previous sections we have assumed a linear relation between the density and curvature perturbations. This is because our main focus was the study of the non-gaussianities of primordial origin. Now we are going to include also the effect of the non-linearities in eq.\,(\ref{eq:NonLinearDelta}). 
In our approach, they enter in two ways. 
First, they 
 modify the relation between the overdensity field and the curvature perturbation; this modification translates into 
 a sort of ``renormalization'' of the value of $\delta_c$ 
 whose net effect is to make the production of black holes harder.

Importantly, non-linearities also change the shape of the profile of the overdensity peaks which eventually collapse into black holes. 
Since the threshold value depends on the shape of the overdensity, non-linearities also affect the way in which $\delta_c$ is computed. 
Ref.\,\cite{Kehagias:2019eil} concludes that $\delta_c$ is robust against non-linearities since its value changes at the percent level. However, ref.\,\cite{Kehagias:2019eil} only considers the idealized case of a $\delta$-function power spectrum.
Deviations from the  $\delta$-function limit are considered in refs.\,\cite{Yoo:2018kvb,Germani:2019zez,Yoo:2020dkz,Kawasaki:2019mbl} using different procedures whose consistency with each other is not yet fully understood. 
We do not aim to clarify this issue in the present work; however, we point out that refs.\,\cite{Yoo:2018kvb,Germani:2019zez,Yoo:2020dkz,Kawasaki:2019mbl} assume that
the curvature perturbation field is gaussian while, at least for the class of models which are relevant for the present analysis, primordial non-gaussianities should be included as well. 

A comprehensive study of the interplay between non-gaussianities of primordial origin and non-linearities in eq.\,(\ref{eq:NonLinearDelta}) is left for future work.
Here we shall focus on the first effect above, and assume that the threshold $\delta_c$ is not significantly affected by the presence of non-linearities.
Under this assumption, in appendix\,\ref{app:NonLin} we show how to compute the PBH abundance accounting for primordial non-gaussianities of local type and non-linearities between density and curvature perturbations.
We find that one can still use eq.\,(\ref{eq:MasterFormula}), but replacing the threshold $x_{\delta}(\bar{\nu})$ with $x^{\rm NL}_{\delta}(\bar{\nu}),$ defined as
\begin{align}\label{eq:Renormalization}
x_{\delta}^{\rm NL}(\bar{\nu}) \equiv \frac{9 (a_m H_m)^2}{4\sigma_2}\frac{\delta_c}{1+2\alpha\sigma_0\bar{\nu}}
\exp\left\{2\sigma_0[\bar{\nu}+\alpha\sigma_0(\bar{\nu}^2-1)]\right\}\,.
\end{align}

\begin{figure}[!htb!]
\begin{center}
\includegraphics[width=.45\textwidth]{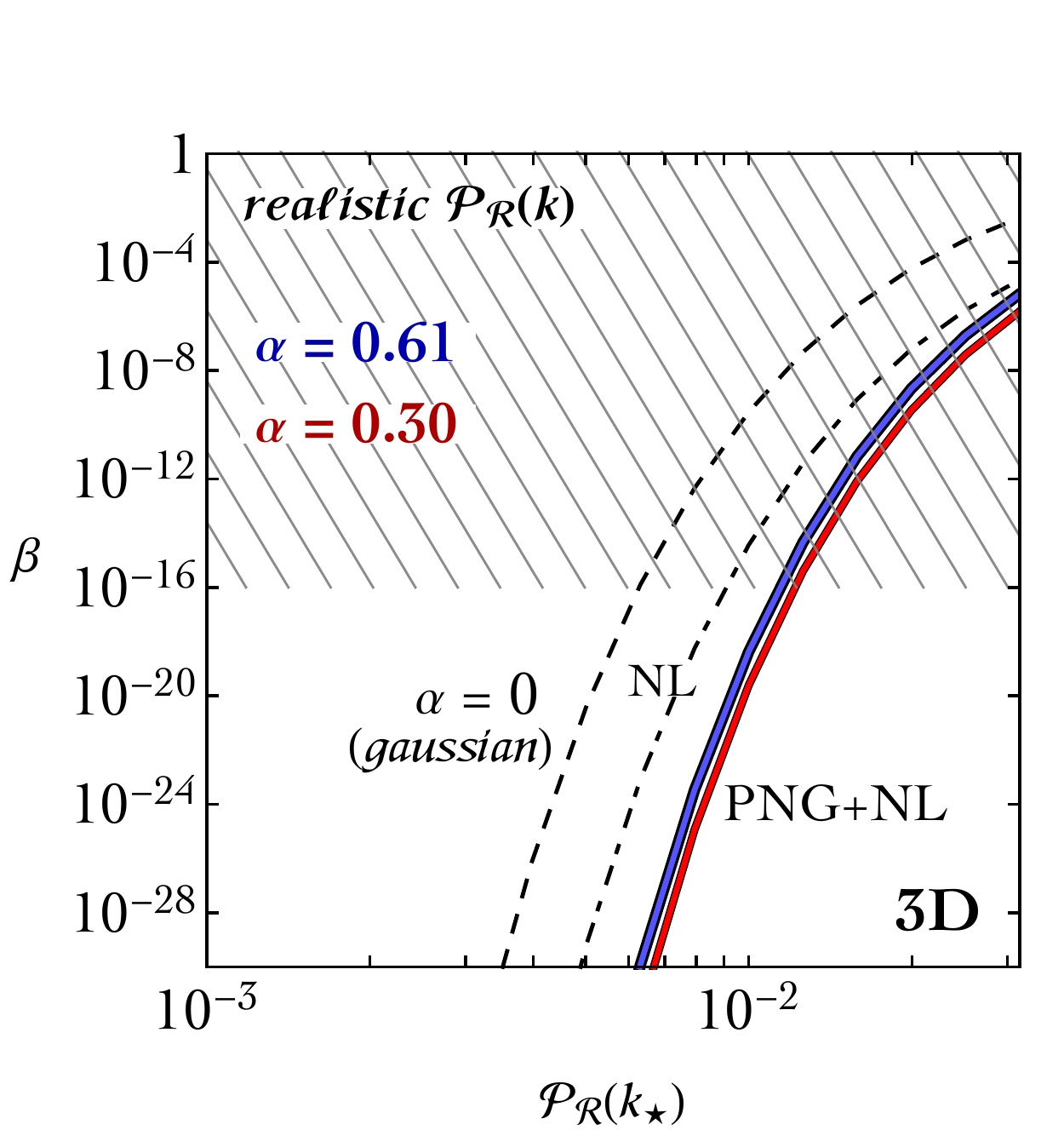}
\qquad
\includegraphics[width=.45\textwidth]{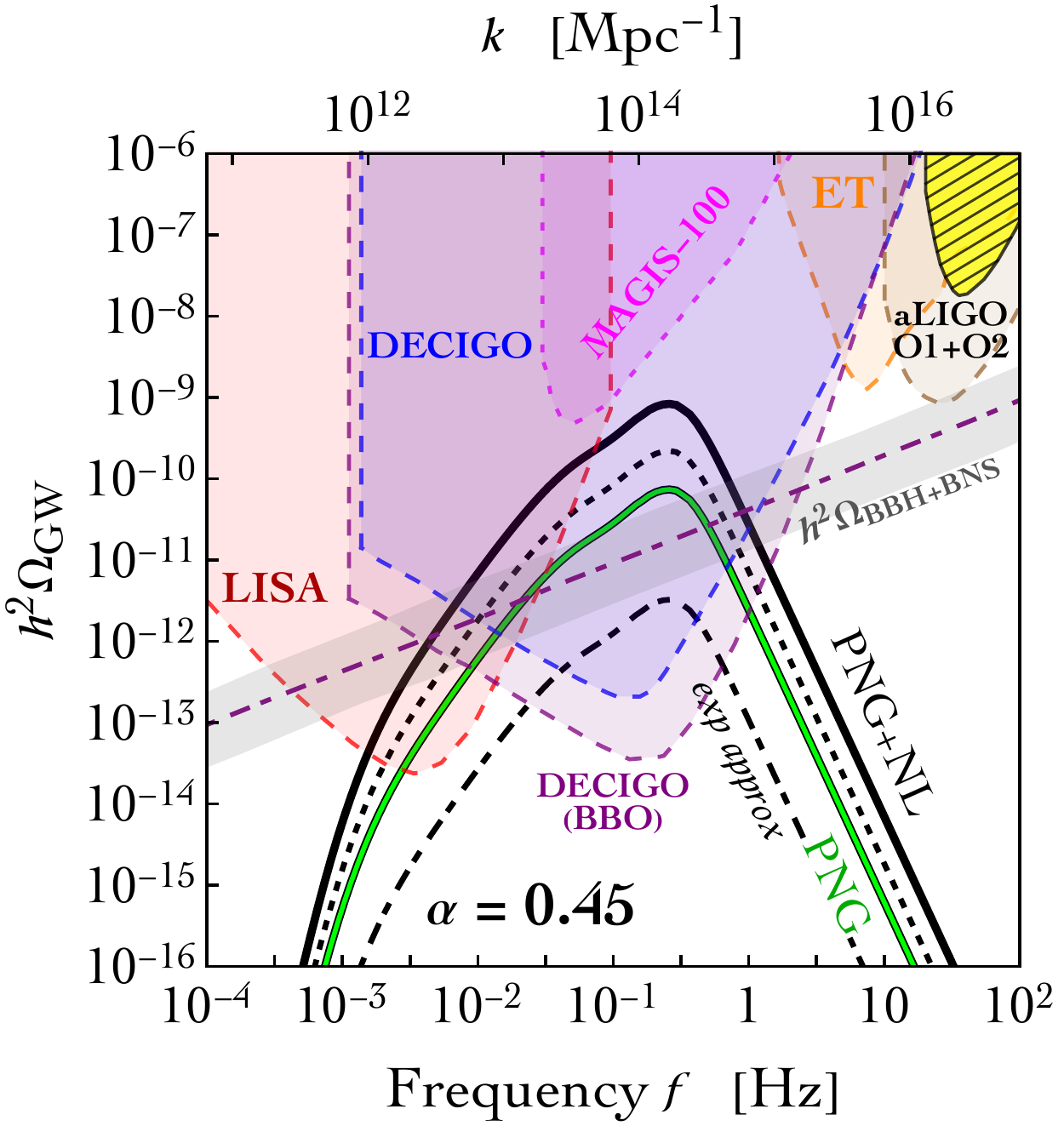}
\caption{\em \label{fig:MasterFormula2}
Left panel. 
Same as in the left panel of fig.\,\ref{fig:MasterFormula}; we compare the gaussian result (dashed black line) 
with {\it i) } the computation that includes non-linearities between curvature and density perturbations (dot-dashed 
line, label ``NL'') and   
{\it ii) } the computation that includes both primordial non-gaussianities and non-linearities 
(solid clored 
lines, label ``PNG+NL''). 
Right panel.  
Fraction of the energy density in gravitational waves relative to the critical energy density of the Universe as a function of the frequency (see eq.\,(\ref{eq:NGGW})). We compare the gaussian result (dashed black line) 
with  the computation that includes non-linearities between curvature and density perturbations 
and  the computation that includes both primordial non-gaussianities and non-linearities.
We also consider for illustration the comparison with the exponential approximation (dot-dashed line, see the right panel of fig.\,\ref{fig:MasterFormula}).
 We take $\alpha = 0.45$. 
We superimpose the signal on the expected sensitivity curves of the future gravitational wave detectors LISA (assuming the C1 configuration\,\cite{Caprini:2015zlo}), DECIGO\,\cite{Yagi:2011wg}, MAGIS-100\,\cite{Coleman:2018ozp} and the Einstein Telescope\,\cite{Maggiore:2019uih}. 
We also show present bound\,\cite{LIGOScientific:2019vic} and future prospect\,\cite{fur} for the 
aLIGO experiment.   
The diagonal gray band is the stochastic gravitational wave background from binary black holes (BBH) and binary neutron stars (BNS), which has been computed following ref.\,\cite{Chen:2018rzo}.
 } 
\end{center}
\end{figure}
\begin{itemize}

\item [$\ast$] Consider first the gaussian case $\alpha \to 0$. 
We have $4\sigma_2 x_{\delta}^{\rm NL}(\bar{\nu})|_{\alpha = 0} 
= 9 (a_m H_m)^2\delta_c e^{2\sigma_0\bar{\nu}}$ 
to be compared with its linear limit 
$4\sigma_2 x_{\delta}(\bar{\nu})|_{\alpha = 0} 
= 9 (a_m H_m)^2\delta_c$. 
The main difference, as already emphasized in ref.\,\cite{DeLuca:2019qsy}, is that non-linearities 
``renormalize'' the value of $\delta_c$ by means of the exponential factor $e^{2\sigma_0\bar{\nu}}$.  
This means that when considering maxima of the curvature field their height, in addition to their curvature, 
becomes important at the non-linear level. 
We remark that these two quantities are statistically correlated, and the amount of correlation is controlled by the power spectrum 
via the a-dimensional parameter $\gamma$ (see appendix\,\ref{app:GaussianPeakTheory}). 
If the power spectrum is very narrow, maxima with large curvature are likely to have also large $\bar{\nu}$; 
consequently, the threshold value $\delta_c$ effectively increases because of the factor $e^{2\sigma_0\bar{\nu}}$. 
This effect decreases the abundance with compared to the gaussian case. 
However, if the power spectrum is not very narrow, the correlation mentioned before is less accentuated, and  maxima with large curvature
are not necessarily characterized by large $\bar{\nu}$; in this case, $\delta_c$ increases by a smaller factor and 
the reduction of the abundance turns out to be less important. Our realistic power spectrum in eq.\,(\ref{eq:RealisticPS}) belongs 
to the last case. 
In the left panel of fig.\,\ref{fig:MasterFormula2} we compare the gaussian computation of $\beta$ in the case
without (dashed black line) and with (dot-dashed black line, label ``NL'') non-linearities. 
As expected, $\beta$ decreases but the change in the peak amplitude of the power spectrum that is needed to 
reproduce the reference value $\beta \simeq 10^{-16}$ is less than a factor of 2 (while very narrow power spectra 
may increase the peak amplitude of the power spectrum up to one order of magnitude\,\cite{DeLuca:2019qsy}).

\item [$\ast$] We take $\alpha \neq 0$. 
As discussed in section\,\ref{eq:Abubnda}, primordial non-gaussianities modify the statistics of the 
overdensity peaks. More peaks form, and, at constants threshold for collapse, the abundance of PBHs increases. 
Non-linearities, however, tend to enhance the value of $\delta_c$ in the sense discussed in the previous paragraph, and this gives the opposite effect 
of reducing the abundance. 
Moreover, when both non-linearities and primordial non-gaussianities are present the 
renormalization of $\delta_c$ also depends on the value of $\alpha$, as shown in eq.\,(\ref{eq:Renormalization}). 
We have, therefore, two competing effects: More overdensity peaks form but their collapse becomes less probable. 
We find that the result is a net decrease of the PBH abundance. 
This is shown in the left panel of fig.\,\ref{fig:MasterFormula2} (solid color lines, label ``PNG+NL''). 
Quantitatively, in order to fit the reference value $\beta \simeq 10^{-16}$ 
the combination of primordial non-gaussianities and non-linearities require a peak amplitude of the power spectrum 
larger compared to its gaussian value. However, we find that $\mathcal{P}_{\mathcal{R}}(k_{\star})$ 
only increases by a factor which is at most 2 in all the realistic cases analyzed in this paper.

\end{itemize}

Before concluding, let us explain the reason why it is 
important to have control on the computation of the PBH abundance at the level of the effects discussed in the present work. 

The only condition $\Omega_{\rm DM}\simeq \Omega_{\rm PBH}$ (which is $\beta\simeq 10^{-16}$ under the assumptions stated at the beginning of this section) is not enough to make the presence of non-gaussianities ``phenomenologically observable''. 
 With this term in quotation marks we mean that 
the presence (or absence) of non-gaussianities can be  simply reabsorbed in a re-tuning of $\mathcal{P}_{\mathcal{R}}(k_{\star})$ 
(which is equivalent to a very modest re-tuning of model parameters, see ref.\,\cite{Taoso:2021uvl})
if the only goal is to reproduce $\beta\simeq 10^{-16}$. 

What changes the rules of the game is the fact that, once $\mathcal{P}_{\mathcal{R}}(k_{\star})$ is fixed, 
the amplitude of the induced gravity wave signal in eq.\,(\ref{eq:NGGW}) is also predicted. 
Importantly, this implies that partial or inaccurate treatments of non-gaussianities will result 
in different values of 
$\mathcal{P}_{\mathcal{R}}(k_{\star})$ (required by imposing the same condition $\beta\simeq 10^{-16}$) and, consequently, different gravitational wave signals.
The (inaccurate) exponential approximation in eq.\,(\ref{eq:MainRiotto}) provides a paradigmatic example. 
As discussed, if taken at face value the exponential approximation in eq.\,(\ref{eq:MainRiotto}) 
may imply a one order-of-magnitude reduction in the value of $\mathcal{P}_{\mathcal{R}}(k_{\star})$ which gives $\beta\simeq 10^{-16}$ 
(see, e.g., fig.\,\ref{fig:MasterFormula} or fig.\,\ref{fig:WrongAppro}). 
Consequently, the amplitude of the induced gravity wave signal in eq.\,(\ref{eq:NGGW}) would be suppressed by two 
orders-of-magnitude. 

In the right panel of fig.\,\ref{fig:MasterFormula2} we compare the value of $h^2\Omega_{\rm GW}$ 
that one gets using peak theory in the gaussian limit (dashed black line) and the 
exponential approximation (dot-dashed black line; we take $\alpha = 0.45$, and consider non-gaussianities at the level of
 the leading part of third-order cumulants). 
 This example is tailored to show that in this case the gravitational wave signal would drop below the expected stochastic background due to mergers 
 of astrophysical black holes and neutron stars (the diagonal gray band).  
 In the same plot we also show the value of $h^2\Omega_{\rm GW}$ that we get in the presence of primordial non-gaussianities (label ``PNG'' 
 this time computed according to eq.\,(\ref{eq:MasterFormula})) and in the presence of both 
 primordial non-gaussianities and non-linearities 
 (label ``PNG+NL'' computed according to our discussion in section\,\ref{sec:nonlin}). 
 As expected, primordial non-gaussianities (the combination of primordial non-gaussianities and non-linearities) decrease (increases) the 
  amplitude of the gravitational wave signal compared to the gaussian result but 
 the net effect in both cases does not exceed the order of magnitude.  
As an order-of-magnitude approximation, therefore, the gaussian limit can be considered trustable (while using the exponential approximation would hide the signal below the stochastic background). 

There is one last point that is worth emphasizing. 
In the gaussian approximation, peak theory gives a PBH abundance which is systematically larger than the one provided by threshold statistics (this is in agreement with the expectation of ref.\,\cite{Young:2014ana} and 
discussed in section\,\ref{sec:Comp}).  
The amplitude of the gravitational wave signal in fig.\,\ref{fig:MasterFormula2} is indeed smaller than the one 
computed with threshold statistics (see ref.\,\cite{Ballesteros:2020qam}) and dangerously closer to the expected stochastic background. 
For this reason, we argue that keeping under control the impact of non-gaussianities could play an important part in the interpretation of possible future detection.

\section{Conclusions and outlook}\label{sec:Con}

We conclude summarizing the main results and novelties of our work.

\begin{itemize}
\item [$\circ$] We have studied the impact of primordial non-gaussianities of local type on the abundance of PBHs.
For this purpose we have focused on the maxima of the the comoving curvature perturbations. 
We have shown that those peaks ``spiky enough'', i.e. with a curvature larger than a certain threshold, are good proxies
for the maxima of the density field which undergo gravitational collapse, and produce PBHs.
Exploiting this observation, we have obtained that the cosmological abundance of PBHs can be computed using eq.\,(\ref{eq:MasterFormula}).
Our calculations extend the gaussian peak theory formalism\,\cite{Bardeen:1985tr} to include the effect of local non-gaussianities.

We have examined the consequences of our results for models of single-field inflation with an ultra slow-roll phase. These scenarios have been investigated for the production of PBHs.
In this context, the impact of primordial non-gaussianities is not dramatic.  
In fact, the desired PBH abundance can be obtained with an amplitude of the power spectrum of curvature perturbations which is only a factor $\lesssim2$ smaller from the value inferred with the gaussian approximation.

\item [$\circ$] In parallel to the exact result presented in eq.\,(\ref{eq:MasterFormula}), we have developed a computational strategy that approximates the effect of primordial non-gaussianities in the form of an expansion in cumulants around the gaussian result.  
In order to have full analytical control, we worked out this part in two (instead of three) spatial dimensions. 
The presence of local non-gaussianities affect cumulants at any order, and we have shown what is their structure.
Furthermore, we have shown that the only inclusion of  third-order cumulants is not sufficient to fully capture the effect of 
local non-gaussianities.

\item [$\circ$] Previous works have studied primordial non-gaussianities in the context of threshold statistics, finding that non-gaussianities exponentially affect the PBH abundance as given by eq.\,(\ref{eq:MainRiotto}).
We have found that that result is not correct and largely overestimate the PBH abundance.

\item [$\circ$] In addition to non-gaussianities of primordial origin, we have also considered non-gaussianities 
that originate from the non-linear relation between curvature and density perturbations. 
In this case we have two competing effects. 
On the one hand, primordial non-gaussianities alter the statistics of peaks enhancing the abundance of PBHs 
as described by eq.\,(\ref{eq:MasterFormula}). 
On the other one, 
primordial non-gaussianities and non-linearities change the condition for collapse. At the linear level and without 
accounting for primordial non-gaussianities, the condition for collapse only involves the 
 laplacian of the curvature perturbation at the position of the peak of the overdensity field while in the presence of 
 non-gaussianities it takes the form of eq.\,(\ref{eq:Renormalization}). 
 This alteration makes  the formation of 
PBHs harder. When combining the two effects, we find that non-gaussianities suppress PBH production but 
the desired PBH abundance can be obtained with an amplitude of the power spectrum of curvature perturbations which is only a factor $\lesssim 2$ larger from the value inferred with the gaussian approximation.

\item [$\circ$] En route, we discuss the difference between threshold statistics and peak theory, 
and we explain under which conditions (and why) peak theory gives a PBH abundance which is larger than the one computed by means of threshold statistics.

\end{itemize}

Finally, let us mention that our treatment of non-gaussianities can be further improved.  
We have computed the threshold for collapse without including the effect of non-gaussianities (both primordial and non-linear) on the shape of the collapsing peak (technically speaking, in eq.\,(\ref{eq:F0}) we used the averaged density profile---while in principle peak theory also gives informations about the 
shape of the peak---and the linear approximation in eq.\,(\ref{eq:NonLinearDeltaTH}))~\footnote{Formalisms which incorporate the variation of the shape and size of the peaks have been proposed in refs.\,\cite{Germani:2019zez,Suyama:2019npc,Young:2020xmk} considering gaussian curvature perturbations.}. 
Including these effects could give an additional modification of $\delta_c$. 
In light of the motivations discussed before, it would be important to keep under control these additional effects. 
We will address this issue in a forthcoming work focused on the interplay between the computation of the black hole abundance and the induced gravitational wave signal.

\begin{acknowledgments}
The research of A.U. is supported in part by the MIUR under contract 2017\,FMJFMW (``{New Avenues in Strong Dynamics},'' PRIN\,2017) 
and by the INFN grant ``SESAMO -- SinergiE di SApore e Materia Oscura.'' 
M.T. acknowledges support from the INFN grant ``LINDARK,'' the research grant ``The Dark Universe: A Synergic Multimessenger Approach No. 2017X7X85'' funded by MIUR, and the project ``Theoretical Astroparticle Physics (TAsP)'' funded by the INFN.
\end{acknowledgments}

\newpage

\appendix

\section{Scalar perturbations during inflation in the presence of ultra slow-roll}\label{app:InflationPrimer}

In this paper we focus on single-field models of inflation. 
We indicate with $\phi$ the canonically normalized inflaton field and with $U=U(\phi)$ its potential. 
The dynamics of $\phi$ can be obtained solving the equation of motion
\begin{align}\label{eq:EoM}
\frac{d^2\phi}{dN^2} + 3\frac{d\phi}{dN} - \frac{1}{2}\left(\frac{d\phi}{dN}\right)^3 
+\left[
3-\frac{1}{2}\left(\frac{d\phi}{dN}\right)^2
\right]\frac{d\log U}{d\phi}=0\,,
\end{align}
with slow-roll initial conditions; $N$ indicates the number of $e$-folds defined by $dN= Hdt$, where 
$H \equiv \dot{a}/a$ is  the Hubble rate, $a$ the scale factor of the Friedmann-Lema\^{\i}tre-Robertson-Walker metric, 
and $t$ the cosmic time (with $\dot{}\equiv~d/dt$). We will also use in the following the 
conformal time $\tau$ defined by means of $dt/d\tau = a$ or, equivalently, $dN/d\tau = aH$ (notice that, in the limit in 
which $H$ is constant, we have the relation $\tau= -1/aH$; the conformal time, therefore, is negative, and 
late times towards the end of inflation can be formally identify with the limit $\tau \to 0^-$). 

Scalar fluctuations can be efficiently described in terms of the Mukhanov-Sasaki field variable $u(\tau,\vec{x})$ which is a gauge-invariant combination of 
both fluctuations of the inflaton field and scalar fluctuations of the background Friedmann-Lema\^{\i}tre-Robertson-Walker geometry.
 The Mukhanov-Sasaki field variable solves the differential equation (for a review, see ref.\,\cite{Riotto:2002yw})
\begin{align}
 \frac{d^2 u}{d\tau^2} & =  \left(\triangle +\frac{1}{z}\frac{d^2z}{d\tau^2}\right)u\,,\label{eq:MS}\\
 \frac{1}{z}\frac{d^2 z}{d\tau^2} & = a^2H^2\left[(1+\epsilon-\eta)(2-\eta)+\frac{1}{aH}\left(\frac{d\epsilon}{d\tau}-\frac{d\eta}{d\tau}\right)\right]\,,\label{eq:TimeDepPot}
\end{align}
with $z\equiv (1/H)(d\phi/d\tau)$ and $\triangle$ the laplacian acting on spatial coordinates.  
The Hubble parameters are defined by $\epsilon \equiv -\dot{H}/H^2$ and  
$\eta \equiv -\ddot{H}/2H\dot{H}$.

The field $u$ can be quantized by defining the operator
\begin{align}\label{eq:OperatorU}
\hat{u}(\tau,\vec{x}) = \int\frac{d^3 \vec{k}}{(2\pi)^{3}}\left[
u_{k}(\tau)a_{\vec{k}}e^{+i\vec{k}\cdot\vec{x}} + u_{k}^{*}(\tau)a^{\dag}_{\vec{k}}e^{-i\vec{k}\cdot\vec{x}}
\right]\,,
\end{align}
with the annihilation and creation operators that satisfy the commutation relations of bosonic fields 
\begin{equation}\label{eq:Anni}
[a_{\vec{k}},a_{\vec{k}^{\prime}}] =[a_{\vec{k}}^{\dag},a_{\vec{k}^{\prime}}^{\dag}] =0\,,~~~~~~~[a_{\vec{k}},a_{\vec{k}^{\prime}}^{\dag}] = (2\pi)^3\delta^{(3)}(\vec{k}-\vec{k}^{\prime})\,,~~~~~~~a_{\vec{k}}|0\rangle = 0\,,
\end{equation}
where the last condition defines the vacuum. 
The equation of motion for each mode $u_k(\tau)$ takes the form of a Schr\"odinger equation
\begin{align}\label{eq:MSmode}
\frac{d^2u_k}{d\tau^2} + \left(k^2-\frac{1}{z}\frac{d^2z}{d\tau^2}\right)u_k = 0\,,
\end{align}
which can be solved imposing Bunch-Davies initial conditions at some initial time when $k \gg aH$. 
This choice defines the initial vacuum state 
as the minimum energy eigenstate for an harmonic oscillator with time-independent frequency 
in a space-time which is locally flat (because we are at length scales much smaller than the de Sitter curvature radius). 
Quantization of scalar perturbations is easier in terms of the Mukhanov-Sasaki field variable $u(\tau,\vec{x})$ 
since the quadratic action corresponding to the equation of motion in eq.\,(\ref{eq:MS}) 
 is the action describing a canonically normalized free field with an effective time-dependent mass. 
 However, the dynamics of the modes $u_k$ in Fourier space is more transparent if we introduce 
 the so-called comoving curvature perturbation $\mathcal{R} = u/z$; in analogy with eq.\,(\ref{eq:OperatorU}), $\mathcal{R}$ can be 
 promoted to a quantum operator 
\begin{align}\label{eq:OperatorR}
\hat{\mathcal{R}}(\tau,\vec{x}) = \int\frac{d^3 \vec{k}}{(2\pi)^{3}}\left[
\mathcal{R}_{k}(\tau)a_{\vec{k}}e^{+i\vec{k}\cdot\vec{x}} + \mathcal{R}_{k}^{*}(\tau)a^{\dag}_{\vec{k}}e^{-i\vec{k}\cdot\vec{x}}
\right]\,,
\end{align} 
with $\mathcal{R}_k = u_k/z$ that satisfies the time-evolution equation 
 \begin{align}\label{eq:Revo}
  \frac{d^2\mathcal{R}_{k}}{dN^2} + \left(
  3+\epsilon - 2\eta
  \right)\frac{d\mathcal{R}_{k}}{dN} + \frac{k^2}{a^2 H^2} \mathcal{R}_{k}=0\,,
\end{align} 
which is the differential equation of a damped harmonic oscillator. 

In the canonical picture of slow-roll inflation (that is for small Hubble parameters $\epsilon,\eta \ll 1$), 
eq.\,(\ref{eq:Revo}) can be solved in two complementary regimes divided by the so-called horizon-crossing condition 
$k = aH$.  
At early times, when $k \gg aH$ (in such case the modes are called sub-horizon), the last term dominates over the friction one, and the solution oscillates.
As time passes by during inflation, the comoving Hubble radius $1/aH$ shrinks, the horizon-crossing condition is met, and 
one eventually enters in the regime characterized by $k \ll aH$ (in such case the modes are called super-horizon); 
the last term in eq.\,(\ref{eq:Revo}) can be neglected and the latter admits the constant solution $d\mathcal{R}_k/dN = 0$. 
In other words, after horizon crossing the mode $\mathcal{R}_k$ freezes to a constant value that it 
maintains in time until the end of inflation (more precisely, until the mode re-enters the conformal Hubble horizon after the end of 
inflation).

In the canonical picture of slow-roll inflation, after horizon crossing quantum fluctuations can be regarded as classical, 
which motivates the description of cosmological perturbations in terms of classical random fields\,\cite{Kiefer:1998jk}. 
We will give a more precise definition of random fields in appendix\,\ref{app:GaussianPeakTheory}. 
At the conceptual level, the previous statement implies that one has the schematic relation 
$\lim_{k/aH \ll 1}\hat{\mathcal{R}}(\tau,\vec{x}) = \mathcal{R}(\vec{x})$ 
meaning that 
for super-horizon modes 
the quantum operator $\hat{\mathcal{R}}$ 
can be interpreted as a classical random field $\mathcal{R}(\vec{x})$; notice that the latter is time-independent because
 the modes $\mathcal{R}_k$ are frozen in time. 
 The previous relation can be formulated in a more 
 precise form by saying that vacuum expectation values on the quantum side are 
 interpreted as statistical averages on the classical side.  
 Formally, in the classical picture 
 we still have the Fourier decomposition 
 \begin{align}\label{eq:StocasticField}
 \mathcal{R}(\vec{x}) = \int\frac{d^3 \vec{k}}{(2\pi)^{3}}\left(
\mathcal{R}_{k} a_{\vec{k}}e^{+i\vec{k}\cdot\vec{x}} + \mathcal{R}_{k}^{*}a^{\dag}_{\vec{k}}e^{-i\vec{k}\cdot\vec{x}}
\right)\,,
\end{align}
 but now 
  creation and annihilation operators are no longer 
 $q$-numbers but stochastic $c$-numbers which are defined by the statistical averages 
 \begin{align}\label{eq:StocaRules}
\langle a_{\vec{k}}a_{\vec{k}^{\prime}}^{\dag}\rangle =  
 \frac{1}{2}(2\pi)^3\delta^{(3)}(\vec{k}-\vec{k}^{\prime}) = \langle  a_{\vec{k}^{\prime}}^{\dag}a_{\vec{k}}\rangle\,,
 ~~~~~~~
 \langle a_{\vec{k}}a_{\vec{k}^{\prime}}\rangle =
 \langle a^{\dag}_{\vec{k}}a^{\dag}_{\vec{k}^{\prime}}\rangle = 0\,.
\end{align}
Notice that $\langle a_{\vec{k}}a_{\vec{k}^{\prime}}^{\dag}\rangle =  \langle  a_{\vec{k}^{\prime}}^{\dag}a_{\vec{k}}\rangle$ 
is only valid for classical fluctuations, since it implies that the stochastic parameters commute.
Furthermore, the 
stochastic parameters $a_{\vec{k}}$, $a_{\vec{k}}^{\dag}$ 
have zero mean value (consequently, $\langle \mathcal{R}(\vec{x})\rangle = 0$) 
in order to match the fact that the quantum operator $\hat{\mathcal{R}}(\tau,\vec{x})$ in eq.\,(\ref{eq:OperatorR}) has zero vacuum expectation value.

The random field $\mathcal{R}(\vec{x})$ is fully specified by the entire hierarchy of 
its correlation functions.  
The simplest one is the two-point correlation function. 
 The latter can be defined by introducing the idea of power spectrum.
The power spectrum $\Delta_{\mathcal{R}}(k)$ is defined by the Fourier transform of the two-point correlation function 
\begin{align}\label{eq:PowerSpectrum}
\langle\mathcal{R}(\vec{x})\mathcal{R}(\vec{x}+\vec{r})\rangle = \int\frac{d^3 k}{(2\pi)^3}e^{i\vec{k}\cdot\vec{r}}\Delta_{\mathcal{R}}(k) 
= \int_0^{\infty}dk 
\frac{\sin(kr)}{kr}
\frac{k^2}{2\pi^2}\Delta_{\mathcal{R}}(k)
\,.
\end{align}
The fact that $\Delta_{\mathcal{R}}(k)$ depends only on $k\equiv |\vec{k}|$ and the explicit angular integrations that we performed in eq.\,(\ref{eq:PowerSpectrum}) are consequences of the assumptions of spatial homogeneity and isotropy (equivalently,  spatial homogeneity and isotropy
imply that $\langle\mathcal{R}(\vec{x})\mathcal{R}(\vec{x}+\vec{r})\rangle$ only depends 
on the relative distance $r\equiv |\vec{r}|$). 
The limit $r\to 0$ in eq.\,(\ref{eq:PowerSpectrum}) defines the variance $\sigma_0^2$ of the comoving curvature perturbation
\begin{align}\label{eq:Variance}
\sigma_0^2 \equiv \langle\mathcal{R}(\vec{x})\mathcal{R}(\vec{x})\rangle = 
\lim_{r\to 0}\langle\mathcal{R}(\vec{x})\mathcal{R}(\vec{x}+\vec{r})\rangle = 
\int_0^{\infty}\frac{dk}{k}\,\frac{k^3}{2\pi^2}\Delta_{\mathcal{R}}(k) 
\equiv \int_0^{\infty}\frac{dk}{k}\,\mathcal{P}_{\mathcal{R}}(k)\,,
\end{align}
where we defined the dimensionless power spectrum $\mathcal{P}_{\mathcal{R}}(k) \equiv (k^3/2\pi^2)\Delta_{\mathcal{R}}(k)$.
This equation gives to the power spectrum an intuitive statistical meaning; $\mathcal{P}_{\mathcal{R}}(k)$ 
represents the contribution to the variance of the field per unit logarithmic bin around the comoving wavenumber $k$.
As stated above, we can also compute eq.\,(\ref{eq:Variance}) by taking the vacuum expectation value of the quantum operator 
$\hat{\mathcal{R}}(\tau,\vec{x})$. We have (using eq.\,(\ref{eq:Anni}))
\begin{align}\label{eq:PS}
\lim_{k/aH \ll 1}\langle \hat{\mathcal{R}}(\tau,\vec{x})\hat{\mathcal{R}}(\tau,\vec{x})\rangle 
= \lim_{k/aH \ll 1}\int_0^{\infty}\frac{dk}{k}\,\frac{k^3}{2\pi^2}|\mathcal{R}_k(\tau)|^2 = 
\int_0^{\infty}\frac{dk}{k}\,\frac{k^3}{2\pi^2}|\mathcal{R}_k|^2\,,
\end{align}
where the last step means that we are considering a sufficiently late time 
(the limit $\lim_{k/aH \ll 1}$, for fixed $k$, represents a time-limit since $aH$ depends on time) such that the mode with comoving       
wavenumber $k$ is frozen to its constant value after horizon crossing (and the time-dependence drops in the last equality). 
Eq.\,(\ref{eq:PS}) gives an operative definition of the power spectrum. 
For a given model of inflation, we can solve eq.\,(\ref{eq:Revo}) (equivalently, eq.\,(\ref{eq:MSmode})) for each $k$ and take a ``late-time limit'' in the sense 
specified above (that is we evaluate $\mathcal{R}_k$ at some late time after it freezes to a constant value). 
The power spectrum is given by $\mathcal{P}_{\mathcal{R}}(k) \equiv (k^3/2\pi^2)|\mathcal{R}_k|^2$ and fully specifies the 
two-point correlator of the random field $\mathcal{R}(\vec{x})$.  
Equivalently, eq.\,(\ref{eq:PS}) can be derived from the computation of $\langle\mathcal{R}(\vec{x})\mathcal{R}(\vec{x})\rangle$ 
by means of the decomposition given in eq.\,(\ref{eq:StocasticField}).

To proceed further, we need to specify the structure of higher-order correlators. 
Under the assumption that the random field $\mathcal{R}(\vec{x})$ is gaussian, however, the power spectrum is enough to 
fully reconstruct higher-order correlators. 
This is because the $N$-point correlation function either vanishes (for odd $N$) or can be expressed in terms of the power spectrum as a consequence of the Isserlis' theorem (for even $N$). 
The assumption that the statistics of the random field $\mathcal{R}(\vec{x})$ is gaussian is well-motivated in 
the context of the canonical picture of slow-roll inflation. 
This is because if one tries to compute, on the quantum-side, the three-point correlator 
$\lim_{k/aH \ll 1}\langle \hat{\mathcal{R}}(\tau,\vec{x})\hat{\mathcal{R}}(\tau,\vec{x})\hat{\mathcal{R}}(\tau,\vec{x})\rangle$ the 
resulting expression turns out to be suppressed---compared to the two-point correlator---by additional powers of the Hubble parameters $\epsilon$ and $\eta$\,\cite{Maldacena:2002vr}. 
This means that non-gaussianities are not relevant during conventional slow-roll dynamics during which the Hubble parameters take $O(\ll 1)$ values (said differently, 
this means that any detection of sizable non-gaussianities at CMB scales will rule out all single field slow-roll models of inflation).

However, we are interested in a situation which deviates from standard slow-roll dynamics.  
We refer to ref.\,\cite{Ballesteros:2020qam} for a detailed (both analytical and numerical) study, and we summarize here the main points.

Standard slow-roll dynamics takes place at large field values. 
However, few $e$-folds before the end of inflation (which ends at the absolute minimum of the potential located at the origin) the 
inflaton field crosses an approximate stationary inflection point. 
The inflaton field almost stops but it possesses just enough inertia to overcome the approximate stationary inflection point. 
During this part of the dynamics the Hubble parameter $\eta$ transits from $\eta \simeq 0$ (that is typical of slow-roll) 
to a large positive value that is maintained for few $e$-folds until the field crosses the approximate stationary inflection point. 
If $\eta \gtrsim 3/2$ (typically one has $\eta \gtrsim 3$), the friction term in eq.\,(\ref{eq:Revo}) becomes negative. This part of the dynamics characterized by the presence of negative friction is dubbed ultra slow-roll.  
During the negative friction phase, the modes $\mathcal{R}_k$---more precisely, their modulus $|\mathcal{R}_k|$---change exponentially fast, and
 can be either enhanced or suppressed depending on the specific value of $k$. 

After the end of ultra slow-roll, $\eta$ transits to a phase during which it takes negative $O(1)$ values. 
The friction term in eq.\,(\ref{eq:Revo}) turns positive, and the modes $\mathcal{R}_k$---after being enhanced or suppressed by the negative friction phase---are now free to 
freeze to their final constant value. 
Since $\mathcal{P}_{\mathcal{R}}(k) \equiv (k^3/2\pi^2)|\mathcal{R}_k|^2$, the negative friction phase that modifies 
exponentially $|\mathcal{R}_k|$ produces a 
distinctive peak in the power spectrum of curvature  perturbations.

This peculiar dynamics has important consequences as far as non-gaussianities are concerned. 
Because of the presence of the negative friction phase, classicalization of the modes do not happens 
after their horizon crossing but is delayed after the end of ultra slow-roll\,\cite{Ballesteros:2020sre}.  
This means that the three-point correlator 
$\langle \hat{\mathcal{R}}(\tau,\vec{x})\hat{\mathcal{R}}(\tau,\vec{x})\hat{\mathcal{R}}(\tau,\vec{x})\rangle$ 
has to be evaluated
after the end of ultra slow-roll. 
Crucially, after the end of ultra slow-roll $\eta$ takes sizable negative $O(1)$ values (while we expect $\epsilon \ll 1$).  
Let us indicate this value with $\eta_0$ (which is a negative number).
This implies that non-gaussianities are no longer negligible since the expected slow-roll suppression is not valid anymore. 
The  explicit computation of the three-point correlator in Fourier space gives, for a triad of comoving wavenumbers $k_1,k_2,k_3$, the so-called local bispectrum\,\cite{Atal:2018neu}
\begin{align}\label{eq:LocalBispectrumFourier}
B_{\mathcal{R}}(k_1,k_2,k_3) \simeq -\eta_0[\Delta_{\mathcal{R}}(k_1)\Delta_{\mathcal{R}}(k_2) + 
\Delta_{\mathcal{R}}(k_1)\Delta_{\mathcal{R}}(k_3) +
\Delta_{\mathcal{R}}(k_2)\Delta_{\mathcal{R}}(k_3)]\,,
\end{align}
where the three-point correlator is given by the Fourier transform
\begin{align}\label{eq:skewness}
\lim_{\hspace{1.5cm}{\rm end\,of\,USR}}\langle \hat{\mathcal{R}}(\tau,\vec{x})\hat{\mathcal{R}}(\tau,\vec{x})\hat{\mathcal{R}}(\tau,\vec{x})\rangle
= \int\frac{d^3\vec{k}_1}{(2\pi)^3}\frac{d^3\vec{k}_2}{(2\pi)^3}
\frac{d^3\vec{k}_3}{(2\pi)^3}(2\pi)^3\delta^{(3)}(\vec{k}_1+\vec{k}_2+\vec{k}_3)
B_{\mathcal{R}}(k_1,k_2,k_3)\,,
\end{align}
and where $\lim_{{\rm end\,of\,USR}}$ indicates explicitly that the three-point correlator has to be evaluated at some late time 
after the end of ultra slow-roll, when the modes $\mathcal{R}_k$ finally set to their final constant value (so that the right-hand
 side of eq.\,(\ref{eq:skewness}) does not depend on time). 
 
The analysis carried out in refs.\,\cite{Atal:2018neu} shows that non-gaussianities in the 
presence of ultra slow-roll are expected to be non-negligible.  
Eqs.\,(\ref{eq:LocalBispectrumFourier},\,\ref{eq:skewness}) represent 
the quantum side of the story, and eq.\,(\ref{eq:LocalBispectrumFourier}) can be obtained by means of the so-called 
``{\it in}-{\it in}'' formalism\,\cite{Maldacena:2002vr}.
It is important to understand the implications from the point of view of the random field $\mathcal{R}(\vec{x})$. 
Consider the non-gaussian random field $\mathcal{R}_{\rm NG}(\vec{x})$ defined by
\begin{align}\label{eq:NONgField}
\mathcal{R}_{\rm NG}(\vec{x}) \equiv \mathcal{R}_{\rm G}(\vec{x}) + 
\frac{(-\eta_0)}{2}\left[
\mathcal{R}_{\rm G}(\vec{x})^2 - \langle \mathcal{R}_{\rm G}(\vec{x})^2\rangle
\right]\,,
\end{align}
where $\mathcal{R}_{\rm G}(\vec{x})$ is a gaussian random field---with variance $\sigma_0^2$ (see eq.\,(\ref{eq:Variance}))---which 
  admits the decomposition given in eq.\,(\ref{eq:StocasticField}); it is a simple exercise to show that 
  the three-point correlator 
$\langle \mathcal{R}_{\rm NG}(\vec{x})\mathcal{R}_{\rm NG}(\vec{x})\mathcal{R}_{\rm NG}(\vec{x})\rangle$ 
has precisely the form given in eq.\,(\ref{eq:skewness}).   
Notice that in eq.\,(\ref{eq:NONgField}) the presence of the constant piece $\langle \mathcal{R}_{\rm G}(\vec{x})^2\rangle$ 
guarantees that the non-gaussian random field $\mathcal{R}_{\rm NG}(\vec{x})$ has zero mean. 
This property is physically motivated by the fact that the background solution is stable.

The physics-case sketched in this appendix motivates the non-gaussianities studied in this paper which are of the form given in eq.\,(\ref{eq:NONgField}).
To avoid cluttering the notation, in the main part of this work we indicate simply with $\mathcal{R}$ the gaussian random field 
$\mathcal{R}_{\rm G}$ in eq.\,(\ref{eq:NONgField}) and with $h$ the non-gaussian one $\mathcal{R}_{\rm NG}$.  
Furthermore, we set $\alpha \equiv (-\eta_0)/2$.

Before proceeding, an important comment is in order. As we have discussed, the structure of the non-gaussian 
random field given in eq.\,(\ref{eq:NONgField}) is motivated by the explicit computation of the 
three-point correlator in eqs.\,(\ref{eq:LocalBispectrumFourier},\,\ref{eq:skewness}) in the presence of ultra slow-roll. 
This computation is based on the ``{\it in}-{\it in}'' formalism in which one expands the interaction Hamiltonian up to the cubic order.  

However, computing only the bispectrum is not the end of the story. 
In principle, one should compute also the trispectrum (that is the connected part of the four-point correlator) and check that 
its contribution does not alter the form of non-gaussianities derived including only cubic interactions. 
Needless to say, the computation of the trispectrum is anything but simple, and we are not aware of explicit results 
in the context of ultra slow-roll. 
From a more pragmatic phenomenological perspective, one possible way to proceed---widely used for the analysis of CMB non-Gaussianity, see ref.\,\cite{Kogo:2006kh}---is the following. 
Instead of eq.\,(\ref{eq:NONgField}), one takes the more general ansatz
\begin{align}\label{eq:NONgField2}
\mathcal{R}_{\rm NG}(\vec{x}) \equiv \mathcal{R}_{\rm G}(\vec{x}) + 
f_2\left[
\mathcal{R}_{\rm G}(\vec{x})^2 - \langle \mathcal{R}_{\rm G}(\vec{x})^2\rangle
\right] + f_3\mathcal{R}_{\rm G}(\vec{x})^3\,,
\end{align} 
where $f_{2,3}$ are free coefficients that parametrize quadratic and cubic deviations from the gaussian limit. 
Notice that eq.\,(\ref{eq:NONgField2}) generalizes eq.\,(\ref{eq:NONgField}) in the sense that 
it introduces cubic corrections but it preserves locality (in the sense that deviations 
from exact gaussianity at $\vec{x}$ are located at the same spatial position).
As discussed before, in the absence of an explicit computation there is no guarantee that ultra slow-roll 
generates deviations from eq.\,(\ref{eq:NONgField}) that have the form given in eq.\,(\ref{eq:NONgField2}). 
Nevertheless, eq.\,(\ref{eq:NONgField2}) can be considered as a phenomenological parametrization to study
deviations from from eq.\,(\ref{eq:NONgField}), as done for instance in ref.\,\cite{Young:2013oia}.
For this reason, in our analysis we will try to set the formalism considering a generic deviation from gaussianity (but always assuming locality) even though 
we will present our results for the motivated case of quadratic non-gaussianities given in eq.\,(\ref{eq:NONgField}).

Finally, let us mention that a resummation of local non-gaussianities at all orders has been presented in ref.\,\cite{Atal:2019cdz} in the context of the $\delta N$ formalism.
The analysis is based on an expansion of the inflaton potential around a local maximum. It would be interesting to go beyond such approximation and extend this analysis for the potentials studied here. Moreover it is also worth investigating the role of higher order contributions leading to non-local non-gaussianities.

\section{CliffsNotes on gaussian peak theory}\label{app:GaussianPeakTheory}

As a warm-up, consider the case of a $n$-dimensional scalar gaussian random field $\mathcal{R}(\vec{x})$ that we identify with the random field associated to curvature perturbations. 
Peak theory in the case of a scalar gaussian random field is well-known\,\cite{Bardeen:1985tr}. 
However, in this appendix we will give a detailed discussion. The reason is that in our approach there is an important 
conceptual difference compared to the standard results of ref.\,\cite{Bardeen:1985tr}. 
Ref.\,\cite{Bardeen:1985tr} computes the number density of peaks of the overdensity field working directly with $\delta(\vec{x},t)$, 
and without any reference to curvature perturbations. 
In this paper, on the contrary, we aim to compute the same quantity but starting from the distribution of 
local maxima of the curvature perturbation. 
Following this alternative route, we will be able to find (see appendix\,\ref{app:NonGaussianPeakTheory}) 
a generalization that accounts for the case in which local non-gaussianities in the definition of $\mathcal{R}(\vec{x})$ are present. 

Let us start from basics. A $n$-dimensional scalar random field $\mathcal{R}(\vec{x})$ is a set of random variables, one for each point $\vec{x}$ in the 
$n$-dimensional real space, equipped with a probability distribution 
$p[\mathcal{R}(\vec{x}_1),\dots ,\mathcal{R}(\vec{x}_m)]d\mathcal{R}(\vec{x}_1)\dots d\mathcal{R}(\vec{x}_m)$
which measures the probability that the function $\mathcal{R}$ has values in the range 
$\mathcal{R}(\vec{x}_j)$ to $\mathcal{R}(\vec{x}_j) + d\mathcal{R}(\vec{x}_j)$  
for each of the $j = 1,\dots,m$, with $m$ an arbitrary integer and $\vec{x}_1,\dots,\vec{x}_m$ arbitrary points.\footnote{Notice that, as in ref.\,\cite{Bardeen:1985tr}, 
all spatial separations and length scales are described in comoving coordinates in the cosmological background. 
This means that the number density that we shall compute at the end of this section in 
eq.\,(\ref{eq:DoubleIntegral}) must be understood as a comoving number density.}
  
We are interested in the behavior of the random field for a point in space that is stationary, and we consider $m=1$ with $\vec{x}_1 = \vec{x}_{\rm st}$. We can expand around this point according to
\begin{align}\label{eq:Stat}
\mathcal{R}(\vec{x}) = \mathcal{R}(\vec{x}_{\rm st}) + \frac{1}{2}\sum_{i,j =1}^{n}\mathcal{R}_{ij}(\vec{x}_{\rm st})(
\vec{x} - \vec{x}_{\rm st}
)_i(
\vec{x} - \vec{x}_{\rm st}
)_j\,,
\end{align}
from which we have
\begin{align}
\mathcal{R}_i(\vec{x}) = \sum_{j =1}^{n}\mathcal{R}_{ij}(\vec{x}_{\rm st})(\vec{x} - \vec{x}_{\rm st})_j\,.
\end{align}
The goal is to obtain the number density of these stationary points in the $n$-dimensional space.
Not to violate
the cosmological principle, we
only want to consider random fields which are statistically homogeneous and
isotropic. Consequently, the specific value of $\vec{x}_{\rm st}$ is irrelevant, and we can always shift to $\vec{x}_{\rm st} = \vec{0}$. 
Let us, therefore, drop the explicit dependence on $\vec{x}_{\rm st}$.

The quantity of central interest for the computation of the number density of stationary points of $\mathcal{R}$ is the joint probability density distribution of the field $\mathcal{R}$, its first and second derivatives. 
This is intuitively obvious, since identifying maxima (or minima) implies a set of conditions on field derivatives, and it is thus mandatory to know what is their probability distribution (derivatives of a random field are also random variables themselves).

Consider the realistic case with $n=3$.
We indicate with $P(\mathcal{R},\mathcal{R}_i,\mathcal{R}_{ij})
d\mathcal{R}
d^3\mathcal{R}_i
d^6\mathcal{R}_{ij}$
the joint probability distribution for the field being in the range $\mathcal{R}$ to $\mathcal{R} + d\mathcal{R}$, 
the field gradient being in the range $\mathcal{R}_i$ to $\mathcal{R}_i +d\mathcal{R}_i$ and the second derivative matrix elements being in the range $\mathcal{R}_{ij}$ to $\mathcal{R}_{ij} + d\mathcal{R}_{ij}$, all at the same point in space. $P(\mathcal{R},\mathcal{R}_i,\mathcal{R}_{ij})$ is the joint ten-dimensional probability density distribution. 

If the point is stationary, we can write the joint probability distribution as
$P(\mathcal{R},\mathcal{R}_i = 0,\mathcal{R}_{ij})\left|{\rm det}(\mathcal{R}_{ij})\right|
d\mathcal{R}
d^3\vec{x}
d^6\mathcal{R}_{ij}$. We set $\mathcal{R}_i = 0$ since the point is stationary, and we used
eq.\,(\ref{eq:Stat}) to change variables in the gradient volume element.
We can, therefore, write the probability distribution to have a stationary point in a volume $d^3\vec{x}$ 
with the field being in the range $\mathcal{R}$ to $\mathcal{R} + d\mathcal{R}$ as
\begin{align}
\label{eq:nmax}
n_{\rm st}(\mathcal{R})d\mathcal{R}d^3\vec{x} \equiv  d\mathcal{R}d^3\vec{x}
\underbrace{\int 
P(\mathcal{R},\mathcal{R}_i = 0,\mathcal{R}_{ij})\left|{\rm det}(\mathcal{R}_{ij})\right|
d^6\mathcal{R}_{ij}}_{\equiv n_{\rm st}(\mathcal{R})}\,,
\end{align}
where the integral is extended to the whole range of variability of the second derivatives since we are considering 
generic stationary points. 
 The probability distribution to have a maximum in a volume $d^3\vec{x}$ with the field being in the range $\mathcal{R}$ to $\mathcal{R} + d\mathcal{R}$ 
 is
\begin{align}\label{eq:nMax}
n_{\rm max}(\mathcal{R})d\mathcal{R}d^3\vec{x} \equiv  d\mathcal{R}d^3\vec{x}
\underbrace{\int_{\rm max}
P(\mathcal{R},\mathcal{R}_i = 0,\mathcal{R}_{ij})\left|{\rm det}(\mathcal{R}_{ij})\right|
d^6\mathcal{R}_{ij}}_{\equiv n_{\rm max}(\mathcal{R})}\,,
\end{align} 
where now the integral is subject to the conditions on the Hessian matrix that define a maximum. 
An equivalent definition holds in the case of a minimum. 
The probability density distribution $n_{\rm max}(\mathcal{R})$ represents the number density of local maxima (where 
``number density'' is defined in a probabilistic sense)
with field value in the range $\mathcal{R}$ to $\mathcal{R} + d\mathcal{R}$.

In order to extract quantitative informations, we need to compute 
$P(\mathcal{R},\mathcal{R}_i,\mathcal{R}_{ij})$, set $\mathcal{R}_i = 0$, and integrate. 
This strategy does not depend on the specific statistics of the random field.

The computation of $P(\mathcal{R},\mathcal{R}_i,\mathcal{R}_{ij})$ drastically simplifies in the gaussian case. This is because
 the joint probability distribution of a gaussian field and its derivatives is a multivariate normal distribution. 
 Consider the simplified case with $n=2$ in which we have more control on analytical 
 formulas;\footnote{The simplest possibility would be $n=1$. 
 However, the case $n=2$ is the simplest setup in which a non-trivial discussion about spatial isotropy is possible.} we have six random variables that we collect in the column vector ${R} \equiv (\mathcal{R},\mathcal{R}_x,\mathcal{R}_y,\mathcal{R}_{xx},\mathcal{R}_{xy},\mathcal{R}_{yy})^{\rm T}$.
 We have
 \begin{align}\label{eq:JointPDFGauss}
 P(\mathcal{R},\mathcal{R}_i,\mathcal{R}_{ij}) = 
 \frac{1}{(2\pi)^{k/2}\sqrt{{\rm det}\,{C}}}
 \exp\left[
 -\frac{1}{2}({R} - \langle {R}\rangle)^{\rm T}({C}^{-1})({R} - \langle {R}\rangle)
 \right]\,,
\end{align}
where $k=6$ is the dimension of $R$, $\langle {R}\rangle$ is the column vector of the expectation values of ${R}$ and 
${C}$ is the covariance matrix defined by 
$C \equiv \langle ({R} - \langle {R}\rangle)({R} - \langle {R}\rangle)^{\rm T} \rangle$ with elements 
 \begin{align}\label{eq:CovarianceMatrixElements}
C_{ij} =  \langle ({R} - \langle {R}\rangle)_i({R} - \langle {R}\rangle)_j \rangle 
=   \langle R_i R_j \rangle - \langle R_i\rangle\langle R_j\rangle\,.
 \end{align}
 From the computation of the covariance matrix one can fully reconstruct the joint probability distribution. 
 
 We restrict the analysis to zero-mean random fields since this assumption is physically motivated (see discussion in 
 appendix\,\ref{app:InflationPrimer}). 
From the explicit computation of the two-point correlators $\langle R_i R_j \rangle$, one finds that the covariance matrix takes the form
\begin{equation}
 C = \bordermatrix{     
            & {\scriptstyle\mathcal{R}}    & {\scriptstyle\mathcal{R}_x}     & {\scriptstyle\mathcal{R}_y} & 
            {\scriptstyle\mathcal{R}_{xx}}  & {\scriptstyle\mathcal{R}_{xy}} & {\scriptstyle\mathcal{R}_{yy}}    \cr
    {\scriptstyle \mathcal{R}}     & \sigma_0^2 & 0 & 0 & -\sigma_1^2/2 & 0 & -\sigma_1^2/2     \cr
    {\scriptstyle \mathcal{R}_x }    & 0 & \sigma_1^2/2 & 0 & 0 & 0 & 0     \cr
   {\scriptstyle  \mathcal{R}_y } & 0 & 0 & \sigma_1^2/2 & 0 &  0& 0   \cr
   {\scriptstyle  \mathcal{R}_{xx} }   & -\sigma_1^2/2 & 0 & 0 & 3\sigma_2^2/8 & 0 & \sigma_2^2/8  \cr
   {\scriptstyle  \mathcal{R}_{xy}  }     & 0  &  0 & 0 &  0 & \sigma_2^2/8 & 0 \cr
   {\scriptstyle  \mathcal{R}_{yy}  } &  -\sigma_1^2/2 & 0 & 0 & \sigma_2^2/8 & 0 & 3\sigma_2^2/8 
}\,,\label{eq:GaussianCorrMatrix}
\end{equation}
and we find that $\mathcal{R}_x$, $\mathcal{R}_y$ and $\mathcal{R}_{xy}$ are completely uncorrelated while 
$\mathcal{R}$, $\mathcal{R}_{xx}$ and $\mathcal{R}_{yy}$ are correlated. 
We introduce the spectral moments 
\begin{equation}\label{eq:sigmaj2}
\sigma_j^2 \equiv \int \frac{dk}{k}\mathcal{P}_{\mathcal{R}}(k)\,k^{2j}\,,
\end{equation}
where $\mathcal{P}_{\mathcal{R}}(k)$ is the dimensionless power spectrum of $\mathcal{R}$. 
Notice that in two spatial dimensions the dimensionless power spectrum of $\mathcal{R}$ is related to the power spectrum 
by means of $\mathcal{P}_{\mathcal{R}}(k) = (k^2/2\pi)\Delta_{\mathcal{R}}(k)$ with $\Delta_{\mathcal{R}}(k) = |\mathcal{R}_k|^2$.

Consider, as an illustrative example, the computation of $\langle \mathcal{R}_{xx}\mathcal{R}_{yy}\rangle$. 
We use the explicit form of $\mathcal{R}$ given in eq.\,(\ref{eq:StocasticField}). 
We find
\begin{align}
\langle \mathcal{R}_{xx}\mathcal{R}_{yy}\rangle & = 
\int\frac{d^2\vec{k}_1}{(2\pi)^2}\frac{d^2\vec{k}_2}{(2\pi)^2}
\langle
\left(
-k_{1,x}^2\mathcal{R}_{k_1}a_{\vec{k}_1}e^{i\vec{k}_1\cdot \vec{x}} 
-k_{1,x}^2\mathcal{R}_{k_1}^*a^{\dag}_{\vec{k}_1}e^{-i\vec{k}_1\cdot \vec{x}} 
\right)
\left(
-k_{2,y}^2\mathcal{R}_{k_2}a_{\vec{k}_2}e^{i\vec{k}_2\cdot \vec{x}} 
-k_{2,y}^2\mathcal{R}_{k_2}^*a^{\dag}_{\vec{k}_2}e^{-i\vec{k}_2\cdot \vec{x}} 
\right)
\rangle\nn \\
 & = \int\frac{d^2\vec{k}_1}{(2\pi)^2}\frac{d^2\vec{k}_2}{(2\pi)^2}
 k_{1,x}^2k_{2,y}^2
 \left[
e^{i(\vec{k}_1 - \vec{k}_2)\cdot \vec{x}}\mathcal{R}_{k_1}\mathcal{R}_{k_2}^*
\langle a_{\vec{k}_1}a^{\dag}_{\vec{k}_2}\rangle 
+
e^{-i(\vec{k}_1 - \vec{k}_2)\cdot \vec{x}}\mathcal{R}_{k_2}\mathcal{R}_{k_1}^*
\langle a^{\dag}_{\vec{k}_1}a_{\vec{k}_2}\rangle
 \right] \nn \\
 & = \int\frac{d^2\vec{k}}{(2\pi)^2} k_x^2k_y^2|\mathcal{R}_{k}|^2 
  = \frac{1}{4\pi^2}\int dk d\varphi k^5\cos^2\varphi\sin^2\varphi\Delta_{\mathcal{R}}(k) 
  = \frac{1}{8}\int \frac{dk}{k}k^4\mathcal{P}_{\mathcal{R}}(k)\,,
\end{align}
where in the last line we just introduced polar coordinates. All the entries in eq.\,(\ref{eq:GaussianCorrMatrix}) can be computed in a similar way.

Interestingly, the pattern of zeros in eq.\,(\ref{eq:GaussianCorrMatrix}) and the relations among different non-zero entries---obtained before by means of a direct  computation---can be understood as a consequence of homogeneity and isotropy.

Homogeneity, that is translational invariance, implies that correlators do not depend on the specific spatial position at which they are computed. 
For instance, this means that the spatial derivative of $\langle \mathcal{R}\mathcal{R}\rangle$ should vanish (this must be true for a generic correlator evaluated at a given spatial point); from this condition, one finds $\partial_i(\langle \mathcal{R}\mathcal{R}\rangle) = 0 \to 
\langle \mathcal{R}\mathcal{R}_i\rangle = 0$ so that $\mathcal{R}$, $\mathcal{R}_x$ and $\mathcal{R}_y$ are uncorrelated. 
Similarly, from $\partial_x(\langle \mathcal{R}_x\mathcal{R}_x\rangle) = 0$ one gets $\langle \mathcal{R}_x\mathcal{R}_{xx}\rangle = 0$ (with similar relations along other directions) so that first and second derivatives  are uncorrelated.  
On the contrary, from $\partial_x(\langle \mathcal{R}\mathcal{R}_x\rangle) = 0$ it follows that $\langle \mathcal{R}\mathcal{R}_{xx}\rangle = - \langle \mathcal{R}_x\mathcal{R}_x\rangle$ as indeed obtained in eq.\,(\ref{eq:GaussianCorrMatrix}). Similarly, $\langle \mathcal{R}\mathcal{R}_{yy}\rangle = - \langle \mathcal{R}_y\mathcal{R}_y\rangle$.

Isotropy, that is rotational invariance, implies that correlators do not depend on a particular direction in space. 
The simplest consequence of isotropy is that $\langle \mathcal{R}_x\mathcal{R}_x\rangle = \langle \mathcal{R}_y\mathcal{R}_y\rangle$ and $\langle \mathcal{R}_x\mathcal{R}_y\rangle = 0$. 
In order to derive these two conditions, a very useful trick (that we shall use also in the non-gaussian computation) is to introduce---instead of the two components $x$ and $y$ of the two-dimensional vector $\vec{x}$---the complex conjugated variables $z\equiv x+iy$ and $z^* = x-iy$ from which we have
 $\partial_z = (\partial_x -i\partial_y)/2$ and $\partial_{z^*} = (\partial_x + i\partial_y)/2$. 
 If we now rotate the two-dimensional vector $\vec{x}$ (without changing its length) the complex number $z$ changes by a phase factor $e^{i\tau}$, 
 that is $z\to e^{i\tau}z$ (and $z^*\to e^{-i\tau}z^*$). Consequently, 
 the derivatives with respect to $z$ and $z^*$ change according to $\partial_z \to e^{-i\tau}\partial_z$ and $\partial_{z^*} \to e^{i\tau}\partial_{z^*}$. 
 If we now consider the correlators $\langle \mathcal{R}_z \mathcal{R}_z \rangle$ and $\langle \mathcal{R}_{z^*} \mathcal{R}_{z^*} \rangle$ they rotate according to 
 $\langle \mathcal{R}_z \mathcal{R}_z \rangle \to e^{-2i\tau}\langle \mathcal{R}_z \mathcal{R}_z \rangle$ and 
 $\langle \mathcal{R}_{z^*} \mathcal{R}_{z^*} \rangle \to e^{2i\tau}\langle  \mathcal{R}_{z^*} \mathcal{R}_{z^*}\rangle$. Because of isotropy of the two-dimensional space, 
 $\langle \mathcal{R}_z \mathcal{R}_z \rangle$ and $\langle \mathcal{R}_{z^*} \mathcal{R}_{z^*} \rangle$ can not depend on $\tau$ and, therefore, they must vanish. 
 Consequently, the system of equations
 \begin{align}
 \langle \mathcal{R}_z \mathcal{R}_z \rangle & = \frac{1}{4}\left(\langle \mathcal{R}_x \mathcal{R}_x \rangle - 2i\langle \mathcal{R}_x \mathcal{R}_y \rangle
  -\langle \mathcal{R}_y \mathcal{R}_y \rangle
 \right) = 0\,,\\
  \langle \mathcal{R}_{z^*} \mathcal{R}_{z^*} \rangle & = \frac{1}{4}\left(\langle \mathcal{R}_x \mathcal{R}_x \rangle + 2i\langle \mathcal{R}_x \mathcal{R}_y \rangle
  -\langle \mathcal{R}_y \mathcal{R}_y \rangle
 \right) = 0\,,
 \end{align}
 admits the solution  $\langle \mathcal{R}_x\mathcal{R}_x\rangle = \langle \mathcal{R}_y\mathcal{R}_y\rangle$ and $\langle \mathcal{R}_x\mathcal{R}_y\rangle = 0$ which are precisely the relations we were looking for. 
 More in general, if we consider the rotation $\partial_z \to e^{-i\tau}\partial_z$ and $\partial_{z^*} \to e^{i\tau}\partial_{z^*}$ a generic correlator will take a phase factor 
 $e^{i\kappa\tau}$ where $\kappa \equiv (\#\,z^*\,{\rm derivatives}) - (\#\,z\,{\rm derivatives})$. 
 If $\kappa \neq 0$, then the correlator must be equal to zero as a consequence of isotropy. 
For instance, we have $\langle\mathcal{R}_{zz}\mathcal{R}_{z^*}\rangle = 0$ (because $\kappa = -1$) but $\langle\mathcal{R}_{z}\mathcal{R}_{z^*} \rangle \neq 0$ (because $\kappa = 0$).

Combining homogeneity and isotropy validates the remaining entries in eq.\,(\ref{eq:GaussianCorrMatrix}). 
For instance, from $\partial_y(\langle\mathcal{R}\mathcal{R}_x\rangle) = \langle\mathcal{R}_x\mathcal{R}_y\rangle + \langle\mathcal{R}\mathcal{R}_{xy}\rangle = 0$ (homogeneity) we find $ \langle\mathcal{R}\mathcal{R}_{xy}\rangle = 0$ since isotropy implies $\langle \mathcal{R}_x\mathcal{R}_y\rangle = 0$.  As a final check, 
we consider the block of the second derivatives in eq.\,(\ref{eq:GaussianCorrMatrix}). 
From the previous argument, isotropy implies that $\langle \mathcal{R}_{zz} \mathcal{R}_{zz}\rangle = 0$ and $\langle \mathcal{R}_{zz^*} \mathcal{R}_{zz}\rangle = 0$ (together with their complex conjugated $\langle \mathcal{R}_{z^*z^*} \mathcal{R}_{z^*z^*}\rangle = 0$ and $\langle \mathcal{R}_{zz^*} \mathcal{R}_{z^*z^*}\rangle = 0$). 
These two relations imply $\langle \mathcal{R}_{xx} \mathcal{R}_{xy}\rangle = \langle \mathcal{R}_{yy} \mathcal{R}_{xy}\rangle = 0$ 
and $\langle \mathcal{R}_{xx} \mathcal{R}_{xx}\rangle = \langle \mathcal{R}_{xx} \mathcal{R}_{yy}\rangle + 2\langle \mathcal{R}_{xy} \mathcal{R}_{xy}\rangle$. 
Both these relations are verified by the entries in eq.\,(\ref{eq:GaussianCorrMatrix}). We can actually do more since it is possible to show that 
$\langle \mathcal{R}_{xx} \mathcal{R}_{yy}\rangle = \langle \mathcal{R}_{xy} \mathcal{R}_{xy}\rangle$. Let us write 
\begin{align}
\langle \mathcal{R}_{xx} \mathcal{R}_{yy}\rangle   & = \langle \partial_{xx}\mathcal{R}(\vec{x})\partial_{yy}\mathcal{R}(\vec{x})\rangle  
=\left. \langle \partial_{x_1x_1}\mathcal{R}(\vec{x}_1)\partial_{y_2y_2}\mathcal{R}(\vec{x}_2)\rangle   \right|_{\vec{x}_1=\vec{x}_2 =\vec{x}} \nn  \\
& = \left. \partial_{x_1x_1}\partial_{y_2y_2}\langle\mathcal{R}(\vec{x}_1)\mathcal{R}(\vec{x}_2)\rangle   \right|_{\vec{x}_1=\vec{x}_2 =\vec{x}} 
 =  \left. \partial_{x_1}\partial_{y_2}\langle\mathcal{R}_{x_1}(\vec{x}_1)\mathcal{R}_{y_2}(\vec{x}_2)\rangle   \right|_{\vec{x}_1=\vec{x}_2 =\vec{x}}\,.\label{eq:A1}
\end{align}
where the first step is just a more explicit definition of  $\langle \mathcal{R}_{xx} \mathcal{R}_{yy}\rangle$ while in the following ones we consider two distinct point $\vec{x}_1 = (x_1,y_1)$ and $\vec{x}_2 = (x_2,y_2)$ that we later set equal again. Now the point is that because of homogeneity of space the correlator $\langle\mathcal{R}_{x_1}(\vec{x}_1)\mathcal{R}_{y_2}(\vec{x}_2)\rangle $ depends only on the distance $|\vec{x}_1 - \vec{x}_2|$; we can, therefore, exchange $1 \leftrightarrow 2$ obtaining 
$\langle\mathcal{R}_{x_2}(\vec{x}_2)\mathcal{R}_{y_1}(\vec{x}_1)\rangle $ without altering the result. From eq.\,(\ref{eq:A1}) this means that we have 
$\langle \mathcal{R}_{xx} \mathcal{R}_{yy}\rangle = \langle \mathcal{R}_{xy} \mathcal{R}_{xy}\rangle$ as indeed verified in eq.\,(\ref{eq:GaussianCorrMatrix}). 
If we combine this result with the previous relation $\langle \mathcal{R}_{xx} \mathcal{R}_{xx}\rangle = \langle \mathcal{R}_{xx} \mathcal{R}_{yy}\rangle + 2\langle \mathcal{R}_{xy} \mathcal{R}_{xy}\rangle$ we find $\langle \mathcal{R}_{xx} \mathcal{R}_{xx}\rangle = 3 \langle \mathcal{R}_{xy} \mathcal{R}_{xy}\rangle$ which is again verified in eq.\,(\ref{eq:GaussianCorrMatrix}).  
Finally, we also note that the condition $\langle \mathcal{R}_{xx} \mathcal{R}_{yy}\rangle = \langle \mathcal{R}_{xy} \mathcal{R}_{xy}\rangle$ implies that 
$\langle \mathcal{R}_{zz^*} \mathcal{R}_{zz^*}\rangle = \langle \mathcal{R}_{zz} \mathcal{R}_{z^*z^*}\rangle$.
These kind of relations based on  homogeneity and isotropy will be useful later in the non-gaussian case.

Using the properties of the exponential function, eq.\,(\ref{eq:JointPDFGauss}) takes the form
 \begin{align}\label{eq:GaussianPDF}
 P(\mathcal{R},\mathcal{R}_i,\mathcal{R}_{ij}) = 
 P(\mathcal{R}_x) P(\mathcal{R}_y) P(\mathcal{R}_{xy}) P(\mathcal{R},\mathcal{R}_{xx},\mathcal{R}_{yy})\,,
 \end{align}
where
 \begin{align}
 P(\mathcal{R}_x) = \frac{1}{\sqrt{\pi \sigma_1^2}}\exp\left(-\frac{\mathcal{R}_x^2}{\sigma_1^2}\right)\,,\hspace{.7cm}
 P(\mathcal{R}_y) = \frac{1}{\sqrt{\pi \sigma_1^2}}\exp\left(-\frac{\mathcal{R}_y^2}{\sigma_1^2}\right)\,, \hspace{.7cm}
  P(\mathcal{R}_{xy}) = \frac{2}{\sqrt{\pi \sigma_2^2}}\exp\left(-\frac{4\mathcal{R}_{xy}^2}{\sigma_2^2}\right)\,,
 \end{align}
 and
   \begin{align}
  P(\mathcal{R},\mathcal{R}_{xx},\mathcal{R}_{yy}) = 
  \frac{1}{(2\pi)^{3/2}\sqrt{{\rm det}\tilde{C}}}\exp\left(
 -\frac{1}{2}\tilde{R}^{\rm T} \tilde{C}^{-1}\tilde{R}
  \right)\,,\hspace{1cm}
\tilde{C} \equiv   
\bordermatrix{     
    & {\scriptstyle\mathcal{R}}    & {\scriptstyle\mathcal{R}_{xx}}     & {\scriptstyle\mathcal{R}_{yy}} & \cr
    {\scriptstyle \mathcal{R}}     & \sigma_0^2 & -\sigma_1^2/2 & -\sigma_1^2/2     \cr
   {\scriptstyle  \mathcal{R}_{xx} }   & -\sigma_1^2/2 & 3\sigma_2^2/8 & \sigma_2^2/8  \cr
   {\scriptstyle  \mathcal{R}_{yy}  } &  -\sigma_1^2/2 & \sigma_2^2/8 & 3\sigma_2^2/8 
}\,,
 \end{align}
 with $\tilde{R} \equiv (\mathcal{R},\mathcal{R}_{xx},\mathcal{R}_{yy})^{\rm T}$. 
 As customary in the gaussian case, 
 from the knowledge of the power spectrum it is possible to fully reconstruct the statistics of the random field. 
From eq.\,(\ref{eq:nMax}) we get the number density of maxima
\begin{align}\label{eq:nMax2}
n_{\rm max}(\mathcal{R}) = \frac{1}{\pi \sigma_1^2}
\int_{\rm max}
d\mathcal{R}_{xx}d\mathcal{R}_{xy}d\mathcal{R}_{yy}
\left|
\mathcal{R}_{xx}\mathcal{R}_{yy}-\mathcal{R}_{xy}^2
\right|P(\mathcal{R}_{xy})P(\mathcal{R},\mathcal{R}_{xx},\mathcal{R}_{yy})\,,
\end{align}
where we used $P(\mathcal{R}_x = 0) = P(\mathcal{R}_y = 0) = 1/\sqrt{\pi \sigma_1^2}$. 
The integration region is defined by the conditions 
${\rm max} = \{\mathcal{R}_{xx}\mathcal{R}_{yy}-\mathcal{R}_{xy}^2 >0 \land \mathcal{R}_{xx}<0 \land \mathcal{R}_{yy}<0\}$.

The change of variables $\{\mathcal{R}_{xx},\mathcal{R}_{yy},\mathcal{R}_{xy}\} \to \{r,s,\theta\}$ defined by
\begin{align}\label{eq:ChangeOfVariable}
r\cos\theta \equiv \frac{1}{2}\left(\mathcal{R}_{xx}-\mathcal{R}_{yy}\right)\,,\hspace{.75cm}
r\sin\theta \equiv \mathcal{R}_{xy}\,,\hspace{.75cm}
s \equiv -\frac{1}{2}\left(\mathcal{R}_{xx} + \mathcal{R}_{yy}\right)\,,
\end{align}
 turns out the be useful. The Jacobian of the transformation is $J=2r$, and we have 
 $\mathcal{R}_{xx}\mathcal{R}_{yy}-\mathcal{R}_{xy}^2 = s^2 - r^2$. 
 The condition  $s^2 - r^2 >0$ becomes $-s< r < s$ 
 with $s>0$ since $\mathcal{R}_{xx}<0$ and $\mathcal{R}_{yy} <0$. Furthermore, we restrict to $r>0$ if 
 $0< \theta < 2\pi$. All in all, we have ${\rm max} = \{0< \theta < 2\pi \land s>0 \land 0<r<s\}$.
 
The parametrization in terms of $\{r,s,\theta\}$ is useful because we have 
\begin{align}
2s = -(\mathcal{R}_{xx}+\mathcal{R}_{yy}) = 
-\triangle \mathcal{R}\,.
  \end{align}
This implies that any condition that restricts the value of the curvature $-\triangle \mathcal{R}$ can be implemented 
through $s$ by imposing $s > s_{\rm min}$ instead of $s>0$. 
This is the case of eq.\,(\ref{eq:SpikyEnough}) with $\alpha = 0$ (thus $h = \mathcal{R}$), which reads
\begin{align}\label{eq:ExplThresh2}
\frac{s_{\rm min}}{\sigma_2} = \frac{9}{8}\frac{(a_m H_m)^2}{\sigma_2}\delta_c\,,
\end{align}

Eq.\,(\ref{eq:nMax2}) becomes
\begin{align}
n_{\rm max}(\mathcal{R},s_{\rm min}) = 
\frac{8}{\pi^2 \sigma_1^2\sigma_2^2\sqrt{\sigma_2^2\sigma_0^2 -\sigma_1^4}}
\int_{s_{\rm min}}^{\infty}ds\int_0^s dr\,
r(s^2 - r^2)\exp\left[
-\frac{(
\mathcal{R}^2 \sigma_2^2 + 4s^2\sigma_0^2 - 4s\mathcal{R}\sigma_1^2
)}{2(\sigma_0^2\sigma_2^2 - \sigma_1^4)} - \frac{4r^2}{\sigma_2^2}
\right]\,,
\end{align} 
where, according to the previous argument, we set to $s_{\rm min}$ the lower limit of integration 
over $s$ and define $n_{\rm max}(\mathcal{R},s_{\rm min}) $ such that $n_{\rm max}(\mathcal{R},s_{\rm min} = 0) = n_{\rm max}(\mathcal{R})$.
The integration over $r$ gives 
\begin{align}\label{eq:nMaxRs}
n_{\rm max}&(\mathcal{R},s_{\rm min}) =\\
& \underbrace{ 
\frac{\sigma_2^2}{4\pi^2 \sigma_1^2\sqrt{\sigma_2^2\sigma_0^2 -\sigma_1^4}}\int_{s_{\rm min}}^{\infty}ds
\left(
\frac{4s^2}{\sigma_2^2} + e^{-\frac{4s^2}{\sigma_2^2}} -1
\right)\exp\left\{
-\frac{\sigma_0^2\sigma_2^2}{2(\sigma_0^2 \sigma_2^2 - \sigma_1^4)}
\left[
\frac{4s^2}{\sigma_2^2} - \frac{4s}{\sigma_2}\left(\frac{\sigma_1^2}{\sigma_2\sigma_0}\right)\frac{\mathcal{R}}{\sigma_0}
 +
 \frac{\mathcal{R}^2}{\sigma_0^2}
\right]
\right\}}_{\equiv \int_{s_{\rm min}}^{\infty}ds\,\bar{n}_{\rm max}(\mathcal{R},s)}\,,\nn
\end{align} 
where 
\begin{align}\label{eq:nMaxRs2}
\bar{n}_{\rm max}(\mathcal{R},s) \equiv \frac{\sigma_2^2}{4\pi^2 \sigma_1^2\sqrt{\sigma_2^2\sigma_0^2 -\sigma_1^4}}
\left(
\frac{4s^2}{\sigma_2^2} + e^{-\frac{4s^2}{\sigma_2^2}} -1
\right)\exp\left\{
-\frac{\sigma_0^2\sigma_2^2}{2(\sigma_0^2 \sigma_2^2 - \sigma_1^4)}
\left[
\frac{4s^2}{\sigma_2^2} - \frac{4s}{\sigma_2}\left(\frac{\sigma_1^2}{\sigma_2\sigma_0}\right)\frac{\mathcal{R}}{\sigma_0}
 +
 \frac{\mathcal{R}^2}{\sigma_0^2}
\right]
\right\}\,,
\end{align} 
 can be interpreted as the number density of 
maxima with field value in the range $\mathcal{R}$ to $\mathcal{R} + d\mathcal{R}$ 
and curvature $-\triangle \mathcal{R}$ in the range $2s$ to $2(s+ds)$. 
We note that the argument of the exponential function in $\bar{n}_{\rm max}(\mathcal{R},s)$ is invariant under the 
 exchange $\mathcal{R}/\sigma_0 \leftrightarrow 2s/\sigma_2$. 

From the structure of the argument of the exponential function in $\bar{n}_{\rm max}(\mathcal{R},s)$, 
it is natural to introduce the dimensionless parameter $\gamma \equiv \sigma_1^2/\sigma_2\sigma_0$ which is 
completely determined, as we shall explain in a moment, by the properties  of the power spectrum. 
Furthermore, it is easy to see that we have $0<\gamma < 1$. In turn, this condition implies $\sigma_2^2\sigma_0^2 - \sigma_1^4 > 0$ so
 that the square root in eq.\,(\ref{eq:nMaxRs2}) is always real valued. 
 Let us verify the non-trivial condition $\sigma_2^2\sigma_0^2 - \sigma_1^4 > 0$; from the definition in eq.\,(\ref{eq:sigmaj2}), we have
\begin{align}
 \sigma_0^2\sigma_2^2 - \sigma_1^4 = 
 \left[\int\frac{dk}{k}\mathcal{P}_{\mathcal{R}}(k)\right]
 \left[\int\frac{dk^{\prime}}{k^{\prime}}\mathcal{P}_{\mathcal{R}}(k^{\prime})k^{\prime\,4}\right] - 
 \left[\int\frac{dk}{k}\mathcal{P}_{\mathcal{R}}(k)k^2\right]^2
 = \int\frac{dk}{k}\frac{dk^{\prime}}{k^{\prime}}
 \mathcal{P}_{\mathcal{R}}(k)
 \mathcal{P}_{\mathcal{R}}(k^{\prime})\left(
 k^{\prime\,4} - k^{\prime\,2}k^2
 \right)\,.
\end{align}
To conclude that $\sigma_0^2\sigma_2^2 - \sigma_1^4 > 0$, we only need to show that $(k^{\prime\,4} - k^{\prime\,2}k^2) > 0$ 
since the power spectrum is positive definite and the integrals over $k$ cover the positive real axis.
In the first double-integral, we can change $k\leftrightarrow k^{\prime}$ without altering the result, and 
this means that we can also substitute $k^{\prime\,4} \to (k^4 + k^{\prime\,4})/2$; if we do this transformation, 
the factor $(k^{\prime\,4} - k^{\prime\,2}k^2)$ becomes $(k^{\prime\,2} - k^2)^2/2$ which is always positive. 
This concludes the proof.

From eq.\,(\ref{eq:nMaxRs2}) we see that the value of $\gamma$ controls the amount of correlation between $\mathcal{R}$ and $s$.  
If we take $\gamma = 0$, the two variables are completely uncorrelated (because $\sigma_1^2 = 0$). 
On the contrary, $\gamma = 1$ corresponds to the case in which they are maximally correlated.
 
To proceed further, we assume the following analytical expression for the power spectrum
\begin{align}\label{eq:ToyPS}
\mathcal{P}_{\mathcal{R}}(k) = \frac{A_g}{\sqrt{2\pi}v}\exp\left[
-\frac{\log^2(k/k_{\star})}{2v^2}
\right]\,.
\end{align}
As far as the spectral moments in eq.\,(\ref{eq:sigmaj2}) are concerned, we find the analytical result 
 \begin{align}\label{eq:SigmajAnal}
 \sigma_j^2 = A_g k_{\star}^{2j}e^{2j^2 v^2}\,,
\end{align} 
which implies $\sigma_0^2 = A_g$ and $\gamma = e^{-2v^2}$, where we see that $0<\gamma<1$ as expected. 
The three parameters $\{A_g,v,k_{\star}\}$ fully specify our problem. The physical picture is the following.
\begin{itemize}
\item [$\circ$] The scale $k_{\star}$ represents the comoving wavenumber at which the power spectrum peaks. 
This quantity is related to the mass of the black holes produced after the collapse of the regions where 
the overdensity field is above threshold. We take $k_{\star} = O(10^{14})$ Mpc$^{-1}$ corresponding to
 $M_{\rm PBH} = O(10^{18})$ g.
\item [$\circ$] The parameter $v$ controls the broadness of the power spectrum and, in turn, the 
broadness of the mass distribution of the black holes. The case $v=0.1$, for instance, corresponds to a very 
narrow power spectrum. This, in turn, will generate a very narrow mass distribution of black holes.
\item [$\circ$] The amplitude of the power spectrum $A_g$ is related to the abundance of dark matter 
in the present-day Universe in the form of black holes, and typical values are of order $A_g/\sqrt{2\pi}v = O(10^{-2})$. 
To fix ideas, to get $\mathcal{P}_{\mathcal{R}}(k_{\star}) = 10^{-2}$  one needs $A_g = 2.5\times 10^{-3}$ for $v = 0.1$.
\end{itemize}
Let us now pause for a moment to clarify the rationale of the computation that we are doing. 
We are interested in regions of space where the field $\mathcal{R}$ has large curvature $-\triangle\mathcal{R}$ since in this case 
the overdensity field (which is proportional to $-\triangle\mathcal{R}$) takes large values.  
In eq.\,(\ref{eq:nMaxRs2}) we can select regions with large curvature by implementing (as done in eq.\,(\ref{eq:nMaxRs})) 
a lower limit of integration over $s$.  
However, eq.\,(\ref{eq:nMaxRs2}) always associates, by construction, regions with large curvature (consequently, peaks of the 
overdensity field) with 
local maxima of $\mathcal{R}$.
This association can be analytically justified as follows.
 Consider the Taylor expansion in eq.\,(\ref{eq:Stat}) which we rewrite in two spatial dimensions taking a local maximum as stationary point of 
 $\mathcal{R}$
 \begin{align}\label{eq:StatM}
\mathcal{R}(\vec{x}) = \mathcal{R}_{\rm M} + 
\frac{1}{2}\sum_{i,j =1}^{2}\mathcal{R}_{ij}(\vec{x}_{\rm M})(
\vec{x} - \vec{x}_{\rm M}
)_i(
\vec{x} - \vec{x}_{\rm M}
)_j\,.
\end{align}
This quadratic equation can be written in a canonical form (that is without cross terms) if we do a coordinate 
transformation $\vec{x}\to \vec{x}^{\,\prime} \equiv (x^{\prime},y^{\prime})$ such that the new axes are aligned along the eigenvectors of
 the matrix $\mathcal{R}_{ij}(\vec{x}_{\rm M})$. 
 We are free to do this rotation because of isotropy. 
 In such case, eq.\,(\ref{eq:StatM}) takes the form  
 \begin{align}\label{eq:StatMPrime}
\mathcal{R}(\vec{x}^{\,\prime}) = \mathcal{R}_{\rm M} -
\frac{1}{2}(\lambda_{1}x^{\prime\,2} + \lambda_{2}y^{\prime\,2})\,,
\end{align} 
where $\lambda_{i=1,2} > 0$ are the (minus) eigenvalues of $\mathcal{R}_{ij}(\vec{x}_{\rm M})$. 
Notice that in eq.\,(\ref{eq:StatMPrime}) we used homogeneity to shift the position of the maximum $\vec{x}_{\rm M}$ to the 
origin of the new coordinate system.
 From eq.\,(\ref{eq:StatMPrime}) we see that an iso-density surface with constant $\mathcal{R}_{\vec{x}^{\prime}} \equiv \mathcal{R}(\vec{x}^{\,\prime})$ 
 is an ellipse with canonical equation
\begin{align}
\frac{\lambda_1}{2(\mathcal{R}_{\rm M} - \mathcal{R}_{\vec{x}^{\prime}})}  x^{\prime\,2} + 
\frac{\lambda_2}{2(\mathcal{R}_{\rm M} - \mathcal{R}_{\vec{x}^{\prime}})}  y^{\prime\,2}  = 1\,,~~~~~~~~~
a_i \equiv \bigg[\frac{2(\mathcal{R}_{\rm M} - \mathcal{R}_{\vec{x}^{\prime}})}{\lambda_i}\bigg]^{1/2}\,,
\end{align}  
and semi-axes $a_{i=1,2}$. 
The key point is that the actual magnitude of the eigenvalues $\lambda_i$ depends on the steepness of the field $\mathcal{R}$ around the position of its maximum.
This follows from eq.\,(\ref{eq:StatMPrime}) if we apply the Laplacian with respect to the coordinates $\vec{x}^{\,\prime}$ since 
we find
\begin{align}
\lambda_1 + \lambda_2 = -\triangle\mathcal{R}(\vec{x}^{\,\prime})\,.
\end{align}
Suppose now that the point $\vec{x}^{\,\prime}$ coincides with a peak of the overdensity field, $\vec{y}_{\rm pk}$. 
Using eq.\,(\ref{eq:SpikyEnough}), we find
\begin{align}\label{eq:Inter}
\lambda_1 + \lambda_2 = -\triangle\mathcal{R}(\vec{y}_{\rm pk}) \gtrsim \frac{9}{4}(aH)^2\delta_c = 
2\sigma_2\underbrace{\frac{9}{8}\frac{(aH)^2}{\sigma_2}\delta_c}_{\gg 1}~~~~~~\Longrightarrow~~~~~~
\frac{\lambda_i}{\sigma_2} \gg 1\,.
\end{align}
In eq.\,(\ref{eq:Inter}) we used the fact that we typically expect $(aH)^2/\sigma_2 \gg 1$; we will comment in more detail 
about this estimate at the end of this section. 
Notice that in eq.\,(\ref{eq:Inter}) we assumed that the two eigenvalues $\lambda_{i}$ have the same magnitude. 
This is because we have separately $\lambda_{1} = - \mathcal{R}_{xx}(\vec{y}_{\rm pk})$ and 
 $\lambda_{2} = - \mathcal{R}_{yy}(\vec{y}_{\rm pk})$, and the second derivatives $\mathcal{R}_{xx}$ ad $\mathcal{R}_{yy}$ 
 have the same covariance (see eq.\,(\ref{eq:GaussianCorrMatrix})).
We can now do the same expansion in eq.\,(\ref{eq:StatM}) but with respect to the first derivatives of $\mathcal{R}$ at 
$\vec{x}_{\rm M}$. The stationary condition $\mathcal{R}_i(\vec{x}_{\rm M}) = 0$ reads
\begin{align}\label{eq:Inter2}
\mathcal{R}_i(\vec{x}) - \sum_{j=1}^2\mathcal{R}_{ij}(\vec{x}_{\rm M})(\vec{x} - \vec{x}_{\rm M})_j = 0
~~~~~~\Longrightarrow~~~~~~
\left\{
\begin{array}{c}
 \mathcal{R}_x(\vec{x}^{\,\prime}) + \lambda_1x^{\prime} = 0    \\
  \\
 \mathcal{R}_y(\vec{x}^{\,\prime}) + \lambda_2 y^{\prime} = 0       
\end{array}
\right.
\end{align}
where in the last line we introduced, as done before, the eigenvalues $\lambda_{i=1,2}$. 
We identify again the point $\vec{x}^{\,\prime}$ with a peak of the overdensity field so that 
we can use the estimate in eq.\,(\ref{eq:Inter}). Furthermore,  $\vec{x}^{\,\prime}$ represents now, by construction, the distance 
between the local maximum of $\mathcal{R}$ and the peak of the overdensity field. 
We can estimate this distance by means of eq.\,(\ref{eq:Inter}) and eq.\,(\ref{eq:Inter2}). 
In eq.\,(\ref{eq:Inter2}), $\mathcal{R}_x(\vec{x}^{\,\prime})$ and $\mathcal{R}_y(\vec{x}^{\,\prime})$ are not 
equal to zero (since we moved away from the local maximum of $\mathcal{R}$) and their magnitude can be estimated (in a probabilistic sense)
by means of the covariance matrix. We have 
$\mathcal{R}_x(\vec{x}^{\,\prime})\approx \mathcal{R}_y(\vec{x}^{\,\prime}) 
\approx \langle \mathcal{R}_x\mathcal{R}_x \rangle^{1/2} \approx \sigma_1$. We find
\begin{align}\label{eq:AnalArgu}
|\vec{x}^{\,\prime}| \approx \frac{\sigma_1}{\sigma_2}\left(\frac{1}{\lambda_i/\sigma_2}\right) \ll 
\frac{\sigma_1}{\sigma_2} = \frac{e^{-3v^2}}{k_{\star}}\,.
\end{align}
As we will see in a moment (see also appendix\,\ref{app:Threshold}), we have that $1/k_{\star}$ is typically of the order of the comoving horizon length $1/aH$ at the time
when the perturbations re-enter the horizon and become causally connected. 
Therefore, we find $|\vec{x}^{\,\prime}| \ll 1/aH$.
This means that high peaks of the overdensity field lie ``close'' (that is within an Hubble radius) to local maxima of the curvature perturbation. 
\begin{figure}[!htb!]
\begin{center}
$$\includegraphics[width=.49\textwidth]{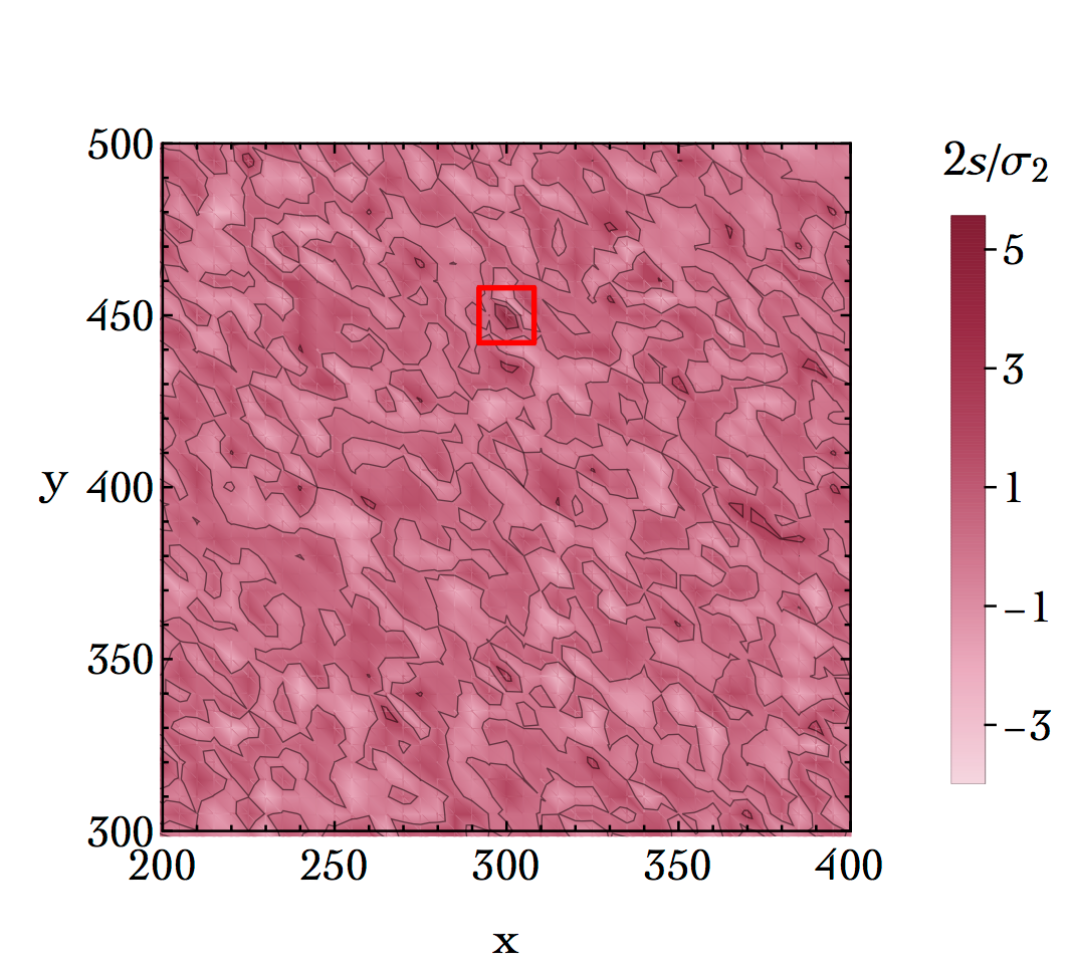}
\qquad\includegraphics[width=.415\textwidth]{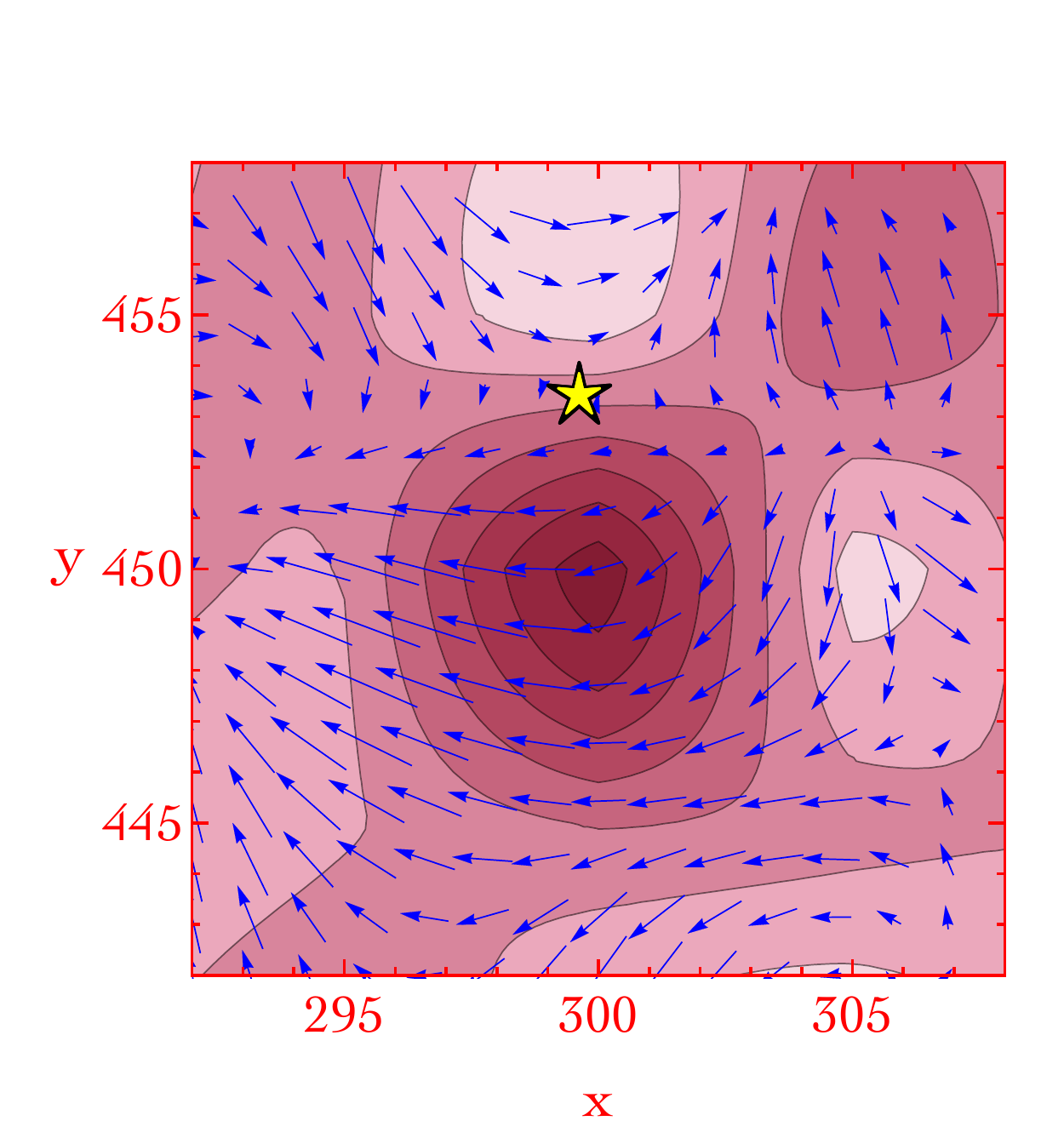}$$
\caption{\em \label{fig:MaxCorrespondence}  
Numerical simulation of the random field $\mathcal{R}$ together with its first and second derivatives. 
At each point in space (discretized in steps $\Delta x = 5$ and $\Delta y = 5$) we associate a vector of values 
$\{\mathcal{R},\mathcal{R}_x,\mathcal{R}_y,\mathcal{R}_{xx},\mathcal{R}_{xy},\mathcal{R}_{yy}\}$ randomly generated from eq.\,(\ref{eq:JointPDFGauss}).
We use the power spectrum in eq.\,(\ref{eq:ToyPS}) to compute the correlation matrix.
We set $v=0.7$, $k_{\star} = 1.5\times 10^{14}$ Mpc$^{-1}$ and $A_g = 2.5\times 10^{-3}$.
Left panel. We show the density plot of the random variable $2s/\sigma_2 = -\triangle\mathcal{R}/\sigma_2$ which is related to the overdensity field.
Right panel. Same as in the left panel but zoomed in the region where the random variable $-\triangle\mathcal{R}/\sigma_2$ has a pronounced peak.  
We superimpose (blue arrows) a vector plot that keeps track of the gradient field $\{\mathcal{R}_x,\mathcal{R}_y\}$. 
The yellow star marks the position of the local maximum of $\mathcal{R}$ that lies close to the peak of $-\triangle\mathcal{R}/\sigma_2$.
 }
\end{center}
\end{figure}
We can validate the analytical approximation by means of a numerical check. 
To this end, we use the full joint probability density distribution in eq.\,(\ref{eq:JointPDFGauss}) to generate a sample of 
random values that we distribute on a two-dimensional grid (see caption of fig.\,\ref{fig:MaxCorrespondence}). 
In other words, at each point on the spatial grid corresponds a value of ${R} = (\mathcal{R},\mathcal{R}_x,\mathcal{R}_y,\mathcal{R}_{xx},\mathcal{R}_{xy},\mathcal{R}_{yy})^{\rm T}$ randomly generated from eq.\,(\ref{eq:JointPDFGauss}). 
In the left panel of fig.\,\ref{fig:MaxCorrespondence} we show the spatial distribution of the random variable 
$2s/\sigma_2 = -\triangle\mathcal{R}/\sigma_2$. 
We focus on a region in which $-\triangle\mathcal{R}$ takes a large value (in units of $\sigma_2$). In the left panel of fig.\,\ref{fig:MaxCorrespondence}, we indicate 
this region with a red contour. We zoom in this part of the plot in the right panel of fig.\,\ref{fig:MaxCorrespondence}. 
The analytical argument explained before suggests that we should find a maximum close to the point where the curvature field peaks. 
We look for this maximum numerically by looking at the behavior of the gradient field $\{\mathcal{R}_x,\mathcal{R}_y\}$ that we plot using blue arrows. 
We indeed find a local maximum that we mark with a yellow star. We checked numerically that at the position of the yellow star where the gradient field vanishes the conditions 
on the second derivatives that define a maximum are verified. 
We note that at the position of the peak of the overdensity field the gradient field $\{\mathcal{R}_x,\mathcal{R}_y\}$ does not vanish but we find that 
its magnitude is of order $O(1)$ (in units of $\sigma_1$) as we argued in eq.\,(\ref{eq:AnalArgu}). 
Furthermore, we checked that the local maximum lies closer to the peak of $-\triangle\mathcal{R}/\sigma_2$ for increasing higher values of the latter.

This numerical result corroborates the validity of the analytical argument in eq.\,(\ref{eq:AnalArgu}), and we conclude, therefore, that eq.\,(\ref{eq:nMaxRs2}) is the right distribution to consider: We count the peaks of $\delta$ by looking at the maxima of $\mathcal{R}$ with large curvature.

\begin{figure}[!htb!]
\begin{center}
$$\includegraphics[width=.45\textwidth]{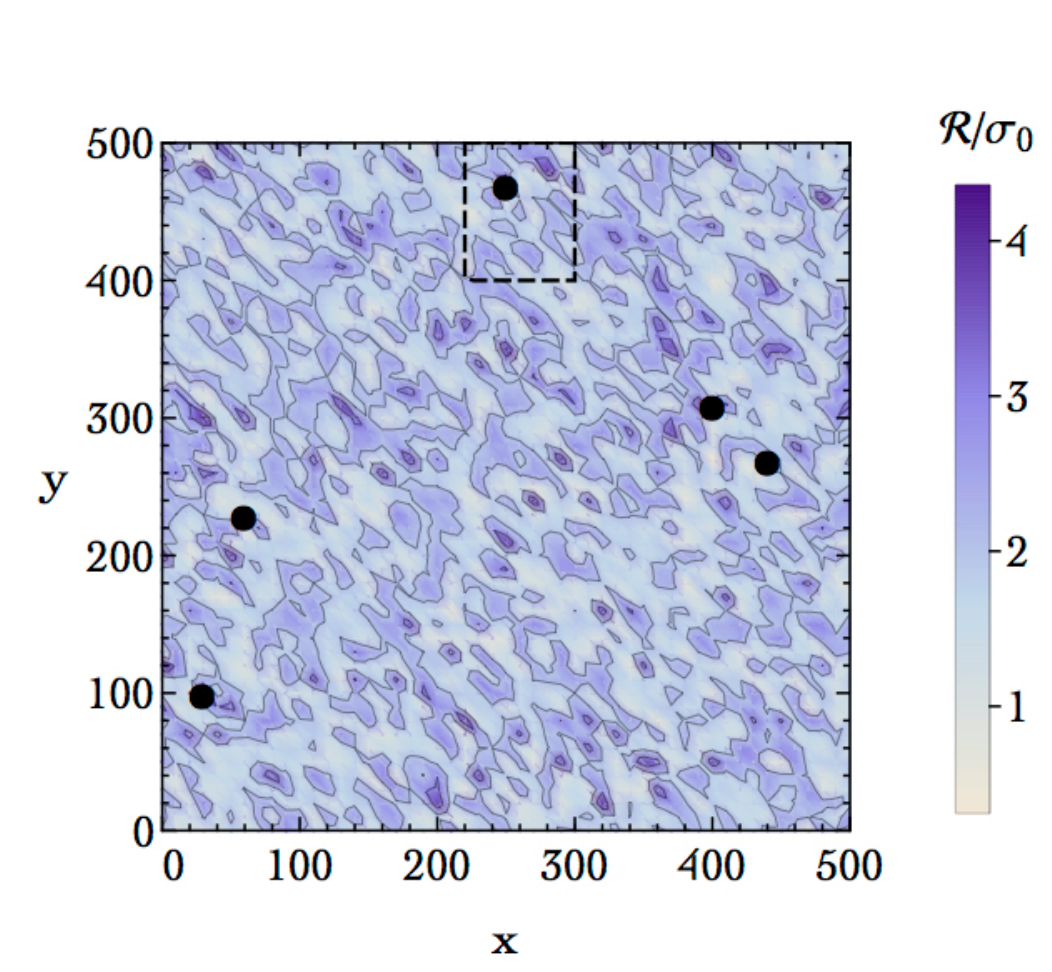}
\qquad\includegraphics[width=.45\textwidth]{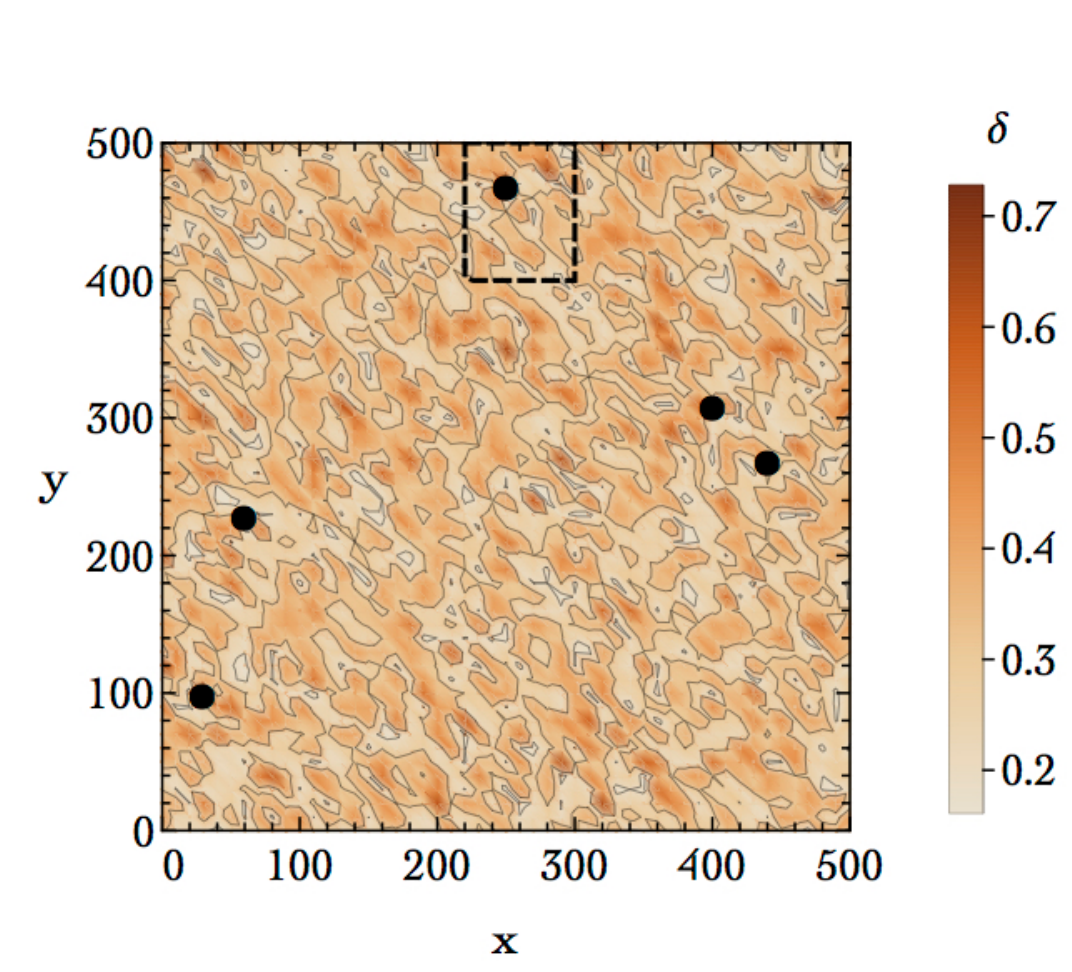}$$
\caption{\em \label{fig:MonteCarlo}  
Numerical simulation of maxima of $\mathcal{R}$ in two spatial dimensions. 
At each point $(x,y)$ in space (discretized in steps $\Delta x = 10$ and $\Delta y = 10$), we extract randomly from the density distribution $\bar{n}_{\rm max}(\mathcal{R},s)$ (defined in eq.\,(\ref{eq:nMaxRs2})) the value of $\mathcal{R}$ and $s = -\triangle\mathcal{R}/2$. The former are shown in the left panel. 
We remark that, by construction, all point generated by means of  $\bar{n}_{\rm max}(\mathcal{R},s)$ are maxima of $\mathcal{R}$. 
As far as the values of $s$ are concerned, we plot on the right panel the corresponding values of $\delta = (8/9)(1/aH)^2s$. 
We use the power spectrum in eq.\,(\ref{eq:ToyPS}), and we set $v=0.1$, $k_{\star} = 1.5\times 10^{14}$ Mpc$^{-1}$ and 
$A_g = 2.5\times 10^{-3}$. 
Furthermore, we take $(k_{\star}/aH)^2 \simeq 0.36$ as suggested by numerical simulations (see appendix\,\ref{app:Threshold}).
For illustrative purposes, we set the threshold $\delta_c = 0.675$ (which is significantly smaller compared to the value expected from 
numerical simulation of gravitational collapse into black holes that is $\delta_c \simeq 1.19$, 
see appendix\,\ref{app:Threshold}; we use a smaller value of $\delta_c$ otherwise events over the threshold would be too rare to be 
simulated in our simplified numerical analysis). 
Points in the simulation with $\delta > \delta_c$ are marked with a black dot.
 }
\end{center}
\end{figure}

\begin{figure}[!htb!]
\begin{center}
$$\includegraphics[width=.4\textwidth]{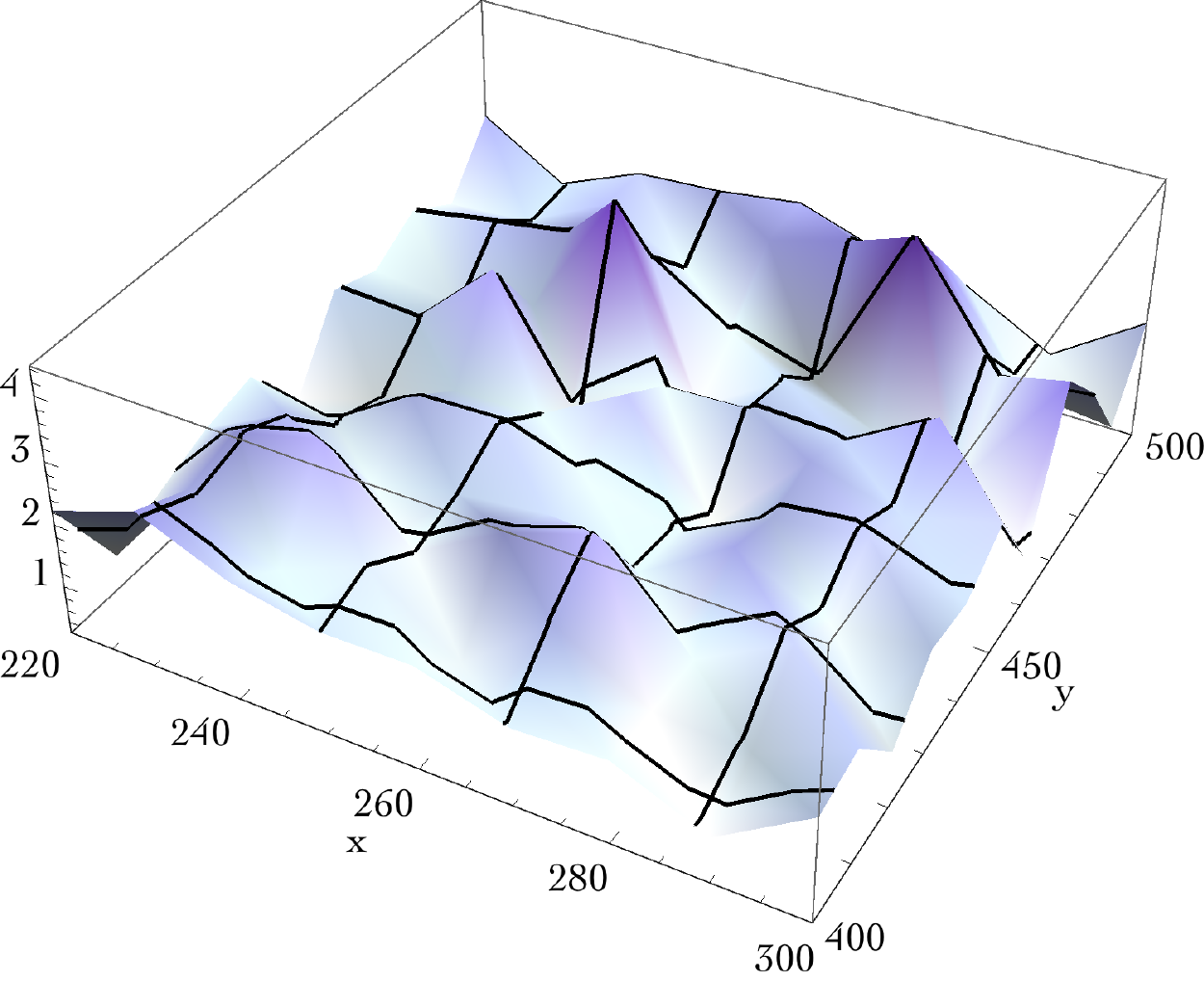}
\qquad\includegraphics[width=.4\textwidth]{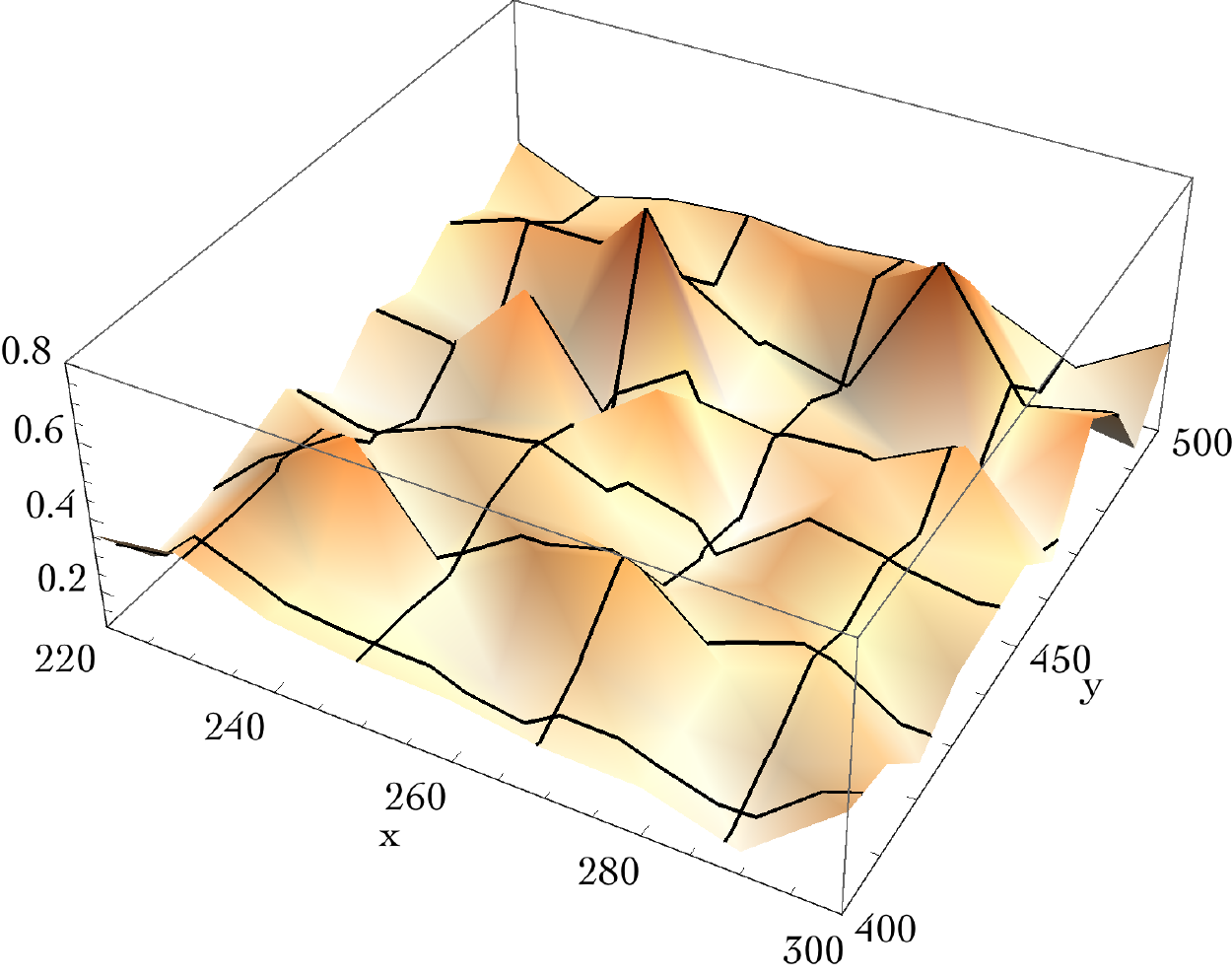}$$
\caption{\em \label{fig:SnapShot}  
Same as in fig.\,\ref{fig:MonteCarlo} but zoomed in the region delimited by dashed lines and displayed in the form of a 
tri-dimensional density plot. 
We see that spiky maxima with large values of $\mathcal{R}$ (left panel) coincide with peaks of $\delta$ (right panel).
 }
\end{center}
\end{figure}

\begin{figure}[!htb!]
\begin{center}
$$\includegraphics[width=.45\textwidth]{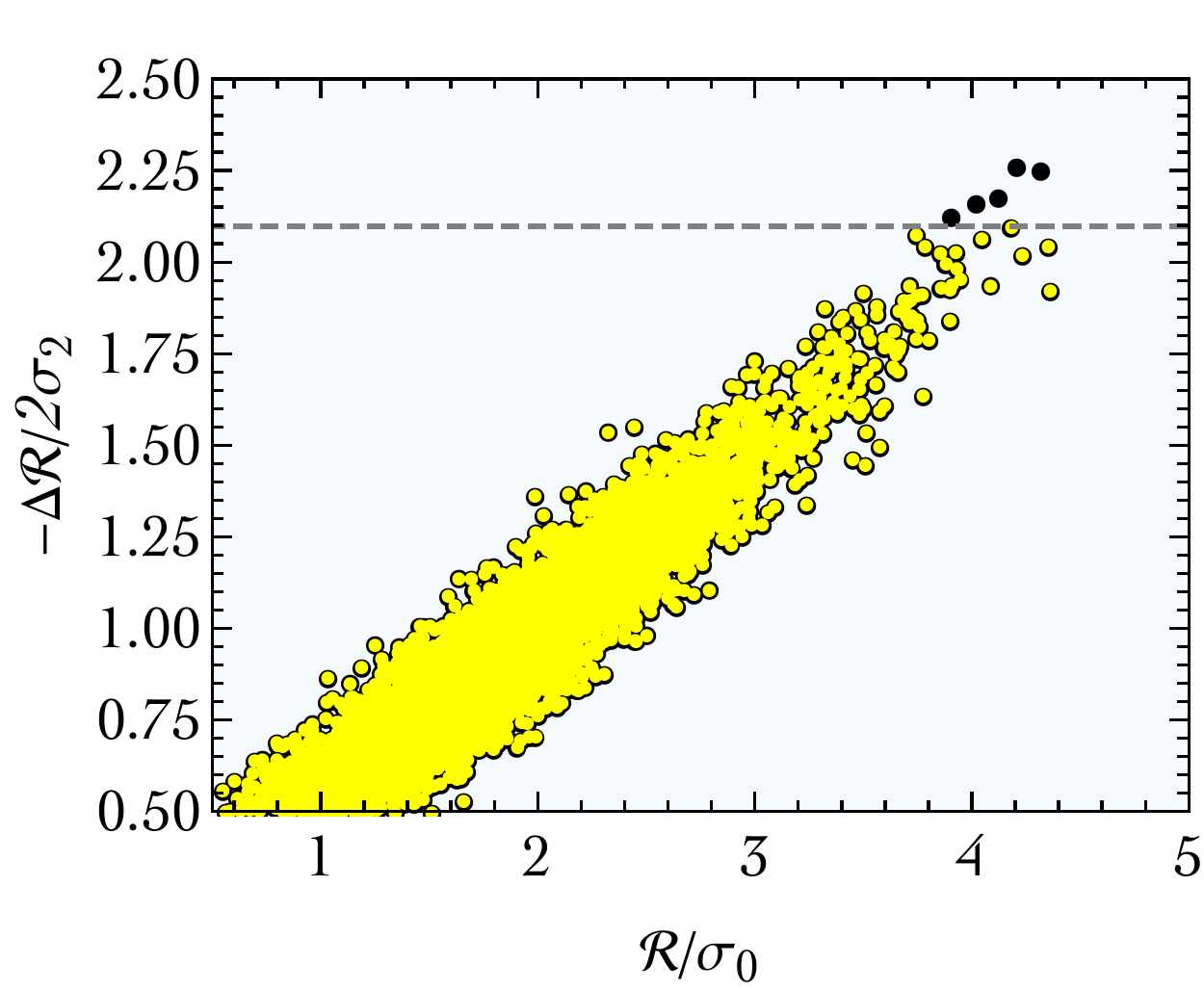}
\qquad
\includegraphics[width=.45\textwidth]{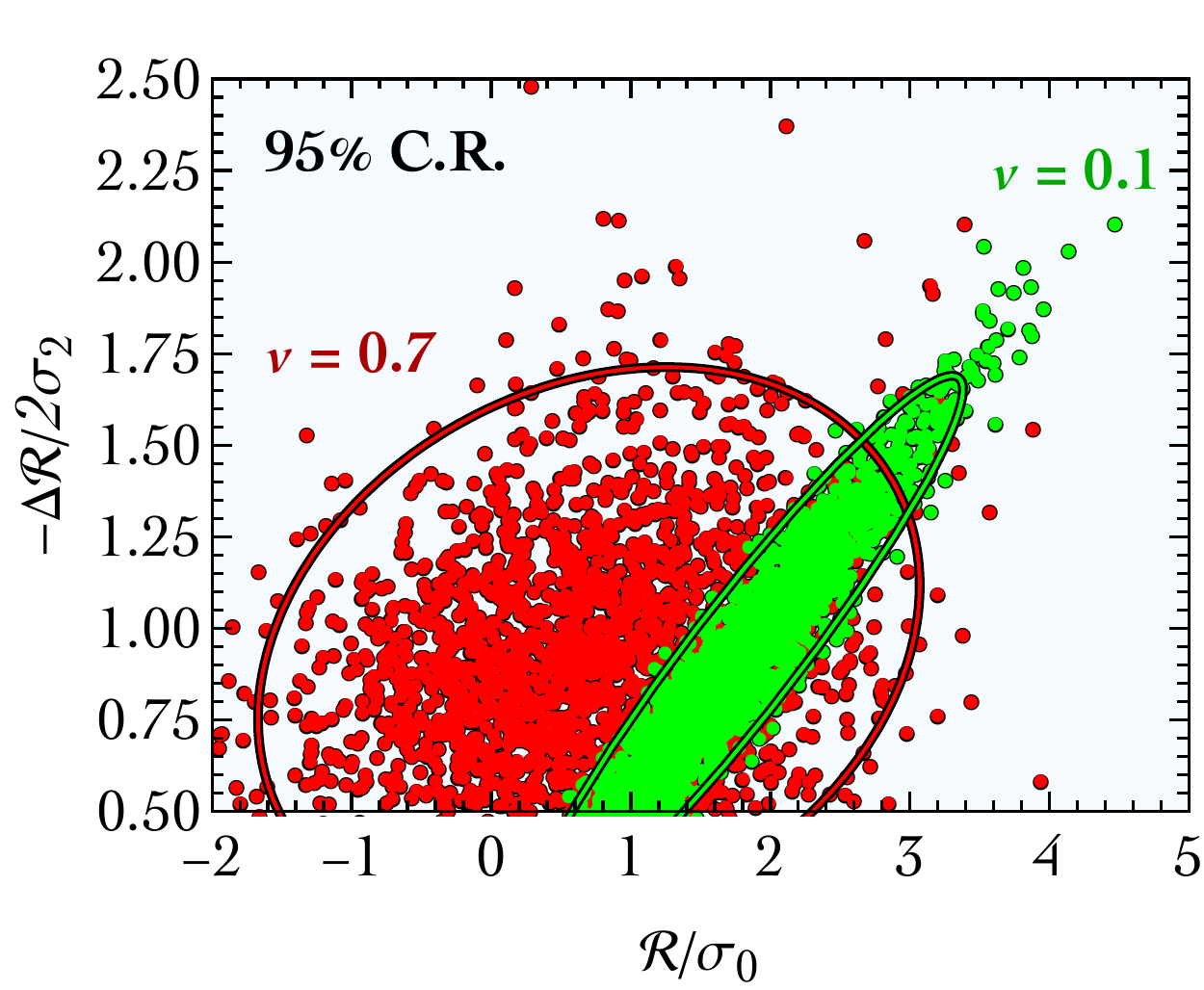}$$
\caption{\em \label{fig:ScanCurvature} 
Left panel. Simulated points in figs.\,\ref{fig:MonteCarlo},\,\ref{fig:SnapShot} shown in the plane $\{\mathcal{R}/\sigma_0, -\triangle\mathcal{R}/2\sigma_2\}$. 
The horizontal dashed line corresponds to the threshold value $\delta_c$ (see caption of fig.\,\ref{fig:MonteCarlo}).
Points above threshold collapse into black holes (marked with black dots here and in figs.\,\ref{fig:MonteCarlo},\,\ref{fig:SnapShot}).
This simulation corresponds to $v=0.1$, and we see that $\mathcal{R}/\sigma_0$ and $-\triangle\mathcal{R}/2\sigma_2$ are strongly correlated. 
Right panel. Comparison between two numerical simulations with $v=0.1$ (green) and $v=0.7$ (red). 
The two ellipses represent the 95\% confidence regions. 
 }
\end{center}
\end{figure}

Next, we ask if there exists some relation between the curvature of a local maximum of $\mathcal{R}$ and the value of $\mathcal{R}$ at the 
maximum.\footnote{The result of this exercise will be useful later to discuss the impact of non-linearities in eq.\,(\ref{eq:Curvature2Density}), see appendix\,\ref{app:NonLin}.}
We can use the probability density distribution $\bar{n}_{\rm max}(\mathcal{R},s)$  
defined in eq.\,(\ref{eq:nMaxRs2}) to generate numerically, in position space, a sample of maxima 
by extracting randomly the value of $\mathcal{R}$ and  curvature $2s  = 
-\triangle \mathcal{R}$. 
We can then use eq.\,(\ref{eq:Curvature2Density}) to extract from the distribution of $s$ the distribution of the overdensity field (of course, with $\mathcal{R}$ instead of $h$ since we are considering here the gaussian case).
The outcome of this exercise is shown in fig.\,\ref{fig:MonteCarlo}, fig.\,\ref{fig:SnapShot} and fig.\,\ref{fig:ScanCurvature} (see captions for details).
 
Let us consider the case in which we take $v=0.1$. We remind that this choice corresponds to a very narrow power spectrum. 
We have $\gamma \simeq 0.98$. As noticed before, in this case we expect a strong correlation 
between $\mathcal{R}$ and $s$.
 This means that 
 regions with large $\mathcal{R}/\sigma_0$ are likely to be also regions with large $2s/\sigma_2$ (and, consequently, regions where the overdensity field peaks).
This is evident in fig.\,\ref{fig:MonteCarlo}, fig.\,\ref{fig:SnapShot} and fig.\,\ref{fig:ScanCurvature} (left panel). 
In particular, in fig.\,\ref{fig:SnapShot} we see that maxima with large values of $\mathcal{R}$ (left panel) maps precisely regions where the 
overdensity field $\delta$ peaks (right panel).
As stated before, this is a consequence of the fact that $\mathcal{R}$ and $s$ (hence $\delta$) are highly correlated.
Furthermore, in the left panel of fig.\,\ref{fig:ScanCurvature} the symmetry of the randomly generated points under the exchange $\mathcal{R}/\sigma_0 \leftrightarrow 2s/\sigma_2$ is evident. 

It is instructive to consider what happens if we take a different value of $v$. Consider, for instance, 
the case with $v=0.7$. We have $\gamma \simeq 0.37$, and $\mathcal{R}$ and $s$  are now much less correlated compared to 
the case with $v=0.1$. This is shown in the right panel of fig.\,\ref{fig:ScanCurvature} (see caption for details). 
For $v=0.7$, spiky maxima (that is maxima with large curvature)  
are seldom characterized also by a large value of $\mathcal{R}$. 
This means that if we aim at deriving a generic formula for the number density of spiky maxima 
we can not restrict eq.\,(\ref{eq:nMaxRs}) to special values of $\mathcal{R}$.
 
Let us summarize our findings so far. 
We have that the number density of local maxima of $\mathcal{R}$ with large curvature gives the number density of regions where the overdensity field peaks. Physically, we are only interested in local maxima with large curvature irrespectively on their value of $\mathcal{R}$. 
Since there is no special relation---for generic values of $v$ (hence $\gamma$)---between $s$ and $\mathcal{R}$, 
eq.\,(\ref{eq:nMaxRs}) must be integrated over the entire range of variability of $\mathcal{R}$.

If we consider the case in which $s_{\rm min}$ does not depend on $\mathcal{R}$, we can integrate eq.\,(\ref{eq:nMaxRs}) analytically. We find the number density 
\begin{align}\label{eq:DoubleIntegral}
\mathcal{N}_{\rm max}(s_{\rm min}) \equiv  \int_{-\infty}^{\infty}d\mathcal{R}
\int_{s_{\rm min}}^{\infty}ds\,\bar{n}_{\rm max}(\mathcal{R},s) = 
\frac{\sigma_2^2}{2\sqrt{2}\pi^{3/2}\sigma_1^2}
\left[
\frac{s_{\rm min}}{\sigma_2}\exp\left(
-\frac{2s_{\rm min}^2}{\sigma_2^2}
\right) + \frac{1}{2}\sqrt{\frac{\pi}{6}}{\rm Erfc}\left(\frac{\sqrt{6}s_{\rm min}}{\sigma_2}\right)
\right]\,.
\end{align}
We can define the dimensionful quantity $R_* \equiv \sqrt{d} \,\sigma_1/\sigma_2$, where $d$ is the number of spatial dimensions. 
From eq.\,(\ref{eq:SigmajAnal}) we find
\begin{align}
R_* = \frac{\sqrt{2}\sigma_1}{\sigma_2} = \frac{\sqrt{2}}{k_{\star}}e^{-3v^2}\,.
\end{align}
In this simplified two-dimensional set-up, we can estimate the mass fraction of black holes by means of the dimensionless quantity 
(for more details, see appendix\,\ref{app:Threshold})
\begin{align}\label{eq:MassFractionGaussian}
R_*^2 \mathcal{N}_{\rm max}(s_{\rm min}) = \frac{1}{\sqrt{2}\pi^{3/2}}
\left[
\frac{s_{\rm min}}{\sigma_2}\exp\left(
-\frac{2s_{\rm min}^2}{\sigma_2^2}
\right) + \frac{1}{2}\sqrt{\frac{\pi}{6}}{\rm Erfc}\left(\frac{\sqrt{6}s_{\rm min}}{\sigma_2}\right)
\right]
\simeq  \frac{1}{\sqrt{2}\pi^{3/2}}\left(
\frac{s_{\rm min}}{\sigma_2}\right)\exp\left(
-\frac{2s_{\rm min}^2}{\sigma_2^2}
\right)
\,,
\end{align}
where the last approximation is valid if $s_{\rm min}/\sigma_2 \gtrsim 1$. 
The mass fraction is controlled by an exponential decaying function with argument 
\begin{align}\label{eq:ExplThresh}
\frac{s_{\rm min}}{\sigma_2} = \frac{9}{8}\frac{(aH)^2}{\sigma_2}\delta_c = 
\frac{9e^{-4v^2}}{8A_g^{1/2}}\left(
\frac{aH}{k_{\star}}
\right)^2\delta_c\,.
\end{align}
The value of $s_{\rm min}/\sigma_2$, which is crucial for the determination of the correct order-of-magnitude of the 
mass fraction, depends on the properties of the power spectrum via the factor $e^{-4v^2}/A_g^{1/2}k_{\star}^2$ and 
the details of the gravitational collapse that leads to black hole formation via the factor  
$(aH)^2\delta_c$.
The comoving horizon length $1/aH$ depends on time. 
The computation of the threshold for black hole production introduces 
the time $t_m$ that is defined by the time when the curvature perturbations cross the horizon and become causally connected. 
This is the time at which eq.\,(\ref{eq:ExplThresh}) has to be computed.\footnote{More precisely, the process of black hole formation involves three different times.
First of all, the number density of peaks that eventually form black holes has to be calculated at some initial time when perturbations are still super-horizon. 
All the analysis done so far is based on this time (even though we do not specify it explicitly) since we are considering comoving curvature perturbations 
which are constant (see appendix\,\ref{app:InflationPrimer}).
Second, we have the time $t_m$ when the perturbations re-enter the horizon and become causally connected. 
Finally, we have the time $t_f$ at which black holes form. 
In general $t_f\neq t_m$, and from simulations of gravitational collapse in numerical relativity we have 
$(t_f/t_m)^{1/2}\simeq 3$\,\cite{Musco:2018rwt}; 
eq.\,(\ref{eq:SpikyEnough}) is usually evaluated at time $t_m$ while the factor $(t_f/t_m)^{1/2}\simeq 3$ can be included 
in the computation of the black hole abundance (see, e.g., discussion in ref.\,\cite{Kalaja:2019uju}).}
The time $t_m$ defines implicitly, by means of the condition $a(t_m)H(t_m)r_m = 1$, the length scale $r_m$. 
This length scale enters in the numerical evaluation of eq.\,(\ref{eq:ExplThresh}), and 
one typically gets $(a_m H_m/k_{\star})^2\delta_c = O(1)$.
A precise numerical evaluation of this factor is needed in order to set the size of the exponential suppression 
in eq.\,(\ref{eq:MassFractionGaussian}). We postpone to appendix\,\ref{app:Threshold} a more detailed discussion about this point.

It is important to check the validity of eq.\,(\ref{eq:MassFractionGaussian}) against the standard result. 
In ref.\,\cite{Bardeen:1985tr}, the number density of peaks of the overdensity field is controlled by the exponential function
$\exp(-\delta_c^2/2\sigma_{\delta}^2)$ where $\sigma_{\delta}^2$ refers to the variance of the overdensity field. 
By means of eq.\,(\ref{eq:Curvature2Density}) (linearized, and with $\mathcal{R}$ instead of $h$), one finds 
$\sigma_{\delta}^2 = (16/81)(1/aH)^4\sigma_2^2$.  
Consequently, $\exp(-\delta_c^2/2\sigma_{\delta}^2)$ matches precisely the exponential function in 
eq.\,(\ref{eq:MassFractionGaussian}). This is another 
indication that computing the number density of peaks of the overdensity field via the number density of maxima of 
$\mathcal{R}$ that are spiky enough leads to sensible results; 
importantly, this is in spite of the fact that we derived eq.\,(\ref{eq:MassFractionGaussian}) in two spatial dimensions.

After this long and detailed discussion about the gaussian case, we are ready to move to the more interesting situation in which 
local non-gaussianities are present. Actually, we are already in the position to make an interesting comment. 
Suppose that we compute the analogue of eq.\,(\ref{eq:nMaxRs2}) with local non-gaussianities  with of course now $h$ and $-\triangle h$ instead of $\mathcal{R}$ and 
$-\triangle\mathcal{R}$. It is crucial to answer the same question that we asked in the gaussian case: 
Is the number density of maxima of $h$ with large curvature $-\triangle h$ a good proxy 
for the number density of peaks of the overdensity field? 
The same analytical argument discussed in eqs.\,(\ref{eq:StatM}-\ref{eq:AnalArgu}) can be repeated with 
$h$ and $-\triangle h$ instead of $\mathcal{R}$ and 
$-\triangle\mathcal{R}$. The estimate of the eigenvalues $\lambda_{i=1,2}$ in eq.\,(\ref{eq:Inter}) remains the same. 
In eq.\,(\ref{eq:AnalArgu}), the only difference is that local non-gaussianities alter the 
entries of the covariance matrix that we used to estimate the magnitude of the gradient field. 
However, we anticipate that we find (see eq.\,(\ref{eq:NonGaussianSecondCumuMatrix})) 
$\langle h_xh_x\rangle = \langle h_yh_y\rangle = \sigma_1^2(1+4\alpha^2\sigma_0^2)/2$ and $\langle h_xh_y\rangle = 0$;
 we conclude that the non-gaussian correction in this case is negligible since we have 
 $\alpha^2\sigma_0^2 = \alpha^2A_g \ll 1$. Remember indeed that the size of $\sigma_0^2$ is controlled 
 by the amplitude of the power spectrum which is $\sigma_0^2 = A_g \ll 1$. This is particularly clear for the simple choice
 of $\mathcal{P}_{\mathcal{R}}$ in eq.\,(\ref{eq:ToyPS}) but remains true in general. 
 
 We conclude that the same argument discussed in eqs.\,(\ref{eq:StatM}-\ref{eq:AnalArgu}) is valid 
 also in the presence of local non-gaussianities. 
 In appendix\,\ref{app:NonGaussianPeakTheoryExact} and appendix\,\ref{app:NonGaussianPeakTheory} we will, therefore, move to compute the number density 
 of maxima of $h$ with curvature above the threshold for black hole production that is needed to generalize eq.\,(\ref{eq:MassFractionGaussian}) to the case $\alpha \neq 0$. 
 We consider two different approaches.
\begin{itemize}

\item [$\ast$] In appendix\,\ref{app:NonGaussianPeakTheoryExact} we come back to the 
formulation of the problem in three spatial dimensions. We will derive an expression analogue to eq.\,(\ref{eq:MassFractionGaussian}) but valid in the case $\alpha \neq 0$ and three spatial dimensions. We dub the result of appendix\,\ref{app:NonGaussianPeakTheoryExact} 
``exact'' because no approximations will be used throughout the computation (apart from the linearization in eq.\,(\ref{eq:Curvature2Density}); the case in which also non-linearities are included will be discussed in appendix\,\ref{app:NonLin}).

\item [$\ast$] In appendix\,\ref{app:NonGaussianPeakTheory} we follow a different route. 
We will consider an expansion in cumulants around the gaussian probability density distribution.  
To make this approach more transparent from the analytic point of view, we will work again in two spatial dimensions. 

\end{itemize}

\section{Peak statistics with local non-gaussianities}
\label{app:NonGaussianPeakTheoryExact}

In appendix\,\ref{app:GaussianPeakTheory} we have shown that the number density of peaks of the overdensity field can be approximated with good accuracy with the number density of maxima of the comoving density perturbation which are spiky enough, i.e. with a Laplacian smaller than a threshold. 
The analysis has been performed assuming
two spatial dimensions, to a have a better control of the analytic expressions, and for ease of visualization of our simulations.
Nevertheless, the same conclusion is also valid in three dimensions, which is the case of physical interest that we are going to consider in this section.
We shall now use these results to compute the  number density of peaks of the overdensity field in presence of local non-gaussianities.
Our starting point is the analogous of eq.\,(\ref{eq:nmax}) for the non-gaussian random field 
$h(\vec{x})$. Furthermore, as explained above, we are interested into maxima of $h$ which are spiky enough.
Explicitly, we have
\begin{align}
\label{eq:peaksh}
n_{\rm pk}(h)\,dh\,d^3x= dh\,d^3x\, \int_{\rm spiky\, max} d^3h_i\,d^6h_{ij}\,P_{\rm NG}(h,h_i,h_{ij})\, \delta^3(h_i)\, |{\rm det}(h_{ij})|\,,
\end{align}
where $P_{\rm NG}(h,h_i,h_{ij})$ is the joint probability distribution function of the non-gaussian field $h,$ the field gradient $h_i$ and the second derivatives $h_{ij}.$ The Dirac delta $\delta^3(h_i)$ enforces the condition that the point is stationary.
We will show in a moment how to restrict the integration volume to the field configurations which represent spiky maxima of $h$.
The comoving number density of peaks is obtained integrating over all the heights of the peaks
\begin{align}
\mathcal{N}_{\rm pk} &= \int_{h_{\rm min}}^{\infty}dh\, n_{\rm pk}(h) = \int_{\rm spiky\, max} dh\,d^3h_i\,d^6h_{ij}\,P_{\rm NG}(h,h_i,h_{ij})\, \delta^3(h_i)\, |{\rm det}(h_{ij})|\,.
\end{align}
Notice that for $\alpha>0$, which is the case relevant for the PBH production (see sec.\,\ref{app:InflationPrimer}), and from the relation
$h(\vec{x})=\mathcal{R}(\vec{x})+\alpha\left( \mathcal{R}(\vec{x})^2-\sigma_0^2\right),$ one realizes that
 $h(x)$ attains a minimum $h_{\rm min}=-(1+4\alpha^2\sigma_0^2)/4\alpha$ for $\mathcal{R}=-1/2\alpha.$ 
 To proceed, it turns out to be useful to consider the gaussian variables $\mathcal{R},$ $\mathcal{R}_i$ and $\mathcal{R}_{ij}$ instead of the non-gaussian fields $h$, $h_{i}$ and $h_{ij}.$
The reason is that the joint probability distribution function of the former variables, that we denote with $P(\mathcal{R},\mathcal{R}_i,\mathcal{R}_{ij})$,  is known and it has a simple analytic expression, see sec.\,\ref{app:GaussianPeakTheory} for the explicit formula in the case of two dimensions.
Using the conservation of the probability in a differential volume $P_{\rm NG}(h,h_i,h_{ij})\,dh\,d^3h_i\,d^6h_{ij} =P(\mathcal{R},\mathcal{R}_i,\mathcal{R}_{ij})\,d\mathcal{R}\,d^3\mathcal{R}_i\,d^6\mathcal{R}_{ij},$ we can perform a change a variable and write
\begin{align}
\label{eq:peakshwithR}
\mathcal{N}_{\rm pk} &= \int_{\rm spiky\, max} d\mathcal{R}\,d^3\mathcal{R}_i\,d^6\mathcal{R}_{ij}\,P(\mathcal{R},\mathcal{R}_i,\mathcal{R}_{ij})\, \delta^3\left[h_i\left(\mathcal{R},\mathcal{R}_i,\alpha\right)\right]\, 
\left|{\rm det}\left[h_{ij}\left( \mathcal{R},\mathcal{R}_i,\mathcal{R}_{ij},\alpha \right) \right] \right|\,.
\end{align}
where the field gradient and the second derivatives of the non gaussian variables are written in terms of the gaussian fields as
\begin{align}
h_i\left(\mathcal{R},\mathcal{R}_i,\alpha\right) =  \mathcal{R}_i \left(1+2\alpha\mathcal{R}\right)\,,~~~~~~~~~~~~~
h_{ij}\left( \mathcal{R},\mathcal{R}_i,\mathcal{R}_{ij},\alpha \right) &= \mathcal{R}_{ij}(1+2 \alpha R)+2\alpha \mathcal{R}_{i} \mathcal{R}_{j}\,.
\end{align}
As noticed in sec.\,\ref{sec:Mot}, stationary points of $\mathcal{R}$ are also stationary points of $h.$ Moreover, from the expressions above, one can see that also the configuration $(1+2 \alpha R)=0$ is a stationary point of $h.$  
However, as mentioned above, it is a minimum of $h$ (for $\alpha>0$), therefore we need to focus only on the configurations with $\mathcal{R}_i=0.$
This means that one can write eq.\,(\ref{eq:peakshwithR}) as
\begin{align}
\label{eq:peakshwithR2}
\mathcal{N}_{\rm pk} &= \int_{\rm spiky\, max} d\mathcal{R}\,d^6\mathcal{R}_{ij}\,P(\mathcal{R},\mathcal{R}_i=0,\mathcal{R}_{ij}) \,\left|{\rm det} \left(h_{ij}\left( \mathcal{R},\mathcal{R}_i=0,\mathcal{R}_{ij},\alpha \right)\right)\right|  \,\frac{1}{\left|1+2\alpha\mathcal{R}\right|^3}\nonumber\\
&= \int_{\rm spiky\, max} d\mathcal{R}\,d^6\mathcal{R}_{ij}\,P(\mathcal{R},\mathcal{R}_i=0,\mathcal{R}_{ij}) \,\left|{\rm det} \left(\mathcal{R}_{ij}\right)\right|\,.
\end{align}
In the equation above all the information about the non-gaussianities is confined in $_{\rm spiky\, max},$ i.e. in the condition to impose on the integration volume in order to restrict on maxima of $h$ spiky enough.
Let us see how these constraints can be implemented explicitly.
The Hessian matrix, for $\mathcal{R}_i=0$, is
\begin{align}
h_{ij}=\mathcal{R}_{ij}\left(1+2\alpha \mathcal{R} \right)\,.
\end{align}
We should, therefore, select maxima (minima) of $\mathcal{R}$ for positive (negative) values of $\left(1+2\alpha \mathcal{R} \right)$.
The calculation can be simplified aligning the coordinate axes along the eigenvectors of the matrix $-\mathcal{R}_{ij}.$ The corresponding eigenvalues are denoted as $\lambda_i$ with $i=1,2,3$ (the same strategy has been adopted in sec.\,\ref{app:GaussianPeakTheory} in the case of two spatial dimensions).
The three eigenvectors, and the three Euler angles defining their orientation, can be used to parametrize the six independent random fields of $\mathcal{R}_{ij}.$
Let us also define the following variables
\begin{align}
\label{eq:defxyz}
\bar{\nu} = \mathcal{R}/\sigma_0\,,~~~~~~
\left\{
\begin{array}{ccc}
\sigma_2 \,x & = &   -\triangle \mathcal{R} = \lambda_1+\lambda_2+\lambda_3 \\
\sigma_2 \,y & = &    (\lambda_1-\lambda_3)/2 \\
\sigma_2 \,z & = &    (\lambda_1-2\lambda_2+\lambda_3 )/2  
\end{array}
\right.
\end{align}
Changing variables and integrating over the Euler angles, from eq.\,(\ref{eq:peakshwithR2}) one obtains\,\cite{Bardeen:1985tr}
\begin{align}
\label{eq:peakshxyz}
\mathcal{N}_{\rm pk} = \int_{\rm spiky\, max} \bar{n}_{\rm pk}(\bar{\nu},x,y,z) d\bar{\nu}\,dx\,dy\,dz\,~~~~~~~~~{\rm with}~~~~~
\bar{n}_{\rm pk}(\bar{\nu},x,y,z) = A\, e^{-Q}\,| F(x,y,z)|\,, 
\end{align}
and
\begin{align}
A & \equiv \sqrt{\frac{5}{3}}\,\frac{\sigma_2^3}{\sigma_1^3}\frac{25}{16\pi^2\,\sqrt{1-\gamma^2}}\,,\\
Q & \equiv \frac{\bar{\nu}^2}{2} + \frac{(x-x_*)^2}{2(1-\gamma^2)} +\frac{5}{2}\left(3y^2+z^2\right)\,,\\
F(x,y,z) & \equiv y\left(y^2-z^2\right)\left[\left(x+z\right)^2-9y^2 \right]\left(x-2z\right)\,,\\
x_* & \equiv \gamma\bar{\nu}\,.\label{eq:AQFx}
\end{align}
We can divide the integration on $\bar{\nu}$ in two parts.
\begin{itemize}
\item [$\circ$] $1+2\alpha\mathcal{R}> 0$. \\
This implies that maxima of $\mathcal{R}$ are also maxima of $h$, and that we are considering $\bar{\nu}>-1/2\sigma_0\alpha.$
 One can choose an ordering for the eigenvalues of the matrix $-\mathcal{R}_{ij}$: $\lambda_1 \geq \lambda_2 \geq \lambda_3.$ Therefore, we can select maxima of $\mathcal{R}$ requiring that $\lambda_3>0.$
Under these conditions, and using eqs.\,(\ref{eq:defxyz}), the domain of integration for the variables $y$ and $z$ reads:
\begin{align}
 \int_0^{x/4} dy\,\int_{-y}^y dz + \int_{x/4}^{x/2} dy\,\int_{3y-x}^y dz\,.
\end{align}
Then, working within the linear approximation in eq.\,(\ref{eq:Curvature2Density}), eq.\,(\ref{eq:SpikyEnough}) implies that only spiky maxima should be selected
\begin{align}
\label{eq:xdelta}
x  > \frac{9 (a_m H_m)^2}{4\sigma_2}\frac{\delta_c}{1+2\alpha\sigma_0\bar{\nu}} \equiv x_{\delta}(\bar{\nu})\,.
\end{align}
In the equation above, the horizon scale $1/a\,H$ has been evaluated at the time $t_m$ when the
curvature perturbations cross the horizon.
Having specified the appropriate domain of integration for all the variables, we can now integrate eq.\,(\ref{eq:peakshxyz}) and multiply by a factor 6 to take into account all the possible orderings of the eigenvalues $\lambda_i$.
We find
\begin{align}
\label{eq:Npk1}
\mathcal{N}_{\rm pk}^{({\rm I})} &= \int_{-\frac{1}{2\alpha\sigma_0}}^{\infty} d\bar{\nu}\, \int_{x_{\delta}(\bar{\nu})}^{\infty} dx\, \bar{n}_{\rm pk}(\bar{\nu},x)\,,
\end{align}
where
\begin{align}
\label{eq:npkxy}
\bar{n}_{\rm pk}(\bar{\nu},x) &= \frac{e^{-\bar{\nu}^2/2}}{(2\pi)^2 R_*^3}\, f(x) \frac{e^{-\frac{(x-x_*)^2}{2(1-\gamma^2)}}}{\sqrt{2\pi(1-\gamma^2)}}\,,
\end{align}
\begin{align}
\label{eq:npkxyfx}
f(x)=\frac{x^3-3x}{2}\left[ {\rm Erf}\left(\sqrt{\frac{5}{2}}\,x\right)+{\rm Erf}\left(\sqrt{\frac{5}{2}}\,\frac{x}{2}\right) \right]+\sqrt{\frac{2}{5\pi}}\left[ \left(\frac{31x^2}{4}+\frac{8}{5}\right) e^{-\frac{5}{8}x^2} +\left(\frac{x^2}{2}-\frac{8}{5}\right) e^{-\frac{5}{2}x^2}
 \right]\,,
\end{align}
and we remind that $R_* \equiv \sqrt{3}\sigma_1/\sigma_2.$

\item [$\circ$] $1+2\alpha\mathcal{R}<0$. \\
In this case minima of $\mathcal{R}$ are maxima of $h.$ Proceeding analogously as before, we have:
\begin{align}
\label{eq:Npk2}
\mathcal{N}_{\rm pk}^{({\rm II})} &= \int_{-\infty}^{-\frac{1}{2\alpha\sigma_0}}d\bar{\nu}\, \int_{-\infty}^{x_{\delta}(\bar{\nu})} dx\, \bar{n}_{\rm pk}(\bar{\nu},x)\,.
\end{align}
\end{itemize}

The comoving number density of peaks of the overdensity field is approximated as the sum of the two terms above: $\mathcal{N}_{\rm pk}=\mathcal{N}_{\rm pk}^{({\rm I})}+\mathcal{N}_{\rm pk}^{({\rm II})}$. 
One can interpret these two contributions along the lines exposed in section\,\ref{sec:Mot}. The term $\mathcal{N}_{\rm pk}^{({\rm II})}$ counts then minima of $\mathcal{R}$ which are maxima of $h.$ Instead $\mathcal{N}_{\rm pk}^{({\rm I})}$ corresponds to maxima of $\mathcal{R}$, and the restriction in the integration range of $\bar{\nu}$ (the lower limit) is designed to subtract those maxima of $\mathcal{R}$ which are minima of $h$. 
Obviously, the gaussian result is recovered for $\alpha\to 0$. In this limit we can compare with the standard expression in ref.\,\cite{Bardeen:1985tr}.
Numerically, and for the power spectra under consideration, we found that the two calculations agree within a factor $\simeq 2$. 
As already mentioned in section\,\ref{app:GaussianPeakTheory} for the case of two spatial dimensions, this confirms that peaks of the overdensity field are well approximated by peaks of the comoving density perturbation which are spiky enough. 
In appendix\,\ref{app:Threshold} we will discuss how to translate $\mathcal{N}_{\rm pk}$ into the primordial 
black hole abundance in eq.\,(\ref{eq:MasterFormula}).

\section{Peak theory with local non-gaussianities: a perturbative approach in 2D}\label{app:NonGaussianPeakTheory}

The idea is to derive the joint probability density distribution for the variables 
$h$, $h_{x}$, $h_{y}$, $h_{xx}$, $h_{xy}$, $h_{yy}$ (considering again, for simplicity, the two-dimensional case). 
Once we get this probability density distribution, we set $h_{x}=h_{y} = 0$ and we integrate over 
$h_{xx}$, $h_{xy}$ and $h_{yy}$ in the domain defining maxima.
This strategy was simple to implement in the case of the gaussian variable $\mathcal{R}$ because the joint 
probability density distribution was a multivariate normal distribution. 
The non-gaussian case is more complicated. 

In order to tackle the problem, we shall use the approach based on the characteristic function. 
In full generality, for a set of $N$ correlated random variables $\xi_i$ the characteristic function 
is the Fourier transform of their joint probability density distribution 
\begin{align}\label{eq:CharaDef}
\chi(\lambda_1,\dots,\lambda_N) &\equiv \int
d\xi_1\dots d\xi_N P(\xi_1,\dots,\xi_N)\exp\left[
i(\xi_1 \lambda_1 + \dots + \xi_N \lambda_N)
\right] \nn\\
& = \int d\xi_1\dots d\xi_N P(\xi_1,\dots,\xi_N)\left[
1+ i(\xi_1 \lambda_1 + \dots + \xi_N \lambda_N) + \frac{i^2}{2!}
(\xi_1 \lambda_1 + \dots + \xi_N \lambda_N)^2 
+\dots
\right]\nn\\
& = 1 + i\sum_{j} \langle \xi_j \rangle\lambda_j  +
\frac{i^2}{2!}\sum_{j_1,j_2}
\langle \xi_{j_1}\xi_{j_2} \rangle \lambda_{j_1}\lambda_{j_2} +
\frac{i^3}{3!}\sum_{j_1,j_2,j_3}
\langle \xi_{j_1}\xi_{j_2}\xi_{j_3} \rangle \lambda_{j_1}\lambda_{j_2}\lambda_{j_3}+ \dots  
\end{align}
where, after a Taylor expansion, we introduced the moments of the joint distribution by means of the integrals 
\begin{align}
\langle \xi_{j_1}\dots\xi_{j_k}\rangle = 
\int
d\xi_1\dots d\xi_N P(\xi_1,\dots,\xi_N)\xi_{j_1}\dots \xi_{j_k}\,.
\end{align}
If we take the natural log of the characteristic function and Taylor expand, we define the cumulants 
\begin{align}\label{eq:DefCumu}
\log\chi(\lambda_1,\dots,\lambda_N) \equiv 
i\sum_j C_1(\xi_j)\lambda_j + 
\frac{i^2}{2!}\sum_{j_1,j_2}C_2(\xi_{j_1},\xi_{j_2})\lambda_{j_1}\lambda_{j_2} + \dots\,.
\end{align}
The relation between moments and cumulants follows from the comparison of the two Taylor expansions. We find the well-known relations
\begin{align}
C_1(\xi_j) & = \langle \xi_{j}\rangle\,,\label{eq:FirstCum}\\
C_2(\xi_{j_1},\xi_{j_2}) & = \langle \xi_{j_1}\xi_{j_2}\rangle -\langle \xi_{j_1}\rangle \langle\xi_{j_2}\rangle\,,\label{eq:SecondCum}\\
C_3(\xi_{j_1},\xi_{j_2},\xi_{j_3}) & = 
\langle \xi_{j_1}\xi_{j_2}\xi_{j_3}\rangle - \langle \xi_{j_1}\rangle\langle \xi_{j_2}\xi_{j_3}\rangle -
\langle \xi_{j_2}\rangle\langle \xi_{j_1}\xi_{j_3}\rangle -
\langle \xi_{j_3}\rangle\langle \xi_{j_1}\xi_{j_2}\rangle 
+2\langle \xi_{j_1}\rangle\langle \xi_{j_2}\rangle\langle \xi_{j_3}\rangle\,,\label{eq:ThirdCum}\\
C_4(\xi_{j_1},\xi_{j_2},\xi_{j_3},\xi_{j_4}) & =  \dots\label{eq:FourthCum}
\end{align} 
and so on.
Working with cumulants instead of moments is more efficient. 
This is particularly true in the gaussian case.
For a set of correlated gaussian variables, 
all cumulants of order higher than two vanish.
In the gaussian case, eq.\,(\ref{eq:DefCumu}) gives 
\begin{align}\label{eq:Babel}
\chi(\lambda_1,\dots,\lambda_N) = \exp\left[
i\sum_j \langle \xi_{j}\rangle \lambda_j 
- \frac{1}{2}\sum_{j_1,j_2}C_2(\xi_{j_1},\xi_{j_2})\lambda_{j_1}\lambda_{j_2}
\right]\,,
\end{align}
and the inverse Fourier transform that gives the probability density distribution reads 
\begin{align}
P(\xi_1,\dots,\xi_N) = \int\frac{d\lambda_1}{(2\pi)}\dots
\frac{d\lambda_N}{(2\pi)}
\exp\left[
i\sum_j\left(\langle \xi_{j}\rangle - \xi_j\right) \lambda_j 
- \frac{1}{2}\sum_{j_1,j_2}C_2(\xi_{j_1},\xi_{j_2})\lambda_{j_1}\lambda_{j_2}
\right]\,.
\end{align}
If we complete the square inside the integrand and compute the resulting multivariate gaussian 
integral, we find precisely eq.\,(\ref{eq:JointPDFGauss}) where the second-order cumulants 
reconstruct the covariance matrix elements in eq.\,(\ref{eq:CovarianceMatrixElements}).
This was precisely the strategy that we followed in the previous section: we computed the elements of the covariance matrix (that are the second-order cumulants) and we (implicitly) performed an 
inverse Fourier transform to get back the probability density distribution.   

In the non-gaussian case, we can try  to apply the same logic. First, we compute the cumulants; second, we reconstruct the probability density distribution by means of an inverse Fourier transform.

The computation of the cumulants require some mathematical tricks that we shall explain in the following.
In full generality, we need to compute the $n^{\rm th}$-order cumulant 
$C_n(\partial_{j} h,\dots,\partial_{l} h)$ where each $\partial_{j}h$ represents a certain number 
of spatial derivatives (zero, one or two) acting on $h$. 
Remember also that the random field $h$ (and its derivatives) is computed at a specific spatial position (say, $\vec{x}$) that will be later identified with a 
stationary point. 
We can write 
\begin{align}
C_n(\partial_{j} h,\dots,\partial_{l} h) & = C_n[\partial_{j} h(\vec{x}),\dots,\partial_{l} h(\vec{x})] \nn \\
& = \left.
C_n[\partial_{j_1} h(\vec{x}_1),\dots,\partial_{l_n} h(\vec{x}_n)]\right|_{\vec{x}_1 = \dots = \vec{x}_n= \vec{x}} 
=  \partial_{j_1}\dots \partial_{l_n}
\left.C_n[h(\vec{x}_1),\dots,h(\vec{x}_n)]\right|_{\vec{x}_1 = \dots = \vec{x}_n= \vec{x}}\,. \label{eq:Cumutrick}
\end{align}
In the first step of eq.\,(\ref{eq:Cumutrick}) we consider each $\partial_{j_k}h$ to act at a different point $\vec{x}_k$, and later we set all points equal again (we already used a similar trick in eq.\,(\ref{eq:A1})).
The advantage of this step is that since now each derivative acts at a
different spatial point, we can bring them outside the cumulant (as done in the last step of eq.\,(\ref{eq:Cumutrick})). 
It is a very simple exercise to check explicitly (for instance, by computing second-order cumulants of 
a gaussian variable with its derivatives) that this procedure is completely legitimate. 
Using our notation $h(\vec{x}_n) = h_n$, the problem reduces to the computation
of $C_n(h_1,\dots,h_n)$ where now the random fields are evaluated at different spatial points. 
After computing $C_n(h_1,\dots,h_n)$, we will take the spatial derivatives and finally set all points equal according 
to the prescription in eq.\,(\ref{eq:Cumutrick}).

To compute $C_n(h_1,\dots,h_n)$, 
we consider the relation $h = \mathcal{R} + \alpha\mathcal{R}^2$. We write in general 
 $h = \mathcal{R}  + f_{\rm NL}(\mathcal{R})$ where $f_{\rm NL}(\mathcal{R})$ can be a generic non-linear function 
 controlled by the parameter $\alpha$. 
 Let us consider the term $\alpha\mathcal{R}^2$ as a perturbation and 
 expand in $\alpha$. We have 
 \begin{align}
C_n(h_1,\dots,h_n) = C_n(\mathcal{R}_1,\dots,\mathcal{R}_n) & + 
C_n[f_{\rm NL}(\mathcal{R}_1),\dots,\mathcal{R}_n] + \dots  +
C_n[\mathcal{R}_1,\dots,f_{\rm NL}(\mathcal{R}_n)] \label{eq:CumuAlpha}  \\
& +\left\{ 
C_n[f_{\rm NL}(\mathcal{R}_1),f_{\rm NL}(\mathcal{R}_2),\dots,\mathcal{R}_n] + \dots \right\} + O(\alpha^3)\,.\label{eq:CumuAlpha2}
\end{align} 
This deconstruction makes clear the fact that we can compute the cumulants $C_n(h_1,\dots,h_n)$ based on 
the cumulants computed for the gaussian random field $\mathcal{R}$.

The problem is now the computation of $C_n[f_{\rm NL}(\mathcal{R}_1),\dots,\mathcal{R}_n]$ 
where we remind that $\mathcal{R}_n = \mathcal{R}(\vec{x}_n)$. 

In order to compute $C_n[f_{\rm NL}(\mathcal{R}_1),\dots,\mathcal{R}_n]$, we need to work out an intermediate result.
First, remember the definition of random field that we gave at the beginning of appendix\,\ref{app:GaussianPeakTheory}: 
The scalar random field $\mathcal{R}(\vec{x})$ is a set of random variables, one for each point $\vec{x}$ in space, equipped with a joint probability density distribution 
$p(\mathcal{R}_1,\dots ,\mathcal{R}_n)$. 
When the random field is gaussian, $p(\mathcal{R}_1,\dots ,\mathcal{R}_n)$ is a
multivariate Gaussian distribution and we can write 
 \begin{align}\label{eq:BasicGau}
 p(\mathcal{R}_1,\dots ,\mathcal{R}_n) = \frac{1}{(2\pi)^{n/2}\sqrt{{\rm det}\sigma}}
 \exp\left[-\frac{1}{2}\sum_{i,j}(\sigma^{-1})_{ij}\mathcal{R}_i\mathcal{R}_j\right]\,,
\end{align} 
where $\sigma_{ij} = \langle\mathcal{R}_i\mathcal{R}_j\rangle$ (as before, we consider zero-mean random field).
 
Consider now a new set of random variables $\{\mathcal{Y}_1,\dots,\mathcal{Y}_n\}$ related to the previous one 
via the transformations $\mathcal{Y}_1 = g_1(\mathcal{R}_1,\dots ,\mathcal{R}_n)$, $\dots$,
$\mathcal{Y}_n = g_n(\mathcal{R}_1,\dots ,\mathcal{R}_n)$ with inverse 
$\mathcal{R}_1 = g_1^{-1}(\mathcal{Y}_1,\dots ,\mathcal{Y}_n)$, $\dots$,
$\mathcal{R}_n = g_n^{-1}(\mathcal{Y}_1,\dots ,\mathcal{Y}_n)$. 
The joint probability density distribution for the transformed variables is given by 
\begin{align}\label{eq:TransformedPDF}
p_{\mathcal{Y}}(\mathcal{Y}_1,\dots ,\mathcal{Y}_n) = 
p[g_1^{-1}(\mathcal{Y}_1,\dots ,\mathcal{Y}_n),\dots ,g_n^{-1}(\mathcal{Y}_1,\dots ,\mathcal{Y}_n)]\left|
{\rm det}J(\mathcal{Y}_1,\dots ,\mathcal{Y}_n)
\right|\,,
\end{align} 
where the Jacobian matrix is 
\begin{align}
J(\mathcal{Y}_1,\dots ,\mathcal{Y}_n) \equiv
\left(
\begin{array}{ccc}
 \frac{\partial\mathcal{R}_1}{\partial\mathcal{Y}_1} & \dots  & \frac{\partial\mathcal{R}_1}{\partial\mathcal{Y}_n}  \\
 \vdots & \ddots  & \vdots  \\
\frac{\partial\mathcal{R}_n}{\partial\mathcal{Y}_1} & \dots  & \frac{\partial\mathcal{R}_n}{\partial\mathcal{Y}_n}  
\end{array}
\right)\,,~~~~~~~~~~{\rm with}~~~~~\frac{\partial\mathcal{R}_j}{\partial\mathcal{Y}_k} = 
\frac{\partial}{\partial\mathcal{Y}_k}\left[g_j^{-1}(\mathcal{Y}_1,\dots ,\mathcal{Y}_n)\right]\,.
\end{align} 
Eq.\,(\ref{eq:TransformedPDF}) is known as Jabobi's multivariate theorem.

We note that in eq.\,(\ref{eq:TransformedPDF}) we can use the relation $\left|
{\rm det}J(\mathcal{Y}_1,\dots ,\mathcal{Y}_n)
\right| = 1/\left|
{\rm det}J(\mathcal{R}_1,\dots ,\mathcal{R}_n)
\right|$ where $J(\mathcal{R}_1,\dots ,\mathcal{R}_n)$ is the Jacobian matrix of the 
inverse transformation with elements 
$\partial\mathcal{Y}_j/\partial\mathcal{R}_k = \partial g_j(\mathcal{R}_1,\dots ,\mathcal{R}_n)/\partial\mathcal{R}_k$. 
This implies that we can rewrite 
\begin{align}\label{eq:TransformedPDF2}
p_{\mathcal{Y}}(\mathcal{Y}_1,\dots ,\mathcal{Y}_n) = 
\int d\mathcal{R}_1\dots d\mathcal{R}_n p(\mathcal{R}_1,\dots ,\mathcal{R}_n)
\delta[\mathcal{Y}_1 - g_1(\mathcal{R}_1,\dots ,\mathcal{R}_n)]\dots
\delta[\mathcal{Y}_n - g_n(\mathcal{R}_1,\dots ,\mathcal{R}_n)]\,,
\end{align} 
and eq.\,(\ref{eq:TransformedPDF}) follows from eq.\,(\ref{eq:TransformedPDF2}) if one applies the transformation property of the multi-dimensional delta function
 \begin{align}\label{eq:DeltaFunctionIdentity}
\delta[\mathcal{Y}_1 - g_1(\mathcal{R}_1,\dots ,\mathcal{R}_n)]\dots
\delta[\mathcal{Y}_n - g_n(\mathcal{R}_1,\dots ,\mathcal{R}_n)] =  
\frac{\delta[\mathcal{R}_1-g_1^{-1}(\mathcal{Y}_1,\dots ,\mathcal{Y}_n)]\dots
\delta[\mathcal{R}_n-g_n^{-1}(\mathcal{Y}_1,\dots ,\mathcal{Y}_n)]}{\left|
{\rm det}J(\mathcal{R}_1,\dots ,\mathcal{R}_n)
\right|}\,,
\end{align}
and integrates over $\mathcal{R}_i$.  
Eq.\,(\ref{eq:TransformedPDF2}) is a very useful formula. Consider the characteristic function of $p_{\mathcal{Y}}(\mathcal{Y}_1,\dots ,\mathcal{Y}_n)$
defined by the Fourier transform
\begin{align}
\chi_{\mathcal{Y}}(\lambda_1,\dots,\lambda_n) = 
\int d\mathcal{Y}_1\dots d\mathcal{Y}_n p_{\mathcal{Y}}(\mathcal{Y}_1,\dots ,\mathcal{Y}_n)
\exp\left[i(\mathcal{Y}_1 \lambda_1 + \dots + \mathcal{Y}_n \lambda_n)\right]\,.
\end{align} 
Using eq.\,(\ref{eq:TransformedPDF2}) and integrating over $\mathcal{Y}_i$ thanks to the delta functions, we find 
\begin{align}
\chi_{\mathcal{Y}}(\lambda_1,\dots,\lambda_n) = 
\int d\mathcal{R}_1\dots d\mathcal{R}_n 
e^{i[g_1(\mathcal{R}_1,\dots ,\mathcal{R}_n) \lambda_1 + \dots + g_n(\mathcal{R}_1,\,\dots\,,\mathcal{R}_n)\lambda_n]}\,
p(\mathcal{R}_1,\dots ,\mathcal{R}_n)\,.
\end{align}
From the definition of cumulants with respect to the characteristic function  $\chi_{\mathcal{Y}}(\lambda_1,\dots,\lambda_n)$, 
we finally find (see eq.\,(\ref{eq:DefCumu}))
\begin{align}
C_n(\mathcal{Y}_1,\dots,\mathcal{Y}_n) & = \left.(-i)^n\frac{\partial}{\partial\lambda_1}\dots
\frac{\partial}{\partial\lambda_n}\log\chi_{\mathcal{Y}}(\lambda_1,\dots,\lambda_n)\right|_{\lambda_1=\,\dots\,=\lambda_n = 0}
\nn\\
& = \left.(-i)^n\frac{\partial}{\partial\lambda_1}\dots
\frac{\partial}{\partial\lambda_n}\log
\int d\mathcal{R}_1\dots d\mathcal{R}_n 
e^{i[g_1(\mathcal{R}_1,\dots ,\mathcal{R}_n) \lambda_1 + \dots + g_n(\mathcal{R}_1,\dots ,\mathcal{R}_n)\lambda_n]}\,
p(\mathcal{R}_1,\dots ,\mathcal{R}_n)\right|_{\lambda_1=\,\dots\,=\lambda_n = 0}\,.
\label{eq:FinalCumu}
\end{align}
This is a generic formula that can be applied to our case. 

Consider the linear order in $\alpha$. From eq.\,(\ref{eq:FinalCumu}), the cumulant 
$C_n[f_{\rm NL}(\mathcal{R}_1),\dots,\mathcal{R}_n]$ 
is given by 
\begin{align}\label{eq:LinearCumu}
C_n[f_{\rm NL}(\mathcal{R}_1),\dots,\mathcal{R}_n] = 
\left.(-i)^n\frac{\partial}{\partial\lambda_1}\dots
\frac{\partial}{\partial\lambda_n}\log
\int d\mathcal{R}_1\dots d\mathcal{R}_n 
e^{i[f_{\rm NL}(\mathcal{R}_1) \lambda_1 + \mathcal{R}_2\lambda_2 + \dots + \mathcal{R}_n\lambda_n]}\,
p(\mathcal{R}_1,\dots ,\mathcal{R}_n)\right|_{\lambda_1=\,\dots\,=\lambda_n = 0}\,.
\end{align}
A similar formula is applicable to all remaining first-order terms in eq.\,(\ref{eq:CumuAlpha}).
Similarly, at order $O(\alpha^2)$ one needs to compute, for instance, cumulants like
\begin{align}
C_n[f_{\rm NL}&(\mathcal{R}_1),f_{\rm NL}(\mathcal{R}_2),\dots,\mathcal{R}_n] = \nn \\
& \left.(-i)^n\frac{\partial}{\partial\lambda_1}\dots
\frac{\partial}{\partial\lambda_n}\log
\int d\mathcal{R}_1\dots d\mathcal{R}_n 
e^{i[f_{\rm NL}(\mathcal{R}_1) \lambda_1 + f_{\rm NL}(\mathcal{R}_2)\lambda_2 + \dots + \mathcal{R}_n\lambda_n]}\,
p(\mathcal{R}_1,\dots ,\mathcal{R}_n)\right|_{\lambda_1=\,\dots\,=\lambda_n = 0}\,.
\label{eq:QuadraticCumu}
\end{align}
All we need to do is computing the integral and taking derivatives with respect to $\lambda_i$. 
The computation of the integrals is simplified by the fact that the variables $\mathcal{R}_i$ are gaussian 
with joint distribution given by eq.\,(\ref{eq:BasicGau}).

Before proceeding, an important remark is in order.
For the sake of simplicity, we derived eq.\,(\ref{eq:TransformedPDF}) assuming that 
 the functions $\mathcal{Y}_{i=1,\dots,n} = g_{i=1,\dots,n}(\mathcal{R}_1,\dots ,\mathcal{R}_n)$ define one-to-one mappings. 
 In this case, there exist unique inverse
functions so that $\mathcal{R}_{i=1,\dots,n} = g_{i=1,\dots,n}^{-1}(\mathcal{Y}_1,\dots ,\mathcal{Y}_n)$.
The case we are interested in, however, is not exactly of this form. 
If we solve $h=\mathcal{R} + \alpha\mathcal{R}^2$ for $\mathcal{R}$, we indeed find $\mathcal{R} = (-1\pm\sqrt{1+4\alpha h})/2\alpha$ 
which has two roots.

However, this is not an insurmountable problem since Jabobi's multivariate theorem in eq.\,(\ref{eq:TransformedPDF}) can be generalized to the case in which the system
 $\mathcal{Y}_{i=1,\dots,n} = g_{i=1,\dots,n}(\mathcal{R}_1,\dots ,\mathcal{R}_n)$
admits at most a countable number of roots. 
Let us indicate these $q=1,\dots,Q$ roots in the form $\mathcal{R}_{q,i=1,\dots,n} = g_{q,i=1,\dots,n}^{-1}(\mathcal{Y}_1,\dots ,\mathcal{Y}_n)$.
Eq.\,(\ref{eq:TransformedPDF}) becomes
 \begin{align}\label{eq:TransformedPDFroots}
p_{\mathcal{Y}}(\mathcal{Y}_1,\dots ,\mathcal{Y}_n) = \sum_{q=1}^{Q}
p[g_{q,1}^{-1}(\mathcal{Y}_1,\dots ,\mathcal{Y}_n),\dots ,g_{q,n}^{-1}(\mathcal{Y}_1,\dots ,\mathcal{Y}_n)]\left|
{\rm det}J_q(\mathcal{Y}_1,\dots ,\mathcal{Y}_n)
\right|\,,
\end{align} 
where now $J_q(\mathcal{Y}_1,\dots ,\mathcal{Y}_n)$ is the Jacobian corresponding to the $q^{\rm th}$ root
\begin{align}
J_q(\mathcal{Y}_1,\dots ,\mathcal{Y}_n) \equiv
\left(
\begin{array}{ccc}
 \frac{\partial\mathcal{R}_{q,1}}{\partial\mathcal{Y}_1} & \dots  & \frac{\partial\mathcal{R}_{q,1}}{\partial\mathcal{Y}_n}  \\
 \vdots & \ddots  & \vdots  \\
\frac{\partial\mathcal{R}_{q,n}}{\partial\mathcal{Y}_1} & \dots  & \frac{\partial\mathcal{R}_{q,n}}{\partial\mathcal{Y}_n}  
\end{array}
\right)\,,~~~~~~~~~~{\rm with}~~~~~\frac{\partial\mathcal{R}_{q,j}}{\partial\mathcal{Y}_k} = 
\frac{\partial}{\partial\mathcal{Y}_k}\left[g_{q,j}^{-1}(\mathcal{Y}_1,\dots ,\mathcal{Y}_n)\right]\,.
\end{align} 
The multi-dimensional delta-function identity in eq.\,(\ref{eq:DeltaFunctionIdentity}) changes accordingly,  and 
the final result in eq.\,(\ref{eq:FinalCumu}) remains unaltered. 

After this digression, we are ready to compute the integral in eq.\,(\ref{eq:FinalCumu}). 
Instead of looking for a generic expression, let us consider specific cases organized for increasing level of difficulty.
\begin{itemize}
\item [$\circ$] The simplest possibility is to truncate the analysis at the linear order in $\alpha$, eq.\,(\ref{eq:LinearCumu}). 
Consider the integral in eq.\,(\ref{eq:LinearCumu}) which we rewrite as
\begin{align}\label{eq:RI1}
\mathcal{I}(\lambda_1,\dots,\lambda_n) \equiv \int d\mathcal{R}_1
e^{if_{\rm NL}(\mathcal{R}_1) \lambda_1}
\bigg[\underbrace{\int d\mathcal{R}_2 \dots d\mathcal{R}_n 
e^{i[\mathcal{R}_2\lambda_2 +\,\dots\,+ \mathcal{R}_n\lambda_n]}\,
p(\mathcal{R}_1,\dots ,\mathcal{R}_n)}_{\equiv\,\mathcal{I}^{\prime}(\mathcal{R}_1,\lambda_2,\dots,\lambda_n)}\bigg]\,.
\end{align} 
The key observation is that the integral inside the square brackets would precisely match the definition of the 
characteristic function $\chi(\lambda_1,\dots,\lambda_n)$ of $p(\mathcal{R}_1,\dots ,\mathcal{R}_n)$
 if it were completed by an additional integration 
over $\mathcal{R}_1$ (see the definition in the first line of eq.\,(\ref{eq:CharaDef})). 
In such case, we could simply use (see eq.\,(\ref{eq:Babel}))
\begin{align}
\int d\mathcal{R}_1 e^{i\lambda_1\mathcal{R}_1} \mathcal{I}^{\prime}(\mathcal{R}_1,\lambda_2,\dots,\lambda_n)= \chi(\lambda_1,\dots,\lambda_n) = \exp\bigg[-\frac{1}{2}\sum_{ij}\sigma_{ij}\lambda_i\lambda_j
\bigg]\,,
\end{align}
 since $\{\mathcal{R}_{1},\dots,\mathcal{R}_{n}\}$ are gaussian. 
However, the fact that integration over  $\mathcal{R}_1$ is missing implies that the integral $\mathcal{I}^{\prime}(\mathcal{R}_1,\lambda_2,\dots,\lambda_n)$
is actually equal to the inverse Fourier transform of $\chi(\lambda_1,\dots,\lambda_n)$ with respect to $\lambda_1$. 
In formulas, eq.\,(\ref{eq:RI1}) becomes 
\begin{align}\label{eq:RI2}
\mathcal{I}(\lambda_1,\dots,\lambda_n) 
& = \int d\mathcal{R}_1
e^{if_{\rm NL}(\mathcal{R}_1) \lambda_1}
\int \frac{d\lambda_1}{(2\pi)}e^{-i\lambda_1\mathcal{R}_1}
\exp\bigg[-\frac{1}{2}\sum_{ij}\sigma_{ij}\lambda_i\lambda_j
\bigg]\nn \\
& =\int d\mathcal{R}_1
e^{if_{\rm NL}(\mathcal{R}_1) \lambda_1}
\exp\bigg[-\frac{1}{2}\sum_{ij\geqslant 2}\sigma_{ij}\lambda_i\lambda_j
\bigg]\int \frac{d\lambda_1}{(2\pi)}
\exp\bigg[
-\frac{1}{2}\sigma_{11}\lambda_1^2 - 
\bigg(
i\mathcal{R}_1 + \sum_{j\geqslant 2}\sigma_{1j}\lambda_j
\bigg)\lambda_1
\bigg]\nn \\
& =\int d\mathcal{R}_1
e^{if_{\rm NL}(\mathcal{R}_1) \lambda_1}
\frac{1}{\sqrt{2\pi\sigma_{11}}}
\exp\bigg[
-\frac{1}{2}\sum_{ij\geqslant 2}\sigma_{ij}\lambda_i\lambda_j 
+\frac{1}{2\sigma_{11}}
\bigg(
i\mathcal{R}_1 + \sum_{j\geqslant 2}\sigma_{1j}\lambda_j
\bigg)^2
\bigg]\,,
\end{align}
where we just reorganized terms before integrating over $\lambda_1$ in the last step. 
If we take the natural log, we find
\begin{align}\label{eq:RI3}
\log[\mathcal{I}(\lambda_1,\dots,\lambda_n) 
] =-\frac{1}{2}\sum_{ij\geqslant 2}\sigma_{ij}\lambda_i\lambda_j 
+\log\int d\mathcal{R}_1
e^{if_{\rm NL}(\mathcal{R}_1) \lambda_1}
\frac{1}{\sqrt{2\pi\sigma_{11}}}
\exp\bigg[\frac{1}{2\sigma_{11}}
\bigg(
i\mathcal{R}_1 + \sum_{j\geqslant 2}\sigma_{1j}\lambda_j
\bigg)^2\bigg]\,.
\end{align}
According to eq.\,(\ref{eq:LinearCumu}), we now need to compute derivatives with respect to 
$\lambda_{i=1,\,\dots\,,n}$ and finally set 
$\lambda_{i=1,\,\dots\,,n} = 0$. We find
\begin{align}\label{eq:RI4}
C_n[f_{\rm NL}(\mathcal{R}_1),\dots,\mathcal{R}_n] = 
\frac{1}{\sqrt{2\pi\sigma_{11}}}\bigg(
\prod_{k=2}^{n}\sigma_{1k}
\bigg)\int d\mathcal{R}_1 f_{\rm NL}(\mathcal{R}_1)g_n(\mathcal{R}_1^{n-1})e^{-\mathcal{R}_1^2/2\sigma_{11}}\,,
\end{align}
where $g_n(\mathcal{R}_1^{n-1})$ is a polynomial of order $\mathcal{R}_1^{n-1}$ whose explicit expression 
is given by\footnote{We can notice that the $g_n$ polynomials are nothing but the Hermite polynomials, except for a pre-factor. In particular:
\begin{equation}
g_n(\mathcal{R}_1^{n-1}) = \frac{1}{(2\sigma_{11})^{n/2}} H_{n-1} \left(\frac{\mathcal{R}_1}{\sqrt{2\sigma_{11}}} \right)
\end{equation}
}
\begin{align}
g_n(\mathcal{R}_1^{n-1}) = (-1)^{n-1} e^{\mathcal{R}_1^2/2\sigma_{11}}\frac{d^{n-1}}{d\mathcal{R}_1^{n-1}}e^{-\mathcal{R}_1^2/2\sigma_{11}}\,.
\end{align}

This result implies that we can just integrate by parts $n-1$ times in eq.\,(\ref{eq:RI4}). 
We find
\begin{align}\label{eq:ExpeDer}
C_n[f_{\rm NL}(\mathcal{R}_1),\dots,\mathcal{R}_n] = 
\bigg(
\prod_{k=2}^{n}\sigma_{1k}
\bigg)\underbrace{
\frac{1}{\sqrt{2\pi\sigma_{11}}}\int d\mathcal{R}_1 f_{\rm NL}^{(n-1)}(\mathcal{R}_1)e^{-\mathcal{R}_1^2/2\sigma_{11}}
}_{=\,\langle f_{\rm NL}^{(n-1)}(\mathcal{R}_1)\rangle}\,.
\end{align}
The leftover integral is nothing but the expectation value of $f_{\rm NL}^{(n-1)}(\mathcal{R}_1)$. 
All in all, we find (remember that $\sigma_{ij} = \langle\mathcal{R}_i\mathcal{R}_j\rangle$)
\begin{tcolorbox}[colframe=navyblue!20,arc=6pt,colback=navyblue!5,width=0.98\textwidth]
\vspace{-.4cm}
\begin{align}
C_n[f_{\rm NL}(\mathcal{R}_1),\dots,\mathcal{R}_n] = 
\langle f_{\rm NL}^{(n-1)}(\mathcal{R}_1)\rangle
\langle\mathcal{R}_1\mathcal{R}_2\rangle\dots\langle\mathcal{R}_1\mathcal{R}_n\rangle\label{eq:CumufNL1}
\end{align}
\end{tcolorbox}

\item [$\circ$] At order $O(\alpha^2)$, consider eq.\,(\ref{eq:QuadraticCumu}) and the integral
\begin{align}\label{eq:RJ1}
\mathcal{J}(\lambda_1,\dots,\lambda_n) \equiv \int d\mathcal{R}_1 d\mathcal{R}_2
e^{if_{\rm NL}(\mathcal{R}_1) \lambda_1 + if_{\rm NL}(\mathcal{R}_2) \lambda_2}
\bigg[\int d\mathcal{R}_3\dots d\mathcal{R}_n 
e^{i[\mathcal{R}_3\lambda_3 +\,\dots\,+ \mathcal{R}_n\lambda_n]}\,
p(\mathcal{R}_1,\dots ,\mathcal{R}_n)\bigg]\,.
\end{align}
The computation goes as before with the only difference that now we need to consider the inverse Fourier transform of the 
characteristic function with respect to both $\lambda_1$ and $\lambda_2$.
For the natural log of $\mathcal{J}(\lambda_1,\dots,\lambda_n)$ we find
 \begin{align}
\log[\mathcal{J}&(\lambda_1,\dots,\lambda_n)] =
-\frac{1}{2}\sum_{ij\geqslant 3}\sigma_{ij}\lambda_i\lambda_j \\ &
 + \log\int d\mathcal{R}_1 d\mathcal{R}_2
\frac{e^{if_{\rm NL}(\mathcal{R}_1) \lambda_1+if_{\rm NL}(\mathcal{R}_2)\lambda_2}}{2\pi\sqrt{\sigma_{11}\sigma_{22}-\sigma_{12}^2}}\exp
\bigg[
\frac{1}{2(\sigma_{11}\sigma_{22}-\sigma_{12}^2)}\bigg(
A_1^2\sigma_{22} + 
A_2^2\sigma_{11} - 2A_1 A_2\sigma_{12}\bigg)
\bigg]\nn\,,
 \end{align}
where $A_1\equiv i\mathcal{R}_1 + \sum_{j\geqslant 3}\sigma_{1j}\lambda_j$ 
and $A_2\equiv i\mathcal{R}_2 + \sum_{j\geqslant 3}\sigma_{2j}\lambda_j$. 
This equation is the analogue of eq.\,(\ref{eq:RI3}) at order $O(\alpha^2)$. 
As before, we need to to compute derivatives with respect to 
$\lambda_{i=1,\,\dots\,,n}$ and finally set 
$\lambda_{i=1,\,\dots\,,n} = 0$. 
Let us consider first the derivatives with respect to $\lambda_1$ and $\lambda_2$. 
We find
\begin{align}\label{eq:QuadraticPartial}
(-i)^2&\frac{\partial}{\partial\lambda_2}
\frac{\partial}{\partial\lambda_1}
\log[\mathcal{J}(\lambda_1,\dots,\lambda_n)]\bigg|_{\lambda_1 = \lambda_2 =0} = \\ &
 + \underbrace{\frac{1}{2\pi\sqrt{\sigma_{11}\sigma_{22}-\sigma_{12}^2}}}_{2\pi\sqrt{{\rm det}\,\tilde{\sigma}}}
\int d\mathcal{R}_1 d\mathcal{R}_2f_{\rm NL}(\mathcal{R}_1)f_{\rm NL}(\mathcal{R}_2)
\exp
\bigg[
\underbrace{\frac{1}{2(\sigma_{11}\sigma_{22}-\sigma_{12}^2)}\bigg(
A_1^2\sigma_{22} + 
A_2^2\sigma_{11} - 2A_1A_2\sigma_{12}\bigg)}_{=\,(1/2)\sum_{i,j=1}^2(\tilde{\sigma}^{-1})_{ij}A_i A_j}
\bigg]\nn \\
&
-\bigg[
\frac{1}{\sqrt{2\pi\sigma_{11}}}
\int d\mathcal{R}_1 f_{\rm NL}(\mathcal{R}_1)e^{A_1^2/2\sigma_{11}}
\bigg]
\bigg[
\frac{1}{\sqrt{2\pi\sigma_{22}}}
\int d\mathcal{R}_2 f_{\rm NL}(\mathcal{R}_2)e^{A_2^2/2\sigma_{22}}
\bigg] 
\,,\nn
 \end{align} 
 where $\tilde{\sigma}$ is the two-by-two sub-matrix of $\sigma$ formed by the first two rows and columns.  
Before proceeding, let us consider a simple example. 
If we are interested to the computation of the second-order cumulant, these two derivatives are the only ones that we need to take. 
Furthermore, in this case we have $A_{j=1,2} = i\mathcal{R}_{j=1,2}$. 
We find the simple result
\begin{tcolorbox}[colframe=navyblue!20,arc=6pt,colback=navyblue!5,width=0.98\textwidth]
\vspace{-.4cm}
\begin{align}\label{eq:CCC}
C_2[f_{\rm NL}(\mathcal{R}_1),f_{\rm NL}(\mathcal{R}_2)] = \langle f_{\rm NL}(\mathcal{R}_1)f_{\rm NL}(\mathcal{R}_2)\rangle - 
 \langle f_{\rm NL}(\mathcal{R}_1)\rangle \langle f_{\rm NL}(\mathcal{R}_2)\rangle
 \end{align} 
\end{tcolorbox} 
 where the expectation values $ \langle f_{\rm NL}(\mathcal{R}_1)\rangle$ and $ \langle f_{\rm NL}(\mathcal{R}_2)\rangle$ are defined as in eq.\,(\ref{eq:ExpeDer}) while 
 $ \langle f_{\rm NL}(\mathcal{R}_1)f_{\rm NL}(\mathcal{R}_2)\rangle$ is computed by means of the joint probability distribution of $\mathcal{R}_1$ and $\mathcal{R}_2$; 
 in formulas, 
 we have 
 \begin{align}
  \langle f_{\rm NL}(\mathcal{R}_1)f_{\rm NL}(\mathcal{R}_2)\rangle = \frac{1}{2\pi\sqrt{{\rm det}\,\tilde{\sigma}}}
  \int d\mathcal{R}_1d\mathcal{R}_2 f_{\rm NL}(\mathcal{R}_1)f_{\rm NL}(\mathcal{R}_2)\exp\bigg[
  -\frac{1}{2}\sum_{i,j=1}^2(\tilde{\sigma}^{-1})_{ij}\mathcal{R}_i \mathcal{R}_j
  \bigg]\,.
  \end{align} 
  For $n \geqslant 3$, we need to compute in eq.\,(\ref{eq:QuadraticPartial}) derivatives with respect to 
$\lambda_{i=3,\,\dots\,,n}$ and finally set 
$\lambda_{i=3,\,\dots\,,n} = 0$.  These derivatives act on the exponential functions in the integrands of the two terms on the right-hand side of eq.\,(\ref{eq:QuadraticPartial}). 
Let us start from the second term; if we compute the $\lambda$-derivatives on the exponential function, we find the following result
 \begin{align}\label{eq:MasterDerivative1}
(-i&)^{n-2}\frac{\partial}{\partial\lambda_n}\dots \frac{\partial}{\partial\lambda_3}\bigg(e^{A_1^2/2\sigma_{11}}e^{A_2^2/2\sigma_{22}}\bigg)
\bigg|_{\lambda_3=\,\dots\,\lambda_n = 0} = \\ &
(-1)^{n-2}\bigg(\prod_{k=3}^{n}\sigma_{1k}\bigg)
\bigg(\frac{d^{n-2}}{d\mathcal{R}_1^{n-2}}e^{-\mathcal{R}_1^2/2\sigma_{11}}\bigg)e^{-\mathcal{R}_2^2/2\sigma_{22}} +
(-1)^{n-2}\bigg(\prod_{k=3}^{n}\sigma_{2k}\bigg)
\bigg(\frac{d^{n-2}}{d\mathcal{R}_2^{n-2}}e^{-\mathcal{R}_2^2/2\sigma_{22}}\bigg)e^{-\mathcal{R}_1^2/2\sigma_{11}}\nn\\
&+(-1)^{n-2}\bigg[
\sum_{a,b>0}^{a+b=n-2}
\bigg(\prod_{k=3+b}^{n}\sigma_{1k}\bigg)\bigg(\prod_{j=3}^{n-a}\sigma_{2j}\bigg)
\bigg(\frac{d^{a}}{d\mathcal{R}_1^{a}}e^{-\mathcal{R}_1^2/2\sigma_{11}}\bigg)
\bigg(\frac{d^{b}}{d\mathcal{R}_2^{b}}e^{-\mathcal{R}_2^2/2\sigma_{22}}\bigg) + 
{\rm perms\,of}\,(3,\dots,n)
\bigg]\,.\nn
  \end{align} 
 To fully understand the content of this equation, few comments are in order. Inside the square brackets, the sum $\sum_{a,b}$ runs over all possible combinations of $a>0$ and $b>0$ such that $a+b = n-2$. For instance, 
  if we take $n=4$ we have only $\{a=1,b=1\}$ while for $n=5$ we have two combinations, namely $\{a=1,b=2\}$ and $\{a=2,b=1\}$ (notice that if $n=3$, there are no combinations that are allowed, and the term in the square brackets does not contribute to the third-order cumulant). 
  Furthermore, after computing the sum over $a$ and $b$, a further sum over all permutations of the indices $(3,\dots,n)$ that give distinct results is required. 
  Let us clarify this point with an explicit example. Consider $n=4$. The sum over $\{a=1,b=1\}$ gives 
  $(e^{-\mathcal{R}_1^2/2\sigma_{11}}e^{-\mathcal{R}_2^2/2\sigma_{22}}\mathcal{R}_1\mathcal{R}_2/\sigma_{11}\sigma_{22})\sigma_{14}\sigma_{23}$; 
  we now need include the term with $(3,4)$ exchanged so that the final result is 
   $(e^{-\mathcal{R}_1^2/2\sigma_{11}}e^{-\mathcal{R}_2^2/2\sigma_{22}}\mathcal{R}_1\mathcal{R}_2/\sigma_{11}\sigma_{22})
   (\sigma_{14}\sigma_{23} + \sigma_{13}\sigma_{24})$. 
   Only permutations that give distinct results need to be included. 
   For instance, take for $n=5$ the term proportional to $\sigma_{14}\sigma_{15}\sigma_{23}$; in this case, the sum over permutations of $(3,4,5)$
   gives only three distinct terms, $\sigma_{14}\sigma_{15}\sigma_{23} \to \sigma_{14}\sigma_{15}\sigma_{23} + 
   \sigma_{13}\sigma_{15}\sigma_{24}+ \sigma_{14}\sigma_{13}\sigma_{25}$ (instead of six---that is the number of permutations of three objects---since the remaining three do not give distinct results).

 As far as the first term on the right-hand side of eq.\,(\ref{eq:QuadraticPartial}) is concerned, we find 
  \begin{align}\label{eq:MasterDerivative2}
&(-i)^{n-2}\frac{\partial}{\partial\lambda_n}\dots \frac{\partial}{\partial\lambda_3}\bigg[
e^{\frac{A_1^2\sigma_{22} + 
A_2^2\sigma_{11} - 2A_1A_2\sigma_{12}}{2(\sigma_{11}\sigma_{22}-\sigma_{12}^2)}}
\bigg]
\bigg|_{\lambda_3=\,\dots\,\lambda_n = 0} = \\ &
(-1)^{n-2}\bigg(\prod_{k=3}^{n}\sigma_{1k}\bigg)
\bigg[\frac{d^{n-2}}{d\mathcal{R}_1^{n-2}}e^{-\frac{\mathcal{R}_1^2\sigma_{22} + 
\mathcal{R}_2^2\sigma_{11} - 2\mathcal{R}_1\mathcal{R}_2\sigma_{12}}{2(\sigma_{11}\sigma_{22}-\sigma_{12}^2)}}\bigg] +
(-1)^{n-2}\bigg(\prod_{k=3}^{n}\sigma_{2k}\bigg)
\bigg[\frac{d^{n-2}}{d\mathcal{R}_2^{n-2}}e^{-\frac{\mathcal{R}_1^2\sigma_{22} + 
\mathcal{R}_2^2\sigma_{11} - 2\mathcal{R}_1\mathcal{R}_2\sigma_{12}}{2(\sigma_{11}\sigma_{22}-\sigma_{12}^2)}}\bigg]\nn\\
&+(-1)^{n-2}\bigg\{
\sum_{a,b>0}^{a+b=n-2}
\bigg(\prod_{k=3+b}^{n}\sigma_{1k}\bigg)\bigg(\prod_{j=3}^{n-a}\sigma_{2j}\bigg)
\bigg[\frac{\partial^{n-2}}{\partial\mathcal{R}_1^{a}\partial\mathcal{R}_2^{b}}
e^{-\frac{\mathcal{R}_1^2\sigma_{22} + 
\mathcal{R}_2^2\sigma_{11} - 2\mathcal{R}_1\mathcal{R}_2\sigma_{12}}{2(\sigma_{11}\sigma_{22}-\sigma_{12}^2)}}
\bigg] + 
{\rm perms\,of}\,(3,\dots,n)
\bigg\}\,,\nn
  \end{align} 
where the sum $\sum_{a,b}$ and the subsequent one over different permutations have the same meaning explained before. 

Eqs.\,(\ref{eq:MasterDerivative1},\,\ref{eq:MasterDerivative2}) allow to compute the integrals in eq.\,(\ref{eq:QuadraticPartial}) by means of $n-2$ integration by parts. 
We find the final result (for $n \geqslant 3$) 
\begin{tcolorbox}[colframe=navyblue!20,arc=6pt,colback=navyblue!5,width=0.98\textwidth]
\vspace{-.4cm}
\begin{align}\label{eq:CumufNL2}
 &C_n[f_{\rm NL}(\mathcal{R}_1),f_{\rm NL}(\mathcal{R}_2),\dots,\mathcal{R}_n] =  \\
 &~~\left[\langle f_{\rm NL}^{(n-2)}(\mathcal{R}_1)f_{\rm NL}(\mathcal{R}_2)  \rangle - 
 \langle f_{\rm NL}^{(n-2)}(\mathcal{R}_1) \rangle\langle f_{\rm NL}(\mathcal{R}_2)  \rangle \right]\langle\mathcal{R}_1\mathcal{R}_3\rangle\dots\langle\mathcal{R}_1\mathcal{R}_n\rangle ~~+\nn \\
 &~~\left[\langle f_{\rm NL}(\mathcal{R}_1)f^{(n-2)}_{\rm NL}(\mathcal{R}_2)  \rangle - 
 \langle f_{\rm NL}(\mathcal{R}_1) \rangle\langle f_{\rm NL}^{(n-2)}(\mathcal{R}_2)  \rangle \right]\langle\mathcal{R}_2\mathcal{R}_3\rangle\dots\langle\mathcal{R}_2\mathcal{R}_n\rangle~~ +\nn \\
 &~~\bigg\{
 \sum_{a,b>0}^{a+b=n-2}\nn
 \left[\langle f^{(a)}_{\rm NL}(\mathcal{R}_1)f^{(b)}_{\rm NL}(\mathcal{R}_2)  \rangle - 
 \langle f_{\rm NL}^{(a)}(\mathcal{R}_1) \rangle\langle f_{\rm NL}^{(b)}(\mathcal{R}_2)  \rangle \right]
  \prod_{k=3+b}^{n}\langle\mathcal{R}_1\mathcal{R}_k\rangle
 \prod_{j=3}^{n-a}\langle\mathcal{R}_2\mathcal{R}_j\rangle
  + 
{\rm perms\,of}\,(3,\dots,n)
 \bigg\}
\end{align} 
\end{tcolorbox}
which generalizes eq.\,(\ref{eq:CumufNL1}) at order $O(\alpha^2)$. We remark that eqs.\,(\ref{eq:CumufNL1},\,\ref{eq:CumufNL2}) are completely generic and do not depend on the functional form of $f_{\rm NL}(\mathcal{R})$.

\item [$\circ$] One can in principle continue the computation and include higher-order $\alpha$-corrections to the cumulants. 
Needless to say, the corresponding expressions quickly become quite unwieldy (even though the computation remains conceptually simple).
One possible strategy is to include only $O(\alpha)$ corrections, and check that $O(\alpha^2)$ terms remain sub-leading. 
In order for this criterium to be satisfactory, one should correctly identify the expansion parameter. 
As we shall see, in the case with no derivatives involved in eq.\,(\ref{eq:Cumutrick}) the latter turns out to be $\alpha^2\sigma_0^2$.

\item [$\circ$]
We can conclude this part with few checks of eqs.\,(\ref{eq:CumufNL1},\,\ref{eq:CumufNL2}). 
For simplicity, let us consider the case in which there are no derivatives in eq.\,(\ref{eq:Cumutrick}). 

{\underline{$n=2$}} 

We compute $C_2(h_1,h_2)$. 
At order $O(\alpha)$, we have 
\begin{align}
C_2(h_1,h_2) & = C_2(\mathcal{R}_1,\mathcal{R}_2) + C_2[f_{\rm NL}(\mathcal{R}_1),\mathcal{R}_2] + C_2[\mathcal{R}_1,f_{\rm NL}(\mathcal{R}_2)] \\
& = \langle \mathcal{R}_1\mathcal{R}_2 \rangle + \langle f_{\rm NL}^{(1)}(\mathcal{R}_1) \rangle \langle \mathcal{R}_1\mathcal{R}_2 \rangle +
  \langle f_{\rm NL}^{(1)}(\mathcal{R}_2) \rangle \langle \mathcal{R}_1\mathcal{R}_2 \rangle\,,
\end{align} 
where we use eq.\,(\ref{eq:SecondCum}) to rewrite the first  term (which is the gaussian one) and eq.\,(\ref{eq:CumufNL1}) to rewrite the last two ones. 
Since we have $f_{\rm NL}(\mathcal{R}) = \alpha\mathcal{R}^2$, we have $\langle f_{\rm NL}^{(1)}(\mathcal{R}_{i=1,2}) \rangle = 0$ since we are considering zero-mean random fields.
This means that there are no corrections at order $O(\alpha)$. We now move to consider corrections at order $O(\alpha^2)$. We use eq.\,(\ref{eq:CCC}). At order $O(\alpha^2)$, we have
\begin{align}\label{eq:CCC2}
C_2(h_1,h_2) = \langle \mathcal{R}_1\mathcal{R}_2 \rangle + \langle f_{\rm NL}(\mathcal{R}_1)f_{\rm NL}(\mathcal{R}_2)\rangle - 
 \langle f_{\rm NL}(\mathcal{R}_1)\rangle \langle f_{\rm NL}(\mathcal{R}_2)\rangle =  \langle \mathcal{R}_1\mathcal{R}_2 \rangle +
 \alpha^2\langle \mathcal{R}_1^2\mathcal{R}_2^2 \rangle -  \alpha^2\langle \mathcal{R}_1^2\rangle\langle \mathcal{R}_2^2\rangle \,.
 \end{align} 
 We now set the two points equal (since we are not considering derivatives of cumulants, there is no need of considering different spatial point).  
 We have $ \langle \mathcal{R}^2 \rangle = \sigma_0^2$ and $\langle \mathcal{R}^4 \rangle = 3 \sigma_0^4$. In conclusion, we find
 \begin{align}\label{eq:SecondOrderCheck}
 C_2(h,h) = \sigma_0^2\left(1+2\alpha^2\sigma_0^2\right)\,.
 \end{align} 
 This is an exact result since corrections of order $O(\alpha^3)$ enter only at higher-orders. 
 As a rule of thumb, 
 for a cumulant $C_n$ of order $n$, only corrections up to order $O(\alpha^n)$ are possible. 
 Eq.\,(\ref{eq:SecondOrderCheck}) shows that the correct expansion parameter is $\alpha^2\sigma_0^2$. 
 In realistic models of inflation we expect $\alpha = O(1)$ while we have $\sigma_0^2 = A_g$ for the specific power spectrum in eq.\,(\ref{eq:ToyPS}). 
 Since one typically has $A_g = O(10^{-3})$, we expect $\alpha^2\sigma_0^2 \ll 1$.  
 Let us also notice that in applications of cosmological interest it is customary to use $f_{\rm NL}(\mathcal{R}) = \alpha(\mathcal{R}^2 - \langle \mathcal{R}^2\rangle)$ since in this way 
 one has a zero-mean non-gaussian field $h$, $\langle h \rangle = 0$. 
 It is simple to see that the constant shift in $f_{\rm NL}(\mathcal{R})$ that is implied by this particular choice  does not affect the computation of cumulants of order equal or higher than two. However, it does change the first order cumulant since $C_1(h) = \langle h \rangle$ (see eq.\,(\ref{eq:FirstCum})). 
 The choice $f_{\rm NL}(\mathcal{R}) = \alpha(\mathcal{R}^2 - \langle \mathcal{R}^2\rangle)$, therefore, leads to $C_1(h) = 0$.

{\underline{$n=3$}} 

Consider now $C_3(h_1,h_2,h_3)$. This cumulant vanishes in the gaussian limit. At order $O(\alpha)$, we use eq.\,(\ref{eq:CumufNL1}) and compute
 \begin{align}
 C_3(h_1,h_2,h_3)  & = C_3[f_{\rm NL}(\mathcal{R}_1),\mathcal{R}_2,\mathcal{R}_3] + 
 C_3[\mathcal{R}_1,f_{\rm NL}(\mathcal{R}_2),\mathcal{R}_3] +C_3[\mathcal{R}_1,\mathcal{R}_2,f_{\rm NL}(\mathcal{R}_3)]\\
 & = 2\alpha\left(
 \langle \mathcal{R}_1\mathcal{R}_2\rangle \langle \mathcal{R}_1\mathcal{R}_3\rangle +
 \langle \mathcal{R}_2\mathcal{R}_1\rangle \langle \mathcal{R}_2\mathcal{R}_3\rangle +
 \langle \mathcal{R}_3\mathcal{R}_1\rangle \langle \mathcal{R}_3\mathcal{R}_2\rangle
 \right)\,.
 \end{align} 
 If we set all points equal, we find
  \begin{align}\label{eq:ThirdOrderCheck}
  C_3(h,h,h)  = 6\alpha\sigma_0^4\,.
   \end{align} 
 We now move to consider the possible presence of a correction at order  $O(\alpha^2)$ that can be computed by means of eq.\,(\ref{eq:CumufNL2}). 
 As already noticed, for $n=3$ the sum $\sum_{a,b}$ does not contribute to the final result. We are left with the two terms
 \begin{align}\label{eq:n3}
   \left[
  \langle f_{\rm NL}^{(1)}(\mathcal{R}_1) f_{\rm NL}(\mathcal{R}_2)\rangle - 
   \langle f_{\rm NL}^{(1)}(\mathcal{R}_1)\rangle
    \langle f_{\rm NL}(\mathcal{R}_2)\rangle
  \right]\langle\mathcal{R}_1\mathcal{R}_3\rangle + \left[
  \langle f_{\rm NL}(\mathcal{R}_1) f_{\rm NL}^{(1)}(\mathcal{R}_2)\rangle - 
   \langle f_{\rm NL}(\mathcal{R}_1)\rangle
    \langle f_{\rm NL}^{(1)}(\mathcal{R}_2)\rangle
  \right]\langle\mathcal{R}_2\mathcal{R}_3\rangle\,.
 \end{align}
 If we set the three points equal and consider explicitly $f_{\rm NL}(\mathcal{R}) = \alpha\mathcal{R}^2$, we find that eq.\,(\ref{eq:n3}) reduces to 
 $4\alpha^2\langle\mathcal{R}^3\rangle\sigma_0^2 $ which is zero because $\langle\mathcal{R}^3\rangle = 0$. 
Eq.\,(\ref{eq:ThirdOrderCheck}), however, is not an exact result since for $n=3$ corrections of order $O(\alpha^3)$ are possible. 
Mimicking eq.\,(\ref{eq:SecondOrderCheck}), we expect
  \begin{align}\label{eq:ThirdOrderCheck2}
  C_3(h,h,h)  = 6\alpha\sigma_0^4[1 + O(\alpha^2\sigma_0^2)]\,.
   \end{align} 
   The correction of order $O(\alpha^3)$ can not be computed by means of eq.\,(\ref{eq:CumufNL2}) but it is expected to be sub-leading since we work under the assumption that $\alpha^2\sigma_0^2 \ll 1$. 
 
{\underline{$n=4$}}

Consider the fourth-order cumulant $C_4(h_1,h_2,h_3,h_4)$. If we limit the analysis to quadratic  non-gaussianities as in $f_{\rm NL}(\mathcal{R}) = \alpha\mathcal{R}^2$, 
it is clear that there are no corrections of order $O(\alpha)$ since $f_{\rm NL}^{(3)}(\mathcal{R}) = 0$. The first non-trivial correction arises at order $O(\alpha^2)$
\begin{align}\label{eq:Temp5}
C_4(h_1,h_2,h_3,h_4) = C_4[f_{\rm NL}(\mathcal{R}_1),f_{\rm NL}(\mathcal{R}_2),\mathcal{R}_3,\mathcal{R}_4] + 5\,{\rm combinations}\,.
\end{align}
If we take eq.\,(\ref{eq:CumufNL2}) it is simple to see that the first two terms on the right-hand side give vanishing contribution. 
In the last line, we have one contribution corresponding to $a=b=1$ that gives two terms since one has to sum $3\leftrightarrow 4$ exchange. 
We have
\begin{align}
 C_4[f_{\rm NL}(\mathcal{R}_1),f_{\rm NL}(\mathcal{R}_2),\mathcal{R}_3,\mathcal{R}_4] & = 
 \left[
 \langle f_{\rm NL}^{(1)}(\mathcal{R}_1) f_{\rm NL}^{(1)}(\mathcal{R}_2)\rangle - 
  \langle f_{\rm NL}^{(1)}(\mathcal{R}_1)\rangle \langle f_{\rm NL}^{(1)}(\mathcal{R}_2)\rangle
 \right](\langle\mathcal{R}_1\mathcal{R}_4\rangle\langle\mathcal{R}_2\mathcal{R}_3\rangle +3\leftrightarrow 4 \rangle)\nn\\
 & = 4\alpha^2\langle\mathcal{R}_1\mathcal{R}_2\rangle(\langle\mathcal{R}_1\mathcal{R}_4\rangle\langle\mathcal{R}_2\mathcal{R}_3\rangle + 
 \langle\mathcal{R}_1\mathcal{R}_3\rangle\langle\mathcal{R}_2\mathcal{R}_4\rangle \rangle)\,.
 \label{eq:C4alpha2}
\end{align}
 If we set all points equal and multiply by six because of eq.\,(\ref{eq:Temp5}), we find
 \begin{align}\label{eq:FourthOrderCheck}
 C_4(h,h,h,h) = 48\alpha^2\sigma_0^6[1 + O(\alpha^2\sigma_0^2)]\,,
 \end{align}
 where we introduced again a sub-leading correction of order $O(\alpha^4)$.
 
 We can check the validity of eqs.\,(\ref{eq:SecondOrderCheck},\,\ref{eq:ThirdOrderCheck2},\,\ref{eq:FourthOrderCheck}) explicitly. 
 The reason is that if we limit the analysis---as done here---to random variables {\it without derivatives}, the computation of the cumulants of $h$ admits an exact analytical solution. 
 We can indeed extract the probability density distribution of the random variable $h$---at a given spatial point, that is what we  need in order to compare with eqs.\,(\ref{eq:SecondOrderCheck},\,\ref{eq:ThirdOrderCheck2},\,\ref{eq:FourthOrderCheck})---by means of the Jacobi's multivariate theorem in eq.\,(\ref{eq:TransformedPDFroots}) using the fact that 
 $\mathcal{R}$ is a gaussian variable with zero mean and variance $\sigma_0^2$. 
 We find
  \begin{align}\label{eq:NoDerivativePDF}
  p(h) = \frac{e^{-\mathcal{R}_+^2/2\sigma_0^2} + e^{-\mathcal{R}_-^2/2\sigma_0^2}}{\sqrt{2\pi}\sigma_0\sqrt{1+4\alpha h}}\,,~~~~~{\rm where}~~
  \mathcal{R}_{\pm} = \frac{-1\pm \sqrt{1+4\alpha h}}{2\alpha}\,,
    \end{align}
with $h \in [-1/4\alpha, \infty)$. By means of the Fa\`a di Bruno's formula, we find the generic $n^{\rm th}$-order 
cumulant\footnote{From the probability density distribution we compute the moments 
$\mu_n = \int_{-1/4\alpha}^{\infty}dh\,h^n\,p(h)$. 
The explicit expression for the $n^{\rm th}$ cumulant in terms of the first $n$ moments 
can be obtained by using Fa\`a di Bruno's formula for higher derivatives of composite functions; explicitly, for $n\geqslant 2$, 
we have 
 $C_n = \sum_{k=1}^{n}(-1)^{k-1}(k-1)!B_{n,k}(0,\mu_2,\dots,\mu_{n-k+1})$, where $B_{n,k}$ are the incomplete Bell polynomials.}
 \begin{align}
C_n(h,\,\dots\,,h) =  2^{n-3}\sigma_0^2(\alpha\sigma_0^2)^{n-2}(n+4\alpha^2\sigma_0^2)(n-1)!\,,
\label{eq:faadibruno}
  \end{align}
  One can check that eqs.\,(\ref{eq:SecondOrderCheck},\,\ref{eq:ThirdOrderCheck2},\,\ref{eq:FourthOrderCheck}) are correctly reproduced in the 
  appropriate limits.
  
\item [$\circ$] Since in the case with no derivatives the probability density distribution can be obtained analytically (see eq.\,(\ref{eq:NoDerivativePDF})), it is instructive to do the following exercise. 
First of all, let us rewrite eq.\,(\ref{eq:NoDerivativePDF}) for the case with $f_{\rm NL}(\mathcal{R}) = \alpha(\mathcal{R}^2 - \langle \mathcal{R}^2\rangle)$ with $\langle \mathcal{R}^2\rangle = \sigma_0^2$. We find 
  \begin{align}\label{eq:NoDerivativePDF2}
  p(h) = \frac{e^{-\mathcal{R}_+^2/2\sigma_0^2} + e^{-\mathcal{R}_-^2/2\sigma_0^2}}{\sqrt{2\pi}\sigma_0\sqrt{1+4\alpha(h+\alpha\sigma_0^2)}}\,,~~~~~{\rm where}~~
  \mathcal{R}_{\pm} = \frac{-1\pm \sqrt{1+4\alpha(h+\alpha\sigma_0^2)}}{2\alpha}\,.
    \end{align}
In this case we have $C_1(h) = 0$ while $C_n(h,\,\dots\,,h)\equiv C_n(h)$ are still given by eq.\,(\ref{eq:faadibruno}). 
We can compute analytically the characteristic function $\log\chi(\lambda) = 
\sum_{n=2}^{\infty}(i^n/n!)C_n(h) \lambda^n$. 
We find  
\begin{align}
\chi(\lambda) =  
\exp\bigg[-\frac{\lambda^2 \sigma_0^2}{2(1-2i\alpha\lambda\sigma_0^2)}-i\alpha\lambda \sigma_0^2\bigg]
\frac{1}{\sqrt{1- 2i\alpha\lambda\sigma_0^2}}\,,
\end{align}   
and one can check for consistency that $p(h)$ in  eq.\,(\ref{eq:NoDerivativePDF2}) is given by 
the inverse Fourier transform 
\begin{align}\label{eq:Dybala}
p(h) = \frac{1}{2\pi}\int_{-\infty}^{\infty}d\lambda\,\chi(\lambda)\,e^{-ih\lambda}\,.
\end{align} 
Once we know $p(h)$, we can compute the integral $\bar{p}(h_c) \equiv \int_{h_c}^{\infty}dh\,p(h)$, 
which is the probability to find $h$ above the threshold $h_c$. 
We find
\begin{align}\label{eq:ExactProb}
\bar{p}_{\rm NG}(h_c) = \frac{1}{2}\bigg[
2 - {\rm Erf}\bigg(\frac{1+\sqrt{1+4h_c\alpha + 4\alpha^2\sigma_0^2}}{2\sqrt{2}\alpha\sigma_0}\bigg)
- {\rm Erf}\bigg(\frac{-1+\sqrt{1+4h_c\alpha + 4\alpha^2\sigma_0^2}}{2\sqrt{2}\alpha\sigma_0}\bigg)
\bigg]\,,
\end{align}
with ${\rm Erf}(x)$ the error function.
This simple case, therefore, can be solved exactly. 
This means that we can use it as a playground to test the following approximation. 
Let us organize the sum over cumulants in a power-series expansion in $\alpha$. We have
\begin{align}\label{eq:ExpCumu}
\sum_{n=2}^{\infty}\frac{i^n}{n!}C_n(h)\lambda^n =  
-\frac{\sigma_0^2\lambda^2}{2} 
-i\alpha \lambda^3 \sigma_0^4 + \alpha^2\sigma_0^4
(-\lambda^2 + 2\lambda^4\sigma_0^2) + O(\alpha^3)\,,
\end{align}
where the first term on the right-hand side corresponds to the quadratic cumulant that reproduces the gaussian limit.
We can truncate eq.\,(\ref{eq:ExpCumu}) at some order in $\alpha$, compute the 
corresponding $\bar{p}(h_c)$ and compare with eq.\,(\ref{eq:ExactProb}). 
At order $\alpha^0$, we find the gaussian result
\begin{align}\label{eq:Gau1}
\bar{p}_{\rm G}(h_c) = \frac{1}{2}{\rm Erf}\left(\frac{h_c}{\sqrt{2}\sigma_0}\right) 
\simeq \frac{1}{\sqrt{2\pi}v_c}\exp\left(-\frac{v_c^2}{2}\right)\,, 
\end{align}
where we define $v_c \equiv h_c/\sigma_0$, and the last approximation corresponds to 
$v_c \gg 1$ which is the limit that is relevant for our analysis.  
Consider now the order $O(\alpha)$. Let us write eq.\,(\ref{eq:ExactProb}) at order $O(\alpha)$ as 
$\bar{p}_{\alpha}(h_c) = \int_{h_c}^{\infty}dh\,p_{\alpha}(h)$ with 
\begin{align}\label{eq:papproxm}
p_{\alpha}(h) & = 
\frac{1}{2\pi}\int_{-\infty}^{\infty}d\lambda\,\exp\bigg[-\frac{\sigma_0^2\lambda^2}{2} 
-i(\alpha\sigma_0) \lambda^3 \sigma_0^3\bigg]\,e^{-ih\lambda} = 
\frac{1}{2\pi}\int_{-\infty}^{\infty}d\lambda\,e^{-\sigma_0^2\lambda^2/2}\,
\bigg[
\sum_{m=0}^{\infty}\frac{(-i\alpha\sigma_0^4)^m}{m!}
\bigg]\,\underbrace{\lambda^{3m}\,e^{-ih\lambda}}_{i^{3m}\frac{\partial^{3m} e^{-ih\lambda}}{\partial h^{3m}}}\nn\\
& = 
\sum_{m=0}^{\infty}\frac{(-i\alpha\sigma_0^4)^m}{m!}i^{3m}
\frac{\partial^{3m}}{\partial h^{3m}}
\underbrace{
\frac{1}{2\pi}\int_{-\infty}^{\infty}d\lambda\,e^{-\sigma_0^2\lambda^2/2}\,
 e^{-ih\lambda}}_{{\rm gaussian\,integral}\,\frac{1}{\sqrt{2\pi}\sigma_0}e^{-h^2/2\sigma_0^2}} 
 \equiv \sum_{m=0}^{\infty}p_{\alpha}^{(m)}(h)\,.
\end{align}
In practice, we can approximate $\bar{p}_{\alpha}(h_c)$ as a sum of terms, 
$\bar{p}_{\alpha}(h_c) = \sum_{m=0}^{\infty}\bar{p}^{(m)}_{\alpha}(h_c)$ each one of them 
defined by means of eq.\,(\ref{eq:papproxm}), that is 
$\bar{p}^{(m)}_{\alpha}(h_c) = \int_{h_c}^{\infty}dh\,p^{(m)}_{\alpha}(h)$. 
A similar decomposition can be defined at higher orders in $\alpha$. 
At order $O(\alpha)$, we find
\begin{align}
\bar{p}^{(1)}_{\alpha}(h_c) & = \frac{1}{\sqrt{2\pi}v_c}\exp\left(-\frac{v_c^2}{2}\right)
(\alpha\sigma_0)v_c^3\bigg(
1-\frac{1}{v_c^2}
\bigg)\,,\label{eq:Approx1}\\
\bar{p}^{(2)}_{\alpha}(h_c) & =
\frac{1}{\sqrt{2\pi}v_c}\exp\left(-\frac{v_c^2}{2}\right)
(\alpha\sigma_0)^2\frac{v_c^6}{2}\bigg(
1 - \frac{10}{v_c^2} + \frac{15}{v_c^4}
\bigg)\,,\label{eq:Approx2}\\
\bar{p}^{(3)}_{\alpha}(h_c) & = \frac{1}{\sqrt{2\pi}v_c}\exp\left(-\frac{v_c^2}{2}\right)
(\alpha\sigma_0)^3\frac{v_c^9}{6}\bigg(
1 - \frac{28}{v_c^2} + \frac{210}{v_c^4} - \frac{420}{v_c^6} + \frac{105}{v_c^8}
\bigg)\,\label{eq:Approx3}\\
\bar{p}^{(4)}_{\alpha}(h_c) & = \dots\nn
\end{align}
Since $v_c \gg 1$, we are tempted to approximate (by neglecting all sub-leading terms in each of the round brackets)  
\begin{align}\label{eq:OrderAlphaResum}
\bar{p}^{(m)}_{\alpha}(h_c) = \frac{1}{\sqrt{2\pi}v_c}\exp\left(-\frac{v_c^2}{2}\right)
\frac{(\alpha\sigma_0 v_c^{3})^m}{m!}~~~~~~\Longrightarrow~~~~~~
\bar{p}_{\alpha}(h_c) = \frac{1}{\sqrt{2\pi}v_c}\exp\left(-\frac{v_c^2}{2} + \alpha\sigma_0 v_c^{3}\right)\,.
\end{align}
We are now in the position to compare 
{\it i)} the gaussian result in eq.\,(\ref{eq:Gau1}), {\it ii)} the exact non-gaussian result in 
eq.\,(\ref{eq:ExactProb}), {\it iii)} the order $O(\alpha)$ approximation 
$\bar{p}_{\alpha}(h_c) = \sum_{m=0}^{\infty}\bar{p}^{(m)}_{\alpha}(h_c)$ 
obtained by truncating the series for increasing values of $m$ and {\it iv)} 
the order $O(\alpha)$ approximation resummed as in eq.\,(\ref{eq:OrderAlphaResum}). 
This comparison is shown in the left panel of fig.\,\ref{fig:SimplifiedCase}.

We can now consider the order $O(\alpha^2)$. We find
\begin{align}
\bar{p}^{(1)}_{\alpha^2}(h_c) & = \frac{1}{\sqrt{2\pi}v_c}\exp\left(-\frac{v_c^2}{2}\right)
(\alpha\sigma_0)\bigg[
v_c^3\bigg(1-\frac{1}{v_c^2}\bigg) + 
2v_c^4\alpha\sigma_0\bigg(1-\frac{5}{2v_c^2}\bigg)
\bigg]\,,\\
\bar{p}^{(2)}_{\alpha^2}(h_c) & =
\frac{1}{\sqrt{2\pi}v_c}\exp\left(-\frac{v_c^2}{2}\right)
(\alpha\sigma_0)^2\times\nn\\
~~~~&\bigg[\frac{v_c^6}{2}\bigg(
1 - \frac{10}{v_c^2} + \frac{15}{v_c^4}
\bigg) + 
2\alpha\sigma_0 v_c^7\bigg(1 - \frac{29}{2v_c^2} + \frac{42}{v_c^4} - \frac{27}{2v_c^6}
\bigg) + 2(\alpha\sigma_0)^2v_c^8\bigg(
1 - \frac{20}{v_c^2} + \frac{381}{4v_c^4} - \frac{363}{4v_c^6}
\bigg)
\bigg]\,,\\
\bar{p}^{(3)}_{\alpha^2}(h_c) & = \dots\nn
\end{align}  
We are again tempted to consider the limit $v_c \gg 1$ in which we keep only the leading term inside the round brackets, meaning 
that we write
\begin{align}
\bar{p}^{(1)}_{\alpha^2}(h_c) & = \frac{1}{\sqrt{2\pi}v_c}\exp\left(-\frac{v_c^2}{2}\right)
(\alpha\sigma_0)\big(
v_c^3 + 
2v_c^4\alpha\sigma_0\big)\,,\\
\bar{p}^{(2)}_{\alpha^2}(h_c) & =
\frac{1}{\sqrt{2\pi}v_c}\exp\left(-\frac{v_c^2}{2}\right)
(\alpha\sigma_0)^2\bigg[\frac{v_c^6}{2} + 
2\alpha\sigma_0 v_c^7 + 2(\alpha\sigma_0)^2v_c^8
\bigg]\,,\\
\bar{p}^{(3)}_{\alpha^2}(h_c) & = \dots\nn
\end{align}
In this case, we find that the series 
$\bar{p}_{\alpha^2}(h_c) = \sum_{m=0}^{\infty}\bar{p}^{(m)}_{\alpha^2}(h_c)$ can be resummed, and we get the 
neat expression
\begin{align}\label{eq:OrderAlpha2Resum}
\bar{p}_{\alpha^2}(h_c) \simeq \frac{1}{\sqrt{2\pi}v_c}\exp\left[-\frac{v_c^2}{2} + 
\alpha\sigma_0 v_c^{3}(1 + 2\alpha\sigma_0 v_c)\right]\,.
\end{align}    
We can do the same comparison we did before, and compare 
{\it i)} the gaussian result in eq.\,(\ref{eq:Gau1}), {\it ii)} the exact non-gaussian result in 
eq.\,(\ref{eq:ExactProb}), {\it iii)} the order $O(\alpha^2)$ approximation 
$\bar{p}_{\alpha^2}(h_c) = \sum_{m=0}^{\infty}\bar{p}^{(m)}_{\alpha^2}(h_c)$ 
obtained by truncating the series at increasing values of $m$ and {\it iv)} 
the order $O(\alpha^2)$ approximation resummed as in eq.\,(\ref{eq:OrderAlpha2Resum}). 
  This comparison is shown in the right panel of fig.\,\ref{fig:SimplifiedCase}.
 \begin{figure}[!htb!]
\begin{center}
$$
\includegraphics[width=.48\textwidth]{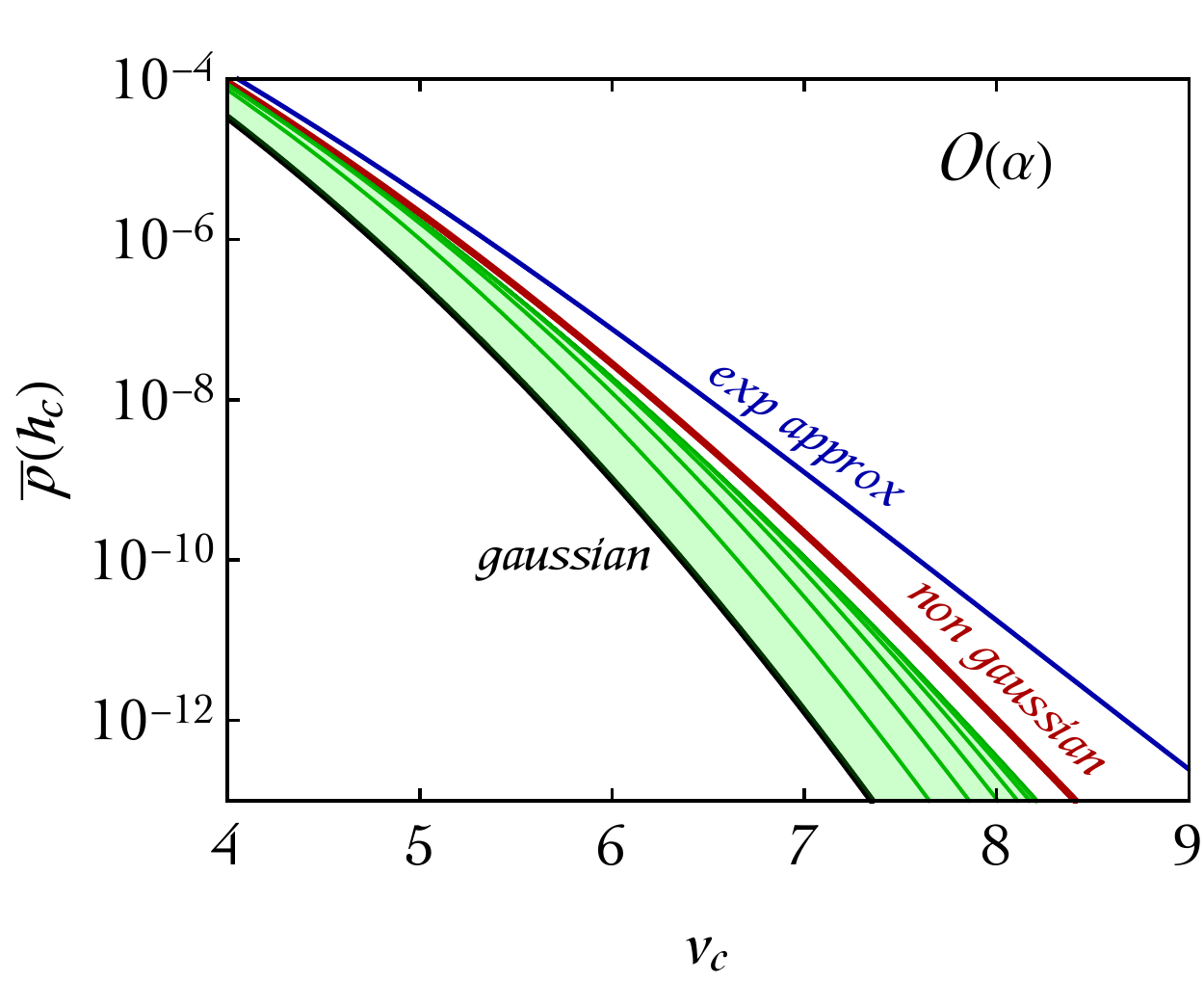}\qquad
\includegraphics[width=.48\textwidth]{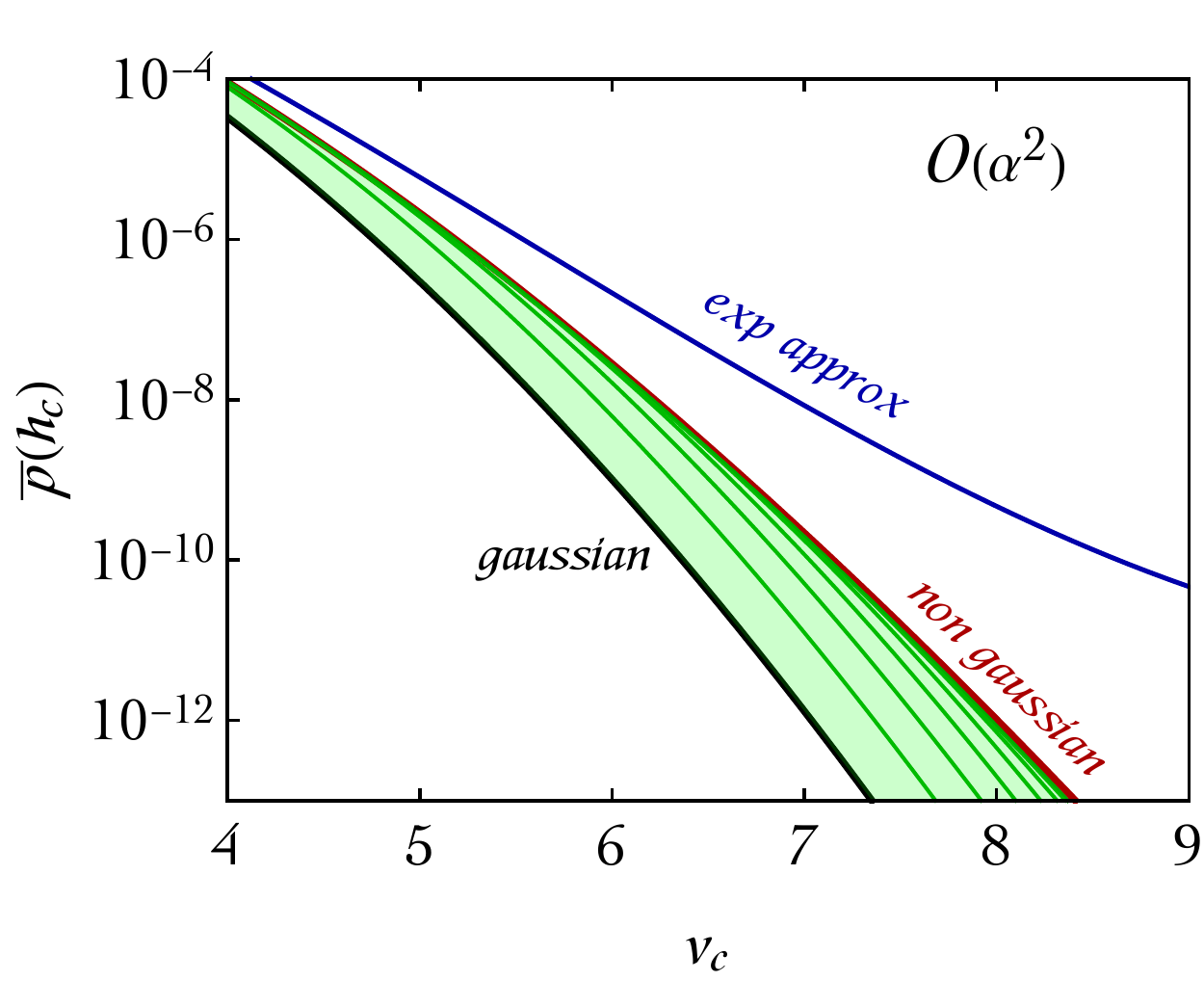}$$
\caption{\em \label{fig:SimplifiedCase}  
Left panel. We compare at order $O(\alpha)$ the 
exact non-gaussian expression for $\bar{p}(h_c)$ (defined below eq.\,(\ref{eq:Dybala})) 
given in eq.\,(\ref{eq:ExactProb}) with: {\it i)} its gaussian limit (eq.\,(\ref{eq:Gau1})), 
ii) its power-series expansion $\bar{p}_{\alpha}(h_c) = \sum_{m=0}^{\infty}\bar{p}^{(m)}_{\alpha}(h_c)$ for increasing values of $m$ (green region) and iii) the resummed expression (labelled ``exp approx'') given in eq.\,(\ref{eq:OrderAlphaResum}).
For illustration, we take $\sigma_0 = 0.1$ and $\alpha = 0.2$. 
Right panel. Same as in the left panel but at order $O(\alpha^2)$.
}
\end{center}
\end{figure}  

The simplified setup studied in this exercise is conceptually different compared to the actual problem we are facing (since we are only considering here the non-gaussian random field $h$ without any information about its derivatives) but, as we shall discuss later, 
we will use an approximation scheme very similar to the one adopted above. 
It is, therefore, instructive to draw some conclusions (this is particularly important in light of the fact that 
in the case without derivatives we know the exact form of the probability density distribution). 
Consider first the $O(\alpha)$ approximation illustrated in the left panel of fig.\,\ref{fig:SimplifiedCase}. 
The sum $\bar{p}_{\alpha}(h_c) = \sum_{m=0}^{\infty}\bar{p}^{(m)}_{\alpha}(h_c)$ (truncated at some finite $m$; in the plot we show the cases with $m \leqslant 9$) already gives a good---even though not optimal---approximation of $\bar{p}_{\rm NG}(h_c)$. 
On the contrary, the resummed expression in eq.\,(\ref{eq:OrderAlphaResum}), more simple to handle, 
tends to overestimate the true result. 
Consider now the $O(\alpha^2)$ approximation illustrated in the right panel of fig.\,\ref{fig:SimplifiedCase}. 
The sum $\bar{p}_{\alpha^2}(h_c) = \sum_{m=0}^{\infty}\bar{p}^{(m)}_{\alpha^2}(h_c)$ 
(truncated at some finite $m$; in the plot we show the cases with $m \leqslant 9$) gives an excellent approximation 
of $\bar{p}_{\rm NG}(h_c)$. 
On the contrary, it is evident that the resummed expression in eq.\,(\ref{eq:OrderAlpha2Resum}) quickly diverges  
from the true result and can not be trusted. 

The reason why the exponential approximation deviates from the exact result can be identified already at order 
$O(\alpha)$. Consider (in units of $\bar{p}_{\rm G}(h_c)$) the first term in eq.\,(\ref{eq:Approx1}) that we kept in our expansion, 
that is $(\alpha\sigma_0)v_c^3$, and compare it with 
one of the terms that we neglected in eq.\,(\ref{eq:Approx3}), for instance the fourth in the round brackets 
$-70(\alpha\sigma_0)^3v_c^3$. These two terms have the same power of $v_c$ but the latter
is parametrically suppressed by $(\alpha\sigma_0)^2$. It is true that we expect 
$(\alpha\sigma_0)^2 \sim 10^{-2} \ll 1$ but we also have an extra numerical factor $-70$ that partially compensate 
the suppression. 
On similar ground, neglecting the term $-(14/3)(\alpha\sigma_0)^3v_c^7$ in eq.\,(\ref{eq:Approx3}) 
compared to the terms that we kept in eq.\,(\ref{eq:Approx1}) and eq.\,(\ref{eq:Approx2}) seems not 
justifiable. 

We conclude that the exponential approximations in eq.\,(\ref{eq:OrderAlphaResum}) and eq.\,(\ref{eq:OrderAlpha2Resum}) 
can not be considered as acceptably good proxies for the exact result since they may overestimate it by many orders of magnitude. 
However, in order to have a good approximation of the exact result it is more proper to 
decompose $\bar{p}_{\alpha}(h_c) = \sum_{m=0}^{\infty}\bar{p}^{(m)}_{\alpha}(h_c)$ (or even better at order $O(\alpha^2)$) and 
sum over $m$ up to some finite value according to the desired accuracy. 
  
\end{itemize}

After this digression, we are finally ready to compute the non-gaussian cumulants for the random variables $h$, $h_{i}$,  $h_{ij}$. 
For simplicity, we start again from the two-dimensional case, and we have six random variables $\{h,h_{x},h_{y},h_{xx},h_{xy},h_{yy}\}$.
We consider the non-gaussian function $f_{\rm NL}(\mathcal{R}) = \alpha(\mathcal{R}^2 - \langle \mathcal{R}^2\rangle)$ since in this case 
we have $C_1(h) = \langle h\rangle = 0$.
Furthermore, in order to fully exploit the properties of homogeneity and isotropy (see discussion below eq.\,(\ref{eq:sigmaj2})), 
it is useful to change basis to the complex conjugated
random variables $\{h,h_z,h_{z^*},h_{zz},h_{zz^*},h_{z^*z^*}\}$.\footnote{
The following relations turn out to be useful. For a generic function $f(x,y):\mathbb{R}^2\to \mathbb{R}$ we have 
$f_z = (f_x - if_y)/2$, $f_{z^*} = (f_x + if_y)/2$ and 
\begin{align}\label{eq:Transformation}
f_{zz^{*}} = \frac{1}{4}\left(f_{xx} + f_{yy}\right)\,,\hspace{0.5cm}
f_{zz} = \frac{1}{4}\left(f_{xx} - f_{yy} - 2if_{xy}\right)\,,\hspace{0.5cm}
f_{z^*z^*} = \frac{1}{4}\left(f_{xx} - f_{yy} + 2if_{xy}\right)\,.
\end{align}
The stationary point condition $f_x = f_y = 0$ becomes $f_z = 0$ (hence $f_{z^*} = 0$). 
The condition $f_{xx}f_{yy} - f_{xy}^2 > 0$ that separates extrema from saddle points becomes 
$f_{zz^*}^2 > f_{zz}f_{z^*z^*} = |f_{zz}|^2$ that is $f_{zz^*} < - |f_{zz}| \lor f_{zz^*} > |f_{zz}|$ (equivalently, 
$|f_{zz^*}| > |f_{zz}|$). Notice that $f_{zz^{*}}$ is real. 
Maxima correspond to the condition $f_{zz^{*}} < 0$ while minima are identified by $f_{zz^{*}} > 0$.
}

All first-order cumulants vanish. This is true for $C_1(h) = \langle h\rangle = 0$ as discussed before. 
As far as the other first-order cumulants are concerned, we have for instance $C_1(h_z) = \langle h_z\rangle = \partial_z\langle h\rangle = 0$ 
which vanishes (even without imposing $C_1(h) = 0$) because $\langle h\rangle$ does not depend, as a consequence of homogeneity, on the specific spatial point 
at which it is computed. Similarly, we have $C_1$ also for the remaining random variables.

As far as the second-order cumulants are concerned, we find the following non-zero entries
\begin{equation}
\bordermatrix{     
            & {\scriptstyle h}    & {\scriptstyle h_z}     & {\scriptstyle h_{z^*}} & 
            {\scriptstyle h_{zz}}  & {\scriptstyle h_{zz^*}} & {\scriptstyle h_{z^*z^*}}    \cr
    {\scriptstyle h}           & C_2(h,h) & 0 & 0 & 0 & C_2(h,h_{zz^*}) & 0    \cr
    {\scriptstyle h_z}         & 0 & 0 & C_2(h_{z},h_{z^*}) & 0 & 0 & 0    \cr
   {\scriptstyle  h_{z^*}}     & 0 & C_2(h_{z},h_{z^*}) & 0 & 0 & 0 & 0    \cr
   {\scriptstyle  h_{zz}}      & 0 & 0 & 0 & 0 & 0 & C_2(h_{zz},h_{z^*z^*})     \cr
   {\scriptstyle  h_{zz^*}}    & C_2(h,h_{zz^*}) & 0 & 0 & 0 & C_2(h_{zz^*},h_{zz^*})  & 0    \cr
   {\scriptstyle  h_{z^*z^*}}  & 0 & 0 & 0 & C_2(h_{zz},h_{z^*z^*})  & 0 & 0     
}\,,\label{eq:NonGaussianSecondCumuMatrix}
\end{equation}
with 
\begin{eqnarray}
C_2(h,h) & = & \sigma_0^2(1+2\alpha^2\sigma_0^2)\,,\label{eq:C21}\\
C_2(h_z,h_{z^*}) & = & \frac{\sigma_1^2}{4}(1+4\alpha^2\sigma_0^2)\,,\label{eq:C22}\\
C_2(h,h_{zz^*}) & = & -\frac{\sigma_1^2}{4}(1+4\alpha^2\sigma_0^2)\,,\label{eq:C23}\\
C_2(h_{zz},h_{z^*z^*}) = C_2(h_{zz^*},h_{zz^*}) & = & \frac{\sigma_2^2}{16}\left[1 + 4\alpha^2\left(\frac{2\sigma_1^4}{\sigma_2^2} + 
\sigma_0^2\right)\right]\,.\label{eq:C24}
\end{eqnarray}
At order $O(\alpha)$, all non-gaussian corrections vanish.
We already computed $C_2(h,h)$ in eq.\,(\ref{eq:SecondOrderCheck}). It is instructive to consider explicitly 
few more cases.  
The relevant equations are eq.\,(\ref{eq:Cumutrick}), eq.\,(\ref{eq:CumufNL1}), eq.\,(\ref{eq:CCC}) 
and eq.\,(\ref{eq:CumufNL2}).

Consider for instance the computation of $C_2(h_z,h_{z^*})$. 
We have 
\begin{align}
& C_2(h_z,h_{z^*})  \overset{\makebox[0pt]{\mbox{\tiny eq.\,(\ref{eq:Cumutrick})}}}{~=~} 
\left.\partial_{z_1}\partial_{z_2^*}C_2(h_1,h_2)\right|_{\vec{x}_1 = \vec{x}_2} \\ & 
= 
\left.\partial_{z_1}\partial_{z_2^*}\left\{
C_2(\mathcal{R}_1,\mathcal{R}_2) + 
C_2[f_{\rm NL}(\mathcal{R}_1),\mathcal{R}_2]+C_2[\mathcal{R}_1,f_{\rm NL}(\mathcal{R}_2)] 
+C_2[f_{\rm NL}(\mathcal{R}_1),f_{\rm NL}(\mathcal{R}_2)]
\right\}\right|_{\vec{x}_1 = \vec{x}_2}\nn \\ & 
=
\partial_{z_1}\partial_{z_2^*}
\big[
\langle\mathcal{R}_1\mathcal{R}_2\rangle + 
\underbrace{\langle f_{\rm NL}^{(1)}(\mathcal{R}_1)\rangle\langle\mathcal{R}_1\mathcal{R}_2\rangle + 
\langle f_{\rm NL}^{(1)}(\mathcal{R}_2)\rangle\langle\mathcal{R}_1\mathcal{R}_2\rangle}_{\tiny {O(\alpha),\,\,\rm eq.\,(\ref{eq:CumufNL1})}} + 
\underbrace{\langle f_{\rm NL}(\mathcal{R}_1)f_{\rm NL}(\mathcal{R}_2) \rangle -
\langle f_{\rm NL}(\mathcal{R}_1)\rangle\langle f_{\rm NL}(\mathcal{R}_2) \rangle}
_{\tiny {O(\alpha^2),\,\,\rm eq.\,(\ref{eq:CCC})}}
\big]\big|_{\vec{x}_1 = \vec{x}_2}\nn \\ &
= \langle\mathcal{R}_z\mathcal{R}_{z^*}\rangle + 
2\langle f_{\rm NL}^{(1)}(\mathcal{R})\rangle\langle\mathcal{R}_z\mathcal{R}_{z^*}\rangle + 
\alpha^2
\partial_{z_1}\partial_{z_2^*}\langle\mathcal{R}_1^2\mathcal{R}_2^2\rangle\big|_{\vec{x}_1 = \vec{x}_2}\label{eq:AA1} \\
& = \langle\mathcal{R}_z\mathcal{R}_{z^*}\rangle + 4\alpha^2\langle \mathcal{R}^2\mathcal{R}_z\mathcal{R}_{z^*}\rangle\,.\label{eq:AA2}
\end{align}
Eq.\,(\ref{eq:AA1}) follows from the fact that the expectation values $\langle f_{\rm NL}(\mathcal{R})\rangle$ and 
$\langle f_{\rm NL}^{(1)}(\mathcal{R})\rangle$ do not depend, because of homogeneity, on the specific spatial point at which they are 
evaluated and, therefore, their spatial derivatives vanish. 
Eq.\,(\ref{eq:AA2}) follows from the fact that $\langle f_{\rm NL}^{(1)}(\mathcal{R})\rangle = 2\alpha\langle \mathcal{R}\rangle = 0$; 
at order $O(\alpha^2)$, we moved the derivatives inside the statistical average and set $\vec{x}_1 = \vec{x}_2$.  
In eq.\,(\ref{eq:AA2}) we have $\langle\mathcal{R}_z\mathcal{R}_{z^*}\rangle =
 (\langle\mathcal{R}_x\mathcal{R}_{x}\rangle+\langle\mathcal{R}_{y}\mathcal{R}_{y}\rangle)/4 =\sigma_1^2/4$ and 
 $\langle\mathcal{R}^2\mathcal{R}_z\mathcal{R}_{z^*}\rangle = 
 \langle \mathcal{R}^2(\mathcal{R}_x^2+\mathcal{R}_y^2)\rangle/4$. 
 This statistical average can be computed by means of the multivariate normal distribution in eq.\,(\ref{eq:GaussianPDF})
 \begin{align}
\langle \mathcal{R}^2(\mathcal{R}_x^2+\mathcal{R}_y^2)\rangle & = 
\int d\mathcal{R}d\mathcal{R}_xd\mathcal{R}_yd\mathcal{R}_{xx}d\mathcal{R}_{xy}d\mathcal{R}_{yy}
  \,\mathcal{R}^2(\mathcal{R}_x^2+\mathcal{R}_y^2)\,
  P(\mathcal{R}_x) P(\mathcal{R}_y) P(\mathcal{R}_{xy}) P(\mathcal{R},\mathcal{R}_{xx},\mathcal{R}_{yy}) \\
 & = 
 \underbrace{\int d\mathcal{R}_xd\mathcal{R}_y(\mathcal{R}_x^2+\mathcal{R}_y^2)P(\mathcal{R}_x) 
 P(\mathcal{R}_y)}_{ = \sigma_1^2}
 \underbrace{\int d\mathcal{R}_{xy}P(\mathcal{R}_{xy})}_{=\,1}
\int d\mathcal{R}\mathcal{R}^2
\underbrace{\int d\mathcal{R}_{xx}d\mathcal{R}_{yy}P(\mathcal{R},\mathcal{R}_{xx},\mathcal{R}_{yy})}_
{=\,\frac{1}{\sqrt{2\pi\sigma_0^2}}\exp(-\mathcal{R}^2/2\sigma_0^2) }\,,
 \nn
\end{align}
and, from the explicit computation of the integrals, we get 
$\langle \mathcal{R}^2(\mathcal{R}_x^2+\mathcal{R}_y^2)\rangle = \sigma_0^2\sigma_1^2$. 
All in all, we find $C_2(h_z,h_{z^*}) = \sigma_1^2(1+4\alpha^2\sigma_0^2)/4$. 
All the remaining non-zero entries in eq.\,(\ref{eq:NonGaussianSecondCumuMatrix}) can be derived in the same way. 

In eq.\,(\ref{eq:NonGaussianSecondCumuMatrix}) the zero entries vanish as a consequence of isotropy. 
We can introduce again the parameter $\kappa \equiv (\#\,z^*\,{\rm derivatives\,in\,}C_2) - (\#\,z\,{\rm derivatives\,in\,}C_2)$  
and set to zero all cumulants with $\kappa \neq 0$ since not invariant under spatial rotations.
The reasoning that we followed in the gaussian case for the computations of the two-point correlators (see discussion below eq.\,(\ref{eq:sigmaj2})) applies also in the case of generic $n^{\rm th}$-order cumulants. This is because cumulants are functions of moments (see eqs.\,(\ref{eq:FirstCum}-\ref{eq:FourthCum})). 
Furthermore, the relation between $C_2(h_z,h_{z^*})$ and $C_2(h,h_{zz^*})$ can be understood as a consequence of homogeneity. 

We now move to consider the third-order cumulants. 
Based on isotropy, we expect only a handful of non-zero cumulants (that are those with $\kappa = 0$) that we list in table\,\ref{tab:1}.
\begin{table}[!h!]
\begin{center}
\begin{adjustbox}{max width=1\textwidth}
\begin{tabular}{|c|c|c|c|}
\multicolumn{1}{c}{\scriptsize No derivatives} & \multicolumn{1}{c}{\scriptsize 2 derivatives} & \multicolumn{1}{c}{\scriptsize 4 derivatives} & \multicolumn{1}{c}{\scriptsize 6 derivatives} \\ 
\hline
\multirow{2}{*}{$C_3(h,h,h)$} & \multirow{2}{*}{$C_3(h,h,h_{zz^*})$} & \multirow{2}{*}{$C_3(h,h_{zz^*},h_{zz^*})$} &  \multirow{2}{*}{$C_3(h_{zz^*},h_{zz^*},h_{zz^*})$}    \\
 {\tiny $\times 1$} & {\tiny $\times 3$} & {\tiny $\times 3$} & {\tiny $\times 1$}   \\ \hline 
\multicolumn{1}{c|}{} & \multirow{2}{*}{$C_3(h,h_{z},h_{z^*})$} &  \multirow{2}{*}{$C_3(h_{z},h_{z^*},h_{zz^*})$} & \multirow{2}{*}{$C_3(h_{zz},h_{z^*z^*},h_{zz^*})$}  \\   
\multicolumn{1}{c|}{} & {\tiny $\times 6$} & {\tiny $\times 6$} & {\tiny $\times 6$}  \\  \cline{2-4}
\multicolumn{1}{c}{} & \multicolumn{1}{c|}{} &  \multirow{2}{*}{$C_3(h,h_{zz},h_{z^*z^*})$} & \multicolumn{1}{|c}{}  \\   
\multicolumn{1}{c}{} & \multicolumn{1}{c|}{} & {\tiny $\times 6$} & \multicolumn{1}{|c}{}   \\  \cline{3-3}
\multicolumn{1}{c}{} & \multicolumn{1}{c|}{} &  \multirow{2}{*}{$C_3(h_z,h_{z},h_{z^*z^*})$} & \multicolumn{1}{|c}{}  \\   
\multicolumn{1}{c}{} & \multicolumn{1}{c|}{} & {\tiny $\times 3$} & \multicolumn{1}{|c}{}   \\  \cline{3-3}
\multicolumn{1}{c}{} & \multicolumn{1}{c|}{} &  \multirow{2}{*}{$C_3(h_{z^*},h_{z*},h_{zz})$} & \multicolumn{1}{|c}{}  \\   
\multicolumn{1}{c}{} & \multicolumn{1}{c|}{} & {\tiny $\times 3$} & \multicolumn{1}{|c}{}   \\  \cline{3-3} 
\end{tabular}
\end{adjustbox}
	\caption{\it 
	Third-order cumulants for the random variables $\{h,h_{z},h_{z^*},h_{zz},h_{zz^*},h_{z^*z^*}\}$ that are non-zero based on isotropy. 
For these cumulants, we have 
$\kappa \equiv (\#\,z^*\,{\rm{\it derivatives\,in\,}}C_3) - (\#\,z\,{\rm{\it derivatives\,in\,}}C_3) = 0$.
We gather together in each column cumulants with the same number of spatial derivatives (cumulants without derivatives in the first column, with two derivatives in the second, four in the third and so on).
For each entry, the tiny numbers in the second row indicate the multiplicity of the corresponding cumulant due to distinct permutations of its arguments.
	}
	\label{tab:1}
\end{center}
\label{default}
\end{table}%

We already computed $C_3(h,h,h) = 6\alpha\sigma_0^4$ in eq.\,(\ref{eq:ThirdOrderCheck2}). As far as the remaining cumulants are concerned, we find
\begin{eqnarray}
C_3(h,h_z,h_{z^*}) & = & \alpha\sigma_0^2\sigma_1^2\,,\label{eq:Cumu1}\\
C_3(h,h,h_{zz^*}) & = & -2\alpha\sigma_0^2\sigma_1^2\,,\label{eq:Cumu1bis}\\
C_3(h,h_{zz^*},h_{zz^*}) & = & \frac{\alpha}{8}\left(3\sigma_1^4 + 2\sigma_0^2\sigma_2^2\right)\,,\label{eq:Cumu2}\\
C_3(h,h_{zz},h_{z^*z^*}) & = & \frac{\alpha}{4}\sigma_0^2\sigma_2^2\,,\label{eq:Cumu3}\\
C_3(h_z,h_{z^*},h_{zz^*}) & = & -\frac{\alpha}{8}\sigma_1^4\,,\label{eq:Cumu4}\\
C_3(h_z,h_z,h_{z^*z^*}) =  C_3(h_{z^*},h_{z^*},h_{zz}) & = & \frac{\alpha}{4}\sigma_1^4\,,\label{eq:Cumu5}\\
C_3(h_{zz^*},h_{zz^*},h_{zz^*}) & = & - \frac{3\alpha}{16}\sigma_1^2\sigma_2^2\,,\label{eq:Cumu6}\\
C_3(h_{zz},h_{z^*z^*},h_{zz^*}) & = & -\frac{\alpha}{16}\sigma_1^2\sigma_2^2\,.\label{eq:Cumu7}
\end{eqnarray}
We find that corrections at order $O(\alpha^2)$ vanish for all third-order cumulants listed before\footnote{This happens because the $O(\alpha^2)$ contribute to $C_3(h_1,h_2,h_3)$ is:
\begin{equation*}
\Delta C_3 (h_1,h_2,h_3) = 2 \alpha^2 \left[\langle \mathcal{R}_1 \mathcal{R}_2\rangle \langle \mathcal{R}_3^2 (\mathcal{R}_1+\mathcal{R}_2)\rangle + \langle \mathcal{R}_2 \mathcal{R}_3\rangle \langle \mathcal{R}_1^2 (\mathcal{R}_2+\mathcal{R}_3)\rangle+\langle \mathcal{R}_3 \mathcal{R}_1\rangle \langle \mathcal{R}_2^2 (\mathcal{R}_3+\mathcal{R}_1)\rangle \right]
\end{equation*}
Thus, when we derive and then we set $\vec{x_1}=\vec{x_2}=\vec{x_3}$, following the prescription (\ref{eq:Cumutrick}) and we return to the $\{x,y\}$ basis, we see that all the terms have the form $\langle \xi_{j_1} \xi_{j_2} \xi_{j_3} \rangle$, where $\xi_{j_k} \in \{\mathcal{R},\mathcal{R}_x,\mathcal{R}_y,\mathcal{R}_{xx},\mathcal{R}_{xy},\mathcal{R}_{yy}\}$. These are third order moments of a gaussian multivariate distribution, which are vanishing.}. 
However, as already discussed in the case of eq.\,(\ref{eq:ThirdOrderCheck2}), eqs.\,(\ref{eq:Cumu1}-\ref{eq:Cumu7}) are not exact because we expect the presence of $O(\alpha^3)$ corrections we did not include.  
It is simple to check that properties of homogeneity  are respected by the explicit expressions in  eqs.\,(\ref{eq:Cumu1}-\ref{eq:Cumu7}). 
For instance, from $\partial_{z^*}C_3(h_z,h_z,h_{z^*}) = 0$ we find $2C_3(h_z,h_{z^*},h_{zz^*}) = - C_3(h_z,h_z,h_{z^*z^*})$ that is indeed verified by eq.\,(\ref{eq:Cumu4}) and eq.\,(\ref{eq:Cumu5}). 
Similarly, from $\partial_{z^*}C_3(h,h,h_z) = 0$ we have $C_3(h,h,h_{zz^*}) = -2C_3(h,h_z,h_{z^*})$ that is indeed verified by eq.\,(\ref{eq:Cumu1}) and eq.\,(\ref{eq:Cumu1bis}).

After computing the cumulants at the desired order in $\alpha$, we are finally ready to take the last step in our computation. 
From the cumulants, we will reconstruct the characteristic function and, via an inverse Fourier transform, the probability density distribution. 
Few technical  remarks are in order because it is crucial to properly identify the variables participating to the Fourier transforms. 
The inverse Fourier transform of the first line in eq.\,(\ref{eq:CharaDef}) is given by 
\begin{align}\label{eq:CharaDefInve}
P(\xi_1,\dots,\xi_N) =  \int\frac{d\lambda_1}{(2\pi)}\dots\frac{d\lambda_N}{(2\pi)}\,\chi(\lambda_1,\dots,\lambda_N)
\exp\left[
-i(\xi_1 \lambda_1 + \dots + \xi_N \lambda_N)
\right]\,.
\end{align}
Both eq.\,(\ref{eq:CharaDef}) and eq.\,(\ref{eq:CharaDefInve}) are valid for real variables. The simplest identification, therefore, would be 
$\{\xi_{i=1,\dots,6}\} = \{h,h_x,h_y,h_{xx},h_{xy},h_{yy}\}$. However, we found more efficient to work with $ \{h,h_z,h_{z^*},h_{zz},h_{zz^*},h_{z^*z^*}\}$ since properties like isotropy become more transparent. Among these variables, $h$ and $h_{zz^*}$ are real while $h_z$ and $h_{zz}$ are complex (with conjugated variables given by $h_{z^*}$ and $h_{z^*z^*}$, respectively). 
In this case a  suitable choice of real variables, therefore,  is $\{\xi_{i=1,\dots,6}\} = \{h,{\rm Re}h_z,{\rm Im}h_z,{\rm Re}h_{zz},h_{zz^*},{\rm Im}h_{zz}\}$ with the obvious relations 
\begin{align}\label{eq:ReImBasics}
{\rm Re}h_z = (h_z + h_{z^*})/2\,,~~~~~{\rm Im}h_z = -i(h_z - h_{z^*})/2\,,~~~~~
{\rm Re}h_{zz} = (h_{zz} + h_{z^*z^*})/2\,,~~~~~{\rm Im}h_{zz} = -i(h_{zz} - h_{z^*z^*})/2\,.
\end{align}
Consequently, we introduce their Fourier counterparts $\{ \lambda_{i=1,\dots,6} \} \equiv \{\lambda, \lambda_{{\rm Re}h_z},
\lambda_{{\rm Im}h_z},\lambda_{{\rm Re}h_{zz}},\lambda_{zz^*},\lambda_{{\rm Im}h_{zz}} \}$ such that the characteristic function in eq.\,(\ref{eq:CharaDef}) is given by the integral
\begin{align}\label{eq.Four}
\int dh\,d{\rm Re}h_z\,d{\rm Im}h_z\,d{\rm Re}h_{zz}\,dh_{zz^*}\,d{\rm Im}h_{zz}
P(h,h_z,h_{z^*},h_{zz},h_{zz^*},h_{z^*z^*})
\exp[i(
h\lambda + {\rm Re}h_z \lambda_{{\rm Re}h_z} + 
 {\rm Im}h_z \lambda_{{\rm Im}h_z} + \dots
)]\,.
\end{align}
Consider, for instance, the two terms ${\rm Re}h_z \lambda_{{\rm Re}h_z} + 
 {\rm Im}h_z \lambda_{{\rm Im}h_z}$ in the argument of the exponential. 
 By means of eqs.\,(\ref{eq:ReImBasics}) we can simply rewrite 
 ${\rm Re}h_z \lambda_{{\rm Re}h_z} + 
 {\rm Im}h_z \lambda_{{\rm Im}h_z} =  h_z \lambda_{z}^* + h_{z^*}\lambda_z$ if we define $\lambda_z \equiv (\lambda_{{\rm Re}h_z} + i\lambda_{{\rm Im}h_z})/2$. 
 Notice that in this case we have $\lambda_{{\rm Re}h_z} = 2{\rm Re}\lambda_z$ and $\lambda_{{\rm Im}h_z} = 2{\rm Im}\lambda_z$. 
 We can, therefore, use $\lambda_z$ as the complex Fourier variable associated to $h_{z^*}$ (and, correspondingly, its conjugated $\lambda_z^*$ associated to $h_z$). 
 Similarly, we introduce $\lambda_{zz} \equiv (\lambda_{{\rm Re}h_{zz}} + i\lambda_{{\rm Im}h_{zz}})/2$, and the exponential function in eq.\,(\ref{eq.Four}) reads 
 $\exp\{
 i[h\lambda + h_{zz^*}\lambda_{zz^*} + (h_z\lambda_z^* + h_{zz}\lambda_{zz}^* + c.c.)]
 \}$. 
 We are now in the position to consider explicitly the inverse Fourier transform of eq.\,(\ref{eq.Four}). As integration variables, 
 instead of $\lambda_{{\rm Re}h_z},
\lambda_{{\rm Im}h_z},\lambda_{{\rm Re}h_{zz}},\lambda_{{\rm Im}h_{zz}}$ we use $\lambda_z$ and $\lambda_{zz}$ introduced before by means of the 
relations $\lambda_{{\rm Re}h_z} = 2{\rm Re}\lambda_z$, $\lambda_{{\rm Im}h_z} = 2{\rm Im}\lambda_z$, 
$\lambda_{{\rm Re}h_{zz}} = 2{\rm Re}\lambda_{zz}$ and $\lambda_{{\rm Im}h_{zz}} = 2{\rm Im}\lambda_{zz}$. 
We have
\begin{align}\label{eq:InegralPDFz}
&P(h,h_z,h_{z^*},h_{zz},h_{zz^*},h_{z^*z^*}) = \\ & \int \frac{d\lambda}{(2\pi)}\frac{d{\rm Re}\lambda_{z}}{\pi}\frac{d{\rm Im}\lambda_{z}}{\pi}
\frac{d\lambda_{zz^*}}{(2\pi)}
\frac{d{\rm Re}\lambda_{zz}}{\pi}\frac{d{\rm Im}\lambda_{zz}}{\pi}
\,\chi(\lambda,\lambda_z,\lambda_{zz},\lambda_{zz^*})\,\exp\left\{
 -i\left[h\lambda + h_{zz^*}\lambda_{zz^*} + \left(h_z\lambda_z^* + h_{zz}\lambda_{zz}^* + c.c.\right)\right]
 \right\}\,,\nn
\end{align}
with of course $\lambda_z = {\rm Re}\lambda_{z} + i {\rm Im}\lambda_{z}$ and $\lambda_{zz} = {\rm Re}\lambda_{zz} + i {\rm Im}\lambda_{zz}$. 
The natural log of the characteristic function $\log\chi(\lambda,\lambda_z,\lambda_{zz},\lambda_{zz^*})$ is written, according to eq.\,(\ref{eq:DefCumu}), as a series expansion 
in terms of the cumulants with respect to the Fourier variables $\{\lambda,\lambda_z,\lambda_{zz},\lambda_{zz^*}\}$ (remember that $\lambda_z$ and $\lambda_{zz}$ are complex variables while $\lambda$ and $\lambda_{zz^*}$ are real).
The first non-gaussian correction arises at order $O(\alpha)$ and reads 
 \begin{align}
 \log\chi(\lambda,&\lambda_z,\lambda_{zz},\lambda_{zz^*}) = \label{eq:ChiGau} \\ &-\frac{1}{2}C_2(h,h)\lambda^2 - C_2(h_z,h_{z^*})|\lambda_z|^2 - C_2(h,h_{zz^*})\lambda\lambda_{zz^*}
 -C_2(h_{zz},h_{z^*z^*})|\lambda_{zz}|^2 - \frac{1}{2}C_2(h_{zz^*},h_{zz^*})\lambda_{zz^*}^2\nn \\
 & -\frac{i}{6}C_3(h,h,h)\lambda^3
 -\frac{i}{2}C_3(h,h,h_{zz^*})\lambda^2\lambda_{zz^*} 
  - iC_3(h,h_z,h_{z^*})\lambda|\lambda_z|^2 
 -\frac{i}{2}C_3(h,h_{zz^*},h_{zz^*})\lambda\lambda_{zz^*}^2 \nn\\
 &- iC_3(h,h_{zz},h_{z^*z^*})\lambda|\lambda_{zz}|^2 -iC_3(h_z,h_{z^*},h_{zz^*})|\lambda_z|^2\lambda_{zz^*} - \frac{i}{2}C_3(h_z,h_z,h_{z^*z^*})(\lambda_z^*)^2\lambda_{zz}
 \nn \\
 &- \frac{i}{2}C_3(h_{z^*},h_{z^*},h_{zz})\lambda_z^2\lambda_{zz}^* -\frac{i}{6}C_3(h_{zz^*},h_{zz^*},h_{zz^*})\lambda_{zz^*}^3
  -iC_3(h_{zz},h_{z^*z^*},h_{zz^*})|\lambda_{zz}|^2\lambda_{zz^*}\,,\nn
 \end{align}
 where the second-order cumulants given in eq.\,(\ref{eq:NonGaussianSecondCumuMatrix}) are taken at order $O(\alpha)$ and, therefore, coincide with their gaussian limit.
 Notice that each term in the series expansion enters with a  coefficient that counts its multiplicity (explicitly written in table\,\ref{tab:1}). 
 This follows from the fact that in the last line in eq.\,(\ref{eq:CharaDef}) there are different terms that give the same contribution. 
 The integration in eq.\,(\ref{eq:InegralPDFz}) gives the probability density distribution $P(h,h_z,h_{z^*},h_{zz},h_{zz^*},h_{z^*z^*})$.  
 After computing $P(h,h_z,h_{z^*},h_{zz},h_{zz^*},h_{z^*z^*})$, it is also possible to reconstruct 
 $P(h,h_x,h_y,h_{xx},h_{xy},h_{yy})$  by means of the Jacobi's multivariate theorem in eq.\,(\ref{eq:TransformedPDF}) using the transformation in eq.\,(\ref{eq:Transformation}). 
 
 It is possible to check this procedure in the gaussian limit $\alpha\to 0$. In the gaussian limit only the second-order cumulants survive (with $\alpha\to 0$ taken in the corresponding 
 expressions given in eq.\,(\ref{eq:NonGaussianSecondCumuMatrix})). 
 The inverse Fourier transform in eq.\,(\ref{eq:InegralPDFz}) can be computed analytically in the gaussian limit, and one gets $P(\mathcal{R},\mathcal{R}_z,\mathcal{R}_{z^*},\mathcal{R}_{zz},\mathcal{R}_{zz^*},\mathcal{R}_{z^*z^*})$ where we used $\mathcal{R}$ instead of $h$ since $\alpha\to 0$.
 The determinant of the Jacobian matrix  is $\left|{\rm det}J(\mathcal{R},\mathcal{R}_x,\mathcal{R}_y,\mathcal{R}_{xx},\mathcal{R}_{xy},\mathcal{R}_{yy})\right| = 1/64$, and Jacobi's multivariate theorem gives  precisely eq.\,(\ref{eq:GaussianPDF}). 
 This is a non-trivial check that our procedure is technically correct.

In the presence of local non-gaussianities, the computation of eq.\,(\ref{eq:InegralPDFz}) is much more complicated because 
$\chi(\lambda,\lambda_z,\lambda_{zz},\lambda_{zz^*})$ in eq.\,(\ref{eq:ChiGau}) is an exponential function whose argument 
is a polynomial of cubic order. 
A possible solution strategy is the following. 
Let us write eq.\,(\ref{eq:InegralPDFz}) in the schematic form (we use for illustration a generic set of variables $\{\lambda_{i=1,\dots,N}\}$  that in the actual computation 
must be replaced with $\{\lambda,\lambda_z,\lambda_{zz},\lambda_{zz^*}\}$)
\begin{align}\label{eq:Trick}
\chi(\lambda_i) = \exp[p_2(\lambda_i) + p_3(\lambda_i)] =
\exp[p_2(\lambda_i)]\exp[p_3(\lambda_i)] = \exp[p_2(\lambda_i)]\sum_{m=0}^{\infty}\frac{1}{m!}p_3(\lambda_i)^m\,,
\end{align}
where $p_2(\lambda_i)$ is the quadratic polynomial in $\{\lambda_i\}$ that corresponds to the gaussian limit $\alpha\to 0$
while $p_3(\lambda_i)$ is the cubic polynomial that contains $O(\alpha)$ deviations. 
At a given order $m$ in the series expansion defined by eq.\,(\ref{eq:Trick}), the inverse Fourier integral in eq.\,(\ref{eq:InegralPDFz}) 
can be computed analytically. 
This is because we can make use of the following identity (the sum over $k$ is understood)
\begin{align}\label{eq:Integral2Derivatives}
\int d\lambda_1\dots\lambda_N (\lambda_i^n\dots\lambda_j^m)\exp[p_2(\lambda_i)]\exp(-i\lambda_k h_k) = 
(i)^n\dots(i)^m \frac{\partial^n}{\partial h_i^n}\dots
\frac{\partial^m}{\partial h_j^m}
\int d\lambda_1\dots\lambda_N\exp[p_2(\lambda_i)]\exp(-i\lambda_k h_k)\,,
\end{align}
where the last integral is nothing but the gaussian integral that can be easily computed. 
We go along with this strategy, that we summarize for ease of reading in the following.
\begin{itemize}
\item [{\it i)}] For fixed $m$, we compute the probability density distribution in eq.\,(\ref{eq:InegralPDFz}) performing the inverse Fourier transform 
by means of the trick in eq.\,(\ref{eq:Trick}) and eq.\,(\ref{eq:Integral2Derivatives}).
\item [{\it ii)}] We transform $P(h,h_z,h_{z^*},h_{zz},h_{zz^*},h_{z^*z^*})$ into $P(h,h_x,h_{y},h_{xx},h_{xy},h_{yy})$ by means of 
Jacobi's multivariate theorem (as illustrated before for the gaussian case).
\item [{\it iii)}] We set $h_x = h_y = 0$, and compute the number density of local maxima $n_{\rm max}(h)$ according to the general definition in 
eq.\,(\ref{eq:nMax}) (with of course $h$ instead of $\mathcal{R}$, and working in two spatial dimensions). 
\item [{\it iv)}] We use the same change of variables proposed in eq.\,(\ref{eq:ChangeOfVariable}) (with again $h$ instead of $\mathcal{R}$), 
and we integrate over $\theta$, $r$, $h$ and $s$ in the domain defining maxima, as discussed in the gaussian case; similarly, we implement the same lower bound of integration 
over $s$ given by the threshold condition for black hole formation in eq.\,(\ref{eq:ExplThresh}). 
Notice that in the following we will set the lower limit of integration over $h$ to be $h_{\rm min} = -\infty$ 
despite the fact that we have $h_{\rm min} = -(1+4\alpha^2\sigma_0^2)/4\alpha$ for 
$h = \mathcal{R}+ \alpha(\mathcal{R}^2 - \sigma_0^2)$ and $\mathcal{R} \in (-\infty,+\infty)$. 
Approximating $h_{\rm min}$ with its gaussian value allows to obtain close analytical formulas. 
We shall justify the validity of this approximation at the end of this section.
\item [{\it v)}]  Finally, we define the mass fraction of black holes as in eq.\,(\ref{eq:MassFractionGaussian}).
\end{itemize}
We now illustrate the result of this procedure. 
Needless to say, the term with $m=0$ reproduces the gaussian result in eq.\,(\ref{eq:MassFractionGaussian}). 
The first correction is given by the term with $m=1$. 
Let us introduce for the sake of simplicity the notation $\beta = \beta_{\rm G} + \sum_{m=1}^{\infty}\beta_{{\rm NG}}^{(\alpha,m)}$ 
with $\beta_{\rm G}$ given by eq.\,(\ref{eq:MassFractionGaussian}).
We find
\begin{align}
\beta_{\rm NG}^{(\alpha,1)} = 
\left(
\frac{\alpha\sigma_1^2}{\sigma_2}
\right)\frac{2\sqrt{2}}{\pi^{3/2}}
e^{-6s_{\rm min}^2/\sigma_2^2}
 \left[
\frac{s_{\rm min}^2}{\sigma_2^2} 
+ e^{4s_{\rm min}^2/\sigma_2^2}\left(
 - \frac{s_{\rm min}^2}{\sigma_2^2} + 4\frac{s_{\rm min}^4}{\sigma_2^4}
\right)
\right]\,.
\end{align}
For $m=2$, we find
\begin{align}
\beta_{\rm NG}^{(\alpha,2)} =
\left(
\frac{\alpha\sigma_1^2}{\sigma_2}
\right)^2\frac{e^{-6s_{\rm min}^2/\sigma_2^2}}{3\pi^{3/2}}\bigg\{&
2\sqrt{3\pi}e^{6s_{\rm min}^2/\sigma_2^2}{\rm Erfc}\bigg(\frac{\sqrt{6}s_{\rm min}}{\sigma_2}\bigg) 
+ 3\sqrt{2}\frac{s_{\rm min}}{\sigma_2}
\bigg[
7 - 40\frac{s_{\rm min}^2}{\sigma_2^2} 
+48\frac{s_{\rm min}^4}{\sigma_2^4}\nn \\
& +e^{4s_{\rm min}^2/\sigma_2^2}\bigg(
-3 + 68\frac{s_{\rm min}^2}{\sigma_2^2} - 176\frac{s_{\rm min}^4}{\sigma_2^4} 
+ 64\frac{s_{\rm min}^6}{\sigma_2^6}
\bigg)\bigg]\bigg\}\,.
\end{align}
For increasing values of $m$ the corresponding expressions of $\beta_{\rm NG}^{(m)}$ 
become more lengthy and less transparent, and we do not report them explicitly.
On the contrary, we make use of the same expansion introduced in 
eq.\,(\ref{eq:MassFractionGaussian}) for $s_{\rm min}/\sigma_2 \gtrsim 1$.
Interestingly, in this limit we find that the generic term $\beta_{\rm NG}^{(\alpha,m)}$ can be written as
\begin{align}\label{eq:BetaAlphaGenericM}
\beta_{\rm NG}^{(\alpha,m)} \simeq \bigg[
\bigg(\frac{s_{\rm min}^3}{\sigma_2^3}\bigg)\frac{\alpha\sigma_1^2}{\sigma_2}
\bigg]^{m} 
\frac{2^{4m-1/2}}{\pi^{3/2}m!}\bigg(\frac{s_{\rm min}}{\sigma_2}\bigg)e^{-2s_{\rm min}^2/\sigma_2^2}\,,
\end{align}
and the sum $\sum_{m=1}^{\infty}\beta_{\rm NG}^{(\alpha,m)}$ admits the following analytical form
\begin{align}
\sum_{m=1}^{\infty}\beta_{\rm NG}^{(\alpha,m)} \simeq 
\frac{1}{\sqrt{2}\pi^{3/2}}\bigg(\frac{s_{\rm min}}{\sigma_2}\bigg)e^{-2s_{\rm min}^2/\sigma_2^2}
\bigg\{
-1 + \exp\bigg[16\left(\frac{s_{\rm min}^3}{\sigma_2^3}\right)
\frac{\alpha \sigma_1^2}{\sigma_2}\bigg]
\bigg\}\,.
\end{align}
The sum $\beta = \beta_{\rm G} + \sum_{m=1}^{\infty}\beta_{\rm NG}^{(\alpha,m)}$ gives
 \begin{align}
\beta \simeq \frac{1}{\sqrt{2}\pi^{3/2}}\left(\frac{s_{\rm min}}{\sigma_2}\right)\exp\left[
-\frac{2s_{\rm min}^2}{\sigma_2^2} + 16\left(\frac{s_{\rm min}^3}{\sigma_2^3}\right)
\frac{\alpha \sigma_1^2}{\sigma_2}
\right]\,.\label{eq:MasterFormulaApp}
\end{align}
  \begin{figure}[!htb!]
\begin{center}
$$
\includegraphics[width=.48\textwidth]{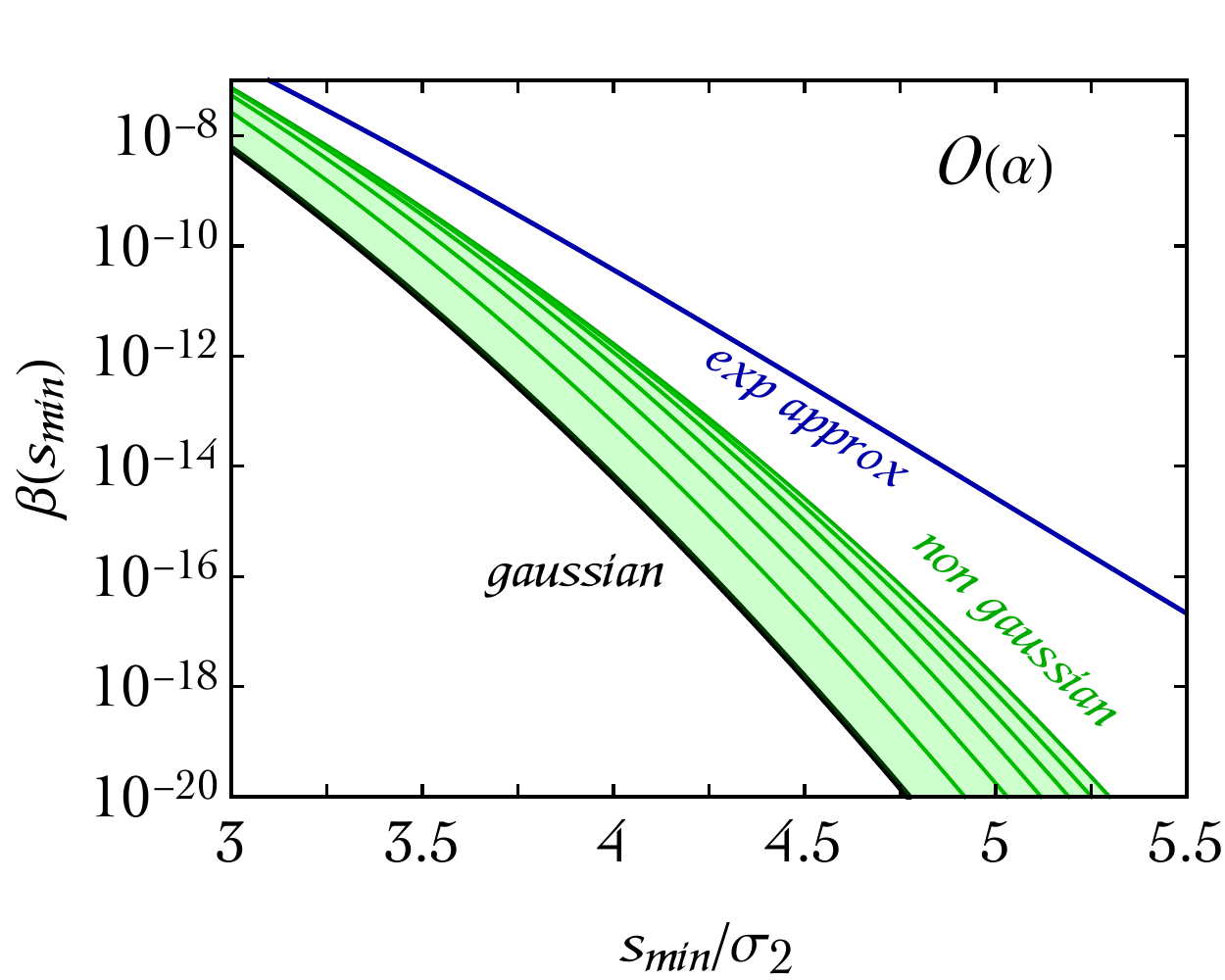}\qquad
\includegraphics[width=.48\textwidth]{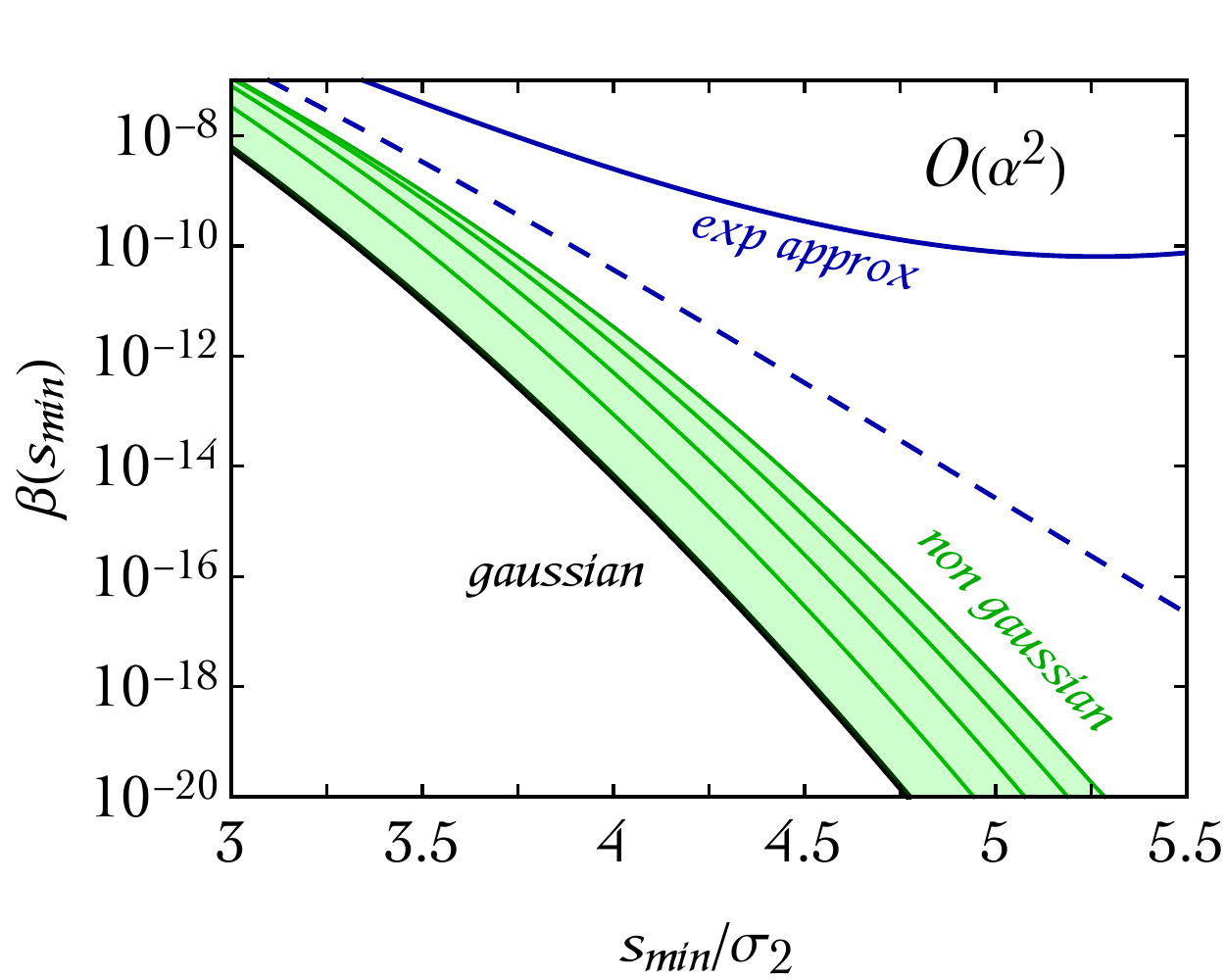}$$
\caption{\em \label{fig:BetaSeries}  
Left panel. 
We compare the gaussian approximation for $\beta$ given in eq.\,(\ref{eq:MassFractionGaussian}) with 
i) its power-series expansion $\beta = \beta_{\rm G} + \sum_{m=1}^{\infty}\beta_{\rm NG}^{(\alpha,m)}$ 
computed at order $O(\alpha)$ for increasing values of $m$ (green region, with the thin solid green lines that label the case $m=1,\dots,6$) and ii) the resummed expression (labelled ``exp approx'') given in eq.\,(\ref{eq:MasterFormulaApp}).
Right panel. Same as in the left panel but at order $O(\alpha^2)$ and up to $m=4$. 
In both panels we consider the log-normal model for the power spectrum, and we use the benchmark values $v=0.6$ and $\mathcal{P}_{\mathcal{R}}(k_{\star}) = 5\times 10^{-3}$. We take $\alpha = 0.2$.
}
\end{center}
\end{figure} 
Notice that the structure of this equation is analogue to the one that we found in eq.\,(\ref{eq:OrderAlphaResum}).

Before proceeding, let us come back to the issue of the lower limit of integration over $h$. 
As anticipated, from the very same definition $h = \mathcal{R}+ \alpha(\mathcal{R}^2 - \sigma_0^2)$ it follows that 
$h> h_{\rm min} = -(1+\alpha^2\sigma_0^2)/4\alpha$. 
To be precise, therefore, this lower limit of integration should be implemented when computing the non-gaussian
contributions $\beta_{\rm NG}^{(\alpha,m)}$. However, implementing the gaussian lower limit 
of integration $h_{\rm min} = -\infty$ does not change qualitatively our results, and the reason is the following.
When computing $\beta$, we only focus on maxima with large curvature $-\triangle h/\sigma_2$ since they correspond to 
peak of the overdensity field. 
The value of $-\triangle h/\sigma_2$ is positively correlated with the value of $h/\sigma_0$, and the amount 
of correlation is controlled by the parameter $\gamma$. This fact was illustrated in appendix\,\ref{app:GaussianPeakTheory} 
in the gaussian case, and it remains valid also in the presence of non-gaussianities 
(for the values of $\alpha$ that are relevant for our analysis). 
When $\gamma \to 1$, the correlation is maximal and regions with large $-\triangle h/\sigma_2$ 
are also regions with positive $h/\sigma_0 \gtrsim 1$. This means that regions where $h$ takes negative values are completely irrelevant 
for our purposes. In the cases that are relevant for our analysis, we typically have $\gamma \sim 0.6$; 
we checked numerically (using for simplicity the log-normal power spectrum, and integrating numerically over $s$) 
that the value of $\beta_{\rm NG}^{(\alpha,m)}$ computed with the proper lower bound 
of integration over $h$ remain equal to those obtained with the gaussian limit $h_{\rm min} = -\infty$. 
In order to see some deviation, one should consider the opposite limit $\gamma \to 0$. 
In this limit $-\triangle h/\sigma_2$  and $h/\sigma_0$ are only weakly correlated and 
it is therefore possible to find regions with large $-\triangle h/\sigma_2$ in which $h$ takes negative values. 
In such situation, of course, implementing the correct lower limit of integration over $h$ becomes important. 
However, cases with $\gamma \to 0$ fall outside the class of models we are considering in this paper.  

We compare in the left panel of fig.\,\ref{fig:BetaSeries} the series $\beta = \beta_{\rm G} + \sum_{m=1}^{m_{\rm max}}\beta_{\rm NG}^{(\alpha,m)}$, truncated at some finite $m_{\rm max}$, with the resummed expression in eq.\,(\ref{eq:MasterFormulaApp}). 
We show the cases $m_{\rm max} = 1,\dots,6$ and we use the log-normal power spectrum in eq.\,(\ref{eq:ToyPS}) 
to compute the spectral parameters. The situation is very similar to what already discussed in fig.\,\ref{fig:SimplifiedCase}, and 
we can exploit the knowledge that we gained from that case to argue the following conclusions.
The series $\beta = \beta_{\rm G} + \sum_{m=1}^{m_{\rm max}}\beta_{\rm NG}^{(\alpha,m)}$ shows a convergence towards 
the order $O(\alpha)$ approximation (that is,  we recall, the approximation in which we only include the leading part of the third-order cumulants) of the exact non-gaussian result while the exponential approximation overestimates the abundance. 
 
We move to consider corrections at order $O(\alpha^2)$.
We already computed in eqs.\,(\ref{eq:C21}-\ref{eq:C24}) the second-order cumulants at order $O(\alpha^2)$.
In addition, we need to compute the fourth-order cumulants at order $O(\alpha^2)$. 
We computed in eq.\,(\ref{eq:FourthOrderCheck}) the simplest one without derivatives, 
$C_4(h,h,h,h) = 48\alpha^2\sigma_0^6$.
\begin{table}[!h!]
\begin{center}
\begin{adjustbox}{max width=1\textwidth}
\begin{tabular}{|c|c|c|c|c|}
\multicolumn{1}{c}{\scriptsize No derivatives} & \multicolumn{1}{c}{\scriptsize 2 derivatives} & \multicolumn{1}{c}{\scriptsize 4 derivatives} & \multicolumn{1}{c}{\scriptsize 6 derivatives} & \multicolumn{1}{c}{\scriptsize 8 derivatives} \\ 
\hline
\multirow{2}{*}{$C_4(h,h,h,h)$} & \multirow{2}{*}{$C_4(h,h,h,h_{zz^*})$} & \multirow{2}{*}{$C_4(h,h,h_{zz},h_{z^*z^*})$} &  \multirow{2}{*}{$C_4(h_z,h_{z^*},h_{zz^*},h_{zz^*})$} &  \multirow{2}{*}{$C_4(h_{zz},h_{zz},h_{z^*z^*},h_{z^*z^*})$}   \\
 {\tiny $\times 1$} & {\tiny $\times 4$} & {\tiny $\times 12$} & {\tiny $\times 12$} & {\tiny $\times 6$}  \\ \hline 
\multicolumn{1}{c|}{} & \multirow{2}{*}{$C_4(h,h,h_z,h_{z^*})$} &  \multirow{2}{*}{$C_4(h,h,h_{zz^*},h_{zz^*})$} & 
\multirow{2}{*}{$C_4(h_z,h_z,h_{z^*z^*},h_{zz^*})$} & \multirow{2}{*}{$C_4(h_{zz^*},h_{zz^*},h_{zz^*},h_{zz^*})$}  \\   
\multicolumn{1}{c|}{} & {\tiny $\times 12$} & {\tiny $\times 6$} & {\tiny $\times 12$}  & {\tiny $\times 1$}  \\  \cline{2-5}
\multicolumn{1}{c}{} & \multicolumn{1}{c|}{} &  \multirow{2}{*}{$C_4(h,h_z,h_{z^*},h_{zz^*})$} & 
\multirow{2}{*}{$C_4(h_{z^*},h_{z^*},h_{zz},h_{zz^*})$} & \multirow{2}{*}{$C_4(h_{zz},h_{z^*z^*},h_{zz^*},h_{zz^*})$}  \\   
\multicolumn{1}{c}{} & \multicolumn{1}{c|}{}  & {\tiny $\times 24$} & {\tiny $\times 12$}  & {\tiny $\times 12$}  \\  \cline{3-5}
\multicolumn{1}{c}{} & \multicolumn{1}{c|}{} &  \multirow{2}{*}{$C_4(h,h_z,h_z,h_{z^*z^*})$} & 
\multirow{2}{*}{$C_4(h,h_{zz^*},h_{zz^*},h_{zz^*})$} & \multicolumn{1}{|c}{}  \\   
\multicolumn{1}{c}{} & \multicolumn{1}{c|}{}  & {\tiny $\times 12$} & {\tiny $\times 4$}  & \multicolumn{1}{|c}{}  \\  \cline{3-4}
\multicolumn{1}{c}{} & \multicolumn{1}{c|}{} &  \multirow{2}{*}{$C_4(h,h_{z^*},h_{z^*},h_{zz})$} & 
\multirow{2}{*}{$C_4(h,h_{zz},h_{z^*z^*},h_{zz^*})$} & \multicolumn{1}{|c}{}  \\   
\multicolumn{1}{c}{} & \multicolumn{1}{c|}{}  & {\tiny $\times 12$} & {\tiny $\times 24$}  & \multicolumn{1}{|c}{}  \\  \cline{3-4}
\multicolumn{1}{c}{} & \multicolumn{1}{c|}{} &  \multirow{2}{*}{$C_4(h_z,h_z,h_{z^*},h_{z^*})$} &  \multirow{2}{*}{$C_4(h_z,h_{z^*},h_{zz},h_{z^*z^*})$} & \multicolumn{1}{c}{}  \\   
\multicolumn{1}{c}{} & \multicolumn{1}{c|}{}  & {\tiny $\times 6$} &  {\tiny $\times 24$}  & \multicolumn{1}{c}{}  \\  \cline{3-4}
\end{tabular}
\end{adjustbox}
	\caption{\it Fourth-order cumulants for the random variables $\{h,h_{z},h_{z^*},h_{zz},h_{zz^*},h_{z^*z^*}\}$ that are non-zero based on isotropy. 
For these cumulants, we have $\kappa \equiv (\#\,z^*\,{\rm{\it derivatives\,in\,}}C_4) - (\#\,z\,{\rm{\it derivatives\,in\,}}C_4) = 0$. 
As in table\,\ref{tab:1}, we gather together in each column cumulants with the same number of spatial derivatives (cumulants without derivatives in the first column, with two derivatives in the second, four in the third and so on).
For each entry, the tiny numbers in the second row indicate the multiplicity of the corresponding cumulant due to distinct permutations of its arguments.}
	\label{tab:2}
\end{center}
\label{default}
\end{table}%
The remaining non-zero cumulants are summarized in table\,\ref{tab:2}. 
We find the following explicit expressions
\begin{eqnarray}
C_4(h,h,h,h_{zz^*}) & = & -18\alpha^2\sigma_0^4\sigma_1^2\,,\label{eq:C4_1}\\
C_4(h,h,h_z,h_{z^*}) & = & 6\alpha^2\sigma_0^4\sigma_1^2\,,\label{eq:C4_2}\\
C_4(h,h,h_{zz},h_{z^*z^*}) & = & \frac{3\alpha^2}{2}\sigma_0^4\sigma_2^2\,,\label{eq:C4_3}\\
C_4(h,h,h_{zz^*},h_{zz^*}) & = &  \frac{\alpha^2}{2}\sigma_0^2(10\sigma_1^4 + 3\sigma_0^2\sigma_2^2)\,,\label{eq:C4_4}\\
C_4(h,h_z,h_{z^*},h_{zz^*}) & = & -\frac{3\alpha^2}{2}\sigma_0^2\sigma_1^4\,,\label{eq:C4_5}\\
C_4(h,h_z,h_{z},h_{z^*z^*}) = C_4(h,h_{z^*},h_{z^*},h_{zz}) & = & \alpha^2\sigma_0^2\sigma_1^4\,,\label{eq:C4_6}\\
C_4(h_z,h_z,h_{z^*},h_{z^*}) & = & 2\alpha^2\sigma_0^2\sigma_1^4\,,\label{eq:C4_7}\\
C_4(h_z,h_{z^*},h_{zz^*},h_{zz^*}) & = & \frac{\alpha^2}{4}\sigma_1^2(\sigma_1^4 + \sigma_0^2\sigma_2^2)\,,\label{eq:C4_8}\\
C_4(h_z,h_z,h_{z^*z^*},h_{zz^*})  = C_4(h_{z^*},h_{z^*},h_{zz},h_{zz^*})  & = & 0\,,\label{eq:C4_9}\\
C_4(h,h_{zz^*},h_{zz^*},h_{zz^*}) & = & -\frac{3\alpha^2}{4}\sigma_1^2(\sigma_1^4 + 2\sigma_0^2\sigma_2^2)\,,\label{eq:C4_10}\\
C_4(h,h_{zz},h_{z^*z^*},h_{zz^*}) & = & -\frac{\alpha^2}{2}\sigma_0^2\sigma_1^2\sigma_2^2\,,\label{eq:C4_11}\\
C_4(h_z,h_{z^*},h_{zz},h_{z^*z^*}) & = & \frac{\alpha^2}{4}\sigma_1^2(\sigma_1^4 + \sigma_0^2\sigma_2^2) \,,\label{eq:C4_11bis}\\
C_4(h_{zz},h_{zz},h_{z^*z^*},h_{z^*z^*}) & = & \frac{\alpha^2}{8}\sigma_0^2\sigma_2^4\,,\label{eq:C4_12}\\
C_4(h_{zz^*},h_{zz^*},h_{zz^*},h_{zz^*}) & = &  \frac{3\alpha^2}{16}\sigma_2^2(3\sigma_1^4 + \sigma_0^2\sigma_2^2)\,,\label{eq:C4_13}\\
C_4(h_{zz},h_{z^*z^*},h_{zz^*},h_{zz^*}) & = & \frac{\alpha^2}{32}\sigma_2^2(3\sigma_1^4 + 2\sigma_0^2\sigma_2^2)\,,\label{eq:C4_14}
\end{eqnarray}
that we derive as before using eq.\,(\ref{eq:Cumutrick}) and eq.\,(\ref{eq:CumufNL2}). 
Conceptually, the rest of the computation follows again points {\it i)}-{\it v)} discussed below eq.\,(\ref{eq:Integral2Derivatives}).
At the technical level, instead of eq.\,(\ref{eq:Trick}) we now have
\begin{align}\label{eq:Trick2}
\chi(\lambda_i) = \exp[p_2(\lambda_i) + p_4(\lambda_i)] =
\exp[p_2(\lambda_i)]\exp[p_4(\lambda_i)] = \exp[p_2(\lambda_i)]\sum_{m=0}^{\infty}\frac{1}{m!}p_4(\lambda_i)^m\,,
\end{align}
where $p_4(\lambda_i)$ is the quartic polynomial that contains $O(\alpha)$ and $O(\alpha^2)$ deviations. 
The computation of the abundance follows the same prescription discussed before, and we introduce the series expansion 
 $\beta = \beta_{\rm G} + \sum_{m=1}^{m_{\rm max}}\beta_{{\rm NG}}^{(\alpha^2,m)}$ 
with $\beta_{\rm G}$ given by eq.\,(\ref{eq:MassFractionGaussian}) and $\beta_{{\rm NG}}^{(\alpha^2,m)}$ which 
corresponds to the series in eq.\,(\ref{eq:Trick2}) truncated at some $m>0$. 
For instance, we find (with $\bar{s}_{\rm min} \equiv s_{\rm min}/\sigma_2$ and $\gamma = \sigma_1^2/\sigma_2\sigma_0$)
\begin{align}
&\beta_{{\rm NG}}^{(\alpha^2,1)} = \\ & ~~~~~~~ \frac{\sqrt{2}\alpha\sigma_0 \bar{s}_{\rm min}}{\pi^{3/2}}e^{-2\bar{s}_{\rm min}^2}
\left[
\alpha(2+\gamma^2)\sigma_0 -2\gamma\bar{s}_{\rm min} - 4\alpha(5+8\gamma^2)\sigma_0 \bar{s}_{\rm min}^2 
+8\gamma\bar{s}_{\rm min}^3 + 16\alpha(1+3\gamma^2)\sigma_0\bar{s}_{\rm min}^4
\right] + O(e^{-6\bar{s}_{\rm min}^2})\,.\nn
\end{align} 
The exact analytic expressions for $\beta_{{\rm NG}}^{(\alpha^2,m)}$ are quite lengthly and we do not report them here explicitly.  
The sum $\sum_{m=1}^{\infty}\beta_{{\rm NG}}^{(\alpha^2,m)}$ admits an analytical expression only if we retain, 
in each one of the $\beta_{{\rm NG}}^{(\alpha^2,m)}$, the highest power of $\bar{s}_{\rm min}$ in the polynomial 
that multiplies the leading exponential suppression. We find 
\begin{align}
\label{eq:betaexpalpha2}
\beta \simeq \frac{1}{\sqrt{2}\pi^{3/2}}\left(\frac{s_{\rm min}}{\sigma_2}\right)\exp\left\{
-\frac{2s_{\rm min}^2}{\sigma_2^2} + 16\left(\frac{s_{\rm min}^3}{\sigma_2^3}\right)
\frac{\alpha \sigma_1^2}{\sigma_2} +
32\left(\frac{s_{\rm min}^4}{\sigma_2^4}\right)\alpha^2\left[
3\left(\frac{\sigma_1^2}{\sigma_2}\right)^2 + \sigma_0^2
\right]
\right\}\,.
\end{align}
The structure of this expression is analogue to eq.\,(\ref{eq:OrderAlpha2Resum}). As shown in fig. \ref{fig:BetaSeries}, the exponential approximation contained in eqs. (\ref{eq:MasterFormulaApp}) and (\ref{eq:betaexpalpha2}) gets worse for increasing $s_{\rm min}/\sigma_2$ and diverges from the actual result, overstimating the abundance. The failure of the exponential approximation is analogue to the one discussed in section \ref{sec:Comp} in the context of threshold statistics.

\section{Towards an exact computation}\label{app:NonPer}

The generic $n$-th order cumulant can be exactly computed if we specify the explicit form of the non-gaussianity. In our case $f_{\mathrm{NL}}(\mathcal{R}) = \alpha (\mathcal{R}^2-\langle\mathcal{R}^2\rangle)$.

The translation only affects the first-order cumulant:
\begin{align}
& C_1[\mathcal{R}+ \alpha (\mathcal{R}^2-\langle\mathcal{R}^2\rangle)]=C_1(\mathcal{R}+ \alpha \mathcal{R}^2)-\alpha\langle\mathcal{R}^2\rangle, \\
& C_n[\mathcal{R}_1+ \alpha (\mathcal{R}_1^2-\langle\mathcal{R}^2\rangle),\dots,\mathcal{R}_n+ \alpha (\mathcal{R}_n^2-\langle\mathcal{R}^2\rangle)] = C_n(\mathcal{R}_1+ \alpha \mathcal{R}_1^2,\dots,\mathcal{R}_n+ \alpha \mathcal{R}_n^2), \quad \forall n>1.
\end{align}
Therefore we can set, without loss of generality, $f_{\mathrm{NL}}(\mathcal{R}) = \alpha \mathcal{R}^2$ in the computation of the cumulants of order higher than one. The quadratic form of the non-linearity allows us to use the Gaussian integral extended to the complex plane thanks to analytic continuation.

We know that, in general $C_n(h_1,\dots,h_n)$ is sum of a series of contributes of different orders of which the last one must be $O(\alpha^n)$ and it is represented by $C_n(\alpha \mathcal{R}_1^2,\dots,\alpha \mathcal{R}_n^2)$.

\subsection{Cancellation of the contributes up to $O(\alpha^{n-2})$}
As a first step, we want to prove that $C_n(\alpha \mathcal{R}_1^2,\dots,\alpha \mathcal{R}_k^2,\mathcal{R}_{k+1},\dots,\mathcal{R}_n)=0$ for $k<n-2$. This implies that $C_n(h_1,\dots,h_n)$ starts at order $O(\alpha^{n-2})$. 
Using eq.\,(\ref{eq:FinalCumu}):
\begin{align}
& C_n(\alpha \mathcal{R}_1^2,\dots,\alpha \mathcal{R}_k^2,\mathcal{R}_{k+1},\dots,\mathcal{R}_n) = \\
&= \left.(-i)^n\frac{\partial}{\partial\lambda_1}\dots
\frac{\partial}{\partial\lambda_n}\log
\int d\mathcal{R}_1\dots d\mathcal{R}_n 
e^{i\sum_{i=1}^k \alpha \mathcal{R}_i^2 \lambda_i+i\sum_{i=k+1}^n \mathcal{R}_i \lambda_i}p(\mathcal{R}_1,\dots ,\mathcal{R}_n)\right|_{\lambda_1=\,\dots\,=\lambda_n = 0} = \\
&= \left.(-i)^n\frac{\partial}{\partial\lambda_1}\dots
\frac{\partial}{\partial\lambda_n}\log
\int d\mathcal{R}_1\dots d\mathcal{R}_{k} 
e^{i\sum_{i=1}^k \alpha \mathcal{R}_i^2 \lambda_i}\underbrace{\left[ \int d\mathcal{R}_{k+1}\dots d\mathcal{R}_{n} e^{i\sum_{i=k+1}^n  \mathcal{R}_i \lambda_i}\,
p(\mathcal{R}_1,\dots ,\mathcal{R}_n) \right]}_{\equiv \mathcal{I}'(\mathcal{R}_1,\dots,\mathcal{R}_{k},\lambda_{k+1},\dots,\lambda_n)}\right|_{\lambda= 0}.
\end{align}
We know that
\begin{align}
\int d\mathcal{R}_1 \dots d\mathcal{R}_k e^{i \sum_{i=1}^{k} \mathcal{R}_i \lambda_i} \mathcal{I}'(\mathcal{R}_1,\dots,\mathcal{R}_{k},\lambda_{k+1},\dots,\lambda_n)= \chi(\lambda_1,\dots,\lambda_n) = e^{-\frac{1}{2}\sum_{i,j=1}^n\sigma_{ij}\lambda_i\lambda_j}\, ,
\end{align}
and so
\begin{align}
\mathcal{I}'(\mathcal{R}_1,&\dots,\mathcal{R}_{k},\lambda_{k+1},\dots,\lambda_n) = \nn \\ & \int \frac{d \lambda_1}{2\pi} \dots \frac{d \lambda_k}{2\pi} \exp \bigg[ -i \sum_{i=1}^{k}\lambda_i \mathcal{R}_i-\frac{1}{2} \bigg(\sum_{i,j=1}^{k}\sigma_{ij}\lambda_i \lambda_j+ +2\sum_{i=1}^{k} \sum_{j=k+1}^{n}\sigma_{ij}\lambda_i \lambda_j+\sum_{i,j=k+1}^{n}\sigma_{ij}\lambda_i \lambda_j \bigg)\bigg]\, .
\end{align}
Let us use the matrix notation, defining the following objects:
\begin{align}
& \bm{\lambda} \equiv (\lambda_1, \dots, \lambda_{k})^{\rm T}=(\lambda_i)_{i=1}^{k}, \quad \bm{\mathcal{R}} \equiv (\mathcal{R}_1, \dots, \mathcal{R}_{k})^{\rm T}=(\mathcal{R}_i)_{i=1}^{k},\\
&\hat{\sigma} \equiv (\sigma_{ij})_{i,j=1}^{k},\qquad \bm{\sigma}_{j} \equiv(\sigma_{1j}, \dots, \sigma_{kj})^{\rm T}= (\sigma_{ij})_{i=1}^{k} \ \mathrm{with} \ j \in \{k+1,\dots,n\}\, .
\end{align}
In this way
\begin{align}
\mathcal{I}'(\bm{\mathcal{R}} , \lambda_{k+1},\dots,\lambda_n) & = e^{-\frac{1}{2}\sum_{i,j=k+1}^n \sigma_{ij}\lambda_i \lambda_j}\int \frac{d^{k} \lambda}{(2\pi)^{k}} e^{-\frac{1}{2} \bm\lambda^{\rm T}\hat{\sigma}\bm\lambda+\bm{\lambda}\cdot\left(-i\bm{\mathcal{R}}-\sum_{j=k+1}^n \lambda_j \bm{\sigma}_j \right)}\\
&=\frac{1}{\sqrt{(2\pi)^{k}\det \hat{\sigma}}} \exp\left[\frac{1}{2}\left(i\bm{\mathcal{R}}+\sum_{j=k+1}^n \lambda_j \bm{\sigma}_j \right)^\mathrm{T}\hat{\sigma}^{-1}\left(i\bm{\mathcal{R}}+\sum_{l=k+1}^n \lambda_l \bm{\sigma}_l \right)-\frac{1}{2}\sum_{i,j=k+1}^n \sigma_{ij}\lambda_i \lambda_j\right]\,.\nn
\end{align}
Now, let us consider the integral
\begin{align}
& \mathcal{I}(\lambda_1, \dots, \lambda_n) = \int d\mathcal{R}_1\dots d\mathcal{R}_{k} e^{i\sum_{i=1}^{k}\alpha\mathcal{R}_{i}^2\lambda_i} \mathcal{I}'(\bm{\mathcal{R}} , \lambda_{k+1},\dots,\lambda_n)\, ;
\end{align}
we can write
\begin{equation}
\sum_{i=1}^{k} \alpha \mathcal{R}^2_i \lambda_i = \sum_{i,j=1}^{k} \alpha \mathcal{R}_i \lambda_i \delta_{ij} \mathcal{R}_j = \alpha \bm{\mathcal{R}}^{\rm T} \Lambda \bm{\mathcal{R}}\, ,
\end{equation}
where we defined the diagonal matrix $\Lambda=\mathrm{diag}(\lambda_1,\dots,\lambda_k)$. In this way, we have
\begin{align}
&\mathcal{I}(\lambda_1, \dots, \lambda_n) \nn \\ & 
= \frac{1}{\sqrt{(2\pi)^{k}\det \hat{\sigma}}} \int d^{k}\mathcal{R} \, \exp\left[i\alpha\bm{\mathcal{R}}^{\rm T}\Lambda\bm{\mathcal{R}}+\frac{1}{2}\left(i\bm{\mathcal{R}}+\sum_{j=k+1}^n \lambda_j \bm{\sigma}_j \right)^\mathrm{T}\hat{\sigma}^{-1}\left(i\bm{\mathcal{R}}+\sum_{l=k+1}^n \lambda_l \bm{\sigma}_l \right) -\frac{1}{2}\sum_{i,j=k+1}^n \sigma_{ij}\lambda_i \lambda_j\right] \\
&=\frac{1}{\sqrt{(2\pi)^{k}\det \hat{\sigma}}}e^{\frac{1}{2}\left(\sum_{j=k+1}^n\lambda_j \bm{\sigma}_j \hat{\sigma}^{-1}\sum_{l=k+1}^n \lambda_l \bm{\sigma}_l-\sum_{i,j=k+1}^n \sigma_{ij}\lambda_i \lambda_j\right)} \int d^k \mathcal{R} \, e^{-\frac{1}{2}\bm{\mathcal{R}}^{\rm T}\left(\hat{\sigma}^{-1}-2i \alpha \Lambda \right)\bm{\mathcal{R}}+i\bm{\mathcal{R}}^{\rm T}\hat{\sigma}^{-1}\sum_{j=k+1}^n\lambda_j \bm{\sigma}_j}\\
&=\frac{1}{\sqrt{\det \left(\mathbb{1}-2i\alpha\hat{\sigma}\Lambda\right)}} \exp\left\{-\frac{1}{2}\sum_{j=k+1}^n \lambda_j \bm{\sigma}_j^{\rm T} \hat{\sigma}^{-1}\left[\left(\mathbb{1}-2i\alpha\hat{\sigma}\Lambda\right)^{-1}-\mathbb{1}\right]\sum_{l=k+1}^n \lambda_l \bm{\sigma}_l-\frac{1}{2}\sum_{i,j=k+1}^n \sigma_{ij}\lambda_i \lambda_j\right\},
\end{align}
where we used that $\det (\hat{\sigma}^{-1}-2i\alpha \Lambda) = \det \hat{\sigma}^{-1} \det(\mathbb{1}-2i\alpha\sigma\Lambda)$ and $\hat{\sigma}^{\rm T}=\hat{\sigma}$ .
Using the following property of the determinant of a matrix:
\begin{align}
\det (\mathbb{1}-2i\alpha \hat{\sigma} \Lambda) = \exp[\mathrm{tr} \log(\mathbb{1}-2i\alpha \hat{\sigma} \Lambda)]= \exp\left[- \mathrm{tr} \sum_{m=1}^\infty\frac{1}{m}(2i\alpha\hat{\sigma}\Lambda)^m\right],
\end{align}
and also the geometric series
\begin{align}
[(\mathbb{1}-2i\alpha\hat{\sigma}\Lambda)^{-1}-\mathbb{1}]= \sum_{m=1}^\infty (2i\alpha\hat{\sigma}\Lambda)^m\, ,
\end{align}
we find
\begin{align}
\mathcal{I}(\lambda_1, \dots, \lambda_n) = \exp \bigg[\frac{1}{2} \mathrm{tr} \sum_{m=1}^\infty\frac{1}{m}(2i\alpha\hat{\sigma}\Lambda)^m -\frac{1}{2}\sum_{j=k+1}^n \lambda_j \bm{\sigma}_j^{\rm T} \hat{\sigma}^{-1} \sum_{m=1}^\infty(2i\alpha\hat{\sigma}\Lambda)^m\sum_{l=k+1}^n \lambda_l \bm{\sigma}_l -\frac{1}{2}\sum_{i,j=k+1}^n \sigma_{ij}\lambda_i \lambda_j\bigg]\,.\nn
\end{align}
Thus, the cumulant is
\begin{align}
&C_n(\alpha\mathcal{R}_1^2,\dots,\alpha\mathcal{R}_{k}^2,\mathcal{R}_{k+1}\dots,\mathcal{R}_n)=(-i)^n\frac{\partial}{\partial\lambda_1}\dots
\frac{\partial}{\partial\lambda_n}\log \mathcal{I}(\lambda_1, \dots, \lambda_n) \bigg|_{\lambda=0}  \label{eq:intermediatecumulant} \\
& = (-i)^n\frac{\partial}{\partial\lambda_1}\dots
\frac{\partial}{\partial\lambda_n}\bigg[\frac{1}{2} \mathrm{tr} \sum_{m=1}^\infty\frac{1}{m}(2i\alpha\hat{\sigma}\Lambda)^m -\frac{1}{2}\sum_{j=k+1}^n \lambda_j \bm{\sigma}_j^{\rm T} \hat{\sigma}^{-1} \sum_{m=1}^\infty(2i\alpha\hat{\sigma}\Lambda)^m\sum_{l=k+1}^n \lambda_l \bm{\sigma}_l -\frac{1}{2}\sum_{i,j=k+1}^n \sigma_{ij}\lambda_i \lambda_j\bigg]\bigg|_{\lambda=0}\,.\nn
\end{align}
A term is non zero only if it contains all the $\lambda_i$'s. For $k<n$ the first term vanishes because it does not contain the $\lambda_i$ with $i>k$. Instead the last term is always vanishing for $n>2$, because the number of derivatives is higher than the number of $\lambda_i$'s present. So, until now
\begin{align}
C_n(\alpha\mathcal{R}_1^2,\dots,\alpha\mathcal{R}_{k}^2,\mathcal{R}_{k+1}\dots,\mathcal{R}_n) = \left.(-i)^n\frac{\partial}{\partial\lambda_1}\dots
\frac{\partial}{\partial\lambda_n}\left[-\frac{1}{2}\sum_{j=k+1}^n \lambda_j \bm{\sigma}_j^{\rm T} \hat{\sigma}^{-1} \sum_{m=1}^\infty(2i\alpha\hat{\sigma}\Lambda)^m\sum_{l=k+1}^n \lambda_l \bm{\sigma}_l \right]\right|_{\lambda=0}.
\end{align}
Now, let us make the derivative with respect to $\lambda_n$
\begin{align}
&C_n(\alpha\mathcal{R}_1^2,\dots,\alpha\mathcal{R}_{k}^2,\mathcal{R}_{k+1}\dots,\mathcal{R}_n)=\\
& (-i)^n\frac{\partial}{\partial\lambda_1}\dots
\frac{\partial}{\partial\lambda_{n-1}}\bigg[-\frac{1}{2}\sum_{j=k+1}^n \delta_{jn} \bm{\sigma}_j^{\rm T} \hat{\sigma}^{-1} \sum_{m=1}^\infty(2i\alpha\hat{\sigma}\Lambda)^m\sum_{l=k+1}^n \lambda_l \bm{\sigma}_l -\frac{1}{2}\sum_{j=k+1}^n \lambda_{j} \bm{\sigma}_j^{\rm T} \hat{\sigma}^{-1} \sum_{m=1}^\infty(2i\alpha\hat{\sigma}\Lambda)^m\sum_{l=k+1}^n \delta_{ln} \bm{\sigma}_l \bigg]\bigg|_{\lambda=0}\,.\nn
\end{align}
The two terms can be summed up, since $\hat{\sigma}^{-1}\left(\hat{\sigma} \Lambda\right)^{m+1}=\Lambda(\hat{\sigma}\Lambda)^{m}$ and this matrix is symmetric:
\begin{align}
\left[\Lambda(\hat{\sigma}\Lambda)^{m}\right]^{\rm T} = \left[(\hat{\sigma}\Lambda)^{m}\right]^{\rm T} \Lambda^{\rm T} = \left[(\hat{\sigma}\Lambda)^{m}\right]^{\rm T} \Lambda = \left[(\hat{\sigma}\Lambda)^{\rm T}\right]^{m} \Lambda = \left(\Lambda^{\rm T} \hat{\sigma}^{\rm T}\right)^{m} \Lambda = \left(\Lambda \hat{\sigma}\right)^{m} \Lambda = \Lambda(\hat{\sigma}\Lambda)^{m}\, ;
\end{align}
so, deriving with respect to $\lambda_{n-1}$
\begin{align}
C_n(\alpha\mathcal{R}_1^2,\dots,\alpha\mathcal{R}_{k}^2,\mathcal{R}_{k+1}\dots,\mathcal{R}_n) & = (-i)^n\frac{\partial}{\partial\lambda_1}\dots
\frac{\partial}{\partial\lambda_{n-1}}\bigg[- \bm{\sigma}_n^{\rm T} \Lambda \sum_{m=1}^\infty(2i\alpha\hat{\sigma}\Lambda)^{m-1}\sum_{l=k+1}^n \lambda_l \bm{\sigma}_l\bigg]\bigg|_{\lambda=0} \nn \\
& = (-i)^n\frac{\partial}{\partial\lambda_1}\dots
\frac{\partial}{\partial\lambda_{n-2}}\bigg[- \bm{\sigma}_n^{\rm T} \Lambda \sum_{m=1}^\infty(2i\alpha\hat{\sigma}\Lambda)^{m-1} \bm{\sigma}_{n-1}\bigg]\bigg|_{\lambda=0}.
\label{eq:genericcumulant}
\end{align}
It is clear that, if $k<n-2$ there are still the derivatives $\frac{\partial}{\partial \lambda_{k+1}} \dots \frac{\partial}{\partial\lambda_{n-2}}$ to apply, but no $\lambda_{k+1},\dots,\lambda_{n-2}$ to meet. Therefore:
\begin{tcolorbox}[colframe=navyblue!20,arc=6pt,colback=navyblue!5,width=1.031\textwidth]
\vspace{-.4cm}
\begin{align}
C_n(\alpha\mathcal{R}_1^2,\dots,\alpha\mathcal{R}_{k}^2,\mathcal{R}_{k+1},\dots,\mathcal{R}_n)= 0, \qquad \forall k<n-2
\end{align}
\end{tcolorbox}
Now let us move to compute the only three orders left.
\subsection{Computation of the $O(\alpha^{n-2})$}
For the leading part of the cumulant, it is enough to set $k=n-2$ in the eq.\,(\ref{eq:genericcumulant}) (let us consider the case $n>2$, so we can get rid of the last term that we canceled in eq.\,\ref{eq:intermediatecumulant}):
\begin{align}
C_n(\alpha\mathcal{R}_1^2,\dots,\alpha\mathcal{R}_{n-2}^2,\mathcal{R}_{n-1},\mathcal{R}_n)=(-i)^n\frac{\partial}{\partial\lambda_1}\dots
\frac{\partial}{\partial\lambda_{n-2}}\bigg[- \bm{\sigma}_n^{\rm T} \Lambda \sum_{m=1}^\infty(2i\alpha\hat{\sigma}\Lambda)^{m-1} \bm{\sigma}_{n-1}\bigg]\bigg|_{\lambda=0},
\end{align}
where now $\Lambda=\mathrm{diag}\left(\lambda_1,\dots\lambda_{n-2}\right)$.

The first observation we can do is that only the $O(\alpha^{n-2})$ matters in the sum over $m$. In fact, for $m<n-2$ the derivatives are too much and they annihilate the term. Instead, for $m>n-2$ the derivatives are not enough and when we put $\lambda=0$ the term vanishes.
\begin{equation}
C_n(\alpha\mathcal{R}_1^2,\dots,\alpha\mathcal{R}_{n-2}^2,\mathcal{R}_{n-1},\mathcal{R}_n)=(2\alpha)^{n-2}\frac{\partial}{\partial\lambda_1}\dots \frac{\partial}{\partial\lambda_{n-2}}\bigg[ \bm\sigma_{n-1}^{\rm T} \Lambda(\hat{\sigma}\Lambda)^{n-3}\bm\sigma_{n} \bigg]\bigg|_{\lambda=0}.
\end{equation}
Now we can see that:
\begin{equation}
\bm\sigma_{n-1}^{\rm T}\left[\Lambda(\hat{\sigma}\Lambda)^{n-3}\right]\bm\sigma_n = \sum_{i,j=1}^{n-2} \sigma_{n-1,i}\left( \sum_{i_1,\dots,i_{n-4}=1}^{n-2} \lambda_i \sigma_{i i_1} \lambda_{i_1} \sigma_{i_1 i_2} \lambda_{i_2} \dots \lambda_{i_{n-4}} \sigma_{i_{n-4}j} \lambda_j \right) \sigma_{jn}\, .
\end{equation}
Of course, because of the presence of the derivatives, in the sum, only the terms with all the $\{\lambda_i\}_{i=1}^{n-2}$ different matters, i.e. we only keep the permutations of $\{1,2,\dots,n-2\}$. So, putting all together
\begin{align}
C_n[\alpha\mathcal{R}_1^2,\dots,\alpha\mathcal{R}_{n-2}^2,\mathcal{R}_{n-1},\mathcal{R}_n] & =(2\alpha)^{n-2}\frac{\partial}{\partial\lambda_1}\dots \frac{\partial}{\partial\lambda_{n-2}}\sum_{i_1,\dots,i_{n-2}=1}^{n-2} \lambda_{i_1} \dots \lambda_{i_{n-2}}\sigma_{n-1, i_1} \sigma_{i_1 i_2} \dots \sigma_{i_{n-3}i_{n-2}}\sigma_{i_{n-2}n} \bigg|_{\lambda=0} \nn \\
&= (2\alpha)^{n-2} \sum_{\substack{\{i_1, \dots, i_{n-2}\}=\\
=\mathrm{perms}\{1,\dots,n-2\}}} \sigma_{n-1, i_1} \sigma_{i_1 i_2} \dots \sigma_{i_{n-3}i_{n-2}}\sigma_{i_{n-2}n}\, .
\end{align}
So, the formula for the $O(\alpha^{n-2})$ contribute to the $n$-th order cumulant is:
\begin{tcolorbox}[colframe=navyblue!20,arc=6pt,colback=navyblue!5,width=1.033\textwidth]
\vspace{-.4cm}
\begin{align}
C_n(\alpha\mathcal{R}_1^2,\dots,\alpha\mathcal{R}_{n-2}^2,\mathcal{R}_{n-1},\mathcal{R}_n)= (2\alpha)^{n-2} \sum_{\substack{\{i_1, \dots, i_{n-2}\}=\\
=\mathrm{perms}\{1,\dots,n-2\}}} \langle \mathcal{R}_{n-1}\mathcal{R}_{i_1} \rangle \langle \mathcal{R}_{i_1}\mathcal{R}_{i_2} \rangle\dots \langle \mathcal{R}_{i_{n-3}}\mathcal{R}_{i_{n-2}} \rangle \langle \mathcal{R}_{i_{n-2}}\mathcal{R}_n \rangle
\end{align}
\end{tcolorbox}
For example, we can check this formula for $n=4$
\begin{align}
& C_4(\alpha\mathcal{R}_1^2,\alpha\mathcal{R}_{2}^2,\mathcal{R}_{3},\mathcal{R}_4)= (2\alpha)^{2} \sum_{\substack{\{i,j\}=\mathrm{perms}\{1,2\}}} \langle \mathcal{R}_{3}\mathcal{R}_{i} \rangle \langle \mathcal{R}_{i}\mathcal{R}_{j} \rangle \langle \mathcal{R}_{j}\mathcal{R}_{4} \rangle \\
& = 4\alpha^2 \left( \langle \mathcal{R}_{3}\mathcal{R}_{1} \rangle \langle \mathcal{R}_{1}\mathcal{R}_{2} \rangle \langle \mathcal{R}_{2}\mathcal{R}_{4} \rangle + \langle \mathcal{R}_{3}\mathcal{R}_{2} \rangle \langle \mathcal{R}_{2}\mathcal{R}_{1} \rangle \langle \mathcal{R}_{1}\mathcal{R}_{4} \rangle\right) = 4\alpha^2 \langle \mathcal{R}_{1}\mathcal{R}_{2} \rangle\left( \langle \mathcal{R}_{3}\mathcal{R}_{1} \rangle \langle \mathcal{R}_{2}\mathcal{R}_{4} \rangle + \langle \mathcal{R}_{3}\mathcal{R}_{2} \rangle \langle  \mathcal{R}_{1}\mathcal{R}_{4} \rangle\right)\,,\nn
\end{align}
that precisely matches the result obtained in eq.\,(\ref{eq:C4alpha2}).
\subsection{Computation of the $O(\alpha^{n-1})$}
For $k=n-1$ we have:
\begin{align}
& \mathcal{I}(\lambda_1, \dots, \lambda_n) = \frac{1}{\sqrt{\det (\mathbb{1}-2i\alpha \hat{\sigma} \Lambda)}} \exp \left\{-\frac{1}{2}\lambda_n^2 \bm\sigma_n^{\rm T} \hat{\sigma}^{-1} \big[(\mathbb{1}-2i\alpha\hat{\sigma}\Lambda)^{-1}-\mathbb{1} \big]\bm\sigma_n-\frac{1}{2}\sigma_{nn}\lambda_{n}^2 \right\},
\end{align}
and so:
\begin{align}
&C_n(\alpha\mathcal{R}_1^2,\dots,\alpha\mathcal{R}_{n-1}^2,\mathcal{R}_n)=(-i)^n\frac{\partial}{\partial\lambda_1}\dots
\frac{\partial}{\partial\lambda_n}\log \mathcal{I}(\lambda_1, \dots, \lambda_n) \bigg|_{\lambda=0}=\\
& = (-i)^n\frac{\partial}{\partial\lambda_1}\dots
\frac{\partial}{\partial\lambda_n}\bigg[\frac{1}{2} \mathrm{tr} \sum_{m=1}^\infty\frac{1}{m}(2i\alpha\hat{\sigma}\Lambda)^m -\frac{1}{2}\lambda_n^2 \bm\sigma_n^{\rm T} \hat{\sigma}^{-1} \sum_{m=1}^\infty(2i\alpha\hat{\sigma}\Lambda)^m \bm\sigma_n-\frac{1}{2}\sigma_{nn}\lambda_n^2\bigg] \bigg|_{\lambda=0}=\\
&= \left.(-i)^n\frac{\partial}{\partial\lambda_1}\dots
\frac{\partial}{\partial\lambda_{n-1}}\left[-\lambda_n \left( \bm\sigma_n^{\rm T} \hat{\sigma}^{-1} \sum_{m=1}^\infty(2i\alpha\hat{\sigma}\Lambda)^m \bm\sigma_n+\sigma_{nn}\right)\right] \right|_{\lambda=0}.
\end{align}
In the last step, everything vanishes since it remains a $\lambda_n$ with no more derivatives with respect to it and in the end we must set $\lambda=0$, so there is no $O(\alpha^{n-1})$ contribution:
\begin{tcolorbox}[colframe=navyblue!20,arc=6pt,colback=navyblue!5,width=1.031\textwidth]
\vspace{-.4cm}
\begin{align}
C_n(\alpha\mathcal{R}_1^2,\dots,\alpha\mathcal{R}_{n-1}^2,\mathcal{R}_n)= 0
\end{align}
\end{tcolorbox}
\subsection{Computation of the $O(\alpha^{n})$}
Now we have $\Lambda=\mathrm{diag}\left(\lambda_1,\dots,\lambda_n\right)$ and the trace part is the only term left, so:
\begin{equation}
\mathcal{I} (\lambda_1,\dots,\lambda_n) =\frac{1}{\sqrt{(2\pi)^n \det\sigma}} \int d^n \mathcal{R} \ e^{-\frac{1}{2}\bm{\mathcal{R}}^{\rm T}\left(\sigma^{-1}-2i\alpha\Lambda\right)\bm{\mathcal{R}}} = \exp \left[-\frac{1}{2} \mathrm{tr} \log \left(\mathbb{1}-2i\alpha\sigma\Lambda\right) \right],
\end{equation}
where now $\sigma=\left(\sigma_{ij}\right)_{i,j=1}^n$ and $\Lambda=\mathrm{diag}\left(\lambda_1,\dots,\lambda_n\right)$.
With the usual steps, knowing that only the $O(\alpha^n)$ matters:
\begin{align}
C_n(\alpha \mathcal{R}_1^2,\dots,\alpha \mathcal{R}_{n}^2) = \left.(-i)^n\frac{\partial}{\partial\lambda_1}\dots
\frac{\partial}{\partial\lambda_n} \left[\frac{1}{2} \mathrm{tr}\sum_{m=0}^\infty\frac{1}{m}(2i\alpha\sigma\Lambda)^m \right]\right|_{\lambda=0} = \frac{2^{n-1}\alpha^n}{n} \frac{\partial}{\partial\lambda_1}\dots
\frac{\partial}{\partial\lambda_n} \mathrm{tr} (\sigma\Lambda)^n \bigg|_{\lambda=0}.
\end{align}
The trace is:
\begin{equation}
\mathrm{tr} (\sigma\Lambda)^n = \sum_{i_1,\dots,i_n=1}^n \lambda_{i_1}\lambda_{i_2}\dots \lambda_{i_n} \sigma_{i_n i_1} \sigma_{i_1 i_2}\dots \sigma_{i_{n-1}i_n} \, ,
\end{equation}
and after the derivation only the permutations of $\{1,2,\dots,n\}$ survive, so:
\begin{tcolorbox}[colframe=navyblue!20,arc=6pt,colback=navyblue!5,width=1.031\textwidth]
\vspace{-.4cm}
\begin{align}
C_n(\alpha\mathcal{R}_1^2,\dots,\alpha\mathcal{R}_{n}^2)= \frac{2^{n-1}\alpha^n}{n} \sum_{\substack{\{i_1, \dots, i_{n}\}=\\
=\mathrm{perms}\{1,\dots,n\}}} \langle \mathcal{R}_{i_n}\mathcal{R}_{i_1} \rangle \dots \langle \mathcal{R}_{i_{n-1}}\mathcal{R}_{i_{n}} \rangle
\end{align}
\end{tcolorbox}

\subsection{Computation of the generic cumulant}
Now that we have all the ingredients, we can compute the generic $n$-th order cumulant up to all orders. First of all, let us focus on the $O(\alpha^{n-2})$ contribute:
\begin{align}
C_n^{O(\alpha^{n-2})}(h_1,\dots,h_n) =(2\alpha)^{n-2} \sum_{\{i,j\}\in A} \sum_{\substack{\{i_1, \dots, i_{n-2}\}=\\
=\mathrm{perms}\{1,\dots,n\} \setminus\{i,j\}}} \langle\mathcal{R}_i \mathcal{R}_{i_1}\rangle \langle\mathcal{R}_{i_1} \mathcal{R}_{i_2}\rangle\dots \langle\mathcal{R}_{i_{n-2}} \mathcal{R}_j\rangle
\end{align}
where $A$ is the set in which all the non-ordered couples of numbers between $\{1,\dots,n\}$ are contained. E.g. for $n=4$, $A=\big\{\{1,2\},\{1,3\},\{1,4\},\{2,3\},\{2,4\},\{3,4\}\big\}$. From combinatorics we know that the permutations of $n-2$ different objects are $(n-2)!$, while the cardinality of $A$ is the number of ways in which we can take 2 objects among $n$ different objects without taking care of their order, which is $\binom{n}{2}=\frac{n(n-1)}{2}$. So, putting these 2 together, the total sum is composed of $\frac{n!}{2}$ elements. If we compute also the exchanged couples, we double the sum, having $n!$ elements, that is the cardinality of $\mathrm{perms}\{1,\dots,n\}$. So we can put the two sums together dividing by 2:
\begin{align}
C_n^{O(\alpha^{n-2})}(h_1,\dots,h_n) = 2^{n-3}\alpha^{n-2} \sum_{\substack{\{i_1, \dots, i_{n}\}=\\
=\mathrm{perms}\{1,\dots,n\}}} \langle\mathcal{R}_{i_1} \mathcal{R}_{i_2}\rangle \langle\mathcal{R}_{i_2} \mathcal{R}_{i_3}\rangle\dots \langle\mathcal{R}_{i_{n-1}} \mathcal{R}_{i_n}\rangle
\end{align}
Now, let us add also the $O(\alpha^n)$ contribution, that we can write as
\begin{align}
C_n(\alpha\mathcal{R}_1^2,\dots,\alpha\mathcal{R}_{n}^2)= \frac{2^{n-1}\alpha^n}{n} \sum_{\substack{\{i_1, \dots, i_{n}\}=\\
=\mathrm{perms}\{1,\dots,n\}}} \langle \mathcal{R}_{i_1}\mathcal{R}_{i_2} \rangle \dots \langle \mathcal{R}_{i_{n}}\mathcal{R}_{i_{1}} \rangle 
\end{align}
And so, summing the two
\begin{tcolorbox}[colframe=navyblue!20,arc=6pt,colback=navyblue!5,width=1.03\textwidth]
\vspace{-.4cm}
\begin{align}
C_n(h_1,\dots,h_n)= 2^{n-3}\alpha^{n-2} \sum_{\substack{\{i_1, \dots, i_{n}\}=\\
=\mathrm{perms}\{1,\dots,n\}}} \langle \mathcal{R}_{i_1}\mathcal{R}_{i_2} \rangle \dots \langle \mathcal{R}_{i_{n-1}}\mathcal{R}_{i_{n}} \rangle \left( 1+\frac{4\alpha^2}{n} \langle\mathcal{R}_{i_1} \mathcal{R}_{i_n} \rangle\right)
\end{align}
\end{tcolorbox}
And we can easily see that, putting $\vec{x}_1=\dots=\vec{x}_n$:
\begin{align}
& C_n(h,\dots,h) = 2^{n-3}\alpha^{n-2} n! \langle \mathcal{R}^2 \rangle^{n-1} \left(1+\frac{4\alpha^2}{n}\langle \mathcal{R}^2 \rangle^{2}\right) = 2^{n-3}(n-1)! \sigma_0^2 \left(\alpha \sigma_0^2 \right)^{n-2} \left(n+4\alpha^2 \sigma_0^2 \right)\,,
\end{align}
and this result exactly reproduces the formula given in eq.\,(\ref{eq:faadibruno}).

\section{Threshold for gravitational collapse into black holes}\label{app:Threshold}

We follow the approach of refs.\,\cite{Germani:2018jgr,Musco:2018rwt} that we generalize to include local non-gaussianities in the curvature perturbation field.\footnote{Recently, ref.\,\cite{Musco:2020jjb} proposed a simplified analytical prescription for the computation of the threshold for PBH formation. 
We note that in ref.\,\cite{Musco:2020jjb} the quantity $\delta_c$ corresponds to $\delta_{\rm th}$ in our notation.
} 
The main steps of the computation are the following

\begin{itemize}
\item [{\it i)}]
Consider eq.\,(\ref{eq:NonLinearDelta}) at the linear order in $h$
\begin{align}\label{eq:NonLinearDeltaTH}
\delta(\vec{x},t) =  
-\frac{4}{9}\left(
\frac{1}{aH}
\right)^2 
\triangle h(\vec{x})\,.
\end{align}
As usual, $\delta(\vec{x},t)$ is a statistical variable and only statistical averages can be used for practical purposes.
Instead of using $\delta(\vec{x},t)$, one then defines the averaged density radial profile $\bar{\delta}(r,t)$ 
by means of 
\begin{align}\label{eq:F0}
\bar{\delta}(r,t) \equiv \frac{\mathcal{F}_0}{(aH)^2}\psi(r)\,,~~~~~~~~~~{\rm with}\,\,\psi(r)= 
\frac{\langle \delta(\vec{x},t)\delta(\vec{x}+\vec{r},t) \rangle}
{\langle \delta(\vec{x},t)\delta(\vec{x},t) \rangle}\,,
\end{align}
where we find
\begin{align}\label{eq:F1}
\langle\delta(&\vec{x},t)\delta(\vec{x}+\vec{r},t)\rangle = \\ & \frac{16}{81}\frac{1}{(aH)^4}
\int dk\,k^3\,\frac{\sin(kr)}{kr}\bigg[
\mathcal{P}_{\mathcal{R}}(k) + \underbrace{\alpha^2k^3\int
\frac{dq}{q}\frac{d\cos\theta}{(k^2 + q^2 -2kq\cos\theta)^{3/2}}
\mathcal{P}_{\mathcal{R}}(q)\mathcal{P}_{\mathcal{R}}(\sqrt{k^2 + q^2 -2kq\cos\theta})}_{\equiv \mathcal{P}_{\rm NG}(k)} 
\bigg]\,,\nn 
\end{align}
while $\langle \delta(\vec{x},t)\delta(\vec{x},t) \rangle$ is given by the limit 
$\lim_{r\to 0}\langle\delta(\vec{x},t)\delta(\vec{x}+\vec{r},t)\rangle$. 
Eq.\,(\ref{eq:F1}) follows from the explicit evaluation (going first in Fourier space) of the two-point correlator;
 $\mathcal{P}_{\rm NG}(k)$ represents the correction to the averaged density profile due to local non-gaussianities, and 
the integral over $q$ arises because of the two convolutions that are needed to compute the term 
$\alpha^2\langle\mathcal{R}(\vec{x})^2\mathcal{R}(\vec{x}+\vec{r})^2\rangle$. 
Remember indeed that the Fourier transform of $\mathcal{R}^2$ is not the square of the Fourier transform of $\mathcal{R}$ but 
it involves a convolution. 

Notice that $\psi(r)$ in eq.\,(\ref{eq:F0}) is time-independent (assuming of course the perturbations to be fully super-horizon thus constant in time), and the only time-dependence of $\bar{\delta}(r,t)$ comes from the overall factor $1/(aH)^2$. 
Furthermore, spherical symmetry is assumed.  
This means that we are neglecting  interactions between adjacent over-densities\,\cite{Bardeen:1985tr}. 
Intuitively, this is a very good assumption since the formation of a PBH is already a rare event, and having two or more 
black holes forming on the same site should be extremely unlikely.
In eq.\,(\ref{eq:F0}) $\mathcal{F}_0$ is a dimensionful parameter which is related to the amplitude of the over-density at the center of the 
radial profile. 
Finally, notice that, by construction, we have $\psi(0) = 1$ and, consequently, $\bar{\delta}(0,t) = \mathcal{F}_0/(aH)^2$.

\item [{\it ii)}] The next step consists in the evaluation of the so-called compaction function $\mathcal{C}(r)$. 
This is the most important quantity because, as explained in refs.\,\cite{Germani:2018jgr,Musco:2018rwt}, PBHs form when local maxima  of the compaction  function exceed a certain threshold value\,\cite{Shibata:1999zs,Helou:2016xyu}. 
The main point of the analysis is the identification of such threshold.

Physically, the compaction  function  is  defined  as  twice  the  local  excess-mass  over  the comoving areal radius, that is (we use explicitly, for the sake of clarity, the reduced Planck mass $\bar{M}_{\rm Pl}^2 = 1/8\pi G_{\rm N}$)
\begin{align}\label{eq:DefC}
\mathcal{C}(r,t) \equiv \frac{2G_{\rm N} \delta M(r,t)}{R(r,t)}\,,
\end{align}
where the areal radius is $R(r,t) \equiv a(t)r$ and the local  excess-mass due to the overdensity field $\delta(r,t)$ within a spherical region of radius $R$ is given by
\begin{align}\label{eq:ExcessMass}
\frac{\delta M(r,t)}{M_b(r,t)} \equiv \frac{1}{V_b(r,t)}\int_0^R 4\pi \delta(r',t)R^{\prime\,2}dR' \,,
\end{align}
where $V_b(r,t) \equiv (4\pi/3)R(r,t)^3$, the background radiation energy density is $\rho_b(t) = 3\bar{M}_{\rm Pl}^2 H(t)^2$ and 
$M_b(r,t) = V_b(r,t)\rho_b(t)$. Using these relation, eq.\,(\ref{eq:DefC}) can be recast in the form 
\begin{align}\label{eq:DefC2}
\mathcal{C}(r,t)= \frac{3(aH)^2}{r}\int_0^r dr' r^{\prime\,2}\delta(r',t)\,.
\end{align}
This is a generic expression. If we now use for the overdensity field the averaged density radial profile $\bar{\delta}(r,t)$ defined in eq.\,(\ref{eq:F0}), we find  
\begin{align}\label{eq:CompactionFunction}
\mathcal{C}(r) \simeq \frac{3\mathcal{F}_0}{r}\int_0^{r} dr^{\prime}\,r^{\prime\,2}\,\psi(r^{\prime})\,,
\end{align}
which is valid in the so-called long-wavelength approximation of ref.\,\cite{Musco:2018rwt}. 
Notice that in this approximation $\mathcal{C}(r)$ does not depend on time.
From $\mathcal{C}(r)$, one computes $r_m$ which is the position of its maximum. The precise computation of $r_m$ for a given power spectrum of curvature perturbation will be discussed below. 

\item [{\it iii)}] After computing $r_m$, we use the fact that, as shown in ref.\,\cite{Musco:2018rwt}, we have 
$\delta_{\rm th} = 3\bar{\delta}(r_m,t_m)$ where $\delta_{\rm th}$ is defined as the threshold value for the local  excess-mass (defined in eq.\,(\ref{eq:ExcessMass}))
associated to  the the averaged curvature perturbations $\bar{\delta}(r,t)$ and evaluated at position $r_m$ and time $t_m$. 
At super-horizon scales, the same computation that led to eq.\,(\ref{eq:CompactionFunction}) gives 
\begin{align}\label{eq:MassCompa}
\frac{\delta M(r,t)}{M_b(r,t)} = \frac{1}{V_b(r,t)}\int_0^R 4\pi \bar{\delta}(r',t)R^{\prime\,2}dR'  \simeq \left(
\frac{1}{aH r}
\right)^2\mathcal{C}(r)\,,
\end{align}
and, known $r_m$, the time $t_m$  is defined implicitly by $a(t_m)H(t_m)r_m = 1$. 
From the previous definition, it follows that $\delta_{\rm th} = \delta M(r_m,t_m)/M_b(r_m,t_m) \simeq \mathcal{C}_c(r_m)$. 

If we knew $\delta_{\rm th}$ it would be possible to use $\delta_{\rm th} = 3\bar{\delta}(r_m,t_m)$ and eq.\,(\ref{eq:F0}) to extract the critical value of $\mathcal{F}_0$, named $\mathcal{F}^c_0$ in the following. 
Correspondingly, we indicate with $\mathcal{C}_c(r)$ the critical compaction function, that is eq.\,(\ref{eq:CompactionFunction}) with 
$\mathcal{F}_0 = \mathcal{F}^c_0$. 

The quantity $\delta_{\rm th}$ is usually extracted from simulations in numerical relativity, and has been found to range in the interval 
$0.4 \lesssim \delta_{\rm th} \lesssim 2/3$. 
The left-side of this interval corresponds to the so-called Harada-Yoo-Kohri limit  that is  the  threshold  for  which  a  very sharply  peaked  over-density  profile collapses  into a zero-mass black hole\,\cite{Harada:2013epa}. 

\item [{\it iv)}]  To proceed further, the simplest way to go is to fix the value of $\delta_{\rm th}$ within the range $0.4 \lesssim \delta_{\rm th} \lesssim 2/3$ and extract $\mathcal{F}_0^c$ from $\delta_{\rm th} = 3\bar{\delta}(r_m,t_m)$.  
We can actually do better and exploit the result of ref.\,\cite{Escriva:2019phb}. 
By using the parameter $q$ defined as $q\equiv -\mathcal{C}^{\prime\prime}(r_m)r_m^2/4\mathcal{C}(r_m)$, the value of  $\delta_{\rm th}$ can be analytically computed as
\begin{align}\label{eq:Germa}
\delta_{\rm th} = \frac{4}{15}e^{-1/q}\left[
\frac{q^{1-5/2q}}{\Gamma(5/2q) - \Gamma(5/2q,1/q)}
\right]\,.
\end{align}
We refer to ref.\,\cite{Escriva:2019phb} for a detailed discussion. Eq.\,(\ref{eq:Germa}) is an extremely neat result and shows 
that the threshold for the compaction function is only sensitive to its curvature at the maximum. 

The strategy, therefore, is the following. 
We compute $q$ from $\mathcal{C}(r)$ and $r_m$, then $\delta_{\rm th}$ from eq.\,(\ref{eq:Germa}).  
From $\delta_{\rm th}$, using $\delta_{\rm th} = 3\bar{\delta}(r_m,t_m)$, we can solve for $\mathcal{F}_0^c$. 
From eq.\,(\ref{eq:F0}), we indeed have $\mathcal{F}_0^c = \delta_{\rm th}/3 r_m^2\psi(r_m)$.

\item [{\it v)}] Finally, we are in the position to extract the value of $\delta_c$ that enters in eq.(\ref{eq:SpikyEnough}), leading to 
eq.\,(\ref{eq:ExplThreshx}) and eq.\,(\ref{eq:ExplThresh2}).

We simply have
\begin{align}\label{eq:ThresholdExpl}
\delta_c = \frac{\delta_{\rm th}}{3\psi(r_m)} = \mathcal{F}_0^c r_m^2\,,
\end{align}
and, from eq.\,(\ref{eq:F0}), we see that $\delta_c$ corresponds to $\bar{\delta}_c(0,t_m)$ that is the value of the critical averaged density profile at the center. 
Furthermore, in eq.\,(\ref{eq:ExplThresh2}) we evaluated the comoving horizon length $1/aH$ at time $t_m$ so that $(a_m H_m)^2 = 1/r_m^2$. 
We remark  that the threshold value for the overdensity field $\delta$ is not set by $\delta_{\rm th}$ (that is the threshold on the compaction function) but rather by $\mathcal{F}_0^c$ which is the threshold at the center of the averaged over-density. 
This point will become more clear in our numerical analysis.

\end{itemize}

We apply this procedure to the case relevant for our analysis. 

First of all, let us comment about the non-gaussian correction in eq.\,(\ref{eq:F1}). For illustration, we consider the power spectrum in eq.\,(\ref{eq:ToyPS}). 
The advantage is that we can do the angular integration that appears in $\mathcal{P}_{\rm{NG}}(k) $ semi-analytically.
We find 
\begin{align}
\mathcal{P}_{\rm{NG}}(k) = 
\alpha^2 k^3\int \frac{dq}{q^2}
\frac{A_g^2}{2\sqrt{2\pi}v k k_{\star}}\exp\left(
\frac{v^4 - \log^2\frac{q}{k_{\star}}}{2v^2}
\right)
\left[
{\rm Erf}\left(
\frac{v^2 + \log\frac{k+q}{k_{\star}}}{\sqrt{2}v}
\right) - 
{\rm Erf}\left(
\frac{v^2 + \log\frac{|k-q|}{k_{\star}}}{\sqrt{2}v}
\right)
\right]\,,\label{eq:EerrorFucnt}
\end{align}
where $-1\leqslant {\rm Erf}(x)\leqslant 1$ is the error function. We can now compare $\mathcal{P}_{\rm{NG}}(k)$ with the leading gaussian term $\mathcal{P}_{\mathcal{R}}(k)$ in eq.\,(\ref{eq:ToyPS}).
\begin{figure}[!htb!]
\begin{center}
$$\includegraphics[width=.4\textwidth]{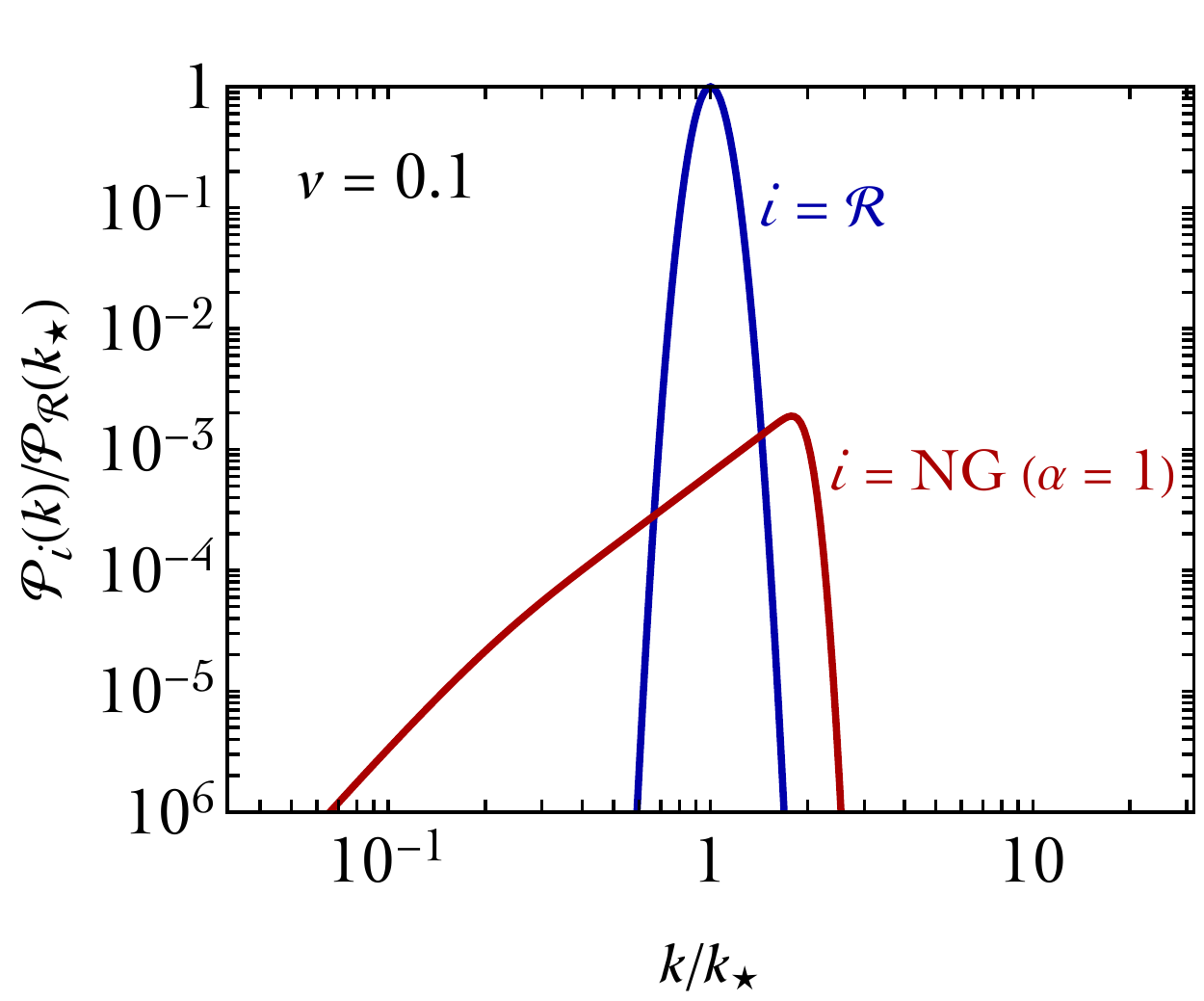}
\qquad\includegraphics[width=.4\textwidth]{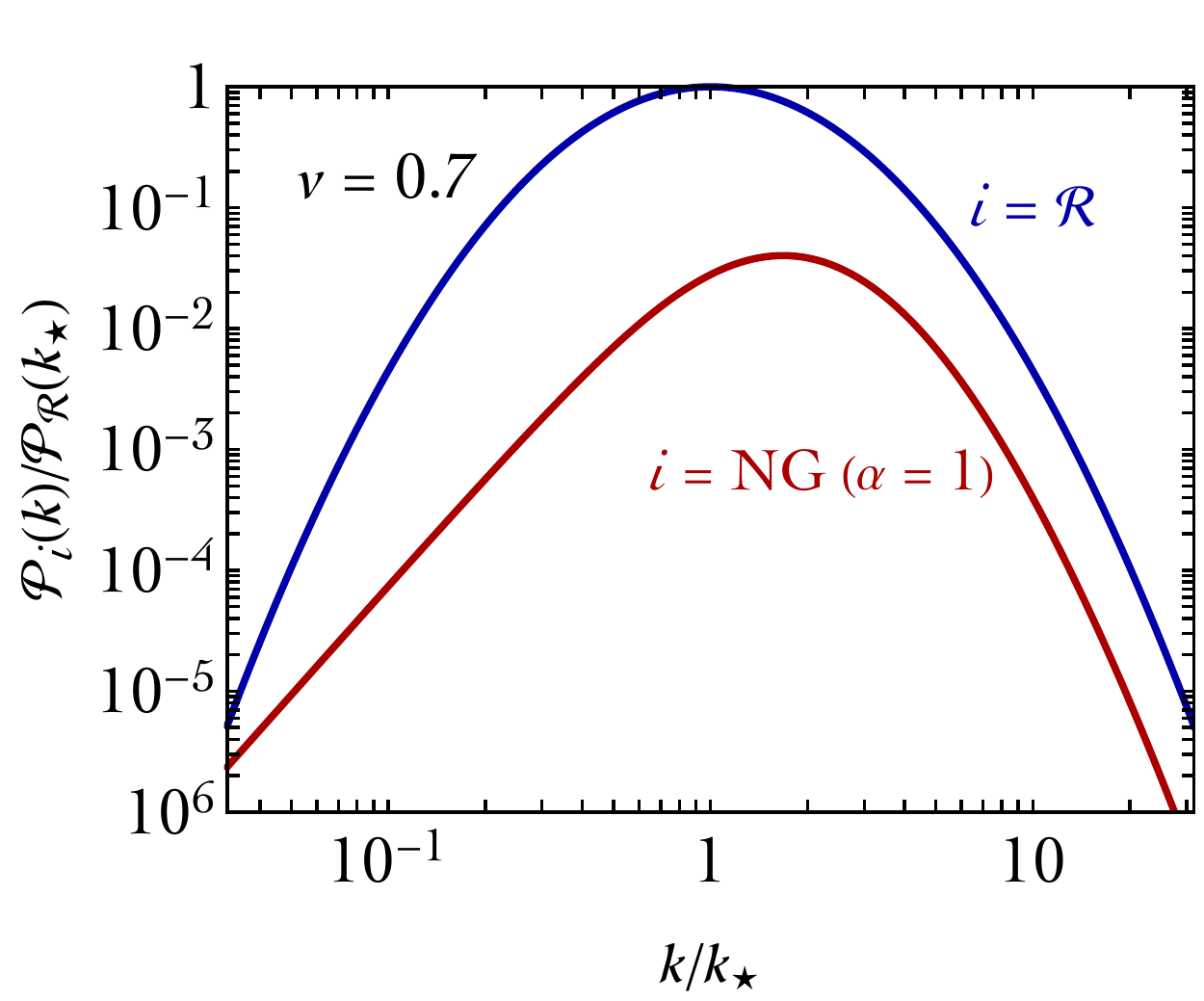}$$
\caption{\em \label{fig:ConvoSpectrum}  
$\mathcal{P}_{\mathcal{R}}(k)$ and  $\mathcal{P}_{\rm{NG}}(k)$ for the power spectrum in eq.\,(\ref{eq:ToyPS}) as a function of the comoving wavenumber $k$ in units of $k_{\star}$. 
We normalize the $y$-axis with respect to $\mathcal{P}_{\mathcal{R}}(k_{\star})$. We set $\alpha = 1$ while $A_g$ is chosen in such a way that $\mathcal{P}_{\mathcal{R}}(k_{\star}) = 10^{-2}$.
 }
\end{center}
\end{figure}
We show this comparison in fig.\,\ref{fig:ConvoSpectrum} for a narrow ($v=0.1$) and a broad ($v=0.7$) spectrum. 
The outcome of this comparison is that the presence of the non-gaussian correction $\mathcal{P}_{\rm{NG}}(k)$ can, in principle, modify the shape of the leading order power spectrum.  
For instance, we note that in the expression for $\mathcal{P}_{\rm{NG}}(k)$ the peak in $k$ is shifted towards slightly higher $k$ compared to $k_{\star}$ as a consequence of the 
convolution.\footnote{This can be easily understood if we look at eq.\,(\ref{eq:EerrorFucnt}).
The dominant contribution to the $q$-integral comes from the region where the argument of the overall exponential vanishes (for $q\gg k_{\star}$ and $q \ll k_{\star}$ we have an exponential suppression), that is for $q\simeq e^{v^2}k_{\star}$. 
If we select this contribution from the $q$-integral, it is easy to see that the resulting expression 
$\mathcal{P}_{\rm{NG}}(k)$ is maximized when $k\simeq k_{\star}e^{v^2}$. This is because 
when $q=e^{v^2}k_{\star}$ the second error function in eq.\,(\ref{eq:EerrorFucnt}) takes the value
$-{\rm Erf}[v^2 + \log(\frac{k-k_{\star}e^{v^2}}{k_{\star}})]$ which is maximized to $1$ when $k=k_{\star}e^{v^2}$,
 that is $-{\rm Erf}[-\infty] = 1$.} 
 However,  the correction $\mathcal{P}_{\rm{NG}}(k)$ is naturally suppressed, in amplitude, since proportional to $A_g^2$ with $A_g \ll 1$. 
 This means that, for $\mathcal{P}_{\rm{NG}}(k)$ to be relevant, one needs to overcome this suppression by taking $\alpha \gg 1$, and 
 this is not what typically happens in the class of models that we have in mind for our analysis (where $\alpha \lesssim 1$). 
 For this reason, in the following we will neglect the non-gaussian correction $\mathcal{P}_{\rm{NG}}(k)$ in eq.\,(\ref{eq:F1}).
Furthermore, since the computation of $\psi(r)$ in eq.\,(\ref{eq:F0}) will be in any case numerical,
we put aside the toy-model in eq.\,(\ref{eq:ToyPS}) and move to consider the more realistic case of the power spectrum introduced in eq.\,(\ref{eq:RealisticPS}). 

As already anticipated in section\,\ref{sec:Res}, in realistic situations in which the power spectrum is not sharply peaked the computation of $\delta_c$ requires some care.
We start by introducing the function 
$\mathcal{P}^{\rm cut}_{\mathcal{R}}(k) \equiv \mathcal{P}_{\mathcal{R}}(k)e^{-k^2/k_{\rm cut}^2}$ 
with some cut-off wavenumber $k_{\rm cut}$.
The reason is that we are interested in the local maximum of the compaction function for which the threshold for collapse in eq.\,(\ref{eq:ExplThreshx}) and eq.\,(\ref{eq:ExplThresh2}) is minimized, thus giving the largest probability for collapse. 
\begin{figure}[!h!]
\begin{center}
\includegraphics[width=.45\textwidth]{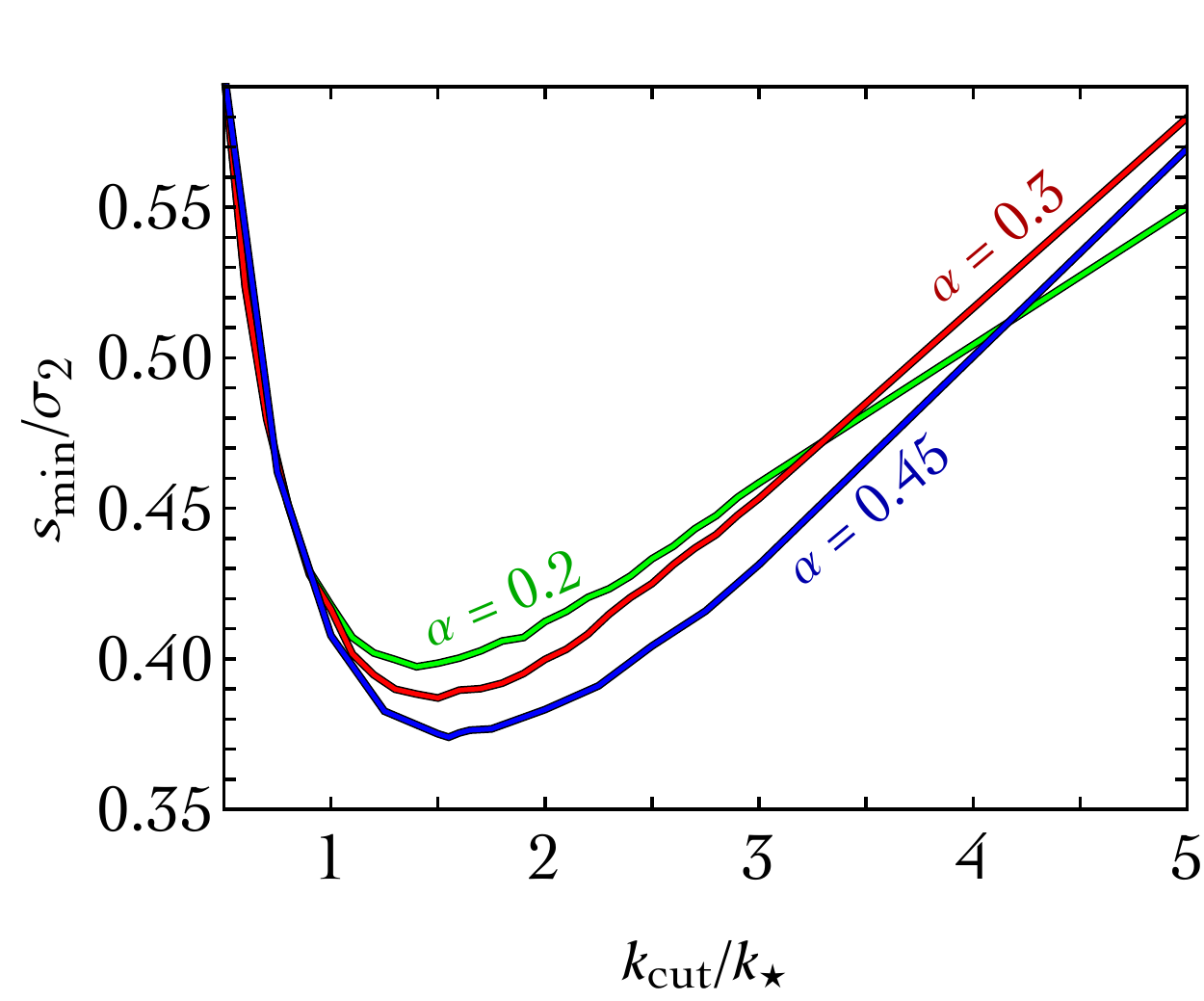}
\caption{\em \label{fig:CutoffAnalysis}  
Ratio $s_{\rm min}/\sigma_2$ defined  in eq.\,(\ref{eq:ExplThresh2}) as a function of the cutoff wavenumber $k_{\rm cut}$ in units of $k_{\star}$. 
More precisely, $s_{\rm min}/\sigma_2$ is computed following 
the procedure sketched in {\it i)}-{\it v)} in appendix\,\ref{app:Threshold} and, for a given $k_{\rm cut}$, 
we used the power spectrum $\mathcal{P}^{\rm cut}_{\mathcal{R}}(k) \equiv \mathcal{P}_{\mathcal{R}}(k)e^{-k^2/k_{\rm cut}^2}$ with 
$\mathcal{P}_{\mathcal{R}}(k)$ given in eq.\,(\ref{eq:RealisticPS}) for three different benchmark values of $\alpha$. 
Notice that $\sigma_2^2 \propto \mathcal{P}_{\mathcal{R}}(k_{\star}) \equiv p_{\star}$ and, consequently, 
for the ratio on the $y$-axis we have $s_{\rm min}/\sigma_2 \propto 1/\sqrt{p_{\star}}$. 
We show the value of $s_{\rm min}/\sigma_2$ in units of $1/\sqrt{p_{\star}}$ meaning that for a given peak amplitude $\mathcal{P}_{\mathcal{R}}(k_{\star})$ and a 
given $k_{\rm cut}$ the corresponding threshold for collapse $s_{\rm min}/\sigma_2$ is given by the value on the $y$-axis times $1/\sqrt{p_{\star}}$.
Since $\mathcal{P}_{\mathcal{R}}(k_{\star}) \ll 1$, we have $s_{\rm min}/\sigma_2 \gg 1$.
 }
\end{center}
\end{figure}
We can look for this maximum by means of a numerical scan over the value $k_{\rm cut}$.  
We show our numerical results in fig.\,\ref{fig:CutoffAnalysis} (see caption for details). 
We find that the value of $s_{\rm min}/\sigma_2$ is minimized at around $k_{\rm cut}/k_{\star} \simeq 1.5$. The precise value depends on the value of $\alpha$. 
This is because, as explained below eq.\,(\ref{eq:RealisticPS}), the spectral index of the power-law falloff of the power spectrum after the peak is controlled by $\alpha$, and different values 
of the latter
give different shapes.   

\begin{figure}[!htb!]
\begin{center}
$$
\includegraphics[width=.48\textwidth]{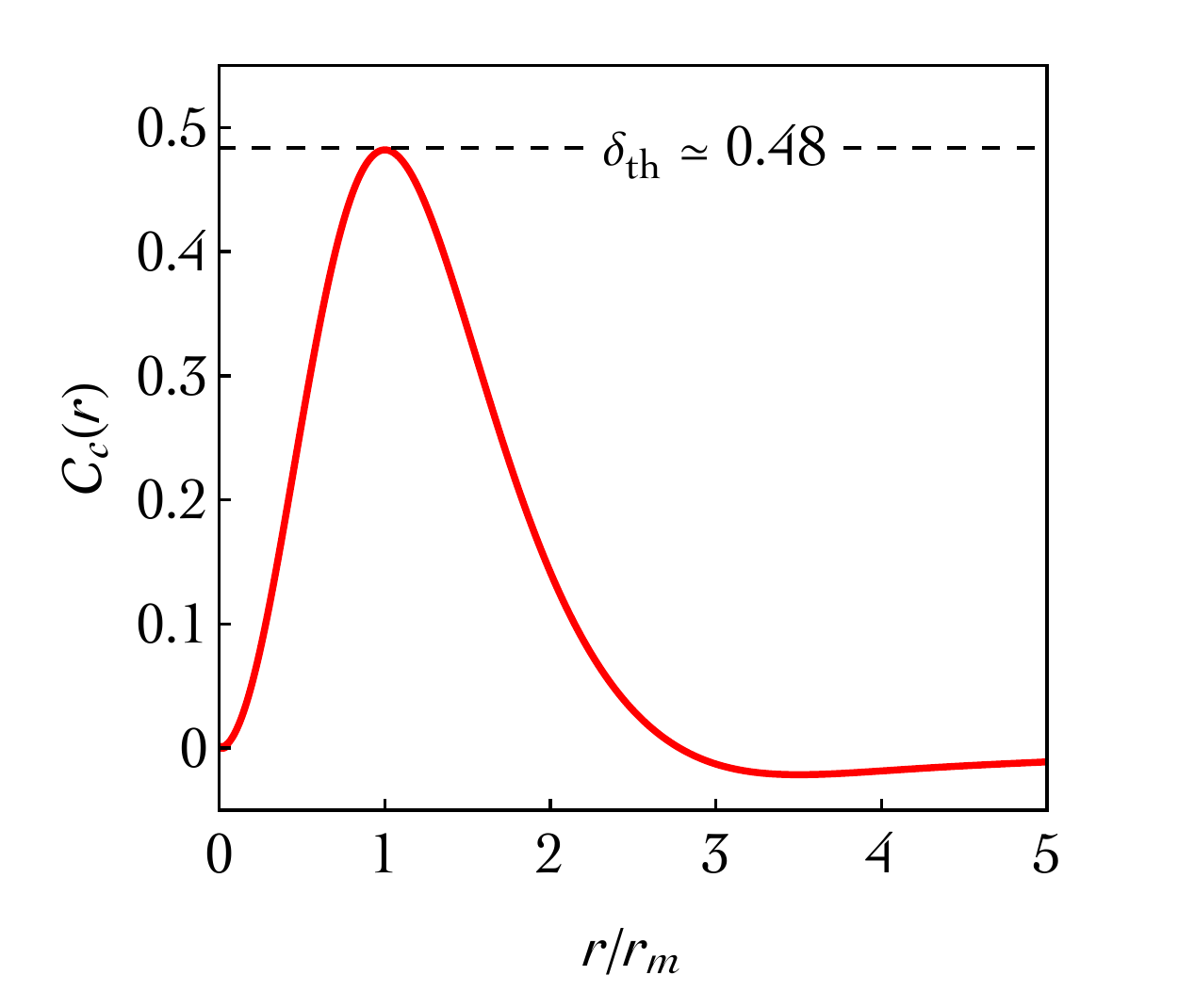}\qquad
\includegraphics[width=.48\textwidth]{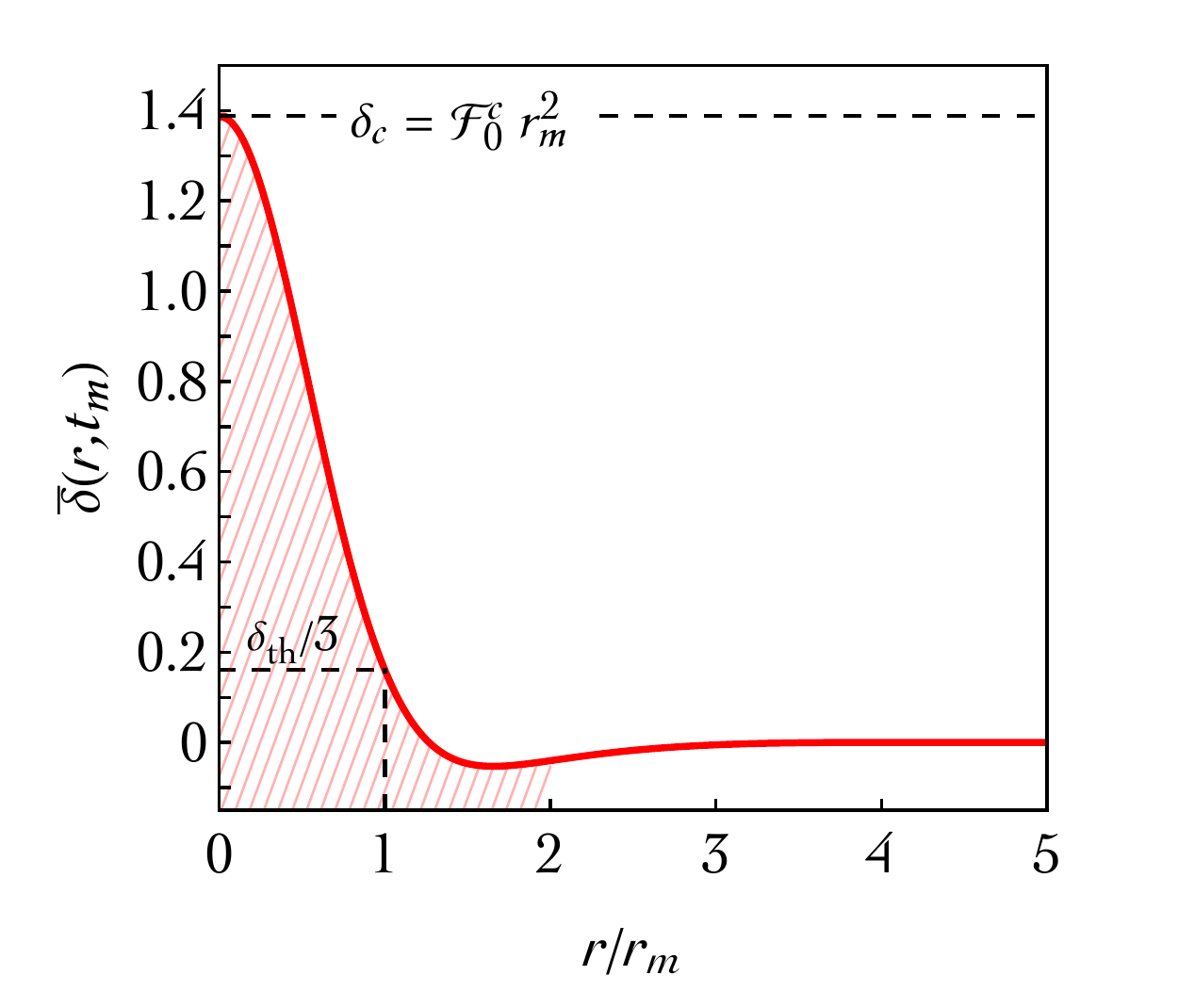}$$
\caption{\em \label{fig:ComparisonPStoy2} 
Left panel. Critical compaction function (that is eq.\,(\ref{eq:CompactionFunction}) evaluated for $\mathcal{F}_0 = \mathcal{F}_0^c$) obtained for $\mathcal{P}^{\rm cut}_{\mathcal{R}}(k)$ with $k_{\rm cut} = 1.4\,k_{\star}$ and $\alpha = 0.2$;
$\delta_{\rm th} \simeq 0.48$ is the threshold computed with eq.\,(\ref{eq:Germa}), and corresponds to the maximum of the critical 
compaction function. 
Right panel. For the same value of $k_{\rm cut}$ and $\alpha = 0.2$, we show the radial profile of the critical averaged over-density.
 }
\end{center}
\end{figure}

In the left panel of fig.\,\ref{fig:ComparisonPStoy2}, we show the critical compaction function 
(which is defined by eq.\,(\ref{eq:CompactionFunction}) with $\mathcal{F}_0 = \mathcal{F}_0^c$) for the case with 
$k_{\rm cut} = 1.4\,k_{\star}$ and $\alpha = 0.2$.
In the right panel of the same figure, we show the radial profile of the critical averaged over-density  
which is given by eq.\,(\ref{eq:F0}) with $\mathcal{F}_0 = \mathcal{F}_0^c$ at time $t=t_m$.
This figure makes clear the difference between $\delta_{\rm th}$ and $\delta_c = \mathcal{F}_0^cr_m^2$ in eq.\,(\ref{eq:ThresholdExpl}). 
The former is the threshold that refers to the value of the compaction function at its maximum while the latter refers to the over-density at the center.

After clarifying the computation of the threshold for gravitational collapse, let us now turn 
to the computation of the PBH abundance. 
At the time $t_f$ of their formation the latter is defined by 
\begin{align}\label{eq:BetaDefinition}
\beta = \frac{1}{\rho_b(t_f)}\int_{\nu_c}^{\infty}d\nu\rho_{\rm PBH}(\nu)\,,
\end{align} 
and it simply corresponds to the ratio between the mass density in PBHs and the background 
energy density at formation time, $\rho_b(t_f) = 3H_f^2/8\pi$. 
In eq.\,(\ref{eq:BetaDefinition}) we use $\nu \equiv \delta/\sigma_{\delta} = 2s/\sigma_2$. 
The parameter $\nu$ controls the hight of the overdensity peak that collapses into a PBH, 
and the integration in  eq.\,(\ref{eq:BetaDefinition}) tells us that only peaks above the threshold contribute to the 
PBH abundance (while perturbations with $\nu < \nu_c$ disperse into the expanding Universe). The mass density of PBHs takes the form 
\begin{align}
\rho_{\rm PBH}(\nu) = M_{\rm PBH}(\nu)n_{\rm max}^{\rm phys}(\nu)\,,
\end{align}
where $M_{\rm PBH}(\nu)$ is the PBH mass 
(usually calculated at horizon crossing time $t_m$) 
and $n_{\rm max}^{\rm phys}(\nu)$ is the number density of peaks of height $\nu$ in the physical space at formation time.
The latter is related to the comoving number density of peaks $n_{\rm max}(\nu)$ by the scaling (in three spatial dimensions) 
$n_{\rm max}^{\rm phys}(\nu) = n_{\rm max}(\nu)/a_f^3$. 
We consider for the moment the realistic case of three spatial dimensions. 
The PBH mass $M_{\rm PBH}(\nu)$ is proportional to the horizon mass at horizon-crossing time, $M_{\rm H}(t_m)$, according to the relation\,\cite{Germani:2018jgr}
\begin{align}\label{eq:PBHmass}
M_{\rm PBH}(\nu) = \mathcal{K}M_{\rm H}(t_m)
\left(\frac{\bar{\sigma}_0 a^2H^2}{a_m^2H_m^2}\right)^{\tilde{\gamma}}(\nu-\nu_{c})^{\tilde{\gamma}}\,.
\end{align}
In this equation the horizon mass at horizon-crossing time is given by the equation 
$M_{\rm H}(t_m) = 1/2H_m$ while the factor $(\nu-\nu_{c})^{\gamma}$ 
is the scaling law for critical collapse (extracted from numerical simulation, and with $\tilde{\gamma}\simeq 0.36$ in the case of radiation\,\cite{Neilsen:1998qc}) while $\mathcal{K} = O(1)$. 
In our notation, $\nu = 2s/\sigma_2$ and $\nu_c = 2s_{\rm min}/\sigma_2$.  
Notice that in eq.\,(\ref{eq:PBHmass}) we introduce, following ref.\,\cite{Germani:2018jgr}, the spectral moments referred to the density perturbation
\begin{align}
\bar{\sigma}_j^2 = \frac{16}{81}\frac{1}{(aH)^4}\int\frac{dk}{k}\mathcal{P}_{\mathcal{R}}(k)k^{2j+4}\,.
\end{align}
All in all, eq.\,(\ref{eq:BetaDefinition}) translates into 
\begin{align}
\label{eq:betadnu}
\beta = \frac{4\pi}{3}\mathcal{K}\left(\frac{\bar{\sigma}_0 a^2H^2}{a_m^2H_m^2}\right)^{\tilde{\gamma}}\left(\frac{1}{a_m H_m}\right)^3\frac{a_f}{a_m}
\int_{\nu_c}^{\infty}d\nu (\nu-\nu_c)^{\tilde{\gamma}}n_{\rm max}(\nu)\,.
\end{align}
We need to specify the peak number density and integrate.  
Let us consider first the gaussian case, in which we have\,\cite{Bardeen:1985tr}\,\footnote{Notice that $R_{*}$ is now defined in terms of the spectral moments $\bar{\sigma}_j$, differently from what done in section\,\ref{app:GaussianPeakTheory} where we have an analogous expression in terms of $\sigma_j$.}
\begin{align}
\label{eq:nmax}
n_{\rm max}(\nu) = \frac{1}{4\pi^2}
\left(\frac{\gamma}{R_{*}}\right)^3
\nu^3\exp(-\nu^2/2)\,,~~~~~~~~~~{\rm with}~~~~\gamma\equiv \frac{\bar{\sigma}_1^2}{\bar{\sigma}_0\bar{\sigma}_2}\,,~~~~
R_{*} \equiv \frac{\sqrt{3}\bar{\sigma}_1}{\bar{\sigma}_2}\,.
\end{align} 
We, therefore, have
\begin{align}
\beta = \frac{\mathcal{K}}{3\pi}\left(\frac{\bar{\sigma}_0 a^2H^2}{a_m^2H_m^2}\right)^{\tilde{\gamma}}
\left(\frac{a_f}{a_m}\right)
\left(\frac{\gamma r_m}{R_{*}}\right)^3
\nu_c^{4+\gamma}\int_{1}^{\infty}dx (x-1)^{\tilde{\gamma}} x^3 e^{-x^2\nu_c^2/2}\,.
\end{align}
Ref.\,\cite{Germani:2018jgr} computed the above integral by means of a saddle point approximation. 
We, instead, note that it admits the following exact expression in terms of generalized hypergeometric functions
\begin{align}
\int_{1}^{\infty}dx (x-1)^{\tilde{\gamma}} x^3 e^{-x^2\nu_c^2/2} = 
\frac{2^{\tilde{\gamma}/2}}{\nu_c^{\tilde{\gamma} + 4}}
\bigg\{&
2\tilde{\gamma}\left(2+\frac{\tilde{\gamma}}{2}\right) 
 {_2}F_2\left[\left\{\frac{1}{2}(1-\tilde{\gamma}),-\frac{\tilde{\gamma}}{2}\right\},\left\{\frac{1}{2},-1-\frac{\tilde{\gamma}}{2}\right\},-\frac{\nu_c^2}{2}\right] \\
&-\sqrt{2}\nu_c\tilde{\gamma}\left(\frac{3+\tilde{\gamma}}{2}\right)
 {_2}F_2\left[\left\{\frac{1}{2}(1-\tilde{\gamma}),1-\frac{\tilde{\gamma}}{2}\right\},\left\{\frac{3}{2},-\frac{1}{2}\left(1-\frac{\tilde{\gamma}}{2}\right)\right\},-\frac{\nu_c^2}{2}\right]
\bigg\}\,.\nn
\end{align}
This result, in turn, allows us to extract the simple approximation 
\begin{align}
\int_{1}^{\infty}dx (x-1)^{\tilde{\gamma}} x^3 e^{-x^2\nu_c^2/2} \simeq 
e^{-\nu_c^2/2}\nu_c^{-2(1+\tilde{\gamma})}\Gamma(1+\tilde{\gamma})\,.
\end{align}
which is valid for $\nu_c/\tilde{\gamma}\gg 1$. 
We checked numerically that this approximation works exquisitely well (at the \% level for $\nu_c =10$ with increasing precision 
for larger thresholds). We find
\begin{align}\label{eq:ExatBeta}
\beta \simeq \underbrace{\left[
\frac{\mathcal{K}}{3\pi}\left(\frac{\bar{\sigma}_0 a^2H^2}{a_m^2H_m^2 \nu_c}\right)^{\tilde{\gamma}}
\left(\frac{a_f}{a_m}\right)
\left(\frac{\gamma r_m}{R_{*}}\right)^3\Gamma(1+\tilde{\gamma})
\right]}_{\sim\,\,O(1)}\,\nu_c^2\,e^{-\nu_c^2/2}
\,,~~~
\resizebox{40mm}{!}{
\parbox{19mm}{
\begin{tikzpicture}[]
\node (label) at (0,0)[draw=white]{ 
       {\fd{2.75cm}{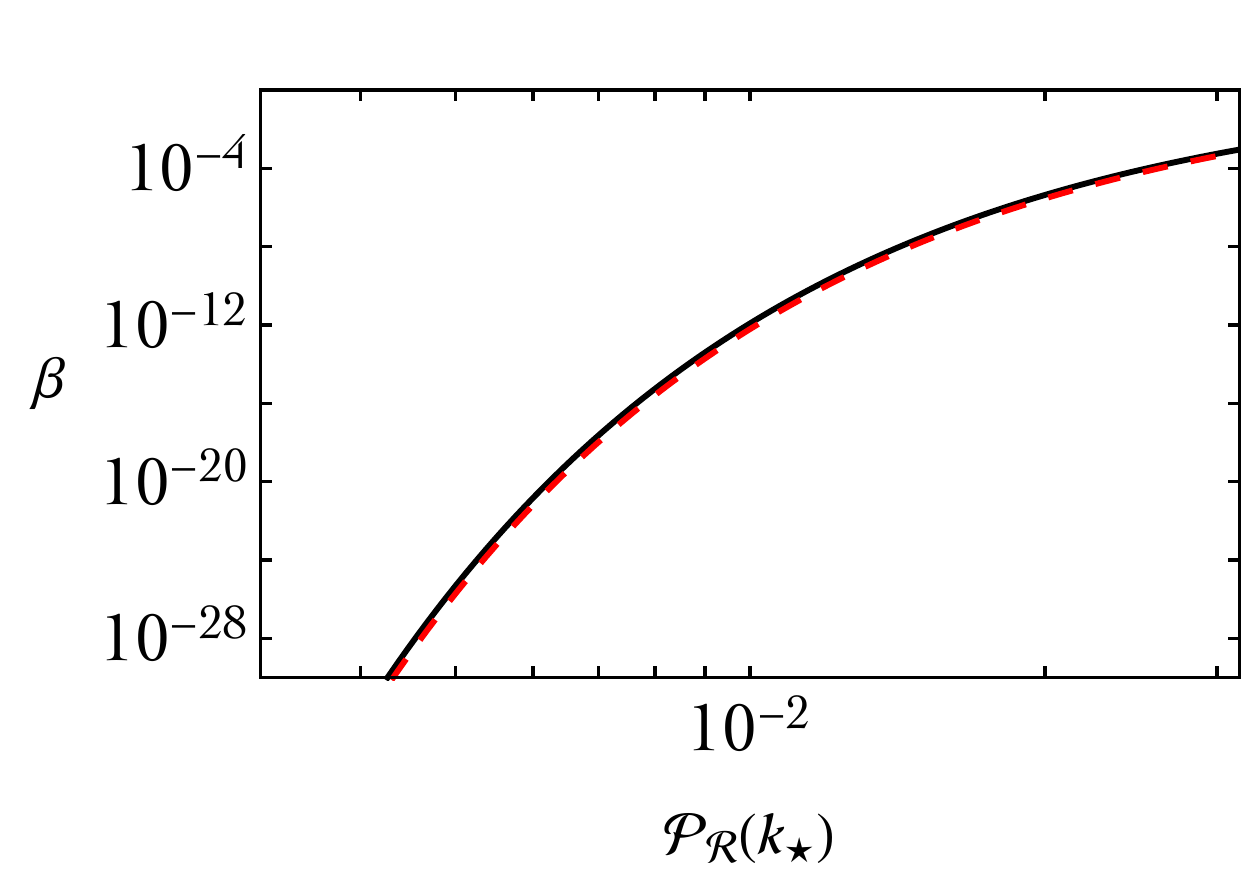}} 
      };
\end{tikzpicture}
}} \hspace{1.75cm}
\end{align}
The numerical value of $\beta$ is controlled by the exponential function $e^{-\nu_c^2/2}$ while 
the pre-factor inside the square brackets in eq.\,(\ref{eq:ExatBeta}) is a dimensionless $O(1)$ number whose exact value only 
plays a sub-leading role in the determination of $\beta$.
To corroborate this statement, in the figure attached to eq.\,(\ref{eq:ExatBeta}) we compare (as function of the peak amplitude of the power spectrum) the abundance $\beta$ computed 
with (black solid line) and without (red dashed line) the pre-factor in the square brackets, and we show that the two computations almost coincide.

\begin{figure}[!h!]
\begin{center}
\includegraphics[width=.45\textwidth]{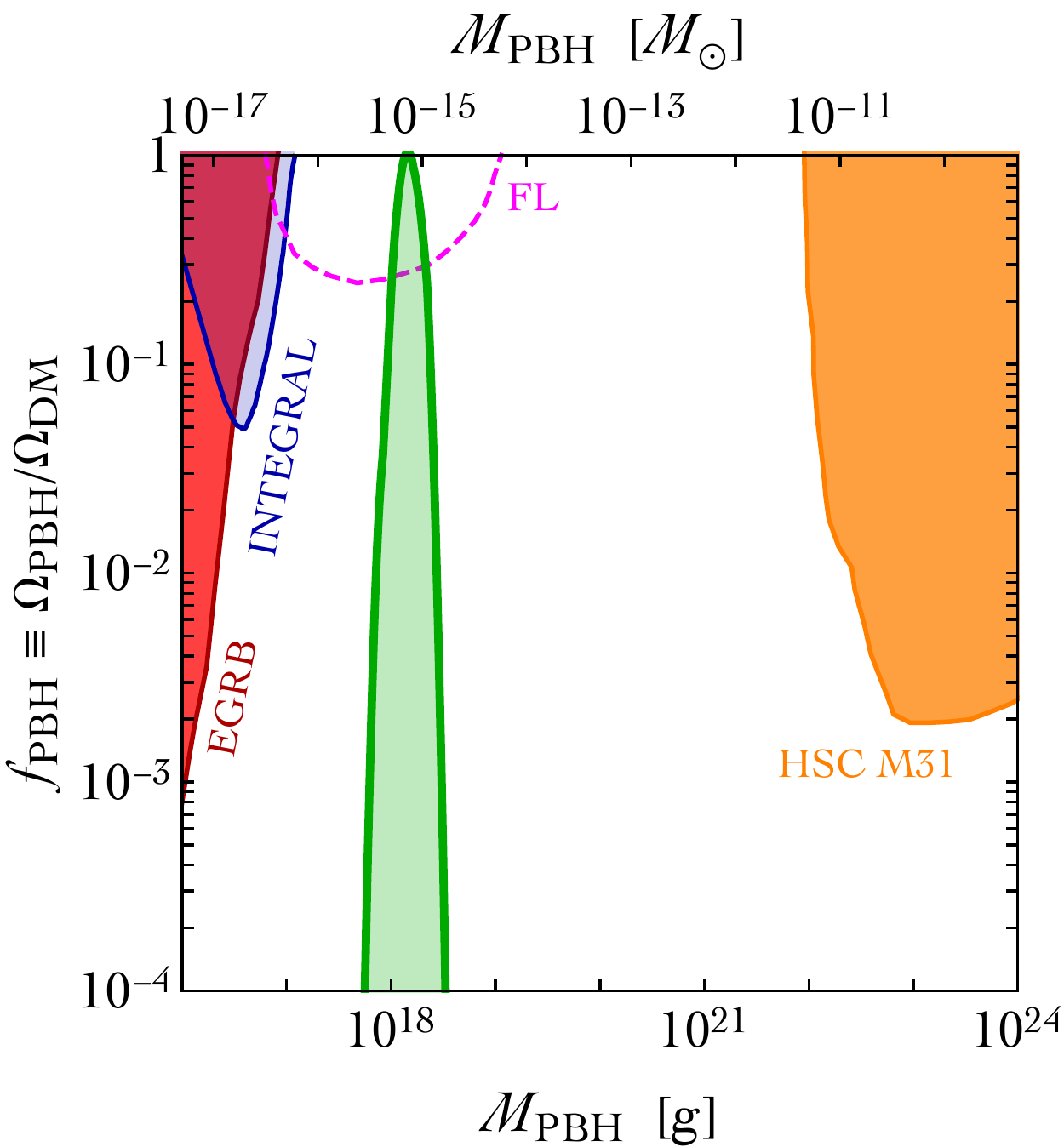}
\caption{\em \label{fig:AbundancePT} PBH abundance in the gaussian approximation for the model in ref.\,\cite{Ballesteros:2020qam}.
We compare the abundance of PBHs with existing bounds from Hawking evaporation from the extragalactic background radiation (EGBR, ref.\,\cite{Carr:2009jm}) and from galactic gamma-ray measurements (INTEGRAL, ref.\,\cite{Laha:2020ivk}).
We also show micro-lensing constraints from observation of the Andromeda galaxy M31 (HSC\,M31, ref.\,\cite{Niikura:2017zjd}),
and future detection prospects using femto-lensing  of gamma-ray bursts (FL, ref.\,\cite{Katz:2018zrn}).
 }
\end{center}
\end{figure}

The computation in eq.\,(\ref{eq:ExatBeta}) shows that in order to estimate the value of $\beta$ in a pragmatic way we can just integrate the comoving number  
density of maxima of the overdensity field and make it dimensionless
\begin{align}\label{eq:ExatBetaAppr}
\beta \simeq 4\pi^2 R_*^3\int_{\nu_c}^{\infty}d\nu\,n_{\rm max}(\nu)\,. 
\end{align}
Eq.\,(\ref{eq:ExatBetaAppr}) gives a perfect approximation of eq.\,(\ref{eq:ExatBeta}).
The corresponding present-day PBHs fractional abundance can then be obtained from eq.\,\ref{eq:PresentDayAbundance}.
Finally, scanning over $k_{\rm cut}$ one can reconstruct the PBHs mass function.
Following the strategy explained above, at each value of $k_{\rm cut}$ we compute the threshold for gravitational collapse, the scale $r_{m}$ and the horizon mass. From these quantities we obtain the PBH mass  and its associated abundance~\footnote{
We estimate the PBH mass for a given $r_m$ as 
$M_{\rm PBH}=\mathcal{K}\Gamma(1+\tilde{\gamma}) \nu_c^{-\tilde{\gamma}} (r_m^2 \bar{\sigma}_0 a^2H^2)^{\tilde{\gamma}} M_{\rm H}(t_m).$
The reason is that using this definition of the PBH mass and considering the number density of peaks above threshold (i.e. integrating eq.\,(\ref{eq:nmax}) above $\nu_c$) one reproduces
eq.\,(\ref{eq:ExatBeta}). In fig.\,\ref{fig:AbundancePT} we adopt $\mathcal{K}=3$.}.
The result of such calculation is presented in fig.\,\ref{fig:AbundancePT} for the 
power spectrum of the model in ref.\,\cite{Ballesteros:2020qam}.
PBHs are produced across some range of masses although the probability for their formation quickly drops when one moves away from the peak, since the critical value $\nu_c$ increases making the formation of PBHs more difficult, see fig.\,\ref{fig:CutoffAnalysis}.

In presence of local non-gaussianities the abundance can be obtained following the same logic that leads to eq.\,(\ref{eq:betadnu}), but computing the number density of peaks as 
explained in sec.\,\ref{app:NonGaussianPeakTheoryExact}, namely identifying this quantity with the number density of spiky maxima of the comoving density perturbations, the sum of eq.\,(\ref{eq:Npk1}) and eq.\,(\ref{eq:Npk2}). We obtain
\begin{align}
\beta = \frac{4\pi}{3}\mathcal{K}&\left(\frac{\bar{\sigma}_0 a^2H^2}{a_m^2H_m^2}\right)^{\tilde{\gamma}}\left(\frac{1}{a_m H_m}\right)^3\frac{a_f}{a_m} \times \nn \\
& \left[
\int_{-\frac{1}{2\alpha\sigma_0}}^{\infty}d\bar{\nu} \int_{x_{\delta}(\bar{\nu})}^{\infty}dx\, [\tilde{\nu}(\bar{\nu},x)-\nu_c]^{\tilde{\gamma}}\,\bar{n}_{\rm max}(\bar{\nu},x) + \int_{-\infty}^{\frac{1}{2\alpha\sigma_0}}d\bar{\nu} \int_{-\infty}^{x_{\delta}(\bar{\nu})}dx\, [\tilde{\nu}(\bar{\nu},x)-\nu_c]^{\tilde{\gamma}}\,\bar{n}_{\rm max}(\bar{\nu},x)
\right]\,,
\end{align}
where $\bar{n}_{\rm max}(\bar{\nu},x)$ is defined in eqs.\,(\ref{eq:npkxy},\,\ref{eq:npkxyfx}). Notice that  $\bar{\nu}=\mathcal{R}/\sigma_0,$ while in eq.\,(\ref{eq:PBHmass}) we have $\nu=\delta/\sigma_\delta.$ In the expression above we have written the latter quantity as $\nu = \tilde{\nu}(\bar{\nu},x)=x\,(1+2\alpha\sigma_0\bar{\nu}).$
Once again, we find that the abundance is well approximated by a simpler expression, obtained by taking the number density of peaks, $\mathcal{N}_{\rm pk}^{\rm (I)}$ + $\mathcal{N}_{\rm pk}^{\rm (II)}$
(eqs.\,(\ref{eq:Npk1},\,\ref{eq:Npk2}) ), and making it dimensionless\,\footnote{In eqs.\,(\ref{eq:beta3Dapprox},\,\ref{eq:2SpatialDimBeta}) $R_{*}$ is defined as in section\,\ref{app:GaussianPeakTheory}, i.e. $R_{*} \equiv \sqrt{d} \,\sigma_1/\sigma_2$.}
\begin{align}
\label{eq:beta3Dapprox}
\beta \simeq \pi^2 R_*^3\, \left[\mathcal{N}_{\rm pk}^{\rm (I)}+\mathcal{N}_{\rm pk}^{\rm (II)}\right].
\end{align}
This equation reproduces eq.\,(\ref{eq:MasterFormula}).

Having looked at the exact definition of $\beta$, let us now consider our simplified two-dimensional model. 
We follow the rationale that led to eq.\,(\ref{eq:ExatBetaAppr}).
The number density of maxima above threshold of the overdensity field is given by $\mathcal{N}_{\rm max}(s_{\rm min})$ (that is, for instance, eq.\,(\ref{eq:DoubleIntegral}) in the gaussian case) and, 
in two spatial dimensions, to make it dimensionless we need to multiply it times $R_*^2$
\begin{align}\label{eq:2SpatialDimBeta}
\beta \simeq R_*^2\mathcal{N}_{\rm max}(s_{\rm min})\,.
\end{align}

\section{Non-gaussianities from non-linearities}\label{app:NonLin}

Consider the relation between the overdensity field and the primordial curvature perturbation
\begin{align}\label{eq:NonLinearDelta3}
\delta(\vec{x},t) =  
-\frac{4}{9}\left(
\frac{1}{aH}
\right)^2 
e^{-2 h(\vec{x})}
\bigg[
\triangle h(\vec{x}) + \frac{1}{2} h_i(\vec{x})
 h_i(\vec{x})
\bigg] \equiv -\frac{4}{9}\left(
\frac{1}{aH}
\right)^2\delta_r(\vec{x})\,.
\end{align}
We have two kinds of non-gaussianities. On the one hand, non-gaussianities of primordial origin---which are the main subject of this paper---that arise from the non-gaussian nature of 
the random field $h$.  
On the other one, non-gaussianities also arise from the presence of non-linear terms in the relation between $h$ and $\delta$ in eq.\,(\ref{eq:NonLinearDelta3}). 
The latter type of non-gaussianities is ineludible in the sense that the
 overdensity field is non-gaussian even if the curvature perturbation is gaussian\,\cite{DeLuca:2019qsy}. 
However, the presence of non-gaussianities of primordial origin is also expected in the context of realistic models of single-field inflation 
which lead to black hole formation; this is because in these models slow-roll conditions (which usually suppress the amount of non-gaussianities) are 
badly violated. 
A comprehensive study in which both effects are included is needed (see ref.\,\cite{Yoo:2019pma} for an attempt in this direction).  
In this appendix, we limit our discussion to a number of  simple qualitative considerations.
Furthermore, as already stated in section\,\ref{sec:Res}, we do not consider the effect of non-linearities in eq.\,(\ref{eq:NonLinearDelta3}) on 
the shape of the peak of the overdensity field (in principle, this may affect the way in which $\delta_c$ is computed; 
as explained in appendix\,\ref{app:Threshold}, we compute $\delta_c$ by taking the linear approximation in eq.\,(\ref{eq:NonLinearDelta3}), and we use the average density profile to describe the shape of the peaks). 
In the following, we will only discuss how non-linearities in eq.\,(\ref{eq:NonLinearDelta3}) are expected to change the probability distribution of the peaks of the overdensity field.

\subsubsection{The case without primordial non-gaussianities}

Let us start from the gaussian case in which $h = \mathcal{R}$ (that is $\alpha = 0$).
Compared to the analysis carried out in appendix\,\ref{app:GaussianPeakTheory}, 
the first thing to check is whether peaks of the overdensity field are still identifiable with local maxima of $\mathcal{R}$ even in the presence of non-linearities. To answer this question, we repeat the numerical analysis that led to fig.\,\ref{fig:MaxCorrespondence}, this time 
using the full relation in eq.\,(\ref{eq:NonLinearDelta3}). A positive answer is found, in agreement with the result of ref.\,\cite{DeLuca:2019qsy}.  
On this basis, 
we can identify $\vec{y}_{\rm pk}$ (the position of the peak of the overdensity field) with $\vec{x}_{\rm M}$ 
(the position of the associated local maximum of $\mathcal{R}$);
consequently, the condition that  
the peak amplitude of the overdensity must be larger than some critical value $\delta_c$ becomes 
\begin{align}\label{eq:NonLinearDelta4}
-\triangle \mathcal{R}(\vec{x}_{\rm M}) 
\gtrsim \frac{9}{4}(aH)^2
e^{2\mathcal{R}_{\rm M}} 
\delta_c\,,
\end{align}
with $\delta_c$ that is now ``renormalized'' by the factor $e^{2\mathcal{R}_{\rm M}}$.  
The bottom line is the following. 
If the value of $\mathcal{R}_{\rm M}$ is so large that $e^{2\mathcal{R}_{\rm M}} \gg 1$ then it may sizably change the 
threshold condition on $-\triangle \mathcal{R}(\vec{x}_{\rm M})$.  
This point was already emphasized in ref.\,\cite{DeLuca:2019qsy}.

In addition, the important remark that we want to make is the following.
At the quantitative level, the importance of this effect strongly depends on the power spectrum. 
If the power spectrum is very peaked $\mathcal{R}_{\rm M}$ and $-\triangle\mathcal{R}(\vec{x}_{\rm M})$ are strongly correlated. 
This point was already highlighted in appendix\,\ref{app:GaussianPeakTheory} (see fig.\,\ref{fig:ScanCurvature} and related discussion). 
This means that maxima with large $-\triangle\mathcal{R}(x_{\rm M})$ are likely to be also regions with large $\mathcal{R}_{\rm M}$. 
In turn, this enhances (because of the factor $e^{2\mathcal{R}_{\rm M}}$) the threshold condition on 
the curvature  $-\triangle \mathcal{R}(\vec{x}_{\rm M})$ compared to the linear case, and 
 we expect a smaller black hole abundance since higher threshold for black hole formation means rarer objects. 
 The net result is that, in order to keep the value of $\beta$ constant, one needs to take larger values of $\mathcal{P}_{\mathcal{R}}(k_{\star})$ to compensate 
 the above effect. This expectation is confirmed in ref.\,\cite{DeLuca:2019qsy} for the case of a very narrow log-normal power spectrum. 
 If we take eq.\,(\ref{eq:ToyPS}) 
 with $v = 0.1$, in particular, ref.\,\cite{DeLuca:2019qsy} finds that one needs to increase $\mathcal{P}_{\mathcal{R}}(k_{\star})$ 
 by more than one order of magnitude.

However, in single-field models of inflation that are relevant in the present work the power spectrum is broader than a very narrow log-normal function. 
This has an important impact.
As argued in appendix\,\ref{app:GaussianPeakTheory} (see fig.\,\ref{fig:ScanCurvature} and related discussion), for a broad power spectrum 
the correlation between $\mathcal{R}_{\rm M}$ and $-\triangle\mathcal{R}(x_{\rm M})$ is much less pronounced (because it is controlled by the parameter 
$0 < \gamma \equiv \sigma_1^2/\sigma_2\sigma_0 < 1$ with $\gamma\to 1$ for a narrow power spectrum and $\gamma \to 0$ for a broad one, see eq.\,(\ref{eq:nMaxRs2})). We expect, therefore, that 
in the realistic case the effect of non-linearities will be far less relevant if compared to what expected based on the case of a narrow  power spectrum: in the realistic case large values of $-\triangle\mathcal{R}(x_{\rm M})$ are not necessarily combined with 
large values of $\mathcal{R}_{\rm M}$, and the enhancement effect in eq.\,(\ref{eq:NonLinearDelta4}) will be less important.
This intuition is confirmed in the left panel of fig.\,\ref{fig:MasterFormula2}. The dot-dashed black line (with label ``NL'') 
represents the computation of $\beta$ in three spatial dimensions in the 
gaussian limit\footnote{This limit corresponds to $\alpha \to 0$ in appendix\,\ref{app:NonGaussianPeakTheoryExact}. 
However, notice that, 
because of our definition of the realistic power spectrum in eq.\,(\ref{eq:RealisticPS}), 
some (very mild) $\alpha$-dependence is present (via the shape of the power spectrum) even if 
 we take the gaussian limit in the computation of the abundance. The dot-dashed line 
 in the left panel of fig.\,\ref{fig:MasterFormula2} corresponds to the case $\alpha = 0.61$. 
} but including the condition  
in eq.\,(\ref{eq:NonLinearDelta4}).
In the analyzed case, we find $\gamma \simeq 0.6$ which is significantly far from the narrow limit $\gamma \to 1$. 
Non-linearities suppress, as expected, the PBH abundance, but the peak amplitude of the power spectrum which is required to get the reference value $\beta \simeq 10^{-16}$ increases only by a factor smaller than 2.  

\subsubsection{The case with primordial non-gaussianities}

We now consider eq.\,(\ref{eq:NonLinearDelta3}) with $h(\vec{x}) = \mathcal{R}(\vec{x}) 
+ \alpha[\mathcal{R}(\vec{x})^2 - \sigma_0^2]$. 
Eq.\,(\ref{eq:NonLinearDelta3}) takes the form
\begin{align}\label{eq:NonLinearNonGauDelta3}
-\triangle \mathcal{R}(\vec{x}_{\rm M}) 
\gtrsim \frac{9}{4}\frac{(aH)^2}{(1+2\alpha\mathcal{R}_{\rm M})}
\exp\left\{2[\mathcal{R}_{\rm M}+\alpha(\mathcal{R}_{\rm M}^2 - \sigma_0^2)]\right\} 
\delta_c\,.
\end{align}
The number density of peaks of the overdensity field can be computed following the same procedure outlined in section\,\ref{app:NonGaussianPeakTheoryExact}.
The only difference is that we should now impose eq.\,(\ref{eq:NonLinearNonGauDelta3}) instead of eq.\,(\ref{eq:SpikyEnough}). This means that eq.\,(\ref{eq:xdelta}) is replaced by the following expression:
\begin{align}
x  > \frac{9 (a_m H_m)^2}{4\sigma_2}\frac{\delta_c}{1+2\alpha\sigma_0\bar{\nu}}
\exp\left\{2\sigma_0[\bar{\nu}+\alpha\sigma_0(\bar{\nu}^2-1)]\right\}  \equiv x_{\delta}^{\rm NL}(\bar{\nu}),
\end{align}
and the function $x_{\delta}^{\rm NL}(\bar{\nu})$ replaces $x_{\delta}(\bar{\nu})$ in the integrals in eqs.\,(\ref{eq:Npk1},\,\ref{eq:Npk2}).
We show our results for the PBH abundance in the left panel of fig.\,\ref{fig:MasterFormula2}.
We find that the additional presence of primordial non-gaussianities makes the formation of PBHs even harder than 
the expectation based solely on the presence of non-linearities.   
However, the enhancement in the value of $\mathcal{P}_{\mathcal{R}}(k_{\star})$ that is required (compared to the gaussian case) 
to fit the reference value $\beta = 10^{-16}$ remains extremely small (less than a factor of $2$ even for $\alpha = 0.61$).



\end{document}